\documentclass{jfm}
\usepackage{graphicx}
\usepackage[usenames, dvipsnames]{xcolor}
\usepackage{amssymb, latexsym}
\usepackage{mathrsfs}
\usepackage{soul}
\usepackage{natbib}
\usepackage{amsbsy}
\usepackage{psfrag}
\usepackage{mathtools}
\usepackage{color}
\usepackage{subfig}
\usepackage{makecell}
\usepackage[makeroom]{cancel}
\definecolor{myBlue}{rgb}{0, 0.4470, 0.7410}
\definecolor{myRed}{rgb}{0.8500, 0.3250, 0.0980}

\usepackage{hyperref}
\hypersetup{
	breaklinks,
	colorlinks=true,
	linkcolor=blue,
  urlcolor=Purple,
	citecolor=Purple,
 }

\def\p{\partial}

\def\beq{\begin{equation}}
\def\eeq{\end{equation}}

\def\p{\partial}

\def\beq{\begin{equation}}
\def\eeq{\end{equation}}

\definecolor{olivegreen}{rgb}{0,0.6,0}

\providecommand\bnabla{\boldsymbol{\nabla}}

\def\spacce#1{\hskip #1pt}
\def\drawline#1#2{\raise 2.5pt\vbox{\hrule width #1pt height #2pt}}
\def\solid{\drawline{24}{.5}\nobreak}

\def\bdash{\hbox{\drawline{5.8}{.5}\spacce{2}}}

\def\dashed{\bdash\bdash\bdash\nobreak}

\def\bdot{\hbox{\drawline{1}{.5}\spacce{2}}}

\def\dotted{\hbox{\leaders\bdot\hskip 24pt}\nobreak}

\def\trian{\raise 1.25pt\hbox{$\scriptstyle\triangle$}\nobreak}

\def\dtrian{\raise 1.25pt\hbox%
{$\scriptscriptstyle\bigtriangledown$}\nobreak}

\def\squar{\raise 1.25pt\hbox{$\scriptstyle\Box$}\nobreak}

\def\diamon{\raise 1.25pt\hbox{$\scriptstyle\diamond$}\nobreak}

\def\p{\partial}

\def\beq{\begin{equation}}
\def\eeq{\end{equation}}

\def\ggg{{\it g}}

\def\citalajim03{Del \'Alamo \& Jim\'enez (2003)}

\newcommand {\bu} {\boldsymbol{u}}
\newcommand {\bU} {\boldsymbol{U}}

\newcommand {\bN} {\boldsymbol{N}}
\newcommand {\bcalN} {\boldsymbol{\mathcal{N}}}
\newcommand {\bxh} {\hat{\boldsymbol{x}}}

\renewcommand {\L} {\mathcal{L}}
\newcommand{\defn}{\ensuremath{\stackrel{\textrm{def}}{=}}}
\newcommand{\e}{\varepsilon}
\newcommand{\bse}{\begin{subequations}}
\newcommand{\ese}{\end{subequations}}
\newcommand {\ba} {\begin {array}}
\newcommand {\ea} {\end {array}}

\usepackage{pifont}
\newcommand{\cmark}{\ding{51}}%
\newcommand{\xmark}{\ding{55}}%

\shorttitle{Cause-and-effect of linear mechanisms in wall turbulence}
\shortauthor{Lozano-Dur\'an et al.}

\title{Cause-and-effect of linear mechanisms sustaining wall turbulence}

\author{Adri\'an~Lozano-Dur\'an\aff{1}\corresp{\email{adrianld@stanford.edu}},
        Navid~C.~Constantinou\aff{2},
        Marios-Andreas~Nikolaidis\aff{3}, 
 	and Michael~Karp\aff{1}}

\affiliation{
\aff{1}Center for Turbulence Research, Stanford University, CA 94305, USA
\aff{2}Research School of Earth Sciences and ARC Centre of Excellence for Climate Extremes,
Australian National University, Canberra ACT 2601, Australia
\aff{3}Department of Physics, National and Kapodistrian University of Athens, Athens 157 72, Greece
}

\begin{document}

\maketitle

\begin{abstract}
   Despite the nonlinear nature of turbulence, there is evidence that
   part of the energy-transfer mechanisms sustaining wall turbulence
   can be ascribed to linear processes. The different scenarios stem
   from linear stability theory and comprise exponential
   instabilities, neutral modes, transient growth from non-normal
   operators, and parametric instabilities from temporal mean-flow
   variations, among others. These mechanisms, each potentially
   capable of leading to the observed turbulence structure, are rooted
   in simplified physical models. Whether the flow follows any or a
   combination of them remains elusive. Here, we evaluate the linear
   mechanisms responsible for the energy transfer from the
   streamwise-averaged mean-flow ($\bU$) to the fluctuating velocities
   ($\bu'$). To that end, we use cause-and-effect analysis based on
   interventions: manipulation of the causing variable leads to
   changes in the effect.  This is achieved by direct numerical
   simulation of turbulent channel flows at low Reynolds number, in
   which the energy transfer from $\bU$ to $\bu'$ is constrained to
   preclude a targeted linear mechanism.  We show that transient
   growth is sufficient for sustaining realistic wall turbulence.
   Self-sustaining turbulence persists when exponential instabilities,
   neutral modes, and parametric instabilities of the mean flow are
   suppressed. We further show that a key component of transient
   growth is the Orr/push-over mechanism induced by spanwise
   variations of the base flow.  Finally, we demonstrate that an
   ensemble of simulations with various frozen-in-time $\bU$ arranged
   so that only transient growth is active, can faithfully represent
   the energy transfer from $\bU$ to $\bu'$ as in realistic
   turbulence. Our approach provides direct cause-and-effect
   evaluation of the linear energy-injection mechanisms from $\bU$ to
   $\bu'$ in the fully nonlinear system and simplifies the conceptual
   model of self-sustaining wall turbulence.
\end{abstract}

\begin{keywords}

\end{keywords}

\section{Introduction}
\label{sec:introduction}

Turbulence is a highly nonlinear phenomenon. Nevertheless, there is
ample agreement that some of the processes sustaining wall-turbulence
can be faithfully represented by linearising the equations of motion
about an appropriate reference flow state, i.e., base
flow~\citep{Malkus1956, Reynolds1967, Hussain1970, Landahl1975,
  Butler1993, Jimenez2013}. One of these processes is the transfer of
kinetic energy from the mean flow to the fluctuating velocities. The
different mechanisms originate from linear stability theory and
constitute the foundations of many control and modelling
strategies~\citep[e.g.][]{Kim2006, Schmid2012, McKeon2017, Rowley2017,
  Zare2020, Jovanovic2020}.  As such, establishing the relevance of a
particular theory is consequential to comprehend, model, and control
the structure of wall-bounded turbulence by linear methods
\citep[e.g.][]{Kim2000, Hogberg2003, Delalamo2006a, Hwang2010, Zare2017,
  Morra2019, Towne2020}.  Despite the ubiquity of linear theories,
their significance in wall turbulence remains outstanding.  One of the
main limitations to assess the role of a concrete linear process in
the flow has been the lack of conclusive cause-and-effect assessment
of the mechanisms in question.  In the present work, we devise a
collection of numerical experiments of turbulent flows over a flat
wall, in which the Navier--Stokes equations are minimally altered to
suppress the causal link entailing the energy-transfer from the mean
flow to the fluctuating velocities via various linear mechanisms.

Before diving into the intricacies of the different linear mechanisms,
one may ask why we should insist on describing this energy transfer
using linear theories if turbulence is undoubtedly a nonlinear
phenomenon. One reason is that the energy source for fluctuations in
wall turbulence is controlled by spatial changes in the mean velocity
(i.e, mean shear) \citep{Batchelor1954, Brown1974, Jimenez2013}. When
the flow is decomposed into a base flow ($\bU$) and fluctuations
($\bu'$), the equations of motion naturally reduce to a system
comprising a linear term and nonlinear term,
\begin{gather} \label{eq:dummy}
\frac{\partial\bu'}{\partial t} =
\underbrace{\mathcal{L}(\bU)\bu'}_{\substack{\textrm{linear}\\\textrm{processes}}} +
\underbrace{\bN(\bu')}_{\substack{\textrm{nonlinear}\\\textrm{processes}}}.
\end{gather}
If $\bU$ is chosen such that the volume integral of $\bu'\cdot \bN$
vanishes (see \S \ref{subsec:baseflow} and \S \ref{sec:nofeeback}),
the linear term in (\ref{eq:dummy}) is the sole source of energy for
$\bu'$, which explains the unceasing surge of interest in linear
theories. Note that constructing (\ref{eq:dummy}) does not require
invoking linearisation about $\bU$ nor assuming that $\bu'$ is
small. We can always partition the flow into $\bU + \bu'$ for an
arbitrary $\bU$, write (\ref{eq:dummy}), refer to the linear
mechanisms supported by $\mathcal{L}(\bU)$, and inquire their
relevance in sustaining turbulence. Hence, we do not challenge here
the validity of a particular linearisation. Instead, the question
raised is whether the linear mechanisms supported by $\bU$ (i.e.,
$\mathcal{L}(\bU)$) are useful in explaining the dynamics of
$\bu'$. It is clear that there exists a myriad of different flow
partitions $\bU + \bu'$, but not all of them are meaningful to explain
the dynamics of the flow. If $\bU$ is chosen wisely, it has been
demonstrated in many occasions that numerous features of the
energy-containing scales can be elucidated from the linear dynamics in
(\ref{eq:dummy})~\citep[e.g.,][]{Reed1996, Cambon1999, Schmid2007,
  Farrell2012, McKeon2017}. This is the case for strongly
inhomogeneous environments, such as wall turbulence with large-scale
pressure or body forces imposed (e.g., in the streamwise direction),
and geophysical flows, in which rotation and stratification impose
strong constraints on the flow~\citep{Farrell2019}. An additional,
less glamorous, motivation for arbitrarily partitioning the flow into
$\bU + \bu'$ (thus enabling the use of linear theories) is a matter of
practicality: our current framework to analyse linear systems is well
beyond the tools to understand nonlinear equations. Hence, inasmuch
the linear equations meaningfully represent the physics of the
problem, linear tools greatly aid the analysis and facilitate the
development of prediction and control strategies.

The rationale behind the formulation and validation of a linear theory
for the energy transfer between flow structures comprises four
elements: (i) the existence in wall turbulence of recurrent fluid
motions (or coherent structures) involved in a self-sustaining
process, (ii) the selection of a base flow which (iii) enables the
prediction of these coherent motions via linear theory, and (iv) a
cause-and-effect framework to evaluate the presence of the linear
mechanism in actual nonlinear turbulence. These four points are
discussed below.

\subsection{Coherent structures and self-sustaining wall turbulence}

Since the experiments by \cite{Klebanoff1962, Kline1967} and
\cite{Kim1971}, it was realised that despite the conspicuous disorder
of wall turbulence, the flow in the vicinity of walls can be
apprehended as a collection of recurrent patterns, usually referred to
as coherent structures~\citep{Richardson1922}. Of particular interest
are those structures carrying most of the kinetic energy and momentum,
further categorised as streaks (regions of high and low velocity
aligned with the mean-flow direction) and rolls/vortices (regions of
rotating fluid) \citep{Robinson1991, Panton2001, Adrian2007,
  Smits2011, Jimenez2012, Jimenez2018}.

Close to the wall in the so-called buffer layer, the current consensus
is that these energy-containing structures are involved in a
quasi-periodic self-sustaining process and that their space-time
structure plays a crucial role in the maintenance of shear-driven
turbulence \citep[e.g.][]{Kim1971, Jimenez1991, Butler1993,
  Hamilton1995, Waleffe1997, Jimenez1999, Schoppa2002, Farrell2012,
  Jimenez2012, Constantinou2014, Farrell2016, Farrell2017}. The self-sustaining process is based on the emergence
of streaks from wall-normal ejections of fluid \citep{Landahl1975}
followed by the meandering and breakdown of the newborn streaks
\citep{Swearingen1987, Hall1991, Waleffe1995, Waleffe1997,
  Schoppa2002, Kawahara2003}. The cycle is restarted by the generation
of new vortices from the perturbations created by the disrupted
streaks. The interwoven relation between vortices and streaks was
demonstrated by \cite{Jimenez1999}, who showed that damping out either
of them inevitably interrupts the turbulence cycle. A similar but more
disorganised scenario is hypothesised to occur for the larger
energy-containing structures further away from the wall within the
logarithmic layer \citep[e.g.][]{Flores2010, Hwang2011, Cossu2017,
  Lozano2019b}, although the focus of the present work is on the
buffer layer (i.e. low Reynolds numbers).  Linear theories have been
instrumental in unfolding and explaining various stages of the
self-sustaining process, and the existence of coherent structures has
aided the selection of particular base flows to linearise the
equations of motion.

The self-sustaining nature of wall turbulence has also been
investigated from the viewpoint of dynamical-systems theory. In this
framework, the spatio-temporal structure of turbulence is thought of
as a low-dimensional manifold around which the dynamical system spends
a substantial fraction of time \citep{Jimenez1987}. According to the
dynamical-systems perspective, the simplest description of turbulence
is then given by a collection of `invariant solutions' (equilibrium
states and periodic orbits) embedded in a high-dimensional turbulent
attractor \citep{Kawahara2012}.  The first dynamical-system
investigations of turbulence in shear flows began with the discovery
of nonlinear equilibrium states, referred to as `exact coherent
structures', of Couette flow \citep{Nagata1990}. Since then, there
have been multiple descriptions of such equilibrium states in shear
flows in channels and pipes, often involving unstable travelling waves
\citep[e.g.,][]{Waleffe2001, Kawahara2001, Wedin2004, Faisst2003,
  Gibson2009, VanVeen2011, Kreilos2012, Park2015,
  Hwang2016}. Particularly relevant for the study of self-sustaining
processes is the discovery of time-periodic solutions by
\cite{Kawahara2001} and later by others \citep[e.g.,][]{Toh2003,
  Viswanath2007, Gibson2008, Kawahara2012, Willis2013}. These
time-periodic solutions were first found for plane Couette flow and
exhibited a full regeneration cycle comprising the formation and
breakdown of streamwise vortices and low-velocity streaks.  The
dynamical-system approach has also provided the grounds to conceive
turbulence as a superposition of invariant solutions and their
manifolds, which would constitute the skeleton of flow trajectories in
turbulence~\citep{Auerbach1987, Cvitanovic1991}. Thus, the simplicity
provided by invariant solutions facilitates the inspection for linear
processes at a given stage in the self-sustaining cycle.  However,
while realistic turbulence does share similarities with these exact
coherent structures, the latter have been restricted to very low
Reynolds numbers.  The actual dynamics of wall turbulence are
significantly more complex and chaotic, and the relationship of
realistic high-Reynolds number turbulent flows with the
exact-coherent-states interpretation remains unsettled. In the present
work, we show that turbulence statistics might be recovered by
ensemble averaging a collection of solutions in the spirit of
\citet{Cvitanovic1991}, although in our case these solution are not
exact coherent structures.

Another theoretical nonlinear framework to describe self-sustaining
processes and transition to turbulence has been proposed by
\citet{Hall1988} and \cite{Hall1991} in terms of vortex--wave
interactions (VWI). The approach has been shown to be the equivalent
high-Reynolds-number representation of the exact coherent structures
discussed above \citep{Wang2007, Hall2010}. VWI theory involves an
intricately delicate balance between a neutrally stable wave, a roll
and a streak. According to VWI, a neutrally stable wave drives a
streamwise-uniform roll by forcing the critical layer of the
streamwise-averaged mean-flow. The roll produces streaks through the
lift-up effect by interacting with a neutrally stable mean-flow
(averaged in streamwise and spanwise directions).  Finally, the
streaks generate a spanwise-varying base-flow that supports the
neutrally stable wave, closing the cycle. Subsequent developments of
the VWI theory include extensions multiscale motions consistent with
the logarithmic layer \citep{Hall2018}. Other descriptions of
self-sustaining turbulence in the vein of vortex-wave interactions
include studies by \citet{Deguchi2013} and \cite{Deguchi2015}, the
high Reynolds number theory by \citet{Ozcakir2016, Ozcakir2019}, and
the semi-analytic model by \cite{Chini2017} and \cite{Montemuro2020};
the latter devoted to the formation and maintenance of uniform
momentum zones and interlaced vortical fissures studied by asymptotic
analysis. While the theories above could provide a plausible
explanation for how turbulence self-sustains, we are still lacking
direct cause-and-effect evidence regarding whether one or a
combination of the above-mentioned mechanisms are actually at work in
realistic turbulent flows.

\subsection{Base flow}
\label{sec:baseflow}

As shown in (\ref{eq:dummy}), formulating a linear theory entails the
partition of the flow into two components: a base flow $\bU$ (which
might be space- and/or time-dependent), and fluctuations (or
perturbations) $\bu'$ about that base flow. In the fluid-stability
community, it is customary to use as base flow a solution of the
Navier-Stokes equations and rigorously linearise the equations about
that state. The resulting analysis is then valid for small-amplitude
perturbations. On the other hand, the turbulence community has
commonly used as base flow a mean velocity defined by some averaging
procedure (which is not a solution of the Navier-Stokes equations) and
then loosely rely on the linear stability theory to analyse the
response of perturbations (which are generally not small in amplitude)
under the assumption of frozen-in-time base flow. This is obviously
far from rigorous and some authors have found questionable the use of
linear stability theory by the turbulence community (further discussed
in \S \ref{sec:literature}). Here, we overcome this hindrance by
considering a cause-and-effect analysis on the full non-linear system
in (\ref{eq:dummy}). First, we refer to base flow $\bU$ as any
arbitrary reference flow state to separate the flow into $\bU + \bu'$.
Second, as discussed above for (\ref{eq:dummy}), we can always
partition the equations for $\bu'$ into a linear and nonlinear
component and inquire the necessity of the linear mechanisms in
$\mathcal{L}(\bU)$ to sustain the flow. The usefulness of the base
flow $\bU$ is measured by to what extent the dynamics of $\bu'$ are
explained by the linear mechanisms supported by $\bU$, which
circumvents the problem of linearisation. Even if the classic
hydrodynamic linear-stability-theory is not rigorously applicable to
our base flows, we still employ the terminology `instability' to the
refer to linear growth provided by $\mathcal{L}(\bU)$.

\begin{figure} 
 \vspace{1cm}
 \begin{center}
   \includegraphics[width=0.85\textwidth]{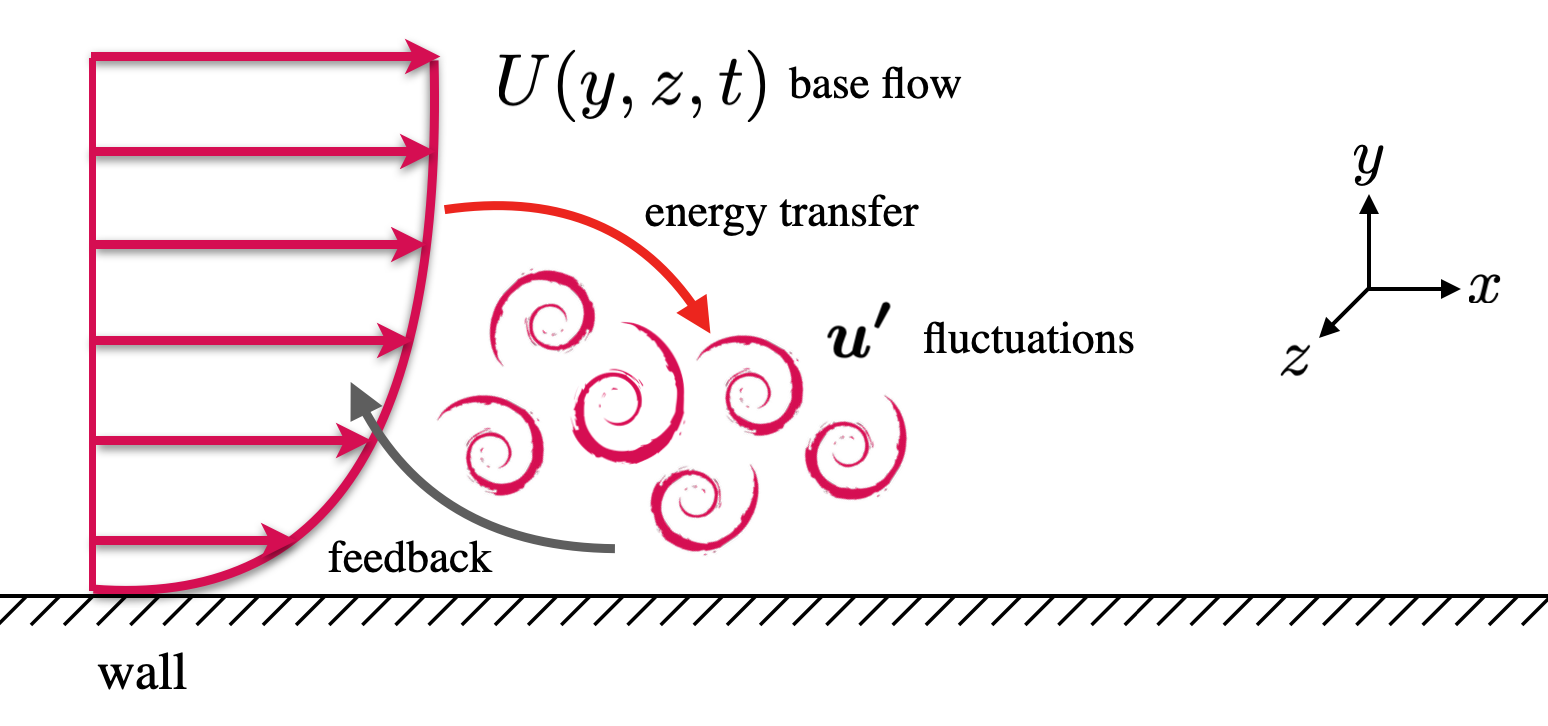} 
 \end{center}
\caption{Schematic of the energy transfer from the base flow $\bU =
  U(y,z,t)\bxh$ to the fluctuating velocities $\bu'$. The energy
  transfer (red arrow) from $\bU$ to $\bu'$ can be investigated via
  the linear dynamics of the governing equation of $\bu'$. The cycle
  is closed by the nonlinear feedback from $\bu'$ back to $\bU$ (gray
  arrow).
  \label{fig:sketch}}
\end{figure}

We now turn our attention to how to choose $\bU$ when the flow is
turbulent.  Historically, the existence of coherent structures in wall
turbulence has motivated the selection of particular base flows, such
that the linear dynamics supported by these base flows is the seed for
the inception of new coherent structures consistent with observations
in real turbulence. The resulting distorted flow might be used again
as a base flow, which describes the generation of new coherent
structures and so forth.  In this manner, the ultimate cause
maintaining turbulence is conceptualised as the energy transfer from
the base flow to the fluctuating flow, as sketched in
figure~\ref{fig:sketch}. The selection of the base flow stands as the
most important decision to formulate linear theory for sustaining
turbulent fluctuations, as the physical mechanisms ascribed to the
linear component of (\ref{eq:dummy}) depend crucially on this choice.
%

Hereafter, we consider the turbulent flow over a flat plate where $x$,
$y$ and $z$ are the streamwise, wall-normal and spanwise directions
respectively; see figure~\ref{fig:sketch}.  Common choices for the
base flow are the average of the streamwise velocity $u$ over
homogeneous directions ($x$ and $z$) and time ($t$), denoted by
$\langle u \rangle_{xzt}$, or only over $x$ or $z$ in some small
(minimal) domain, denoted by $\langle u \rangle_{x}$ and $\langle u
\rangle_{z}$, respectively. The notation $\langle u \rangle_{ij...}$
denotes averaging over the coordinates $i,j,...$, and it is formally
introduced in \S \ref{sec:numerical}. In turbulent boundary layers and
channels, the $y$-dependent base flow $\langle u \rangle_{xzt}$ has
been successful in predicting the formation of streaks, a process
sometimes referred to as primary instability (or more generally
primary linear process).  The resulting streaky flow (now represented
by $\langle u \rangle_{x}$) can be utilised as the new base flow to
generate the subsequent vortices or, more generally, disorganised
fluctuations. This process is usually referred to as secondary
instability (or secondary linear process). We next survey the main
linear theories associated with these two sets of linear processes.

\subsection{Linear theories of self-sustaining wall turbulence}

Several linear mechanisms have been proposed as plausible scenarios to
rationalise the transfer of energy from the large-scale mean flow to
the fluctuating velocities. We discuss below the linear processes
ascribed to two of the most widely used base flows, namely, the
$y$-dependent streakless mean velocity profile $\langle u
\rangle_{xzt}$, and the $y-z$-dependent time-varying streaky base flow
$\langle u \rangle_{x}$. The predictions by the two base flows should
not be considered contradictory but rather complementary, as the
former might be thought as the cause of the latter.

In the primary linear process, it is generally agreed that the linear
dynamics about $\langle u \rangle_{xzt}$ is able to explain the
formation of the streaks $U_{\mathrm{streak}}=\langle u \rangle_{x} -
\langle u \rangle_{xzt}$. The process involves the redistribution of
fluid near the wall by streamwise vortices leading to the formation of
streaks through a combination of the lift-up
mechanisms~\citep{Landahl1975, Lee1990, Farrell1993, Butler1993,
  Kim2000, Jimenez2012}.  In this case, the base flow, while being
exponentially stable, supports the growth of perturbations for a
period of time due to the non-normality of the linear operator about
that very base flow; a process referred to as non-modal transient
growth \citep[e.g.][]{Farrell1988, Gustavsson1991, Trefethen1993,
  Butler1993, Farrell1996, Delalamo2006a, Schmid2007, Cossu2009}.
Other studies suggest that the generation of streaks is due to the
structure-forming properties of the linearised Navier--Stokes
operator, independent of any organised vortices
\citep{Chernyshenko2005}, or due to the interaction of the background
free-stream turbulence and the roll-streak structures
\citep{Farrell2017b}, but the non-modal nature of the linear operator
is still crucially invoked.
Input--output analysis of the linearised Navier--Stokes equations has
also been successful in characterising the non-modal response of the
base flow $\langle u \rangle_{xzt}$ \citep{Farrell1993b,
  Jovanovic2005, Hwang2010a, Zare2017, Ahmadi2019, Jovanovic2020}. The
approach combines the linearised Navier--Stokes equations with
harmonic or stochastic forcing (white or coloured in time) to
qualitatively predict structural features of turbulent shear
flows. Similarly, resolvent analysis \citep{McKeon2010, McKeon2017}
provides pairs of response and nonlinear-forcing modes consistent with
the linear Navier--Stokes operator with respect to the base flow
$\langle u \rangle_{xzt}$ and enables the identification of the most
amplified energetic motions in wall turbulent flows.  A key aspect of
the latter energy transfer is the formation of critical layers or
regions where the wave velocity is equal to the base flow \citep[see
  also][]{Moarref2013}. Both input--output and resolvent analysis
formulate the problem in the frequency domain, and the sustained
response of the perturbations should be understood through a
persistent forcing in time. These amplification mechanisms can be
classified as resonant or pseudoresonant, depending on whether the
amplification of perturbations is associated with modal instabilities
or non-normality of the operator, respectively. Interestingly, the
flow structures responsible for the energy transfer obtained in the
frequency-domain are remarkably similar to the structures identified
with non-normal transient growth posed as an initial value
problem~\citep{Hwang2010b, Symon2018}, i.e. the genesis of streaks
from cross flow perturbation via lift-up mechanism.  In the present
work, we favour the time-domain formulation over the frequency-domain
approach as the former is easily understood as a sequence of events,
which facilitates the cause-and-effect analysis of self-sustaining
turbulence pursued here.

The scenarios described in the paragraph above pertain to the study of
$y$-dependent base flows and, as such, are concerned with primary
linear processes.  The summary of studies in the left column of table
\ref{table:literature} shows that, except for a handful of studies
performed under very particular conditions, most investigations
advocate for transient growth as the main cause for the genesis of the
streamwise streaks via energy transfer from $\langle u \rangle_{xzt}$
to $U_{\mathrm{streak}}$. Indeed, the few works which do not support
the transient growth are from the 1950's or performed in a different
context, such as laminar--turbulent transition. There is hardly any
controversy regarding the formation of the streaks, and here we focus
on the linear mechanisms underpinned by $\langle u \rangle_{x}$ once
the streak is formed, i.e., secondary linear processes.

Motivated by the streamwise-elongated structure of the streaks, we
take our base flow $\bU$ to consist of the instantaneous
streamwise-averaged velocity $U(y,z,t) = \langle u \rangle_{x}$ in the
streamwise direction of a minimal channel domain (see \S
\ref{sec:numerical}) with zero wall-normal and spanwise flow, i.e.,
$\bU = (U,0,0)$.  Our choice is supported by previous studies in the
literature, and most of the works reported on table
\ref{table:literature} (right column) conducted their analysis by
linearising the equations of motion about $U(y,z,t)$.  Yet, other
alternative base flows might be also justified \emph{a priori}, and
one of the goals here is to investigate whether $U(y,z,t)$ is a
meaningful choice to describe the energy transfer from the large
scales to the fluctuating flow.

The linear mechanisms supported by $U(y,z,t)$ can be categorised into
three groups: (i)~modal instability of the mean streamwise flow,
(ii)~non-modal transient growth, and (iii)~non-modal transient growth
assisted by parametric instability of the time-varying base
flow. Other classifications are possible, and ours is motivated by the
terminology adopted in previous works.  Table \ref{table:literature}
(right columns) compiles the literature in favour of one or other
mechanism. The table, while not an exhaustive compilation of existing
works on the topic, hints at a lack of consensus on which is the
prevailing linear mechanism responsible for the energy transfer from
the streaky mean-flow to the fluctuations, or if any, it implies that
the dominant idea is that exponential instability is the one
responsible. We show in this work that the latter is not the case;
modal instabilities of the mean streamwise flow are not crucial for
self-sustaining turbulence. Next, we briefly describe mechanisms (i),
(ii) and (iii).
%
 \begin{table}
   {  \scriptsize
 \begin{center}
   \begin{tabular}{ll | lll}
     Reference   &  \makecell{Linear mechanism \\ for $y$-dependent \\ base flow $\langle u \rangle_{xzt}$} & Reference
     &  \makecell{Linear mechanism \\ for $(y,z)$-dependent \\ base flow $\langle u \rangle_{x}$} \\[1ex] \hline
\cite{Malkus1956}	 	 & NEU              & \cite{Schoppa2002}	   & TG \\
\cite{Kim1971}     	 	 & EXP (V)          & \cite{Hopffner2005}          & TG \& EXP (S) \\
\cite{Skote2002}	 	 & EXP (V)          & \cite{Giovanetti2017}	   & TG \& EXP (S) \\
\cite{Jovanovic2005}             & EXP/TG           & \cite{Cassinelli2017}	   & TG \& EXP (S) \\
\cite{Farrell1988}               & TG               & \cite{Farrell2012}	   & TG PARA  \\
\cite{Landahl1990} 	 	 & TG               & \cite{Thomas2015}	           & TG PARA  \\  
\cite{Lee1990}                   & TG               & \cite{Farrell2016}	   & TG PARA  \\ 
\cite{Farrell1993b}              & TG               & \cite{Farrell2017c}          & TG PARA  \\
\cite{Reddy1993}	       	 & TG               & \cite{Nikolaidis2016}	   & TG PARA  \\   
\cite{Butler1993}	 	 & TG               & \cite{Swearingen1987}	   & EXP (S)  \\  
\cite{Trefethen1993} 	 	 & TG               & \cite{Hall1991}	           & EXP (S)  \\
\cite{Kim2000}	                 & TG               & \cite{Yu1991}	           & EXP (S)  \\
\cite{Chernyshenko2005}	  	 & TG               & \cite{Yu1994}       	   & EXP (S)  \\  
\cite{Delalamo2006a}          	 & TG               & \cite{Li1995}		   & EXP (S/V) \\  
\cite{Cossu2009}                 & TG               & \cite{Park1995}	           & EXP (S)  \\  
\cite{Pujals2009} 	         & TG               & \cite{Hamilton1995}	   & EXP (S)  \\ 
\cite{Hwang2010b}                & TG               & \cite{Bottaro1996}	   & EXP (S)  \\  
\cite{Hwang2010a}                & TG               & \cite{Waleffe1997}	   & EXP (S)  \\  
\cite{McKeon2010}                & TG               & \cite{Reddy1998}	           & EXP (S)  \\ 
\cite{Jimenez2013}   	         & TG               & \cite{Andersson2001}	   & EXP (S)  \\  
\cite{Alizard2015} 	 	 & TG               & \cite{Asai2002}	           & EXP (S)  \\  
\cite{Jimenez2015}               & TG               & \cite{Kawahara2003}	   & EXP (S)  \\  
\cite{Encinar2020}	 	 & TG               & \cite{Hall2010}	           & EXP (S)  \\
                                 &                  & \cite{Marquillie2011}	   & EXP (S/V)\\ 
                                 &                  & \cite{Park2011}	           & EXP (S)  \\ 
	 	                 &                  & \cite{Alizard2015}	   & EXP (S)  \\ 
                                 &                  & \cite{Chini2017}             & EXP (V)  \\
                                 &                  & \cite{Hack2018}	           & EXP (V)  \\
                                 &                  & \cite{Montemuro2020}         & EXP (V)  \\
                                 &                  & \cite{Wang2007}              & EXP (S)  \\  
                                 &                  & \cite{Hall2010}              & NEU      \\  
                                 &                  & \cite{Deguchi2013}           & NEU      \\
                                 &                  & \cite{Deguchi2015}           & NEU      \\
                                 &                  & \cite{Hall2018}              & NEU      \\ \hline 
\end{tabular}
\end{center}
}
\caption{ \small Proposed linear mechanisms responsible for the energy
  transfer from the base flow to fluctuations for: left-columns,
  $y$-dependent base flows (primary linear process); right-columns,
  $(y,z)$-dependent base flows (secondary linear process). Mechanisms
  are abbreviated as: EXP, exponential instability; TG, transient
  growth; TG PARA, transient growth assisted by parametric
  instability; NEU, modally neutral.  For EXP, V and S refer to
  varicose and sinuous instabilities, respectively.  The work by
  \cite{Hack2018} considered a $(x,y,z)$-dependent base flow, but it
  was included in the right-columns as it is devoted to the study of
  secondary instability. The label TG for studies formulated in the
  frequency-domain should be understood as pseudoresonant
  amplification of perturbations due to non-normality of the linear
  operator.  \cite{Swearingen1987, Yu1991, Yu1994, Hall1991,
    Bottaro1996, Li1995, Park1995} study the secondary instability in
  Taylor-G\"ortler vortices. \cite{Asai2002, Bottaro1996, Park1995,
    Reddy1993, Hopffner2005, Jovanovic2005, Wang2007} investigate
  laminar-to-turbulent transition and suggest that the mechanism might
  be at play in the turbulent regime. The works by \cite{Kim1971,
    Swearingen1987, Bottaro1996} and \cite{Asai2002} are laboratory
  experiments, whereas the remainder are numerical
  investigations. \cite{Farrell2012, Thomas2015, Farrell2016,
    Nikolaidis2016} are carried out in the context of Restricted
  NonLinear Navier--Stokes. Additionally, some works focus on the
  buffer layer, logarithmic layer, or outer layer. The table
  highlights one or two linear mechanisms from each reference, but
  many works acknowledge the presence of other mechanisms which are
  not mentioned in the table. The reader is referred to each
  particular work for details.
\label{table:literature}}
\end{table}

In mechanism (i), it is hypothesised that the energy is transferred
from the mean flow $U(y,z,t)$ to the fluctuating flow through modal
instability in the form of strong inflectional variations in the
spanwise direction~\citep{Hamilton1995, Waleffe1997, Karp2017} or
wall-normal direction~\citep{Chini2017, Montemuro2020}, corrugated
vortex sheets~\citep{Kawahara2003}, or intense localised patches of
low-momentum fluid~\citep{Andersson2001, Hack2018}.  These exponential
instabilities are markedly robust at all
times~\citep{Lozano_brief_2018b} and, therefore, their excitation has
been proposed to be the mechanism that replenishes the perturbation
energy of the turbulent flow~\citep{Hamilton1995, Waleffe1997,
  Andersson2001, Kawahara2003, Marquillie2011, Hack2014, Hack2018}.
Other studies have speculated that the streaky base flow $U(y,z,t)$
might originate from the primary Taylor-G\"ortler instability. In this
case, the varying wall shear induced by large scales structures gives
rise to sufficient streamline curvature in $x$ to trigger the
instability~\citep{Brown1977, Phillips1996, Saric2003}. Consequently,
it has also been hypothesised that the following secondary exponential
instability of the Taylor-G\"ortler base-flow is the mechanisms to
generate turbulence fluctuations~\citep{Swearingen1987, Yu1991,
  Yu1994, Hall1991, Bottaro1996, Li1995, Park1995, Karp2018}.
Exponential instabilities above are commonly classified according to
their symmetries as varicose and sinuous.  The varicose instability
(symmetric in the streamwise and wall-normal velocities) is commonly
associated with inflection points in the base flow along the
wall-normal direction, while the sinuous instability (symmetric in the
spanwise velocity) relates to inflection points in the spanwise
directions~\citep{Park1995, Schmid2012}. In all of the scenarios
above, the exponential instability of the streak is thought to be
central to the maintenance of wall turbulence. Additionally, the modal
character of the base flow also plays a key role in the VWI theory but
in this case it is not necessary for base flows to be unstable for
nonlinear states to develop. Instead, it is postulated that the
regeneration cycle is supported by the interaction of a roll with the
neutrally stable mean streamwise-flow \citep{Hall1991, Deguchi2013,
  Hall2018}.

Mechanism (ii), transient growth, involves the redistribution of
energy from the streak to the fluctuations via transient algebraic
amplification.  The transient growth scenario of the streaky base flow
$U(y,z,t)$ (not to be confused with the transient growth of $\langle u
\rangle_{xzt}(y)$ discussed above) gained popularity since the work by
\citet{Schoppa2002}, who argued that transient growth may be the most
relevant mechanism not only for streak formation but also for their
eventual breakdown. \citet{Schoppa2002} showed that most streaks
detected in actual wall-turbulence simulations are indeed
exponentially stable for the set of wavenumbers considered. Hence, the
loss of stability of the streaks would be better explained by the
transient growth of perturbations that would lead to vorticity sheet
formation and nonlinear saturation.  The findings by
\citet{Schoppa2002} have long been criticised, and the absence of
unstable streaks can be also interpreted as an indication that
instability is important, as the unstable streaks break fast and are
harder to observe. Other criticism argues that, far from the wall,
streaks might not provide a reservoir of energy large enough to
sustain the flow fluctuations \citep{Jimenez2018}. Some authors have
further argued that distinguishing between streak transient growth and
streak modal instability would be virtually impossible, as both emerge
almost concurrently during the streak breakdown \citep{Hopffner2005,
  Giovanetti2017, Cassinelli2017}, and hence both are driving
mechanisms of self-sustaining turbulence.

Finally, mechanism~(iii) has been advanced in recent years by Farrell,
Ioannou and coworkers~\citep{Farrell1999, Farrell2012, Farrell2016,
  Nikolaidis2016, Farrell2017, Farrell2017c, Bretheim2018}. They
adopted the perspective of statistical state dynamics (SSD) to develop
a tractable theory for the maintenance of wall turbulence. Within the
SSD framework, the perturbations are maintained by an essentially
time-dependent, parametric instability of the base flow. The concept
of ``parametric instability'' refers here to perturbation growth that
is inherently caused by the time-dependence of the base flow $U$. The
self-sustaining mechanism proposed by SSD still relies on the highly
non-normal streamwise roll and streak structure. However, it differs
from other mechanisms above in that it requires the time-variations of
$U$ for the growth of perturbations to be supported. Furthermore, it
implies that mechanisms based on critical layers
\citep[e.g.][]{Hall1988, Hall1991, Hall2010} and modal or non-modal
growth processes alone \citep[e.g.][]{Waleffe1997, Schoppa2002} are
not responsible for most of the energy transfer from $U$ to
$\boldsymbol{u}'$, as they ignore both the intrinsic time-dependence
of the base flow or the non-normal aspect of the linear dynamics.

\subsection{Cause-and-effect of linear mechanisms}

The scenarios (i), (ii), and (iii), although consistent with the
observed turbulence structure~\citep{ Robinson1991, Panton2001,
  Jimenez2018}, are rooted in simplified theoretical arguments. It
remains to establish whether self-sustaining turbulence follows
predominantly one of the above mentioned mechanisms, or a combination
of them.  One major obstacle to assess linear theories arises from the
lack of tools in turbulence research that resolve the cause-and-effect
dilemma and unambiguously attributes a set of observed dynamics to
well-defined causes. This brings to attention the issue of causal
inference, which is a central theme in many scientific disciplines but
is barely discussed in turbulence research with the exception of a
handful of works \citep{Tissot2014, Liang2017, Bae2018a, Lozano2019b}.
It is via cause-and-effect relationships that we gain a sense of
understanding of a given phenomenon, namely, that we are able to shape
the course of events by deliberate actions or policies
\citep{Pearl2009}. It is for that reason that causal thinking is so
pervasive. Typically, causality is inferred from \emph{a priori}
analysis of frozen flow snapshots or, at most, by time-correlation
between pairs of signals extracted from the flow.  However,
elucidating causality, which inherently occurs over the course of
time, is challenging using a frozen-analysis approach, and
time-correlations lack the directionality and asymmetry required to
guarantee causation (i.e., correlation does not imply causation)
\citep{Beebee2012}. Recently, \cite{Lozano2019b} introduced a
probabilistic measure of causality to study self-sustaining wall
turbulence based on the Shannon entropy that relies on a non-intrusive
framework for causal inference.  In the present work, we provide a
complementary `intrusive' viewpoint.

Here, we evaluate the contribution of different linear mechanisms via
direct numerical simulation of channel flows with constrained energy
extraction from the streamwise-averaged mean-flow.  To that end, we
modify the Navier--Stokes equations to suppress the causal link for a
targeted linear mechanism, while maintaining a fully nonlinear
system. This approach falls within the category of ``instantiated''
causality, i.e., intrusively perturbing a system (cause) and observing
the consequences (effect) \citep{Pearl2009}. In our case, altering the
system has the benefit of providing a clear cause-and-effect
assessment of the importance of each linear mechanism implicated in
sustaining the flow.  These `conceptual numerical experiments' have
been long practised in turbulence research and many notorious examples
can be found in the literature. However, the connection between
conceptual numerical experiments and causality has been loose. In the
present work, we aim to promote the formalisation and systematic use
of cause-and-effect analysis to solve new and long-standing unsettled
problems in Fluid Mechanics.

The study is organised as follows: \S~\ref{sec:numerical} contains the
numerical details of the turbulent channel flow simulations.  The
statistics of interest for wall turbulence are reviewed in
\S~\ref{sec:regular}.  In \S~\ref{sec:theories}, we briefly outline
the linear theories of self-sustaining wall turbulence and evaluate
\emph{a priori} their potential relevance for sustaining the flow.
In \S~\ref{sec:interventions} we discuss the discovery of
cause-and-effect relationships by interventions in the system. The
actual relevance of different linear mechanisms from a
cause-and-effect perspective is investigated in
\S~\ref{sec:constrain}. The section is further subdivided into
subsections devoted to the cause-and-effect of exponential
instabilities and transient growth with and without parametric
instability. Finally, we conclude in \S~\ref{sec:conclusions}.

\section{Minimal turbulent channel flows units}
\label{sec:numerical}

\subsection{Numerical experiments}

To investigate the role of different linear mechanisms, we perform
direct numerical simulations of incompressible turbulent channel flows
driven by a constant mean pressure gradient. Hereafter, the
streamwise, wall-normal, and spanwise directions of the channel are
denoted by $x$, $y$, and $z$, respectively, the corresponding flow
velocity components by $u$, $v$, $w$, and pressure by $p$. The density
of the fluid is $\rho$, the kinematic viscosity of the fluid is $\nu$,
and the channel height is $h$.  The wall is located at $y=0$, where
no-slip boundary conditions apply, whereas free stress and no
penetration conditions are imposed at $y=h$. The streamwise and
spanwise directions are periodic.

The simulations are characterised by the friction Reynolds number,
$\mathrm{Re}_\tau$, defined as the ratio of the channel height to the
viscous length-scale $\delta_v = \nu/u_\tau$, where $u_\tau$ is the
characteristic velocity based on the mean skin friction at the wall
$u_\tau^2 \equiv \nu \langle \p u(x,0,z,t)/ \p y\rangle_{xzt}$. Here,
the Reynolds number is $\mathrm{Re}_\tau = h/\delta_v \approx
180$. The streamwise, wall-normal, and spanwise sizes of the
computational domain are $L_x^+ \approx 337$, $L_y^+ \approx 186$, and
$L_z^+ \approx 168$, respectively, where the superscript $+$ denotes
quantities scaled by~$\nu$ and~$u_\tau$. \citet{Jimenez1991} showed
these `minimal channels' contains an elementary turbulent flow unit
comprised of a single streamwise streak and a pair of staggered
quasi-streamwise vortices, that reproduce the dynamics of the flow in
larger domains. Hence, the current numerical experiment isolates the
few, most relevant, coherent structures involved in self-sustaining
turbulence in the buffer layer.  It also provides an ideal testbed for
studying linear mechanisms, as it enables the identification of a
meaningful base flow for these elementary coherent structures. In
Appendix \ref{sec:appendix_2Lx}, we assess the sensitivity of the key
results presented in this study to changes in the domain extent ($L_x$
and $L_z$). We find that our conclusions still hold when the size of
the computational domain is doubled in each direction.

We integrate the incompressible Navier--Stokes equations 
\begin{subequations}\label{eq:NS_original}
\begin{gather}
  \frac{\partial\bu}{\partial t} = - \bu \cdot
  \bnabla \bu - \frac{1}{\rho}\bnabla p + \nu \nabla^2
  \bu + \boldsymbol{f}, \\
  \bnabla \cdot \bu = 0,
\end{gather}
\end{subequations}
with $\bu \defn (u,v,w)$ and $\boldsymbol{f} = (u_\tau^2/h,0,0)$.

The simulations are performed with a staggered, second-order, finite
differences scheme \citep{Orlandi2000} and a fractional-step method
\citep{Kim1985} with a third-order Runge-Kutta time-advancing scheme
\citep{Wray1990}.  The solution is advanced in time using a constant
time step chosen appropriately so that the Courant--Friedrichs--Lewy
condition is below 0.5. The code has been presented in previous
studies on turbulent channel flows \citep{Lozano2016_Brief, Bae2018b,
  Bae2018c}. In addition, we performed various numerical experiments
(summarised in the second column of table \ref{table}) in which we
time-advance two sets of equations: one for the base flow $\bU$ and
one for the fluctuations $\bu'$. In this manner, we are able to
independently control the dynamics of $\bU$ and $\bu'$. We discuss in
detail these additional experiments in \S \ref{sec:constrain}.

The streamwise and spanwise grid resolutions are $\Delta x^+\approx
6.5$ and $\Delta z^+\approx3.3$, respectively, and the minimum and
maximum wall-normal resolutions are $\Delta
y_{\mathrm{min}}^+\approx0.2$ and $\Delta
y_{\mathrm{max}}^+\approx6.1$. The corresponding number of grid points
in $x$, $y$, and $z$ are $64 \times 90 \times 64$, respectively. All
the simulations presented here were run for at least $300h/u_\tau$
units of time after transients. This time-period is orders of
magnitude longer than the typical lifetime of individual
energy-containing eddies~\citep{Lozano2014b}, and allows us to
collect meaningful statistics of the self-sustaining cycle.

\subsection{Base flow}
\label{subsec:baseflow}

\begin{figure}
 \vspace{1cm}
 \begin{center}
   \includegraphics[width=0.95\textwidth]{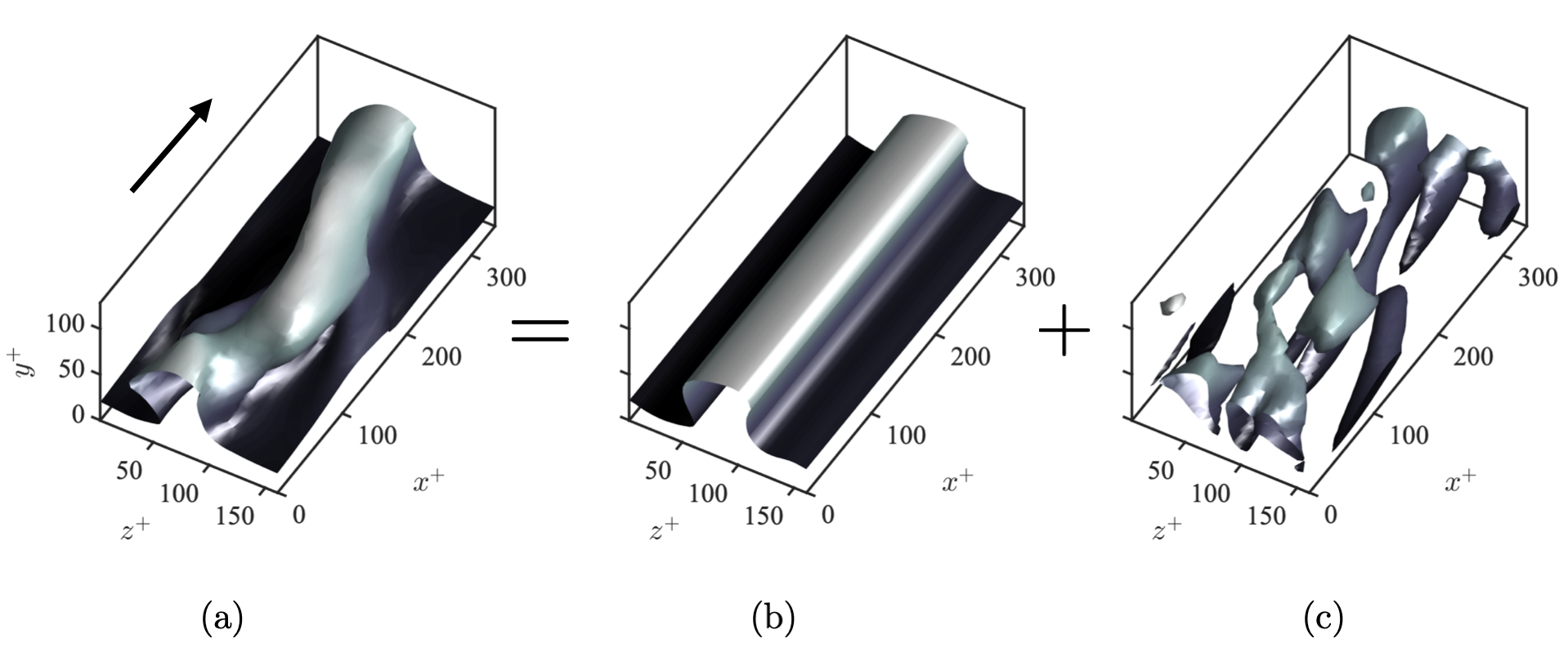}  
 \end{center}
\caption{ Decomposition of the instantaneous flow into a streamwise
  mean base flow and fluctuations.  Instantaneous isosurface of
  streamwise velocity for (a)~the total flow $u$, (b)~the streak base
  flow $U$, and (c)~the absolute value of the fluctuations $|u'|$.
  The values of the isosurfaces are 0.6 (a and b) and 0.1 (c) of the
  maximum streamwise velocity.  Shading represents the distance to the
  wall from dark ($y=0$) to light ($y=h$). The arrow in panel~(a)
  indicates the mean flow direction. Results for case R180.
  \label{fig:snaphots}}
\end{figure}

\begin{figure} 
 \vspace{1cm}
  \begin{center}
    \subfloat[]{\includegraphics[width=0.34\textwidth]{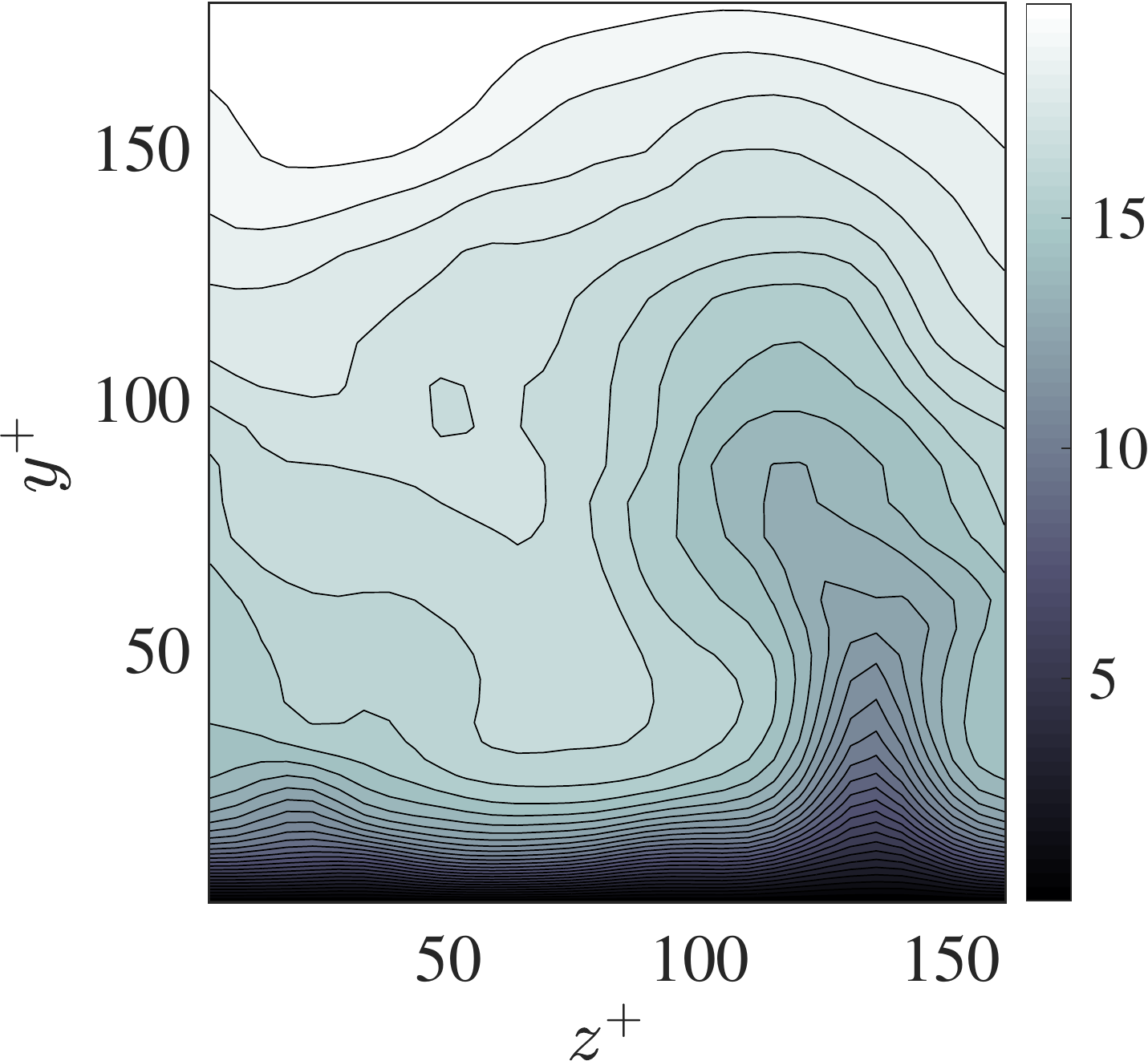}} 
    \subfloat[]{\includegraphics[width=0.34\textwidth]{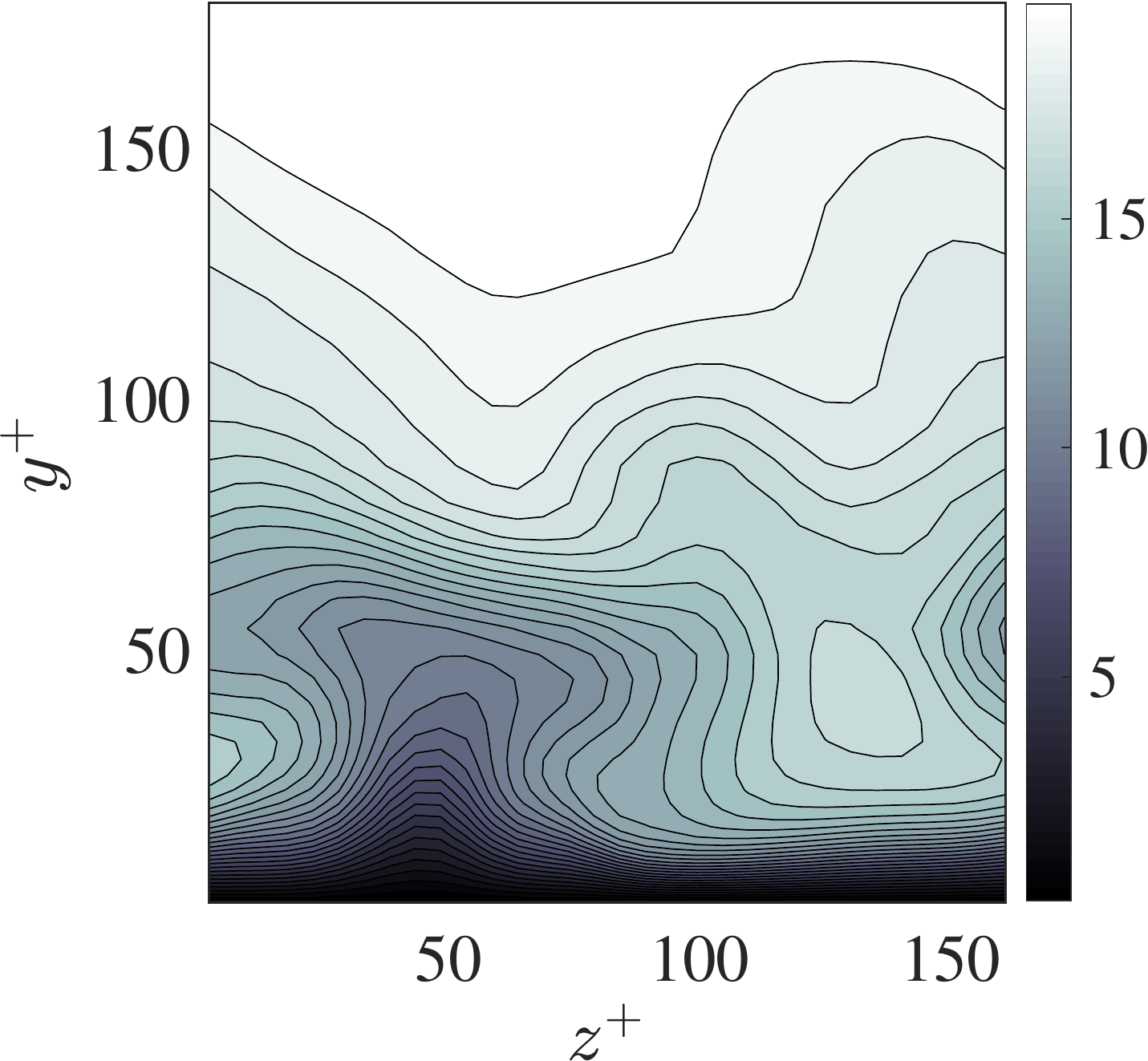}} 
    \subfloat[]{\includegraphics[width=0.34\textwidth]{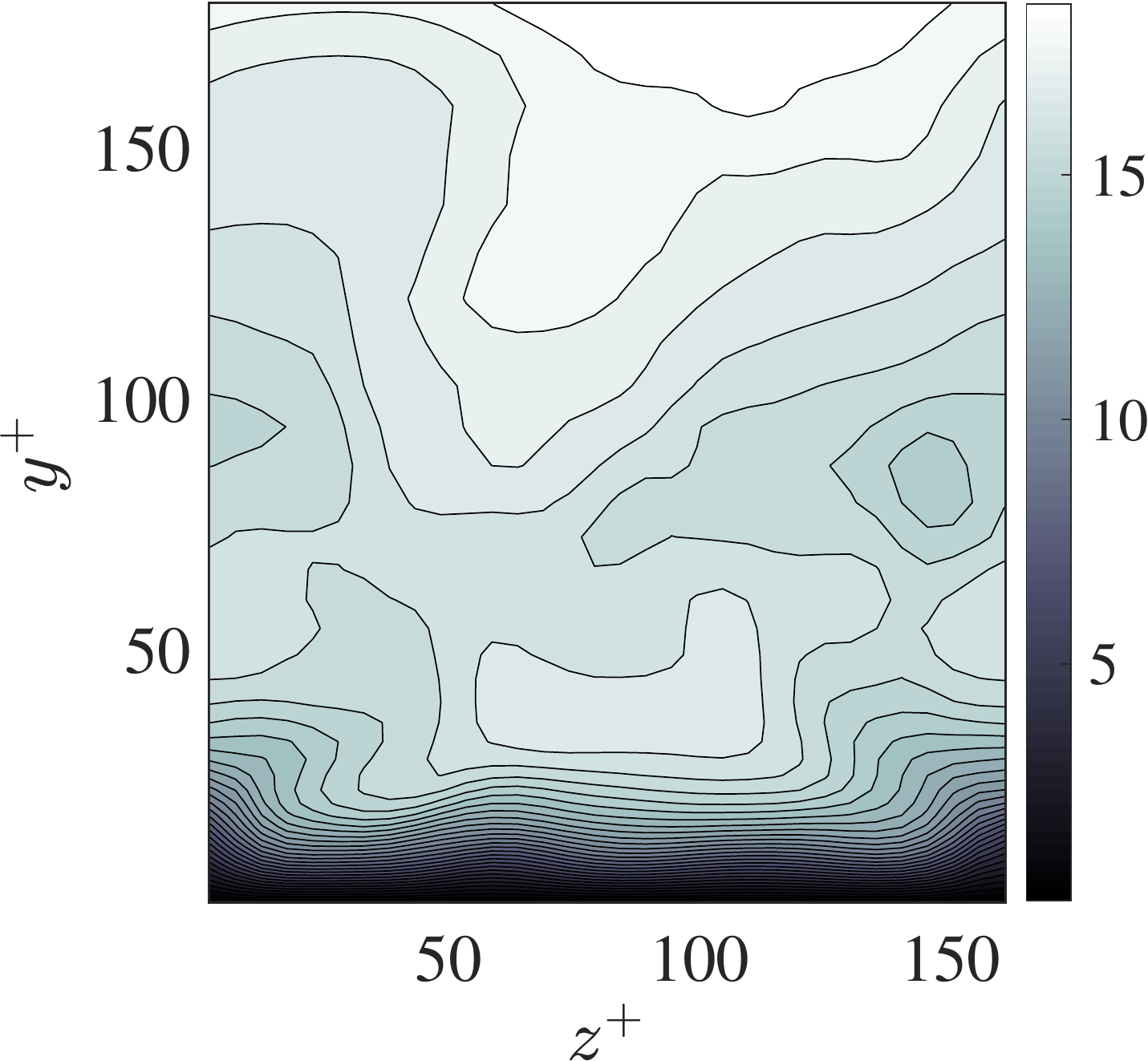}} 
 \end{center}
\caption{ Examples of base flow, defined as $U(y,z,t) \defn \langle u
  \rangle_x$, for a turbulent channel flow at $\mathrm{Re}_\tau\approx
  180$ (case R180 from \S~\ref{sec:regular}). The examples are
  representative instances with (a,b) strong streak activity (c) and
  quiescent times.
  \label{fig:baseflow}}
\end{figure}

We partition the flow into fluctuating velocities $\bu' \defn (u', v',
w')$ and base flow $\bU$, defined as the time-varying mean streamwise
velocity $\bU \defn (U, 0, 0)$, where
\begin{equation}\label{eq:baseflow}
  U(y, z, t) \defn \langle u \rangle_x
  = \frac{1}{L_x} \int_{0}^{L_x} u(x,y,z,t) \,\mathrm{d}x,
\end{equation}
such that $u' \defn u - U$, $v' \defn v$, and $w' \defn
w$. Hereafter, $\langle \,\boldsymbol{\cdot}\, \rangle_{ijk...}$ denotes averaging over
the directions (or time) $i$, $j$, $k$,..., for example,
\begin{equation}
  \langle u \rangle_{xzt}
  = \frac{1}{L_x L_z T_s} \int_{0}^{L_x}\int_{0}^{L_z}\int_{0}^{T_s} u(x,y,z,t) \,
  \mathrm{d}t\mathrm{d}z\mathrm{d}x,
\end{equation}
where $T_s$ is a time-horizon long enough to remove any fluctuations.
Figure~\ref{fig:snaphots} illustrates this flow decomposition and
figure~\ref{fig:baseflow} depicts three typical snapshots of the base
flow defined in~\eqref{eq:baseflow}. Note that because we are using a
minimal box for the channel, only a single energy-containing eddy fits
in the domain. Hence, $U$ computed in minimal boxes is a meaningful
base flow ``felt'' by individual flow structures. This would not be
the case in larger domains in which the contribution of the multiple
structures present in the flow cancels out and does not contribute to
$U$.

We have not included in the base flow~\eqref{eq:baseflow} the
contributions from the streamwise averages of~$v$~and~$w$ components,
$V \defn \langle v \rangle_x$ and $W \defn \langle w \rangle_x$, as
these are not traditionally included in the study of stability of the
streaky flow. Indeed, the vast majority of studies reported in table
\ref{table:literature} do not account for $V$ and $W$ in the
analysis. The results obtained using $(U,0,0)$ as base flow were
repeated for a base flow consisting of $(U,V,W)$, and a concise
overview of the findings can be found in Appendix
\ref{sec:appendix_UVW}. In summary, the conclusions drawn for base
flows $(U,0,0)$ or $(U,V,W)$ are similar, and thus we focus on the
former for simplicity.
%
%

The equation of motion for the base flow
  $\bU=(U,0,0)$ is obtained by averaging the Navier--Stokes
  equations~\eqref{eq:NS_original} in the streamwise direction,
\begin{subequations}\label{eq:NS_U}
\begin{gather}
  \frac{\partial \bU}{\partial t} + \bU \cdot
 \bnabla \bU =
 -\mathcal{D}\langle \bu' \cdot
  \bnabla \bu' \rangle_x
 - \frac{\mathcal{D}}{\rho}\bnabla \langle p \rangle_x + \nu \nabla^2
 \bU + \boldsymbol{f}, \\
 \bnabla \cdot \bU = 0,
\end{gather}
\end{subequations}
where the operator $\mathcal{D}$ sets the $y-$ and $z-$components of
the nonlinear terms and pressure to zero for consistency with $\bU =
(U,0,0)$ (see Appendix \ref{sec:appendix_UVW}).
Subtracting~\eqref{eq:NS_U} from~\eqref{eq:NS_original} we get that
the fluctuating flow $\bu'$ is governed by
\begin{gather} \label{eq:NS}
\frac{\partial\bu'}{\partial t} =  
\underbrace{\mathcal{L}(U)\bu'}_{\substack{\textrm{linear}\\\textrm{processes}}} +
\underbrace{\bN(\bu')}_{\substack{\textrm{nonlinear}\\\textrm{processes}}},
\end{gather}
where $\mathcal{L}(U)$ is the linearised Navier--Stokes operator for the
fluctuating state vector about the instantaneous $\bU$ (see
figure~\ref{fig:snaphots}b) such that
\begin{equation}\label{eq:L}
  \mathcal{L}(U)\bu' =
  \mathcal{P}\left[ -\bU\cdot \bnabla \bu' - \bu' \cdot \bnabla \bU
  + \nu \nabla^2 \bu' \right].
\end{equation}
The operator $\mathcal{P}$ accounts for the kinematic divergence-free
condition, $\bnabla \cdot \bu' = 0$. Conversely, $\bN(\bu')$
collectively denotes the nonlinear terms, which are quadratic with
respect to fluctuating flow fields,
\begin{equation}\label{eq:N}
\bN(\bu') = \mathcal{P}\left[ -\bu' \cdot \bnabla \bu'
+\mathcal{D}\langle \bu' \cdot \bnabla \bu' \rangle_x \right].
\end{equation}
We are interested on the dynamics of $\bu'$ governed by
(\ref{eq:NS}). Note that the flow partition $\bU + \bu'$ implies that
the energy injection into the velocity fluctuations is ascribed to
linear processes from $\mathcal{L}(U)$, since the term $\bN(\bu')$ is
only responsible for redistributing the energy in space and scales
among the fluctuations, i.e., the domain-integral of $\bu'\cdot\bN$
vanishes identically and thus 
\begin{equation}
    \frac{\p }{\p t} \left\langle E \right\rangle_{xyz}
  = \left\langle \bu' \cdot \mathcal{L}(U)\bu'\right\rangle_{xyz},
\end{equation}
where $E \defn \tfrac1{2}|\bu'|^2$ is the fluctuating turbulent
kinetic energy. 

\begin{table}
\begin{center}
\begin{tabular}{lcccc}
  \makecell{ Case \\ \hline Sustained?}   & Equation for $\bu'$  & Equation for $\bU$ & \makecell{Feedback from \\
    $\bu'\rightarrow \bU$} &
  \makecell{Active linear mechanisms for \\ energy transfer from $\bU\rightarrow \bu'$}\\[1ex] \hline
  \makecell{ R180 \\ \hline \cmark } & (\ref{eq:R180_1}) & \makecell{$U(y,z,t)$ \\ from (\ref{eq:R180_2})} & \cmark & \makecell{Exponential instabilities \\ Transient growth \\ Parametric instabilities} \\  \hline
 \makecell{ NF180 \\ \hline \cmark } & (\ref{eq:NF180}) & \makecell{Precomputed \\ $U(y,z,t)$ \\ from R180} & \xmark & \makecell{Exponential instabilities \\ Transient growth \\ Parametric instabilities} \\ \hline
 \makecell{ NF-SEI180 \\ \hline \cmark } & (\ref{eq:NF-SEI180}) & \makecell{Precomputed \\ $U(y,z,t)$ \\ from R180} & \xmark & \makecell{Transient growth \\ Parametric instabilities} \\ \hline
 \makecell{ R-SEI180 \\ \hline \cmark } & (\ref{eq:R-SEI180_1}) & \makecell{$U(y,z,t)$ \\ from (\ref{eq:R-SEI180_2})} & \cmark & \makecell{Transient growth \\ Parametric instabilities} \\ \hline
 \makecell{ NF-TG180 \\ \hline \cmark/\xmark } & (\ref{eq:NF-TG180_1}) & \makecell{ Precomputed \\ $U(y,z,t_0)$ \\ from R180\\ at a frozen $t_0$} & \xmark & \makecell{Transient growth} \\ \hline
 \makecell{ NF-NLU180 \\ \hline \cmark } & \makecell{ (\ref{eq:linear_details_1_mod1}), (\ref{eq:linear_details_2_mod}), \\ (\ref{eq:linear_details_3}), (\ref{eq:linear_details_4})}
 & \makecell{ Precomputed \\ $U(y,z,t)$ \\ from R180} & \xmark & \makecell{Exponential instabilities \\ Transient growth \\ without lift-up \\ Parametric instabilities} \\ \hline
 \makecell{ NF-NPO180 \\ \hline \xmark }& \makecell{(\ref{eq:linear_details_1_mod2}), (\ref{eq:linear_details_2}), \\ (\ref{eq:linear_details_3}), (\ref{eq:linear_details_4})}
 & \makecell{ Precomputed \\ $U(y,z,t)$ \\ from R180} & \xmark & \makecell{Exponential instabilities \\ Transient growth \\ without push-over \\ Parametric instabilities} \\ \hline
 \makecell{ NF-NO180 \\ \hline \xmark } & \makecell{(\ref{eq:linear_details_1}), (\ref{eq:linear_details_2_mod}), \\ (\ref{eq:linear_details_3}), (\ref{eq:linear_details_4})} & \makecell{ Precomputed \\ $U(y,z,t)$ \\ from R180} & \xmark & \makecell{Exponential instabilities \\ Transient growth \\ without Orr \\ Parametric instabilities} \\ \hline
\end{tabular}
\end{center}
\caption{ List of cases of turbulent channel flows with and without
  constrained linear mechanisms. The friction Reynolds number is
  $\mathrm{Re}_\tau \approx 180$ for all cases. The cases are labelled
  following the nomenclature: R, regular wall turbulence with feedback
  $\bU\rightarrow \bu'$ allowed; NF, no-feedback from $\bU\rightarrow
  \bu'$ allowed; SEI, suppressed exponential instabilities; TG, only
  transient growth without exponential nor parametric instabilities;
  NLU, no linear lift-up of the streak; NPO, no linear push-over of
  the streak; NO, no linear Orr of the streak.
\label{table}}
\end{table}
%
In the rest of the paper, in addition to solutions of the
Navier-Stokes equations~\eqref{eq:NS_original}, we modify
~\eqref{eq:L} to preclude the energy transfer from $\bU$ to $\bu'$ for
targeted linear mechanisms. The simulations carried out are summarised
in table~\ref{table}, which includes the active linear mechanisms for
energy transfer from $\bU$ to $\bu'$ and whether the cases are capable
of sustaining turbulent fluctuations. The details on how the equations
of motion are modified for each case are discussed in the remainder of
the paper.

\section{Regular wall turbulence}
\label{sec:regular}

First, we solve the Navier--Stokes equations without any modification,
so that all mechanisms for energy transfers from the base flow to the
fluctuations are naturally allowed. We refer to this case as the
``regular channel''~(R180).  We provide an overview of the
self-sustaining state of the flow and one-point statistics for
R180. The results are used as a reference solution in forthcoming
sections.  The governing equations for the regular channel flow
are~\eqref{eq:NS_U} and~\eqref{eq:NS}:
\begin{subequations}
  \label{eq:R180}
\begin{gather} \label{eq:R180_1}
\frac{\partial\bu'}{\partial t} = 
\mathcal{L}(U)\bu'+ \bN(\bu'),\\
  \frac{\partial \bU}{\partial t} =
 -\bU \cdot
 \bnabla \bU
 -\mathcal{D}\langle \bu' \cdot
  \bnabla \bu' \rangle_x
 - \frac{\mathcal{D}}{\rho}\bnabla \langle p \rangle_x + \nu \nabla^2
 \bU + \boldsymbol{f}, \quad 
 \bnabla \cdot \bU = 0. \label{eq:R180_2}
\end{gather}
\end{subequations}

\begin{figure} 
 \begin{center}
   \subfloat[]{\includegraphics[width=0.45\textwidth]{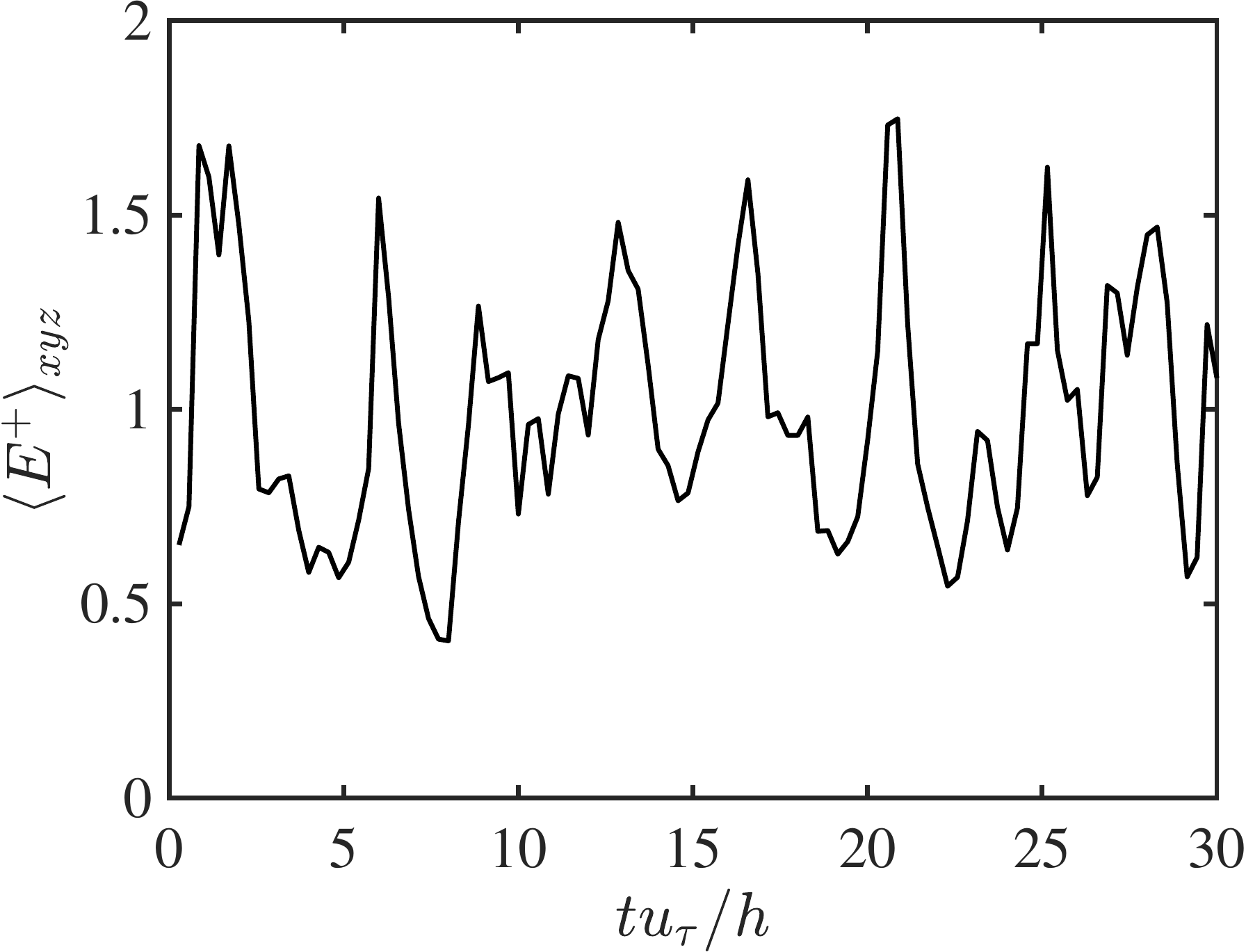}} 
   \hspace{0.1cm}
   \subfloat[]{\includegraphics[width=0.45\textwidth]{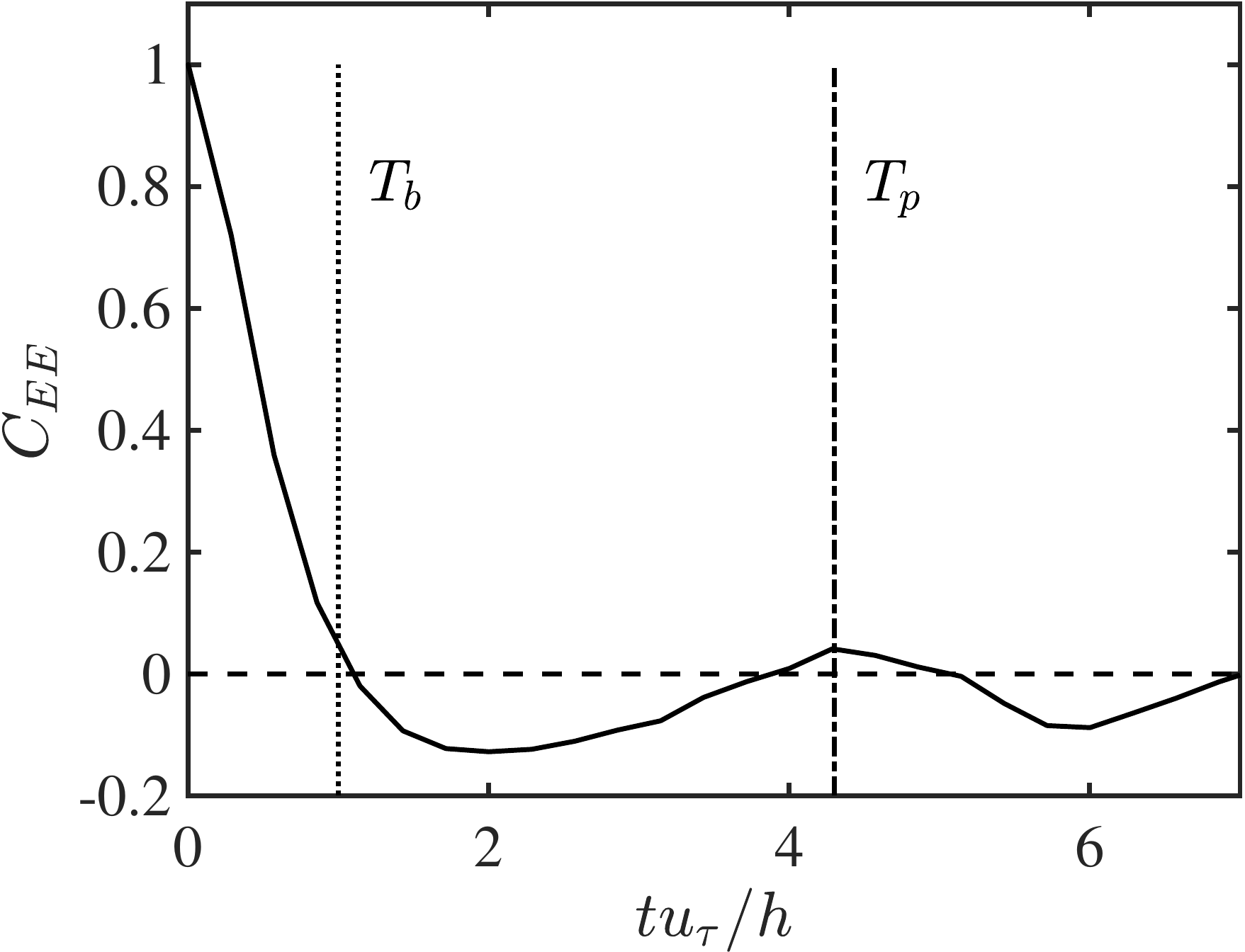}} 
 \end{center}
 \caption{ (a)~The history of the domain-averaged turbulent kinetic
   energy of the fluctuations $\langle E \rangle_{xyz}$. Note that only
   $30 h/u_\tau$ units of time are shown in the panel but the
   simulation was carried out for more than $300 h/u_\tau$.  (b) The
   time-autocorrelation of $\langle E \rangle_{xyz}$. The vertical
   dotted and dash-dotted lines are $t=h/u_\tau\approx T_b$ (burst
   duration) and $t=4.3h/u_\tau \approx T_p$ (time between bursts),
   respectively. Results for regular channel flow R180.
 \label{fig:TKE_regular}}
\end{figure}

\begin{figure}
 \begin{center}
  \includegraphics[width=1\textwidth]{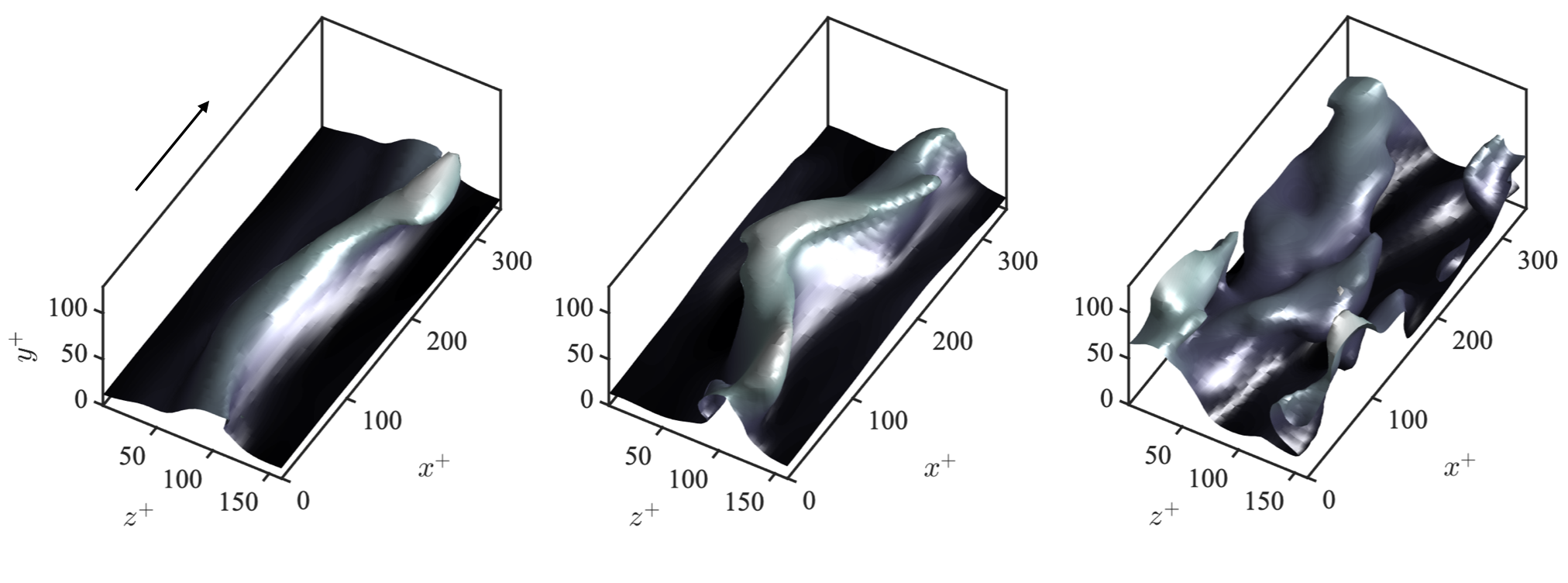} 
 \end{center}
\caption{ Instantaneous isosurface of the streamwise velocity $u$ at
  different times for R180.  The value of the isosurface is 0.65 of
  the maximum streamwise velocity.  Shading represents the distance to the
  wall from dark ($y=0$) to light ($y=h$). The arrow indicates the mean flow
  direction. \label{fig:snapshots_regular}}
\end{figure}

The history of the domain-averaged turbulent kinetic energy, $\langle
E \rangle_{xyz}$, is shown in figure~\ref{fig:TKE_regular}(a). The
evolution of $\langle E \rangle_{xyz}$ reveals the widely documented
intermittent behaviour of the turbulent kinetic energy: relatively low
turbulent kinetic energy states followed by occasional spikes usually
ascribed to the regeneration and bursting stages of the
self-sustaining cycle. As an example,
figure~\ref{fig:snapshots_regular} contains the streamwise velocity at
three instants with different degrees of turbulence activity.
%
%
If we interpret bursts events as moments of intense turbulent kinetic
energy, the time-autocorrelation of $\langle E \rangle_{xyz}$ allows
us to define a characteristic burst duration ($T_b$), and the period
between two consecutive bursts ($T_p$).
Figure~\ref{fig:TKE_regular}(b) shows that $T_b \approx h/u_\tau$
measured as the time for zero correlation, while $T_p \approx
4h/u_\tau$ given by the time-distance between two consecutive
maxima. Later on, we compare this burst period $T_b$ with the
characteristic time-scales for energy-injection into $\bu'$.
%
%

\begin{figure}
 \begin{center}
   \subfloat[]{\includegraphics[width=0.45\textwidth]{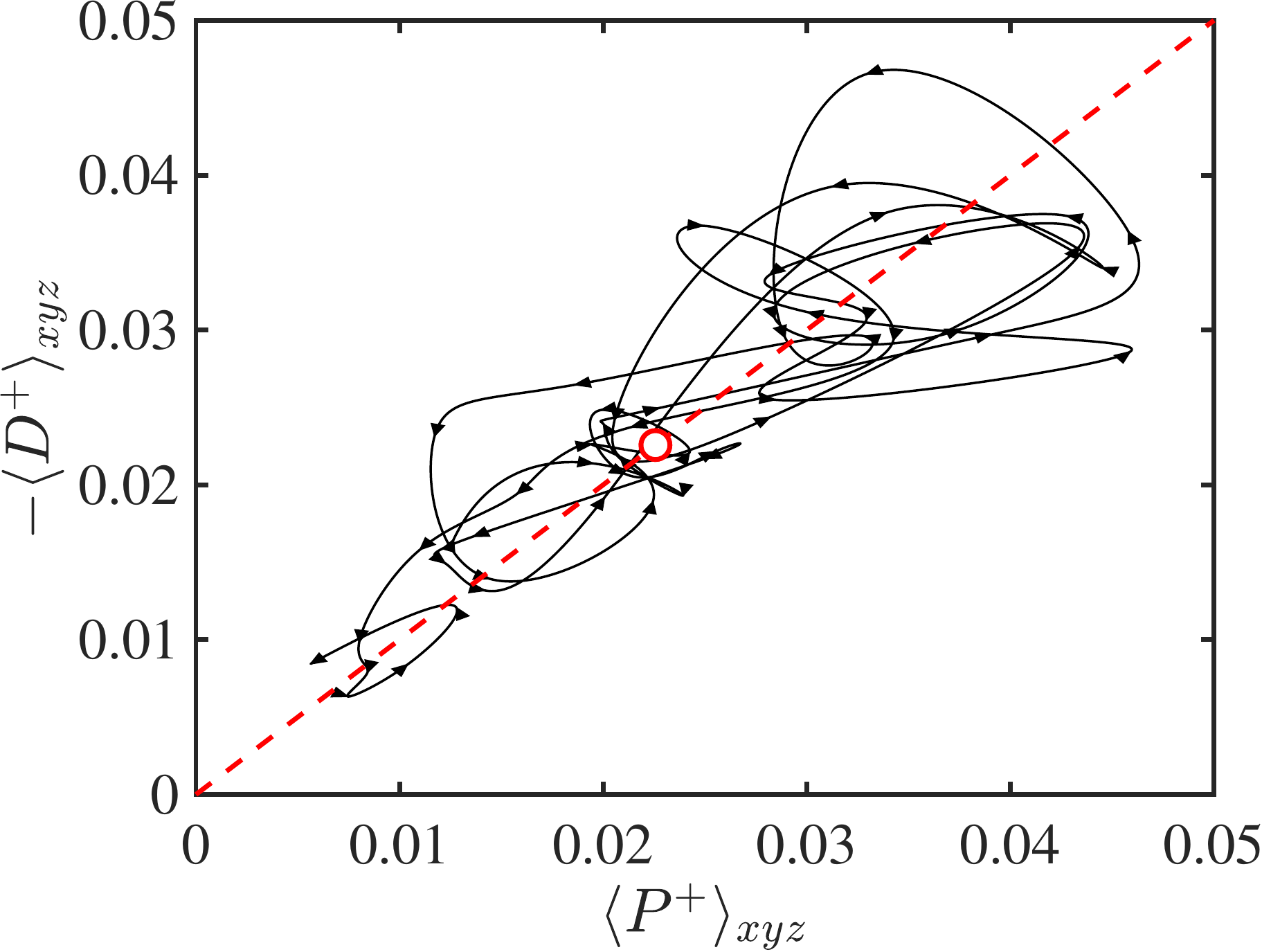}}   
   \subfloat[]{\includegraphics[width=0.45\textwidth]{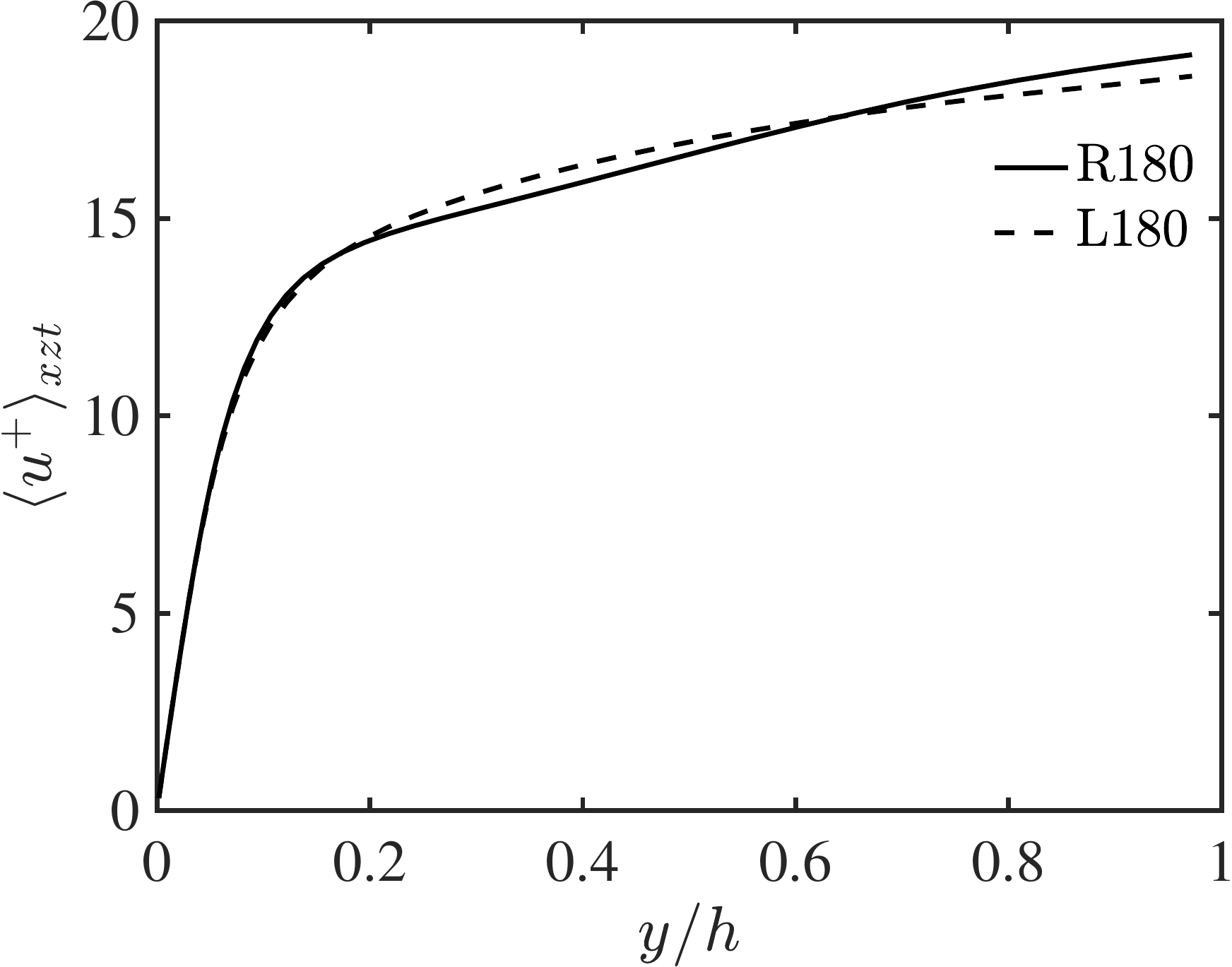}} 
 \end{center}
 \begin{center}
   \subfloat[]{\includegraphics[width=0.33\textwidth]{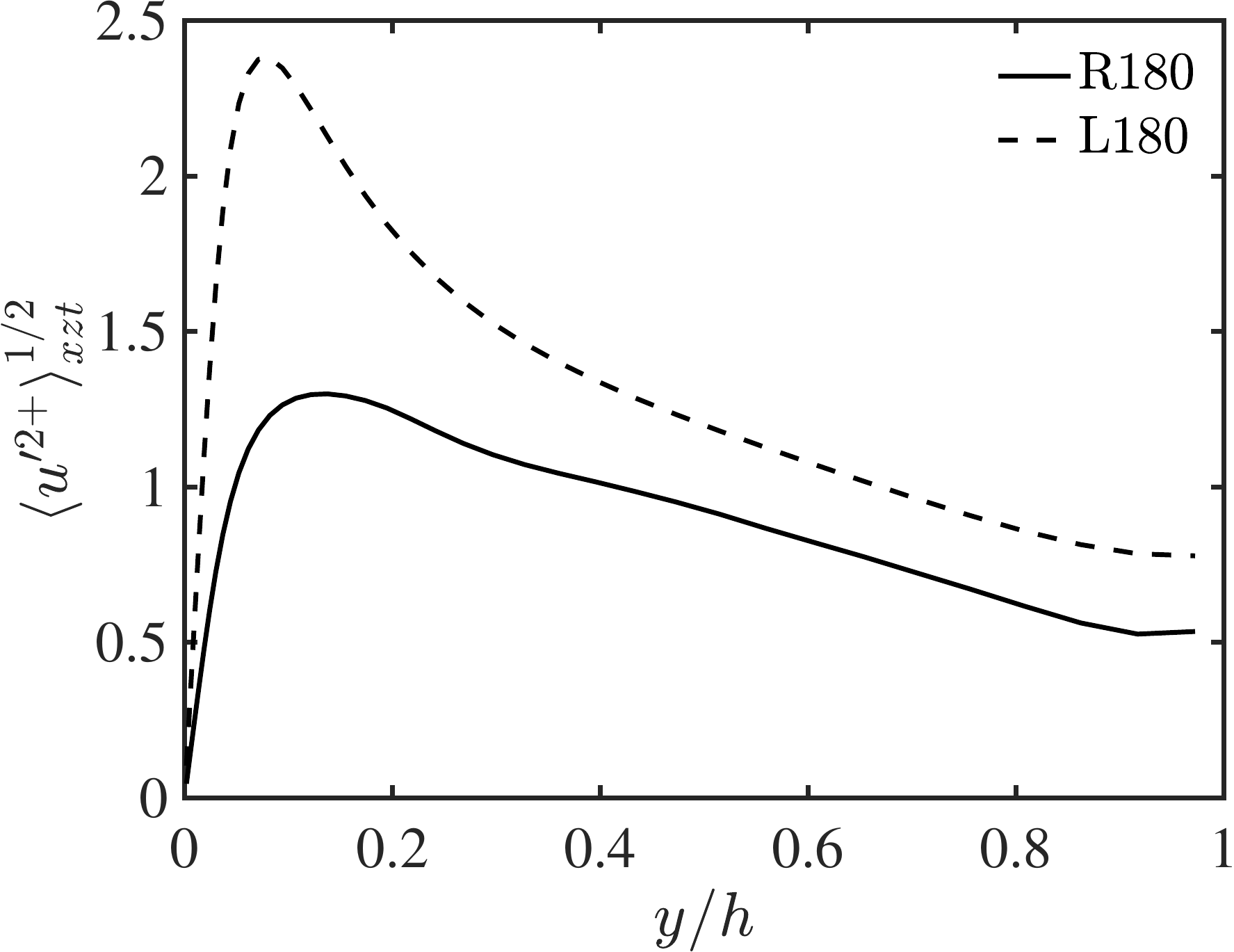}} 
   \subfloat[]{\includegraphics[width=0.33\textwidth]{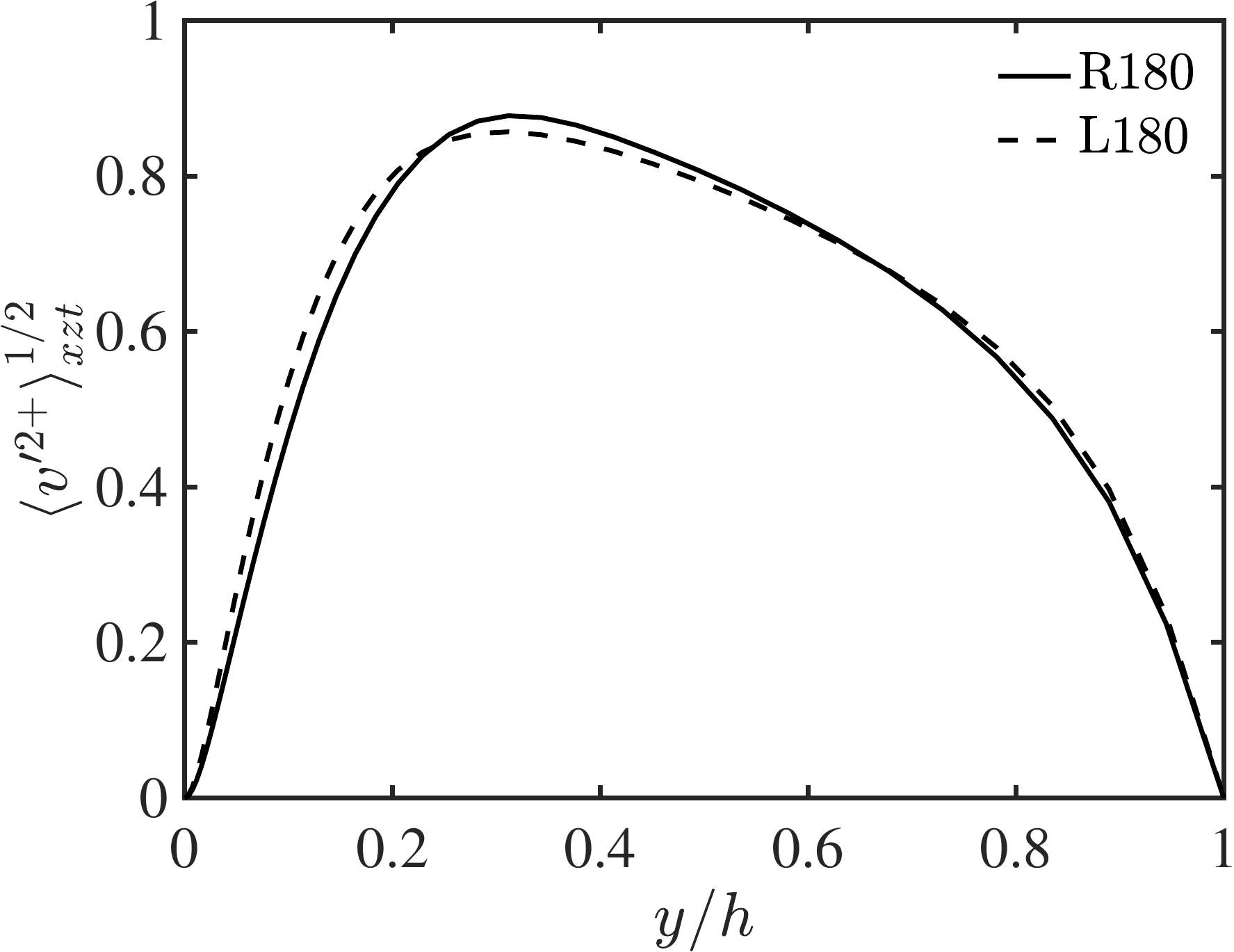}} 
   \subfloat[]{\includegraphics[width=0.33\textwidth]{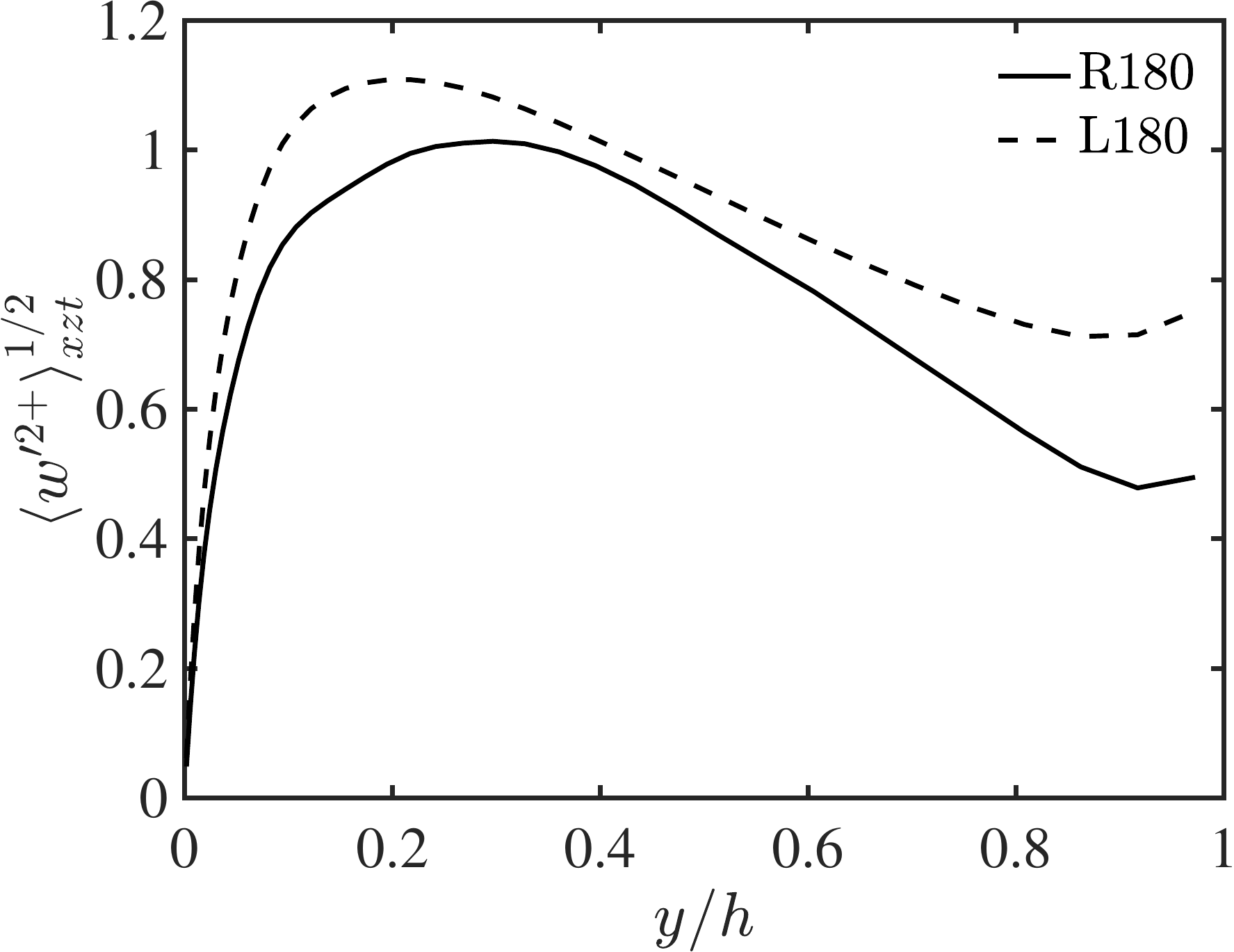}} 
 \end{center}
\caption{ (a)~Projection of the flow trajectory for R180 onto the
  average production rate $\langle P \rangle_{xyz}$ and dissipation
  rate $\langle D \rangle_{xyz}$ plane. The arrows indicate the time
  direction of the trajectory, which on average rotates
  counter-clockwise.  The red dashed line is $\langle P \rangle_{xyz}
  = -\langle D \rangle_{xyz}$ and the red circle $\langle P
  \rangle_{xyzt} = -\langle D \rangle_{xyzt}$. The trajectory
  projected covers $15 h/u_\tau$ units of time.  (b) Streamwise mean
  velocity profile and (c) streamwise, (d) wall-normal, and (e)
  spanwise root-mean-squared fluctuating velocities as a function of
  the wall-normal distance for R180 (\solid) and equivalent
  non-minimal channel L180 (\dashed) with $L_x^+ \times L_z^+= 2312
  \times 1156$ ($L_x \times L_z \approx 12.5h \times 6.3h$).
 \label{fig:PD_stats_regular}}
\end{figure}

A useful representation of the high-dimensional dynamics of the
solution is obtained by projecting the instantaneous flow trajectory
onto the two-dimensional space defined by the domain-averaged
production and dissipation rates
\begin{align}
\langle P \rangle_{xyz} & \defn \left\langle -u'v'\frac{\partial U}{\partial y} -u'w'\frac{\partial U}{\partial z} \right\rangle_{xyz}, \\
\langle D \rangle_{xyz} & \defn \left\langle -2\nu \mathcal{S}:\mathcal{S} \right\rangle_{xyz},
\end{align}
where $\mathcal{S}$ is the rate of strain tensor for the fluctuating
velocities, and the colon denotes double inner product. The
statistically stationary state of the system requires $\langle P
\rangle_{xyzt} =- \langle D \rangle_{xyzt}$. The results, plotted in
figure~\ref{fig:PD_stats_regular}(a), show that the projected solution
revolts around $\langle P \rangle_{xyzt} = -\langle D \rangle_{xyzt}$
and is characterised by excursions into the high dissipation and high
production regions, consistent with previous
works~\citep[e.g.][]{Jimenez2005, Kawahara2012}.
%

The mean velocity profile and root-mean-squared (rms) fluctuating
velocities for the regular channel are shown in
figures~\ref{fig:PD_stats_regular}(b--e). The results are compiled for
the statistical steady state after initial transients. These have also
been reported in the literature, with the worth noting difference that
here the streamwise fluctuating velocity is defined as $u' = u -
\langle u \rangle_x$, while in previous studies is common to choose
$u''=u - \langle u \rangle_{xzt}$
instead. Figures~\ref{fig:PD_stats_regular}(b--e) also contain the
one-point statistics for a non-minimal channel flow with $L_x^+ \times
L_z^+= 2312 \times 1156$ ($L_x \times L_z \approx 12.5h \times 6.3h$)
denoted by L180. The mean profile and cross-flow fluctuations in
larger unconstrained domains are essentially captured in the minimal
box, while $u'$ is underpredicted. The missing $u'$ is due to
larger-scale motions that do not participate in the buffer layer
self-sustaining cycle \citep{Jimenez1991, Flores2010}. A large amount
of $u'$ is recovered when minimal channel domain is enlarged in the
streamwise direction and Appendix \ref{sec:appendix_2Lx} shows that
our conclusions still hold when the minimal channel streamwise length
is doubled.

\section{Linear theories of self-sustaining wall turbulence: \emph{a priori} non-causal analysis}
\label{sec:theories}

The expected scenario of the full self-sustaining cycle in wall
turbulence is the linear amplification of $\boldsymbol{u}'$ induced by
the operator $\mathcal{L}(U)$ followed by nonlinear saturation of
$\boldsymbol{u}'$, scattering and generation of new disturbances by
$\boldsymbol{N}$. We focus here on the linear component of
(\ref{eq:NS}),
\begin{gather} \label{eq:NS_linear}
\frac{\partial\bu'_{\mathrm{linear}}}{\partial t} =
\mathcal{L}(U)\bu'_{\mathrm{linear}}.
\end{gather}
The most general solution to (\ref{eq:NS_linear}) is given by the
Peano-Baker series (see \S~\ref{subsec:theories_parametric}), which
accounts simultaneously for exponential growth, non-modal transient
growth, and non-modal transient growth assisted by parametric
instability. However, we dissect (\ref{eq:NS_linear}) and revisit
separately the different linear mechanisms that can transfer energy
from the base flow to the fluctuating velocities.  The plausibility of
each mechanism in $\mathcal{L}(U)$ as a contender to transfer energy
from $\bU$ to $\bu'$ is investigated in a non-intrusive manner by
interrogating the data from R180. This constitutes a non-causal
analysis, as we are neglecting the nonlinear terms in
(\ref{eq:NS_linear}), whereas the actual system (\ref{eq:NS}) is
non-linear.  This is not a minor point as the non-linear term $\bN$
can immediately counteract the linear growth by
$\mathcal{L}(U)\bu'$. Thus, this section only provides an assessment
on the plausibility of different linear growths.  The actual relevance
of the linear mechanisms is assessed in the cause-and-effect analysis
in \S \ref{sec:constrain}.

\subsection{Energy transfer via exponential instability}
\label{subsec:theories_modal}

The first mechanism considered is modal instability of the
instantaneous base flow.  At any given time, the exponential
instabilities are obtained by eigen-decomposition of the matrix
representation of the linear operator $\mathcal{L}(U)$
in~\eqref{eq:NS},
\begin{equation} \label{eq:eigen_LU}
\mathcal{L}(U) = \mathcal{Q} \Lambda \mathcal{Q}^{-1},
\end{equation}
where $\mathcal{Q}$ consists of the eigenvectors organised in columns,
$\mathcal{Q}^{-1}$ is the inverse of $\mathcal{Q}$, and $\Lambda$ is
the diagonal matrix of associated eigenvalues, $\lambda_j +
i\omega_j$, with $\lambda_j, \omega_j \in \mathbb{R}$.  Equation
(\ref{eq:NS_linear}) admits solutions of the form
$\bu'_{\mathrm{linear}} \sim \boldsymbol{c} \exp{[(\lambda_j +
    i\omega_j)t]}$, with $\boldsymbol{c}$ a constant. Hence, we say
that the base flow is unstable if any of the growth rates $\lambda_j$
is positive.  More details on the stability analysis are provided in
Appendix~\ref{sec:appendix_details} along with the validation of our
implementation in Appendix~\ref{sec:appendix_validation}.
Figure~\ref{fig:modal_example} shows a representative example of the
streamwise velocity for an unstable eigenmode. The predominant
eigenmode has typically a sinuous structure of positive and negative
patches of velocity flanking the velocity streak side by side, which
may lead to its subsequent meandering and eventual
breakdown. Varicose-type modes are also observed but they are less
frequent.
%
\begin{figure}
  \begin{center}
    \includegraphics[width=0.95\textwidth]{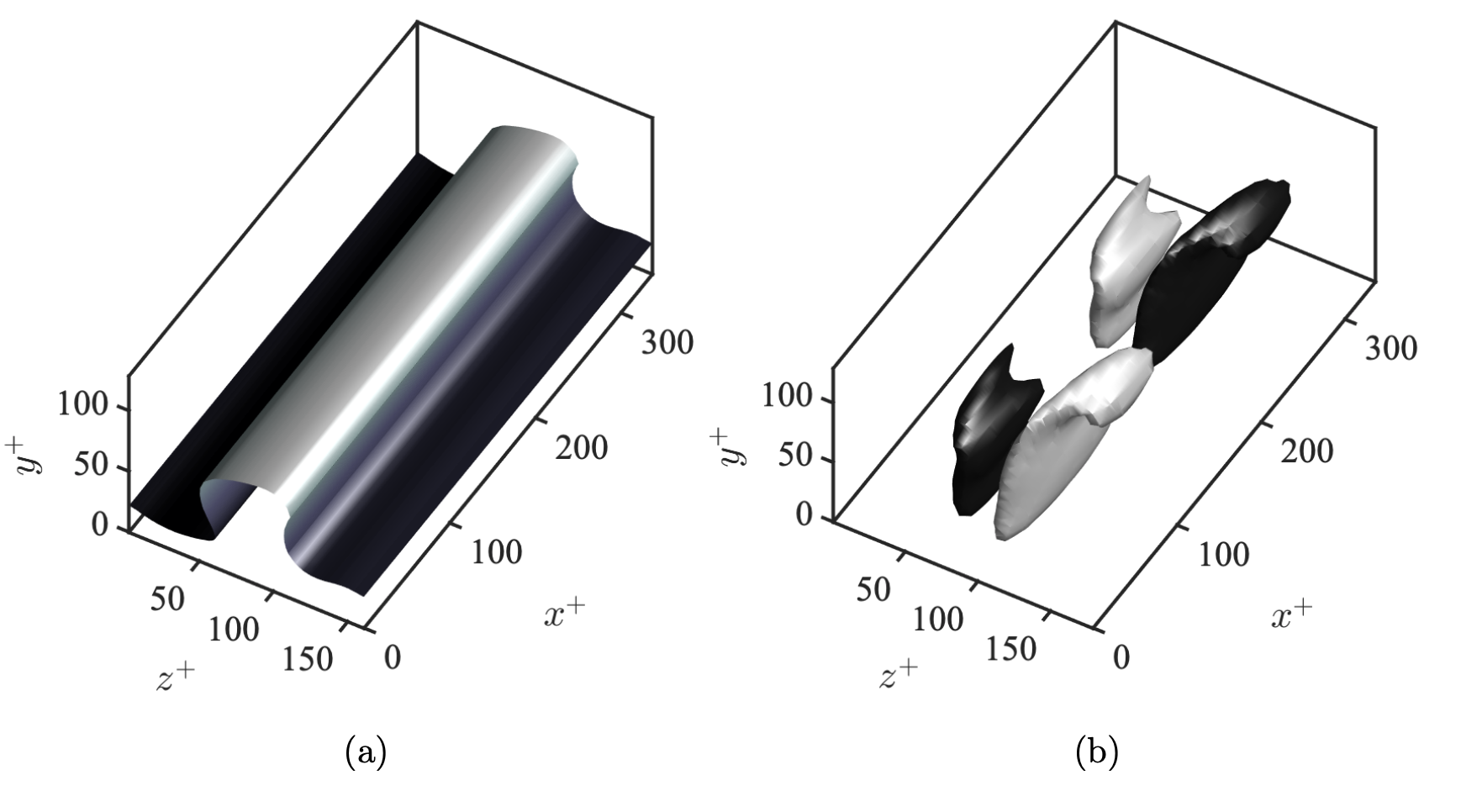} 
 \end{center}
\caption{Representative exponential instability of the streak. (a)
  Instantaneous isosurface of the base flow $U$. The value of the
  isosurface is $0.6$ of the maximum and the colour represents the
  distance to the wall. (b) Isosurface of the instantaneous streamwise
  velocity for the eigenmode associated with the most unstable
  eigenvalue $\lambda_\mathrm{max} h/u_\tau \approx 3$ at $t=5.1
  h/u_\tau$. The flow structure of the eigenmode is consistent with a
  sinuous instability. The values of the isosurface are $-0.5$ (dark)
  and $0.5$ (light) of the maximum streamwise
  velocity. \label{fig:modal_example} }
\end{figure}

\begin{figure}
  \begin{center}
   \subfloat[]{\includegraphics[width=0.32\textwidth]{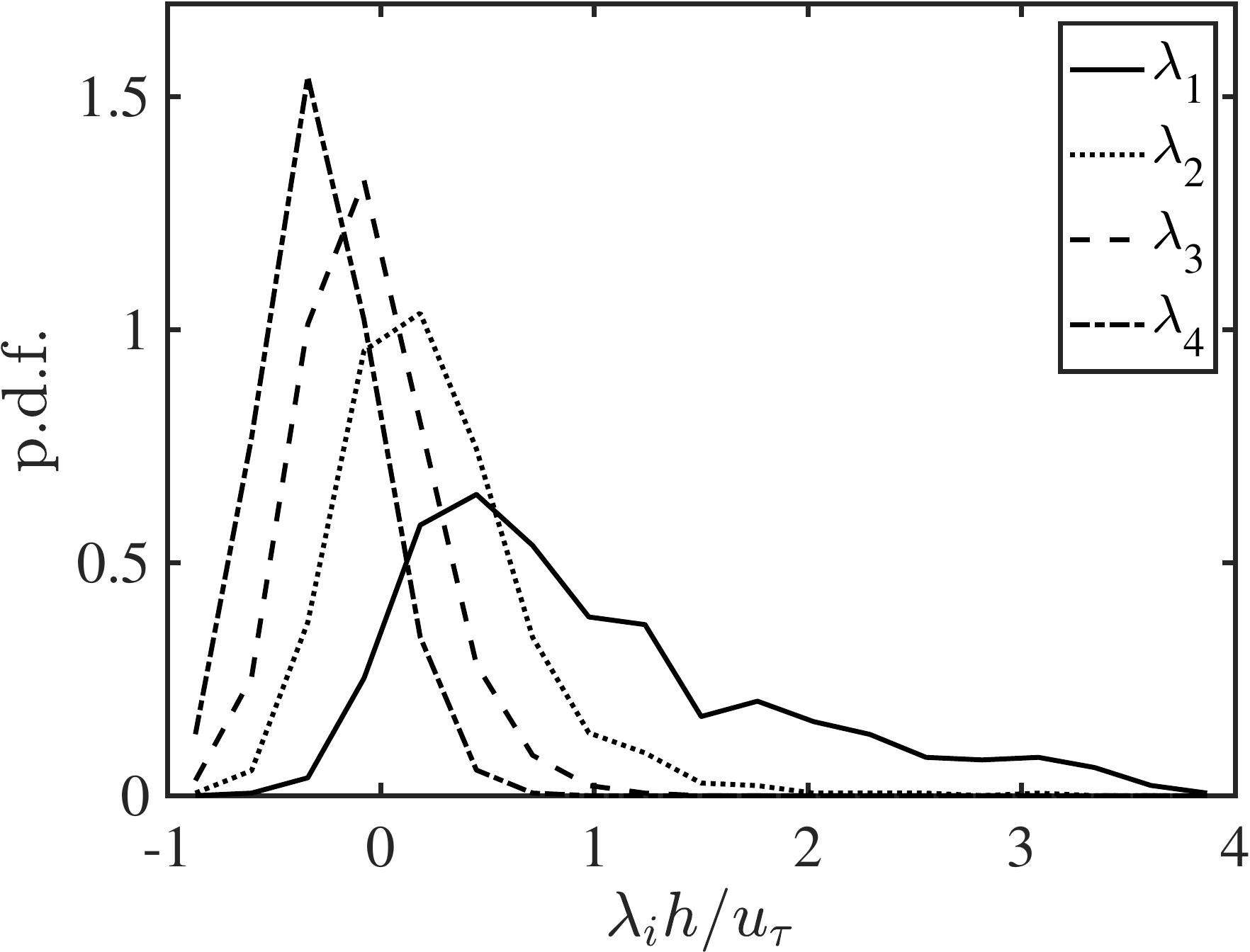}} 
   \hspace{0.07cm}
   \subfloat[]{\includegraphics[width=0.32\textwidth]{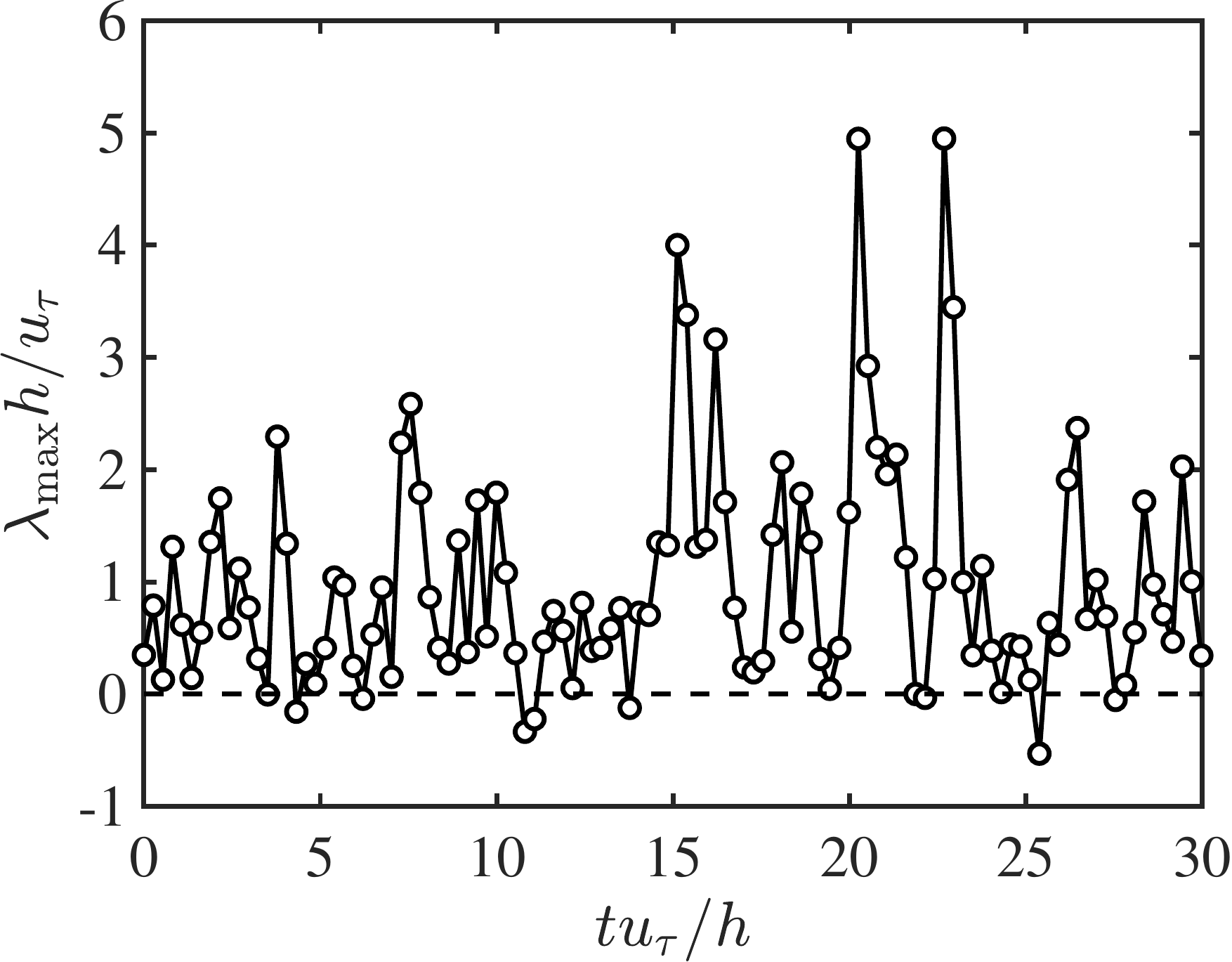}} 
   \hspace{0.07cm}
   \subfloat[]{\includegraphics[width=0.32\textwidth]{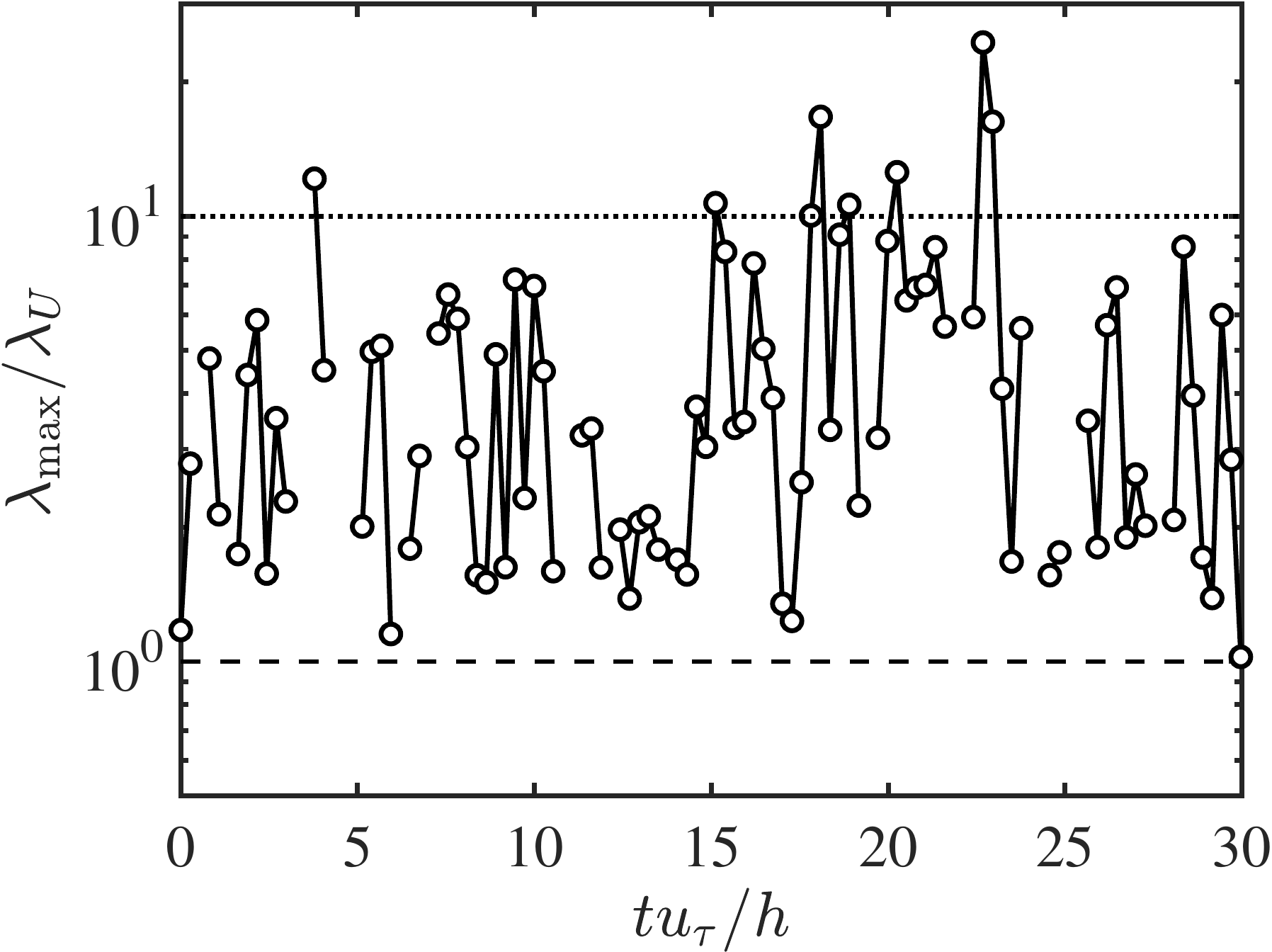}} 
 \end{center}
  \caption{ (a) Probability density functions of the growth rate of
    the four least stable eigenvalues of $\mathcal{L}(U)$,
    $\lambda_1>\lambda_2>\lambda_3>\lambda_4$. (b) The time-series of
    the most unstable eigenvalue~$\lambda_{\rm max}=\lambda_1$ of
    $\mathcal{L}(U)$. (c) The time-series of the ratio of
    $\lambda_{\rm max}/ \lambda_U$, where $\lambda_U$ is the growth
    rate of the base flow given by~(\ref{eq:lambdaU}). The horizontal
    dashed and dotted lines are $\lambda_{\rm max}/ \lambda_U=1$ and
    $\lambda_{\rm max}/ \lambda_U=10$, respectively.  Results for
    regular channel R180. \label{fig:P_eig_regular}}
\end{figure}

Figure~\ref{fig:P_eig_regular}(a) shows the probability density
functions of the growth rate of the four least stable eigenvalues of
$\mathcal{L}(U)$.  On average, the operator $\mathcal{L}(U)$ contains
2 to 3 unstable eigenmodes at any given instant. Denoting the Fourier
streamwise wavenumber as $k_x$, the most unstable eigenmode usually
corresponds to $k_x=2\pi/L_x$, although occasionally modes with
$k_x=2\pi/(2L_x)$ become the most unstable. The sensitivity of our
results to $L_x$ is further discussed in Appendix
\ref{sec:appendix_2Lx}. The history of the maximum growth rate
supported by $\mathcal{L}(U)$, denoted by $\lambda_1 =
\lambda_{\mathrm{max}}$, is shown in
figure~\ref{fig:P_eig_regular}(b). The flow is exponentially unstable
($\lambda_{\mathrm{max}}>0$) more than 90\% of the time.  The average
e-folding time for an exponentially unstable perturbation is roughly
$h/u_\tau$, which is comparable to the bursting duration $T_b$.
%

The ansatz underlying modal instability is that the spatial structure
of the base flow remains constant in time. Therefore, we
expect the above linear instability to manifest in the flow only when
$\lambda_{\mathrm{max}}$ is much larger than time rate of change of
the base flow~$U$, defined as
\begin{equation}
  \label{eq:lambdaU}
  \lambda_U \defn \left\langle
  \frac{1}{2} \frac{|\mathrm{d} E_U  /\mathrm{d}t|}{E_U} \right\rangle_{yz},
\end{equation}
where $E_U \defn \tfrac{1}{2} U^2$ is the energy of the base flow.
The ratio $\lambda_{\mathrm{max}}/\lambda_U$ for
$\lambda_{\mathrm{max}}>0$, shown in
figure~\ref{fig:P_eig_regular}(c), is about 5 on average and
occasionally achieves values above 20, i.e., the time-changes of $U$
might be 5 to 20 times slower than e-folding time of the most unstable
eigenmode. The growth of the modal instabilities is not overwhelmingly
faster than the changes on the base flow. However, considering that
the exponential growth of disturbances is supported for a
non-negligible fraction of the flow history (roughly 90\% of the time
as shown before), modal instability of $\mathcal{L}(U)$ still stands
as a potential mechanism sustaining the fluctuations. Note that the
argument above does not imply that exponential instabilities are
necessarily relevant for the flow when
$\lambda_{\mathrm{max}}/\lambda_U$ is large, but only that they could
manifest based on their characteristic time-scales. In fact, we show
in \S \ref{subsec:nonmodal_Upred} that exponential instabilities are
not a requisite to sustain turbulence.

\subsection{Energy transfer via transient growth}
\label{subsec:theories_nonmodal}

The second linear mechanism considered is the non-modal transient
growth of the fluctuations.  The linear dynamics
of~\eqref{eq:NS_linear} can be formally written as:
\begin{equation}\label{eq:linearprop}
\bu'_{\mathrm{linear}}(t+T) = \Phi_{t \rightarrow t+T}\, \bu'_{\mathrm{linear}}(t).
\end{equation}
The propagator $\Phi_{t \rightarrow t+T}$ maps the fluctuating flow
from time $t$ to time $t+T$ and represents the cumulative effect of
the linear operator $\mathcal{L}(U)$ during the period from $t$ to
$t+T$. If the base flow remains constant for $t\in[t_0, t_0+T]$, then
the propagator simplifies to
\begin{equation}\label{eq:P}
\Phi_{t_0 \rightarrow t_0+T} = \exp \left( \mathcal{L}_0\,T \right),
\end{equation}
where $\mathcal{L}_0$ denotes $\mathcal{L}(U(y, z, t_0))$.

Equation (\ref{eq:linearprop}) accounts for both the modal and
non-modal growth of $\bu'$ for $t\in[t_0, t_0+T]$. The exponential
growth of the fluctuating velocities was already quantified in
\S~\ref{subsec:theories_modal}. Here we are concerned with the
transient growth of $\bu'$ supported by $\L_0$. To that end, we
exclude from the analysis any growth of fluctuations due to the modal
instabilities of $\L_0$. This is achieved by the modified operator
$\tilde{\mathcal{L}}_0$,
\begin{equation}\label{eq:Atilde}
\tilde{\mathcal{L}}_0 \defn \mathcal{Q} \tilde{\Lambda} \mathcal{Q}^{-1},
\end{equation}
where $\tilde{\Lambda}$ is the stabilised counterpart of $\Lambda$
in~\eqref{eq:eigen_LU} obtained by setting the real part ($\lambda_j$)
of all unstable eigenvalues of $\Lambda$ equal to $-\lambda_j$, while
their phase speed and eigenmode structure are left unchanged. We
assessed that the transient growth properties of
$\tilde{\mathcal{L}}_0$ are mostly insensitive to the amount of
stabilisation introduced in $\Lambda$ when $\lambda_j>0$ are replaced
by $-a\lambda_j$ with $a\in[1/10,10]$. The potential effectiveness of
transient growth due to a base flow $U(y,z,t_0)$ is then characterised
by the energy gain $G$ over some time-period $T$, defined as
\beq\label{eq:G_def}
  G(t_0, T, \bu'_{0}) \defn \frac{\big \langle \bu'_T \cdot\bu_T' \big \rangle_{xyz}}{\big \langle \bu'_0\cdot\bu_0' \big \rangle_{xyz}},
  \eeq
where $\bu'_{T} \defn \bu'_{\mathrm{linear}}(x,y,z,t_0+T)$,
$\bu'_0 \defn \bu'_{\mathrm{linear}}(x,y,z,t_0)$ and $T$ is the
time-horizon for which the gain $G$ is computed.

The energy, being a bilinear form, can be expressed as an inner
product, e.g.,
\beq\label{eq:innerproduct}
  (\bu'\,,\,\bu') \defn \big\langle \bu'\cdot\bu' \big\rangle_{xyz}.
\eeq
Using the definition~\eqref{eq:innerproduct} and the form of the
propagator~\eqref{eq:P} for the frozen linear operator $\tilde \L_0$,
the energy gain is rewritten as:
\begin{align}\label{eq:G_0}
  G(t_0, T, \bu'_0)  & = 
  \frac{\big ( \bu'_T \, , \, \bu'_T \big )}{\big ( \bu'_0 \, , \, \bu'_0 \big )}  = \frac{\big ( e^{\tilde \L_0 T} \bu'_0 \, , \, e^{\tilde \L_0 T}\bu'_0 \big )}{\big ( \bu'_0 \, , \, \bu'_0 \big )} 
  = \frac{\big ( \bu'_0 \, , \, e^{\tilde \L_0^\dagger T}e^{\tilde \L_0 T}\bu'_0 \big )}{\big ( \bu'_0 \, , \, \bu'_0 \big )} .
\end{align}
In the last equality, dagger $\dagger$ denotes the adjoint operator.
Note that, for $T \rightarrow \infty$, the energy gain (\ref{eq:G_0})
tends to $0$, since the operator $\tilde{\mathcal{L}}_0$ is
exponentially stable.  The maximum gain over all initial conditions
$\bu'_0$, denoted by $G_\mathrm{max}(t_0,T) =
\mathrm{sup}_{\bu'_0}(G)$, is given by the square of the largest
singular value of the stabilised linear propagator $\tilde \Phi_0$
\citep{Butler1993, Farrell1996},
\begin{align}
\tilde{\Phi}_{t_0 \rightarrow t_0+T} & = \exp(\tilde{\mathcal{L}}_0 T) \\ \label{eq:tildePhi0}
& = \mathcal{M} \Sigma \mathcal{N}^\dagger,
\end{align}
where $\Sigma$ is a diagonal matrix, whose positive entries $\sigma_j$
are the singular values of $\exp(\tilde{\mathcal{L}}_0 T)$ and the
columns of $\mathcal{M}$ and of $\mathcal{N}$ are the output modes (or
left-singular vectors) and input modes (or right-singular vectors) of
$\exp(\tilde{\mathcal{L}}_0 T)$, respectively.

\begin{figure}
  \begin{center}
  \hspace{0.1cm}
  \subfloat[]{\includegraphics[width=0.475\textwidth]{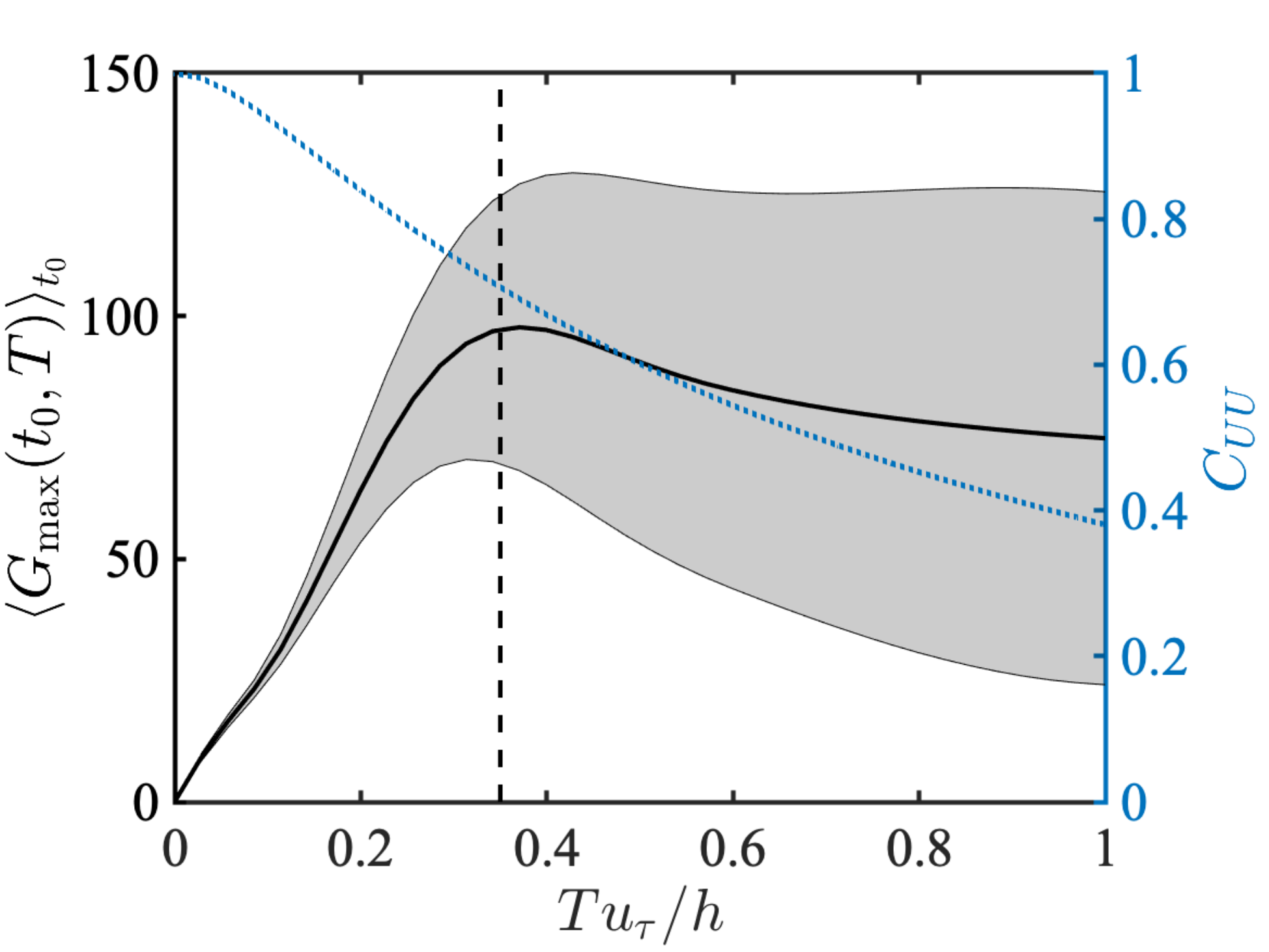}}   
  \hspace{0.1cm}
  \subfloat[]{\includegraphics[width=0.435\textwidth]{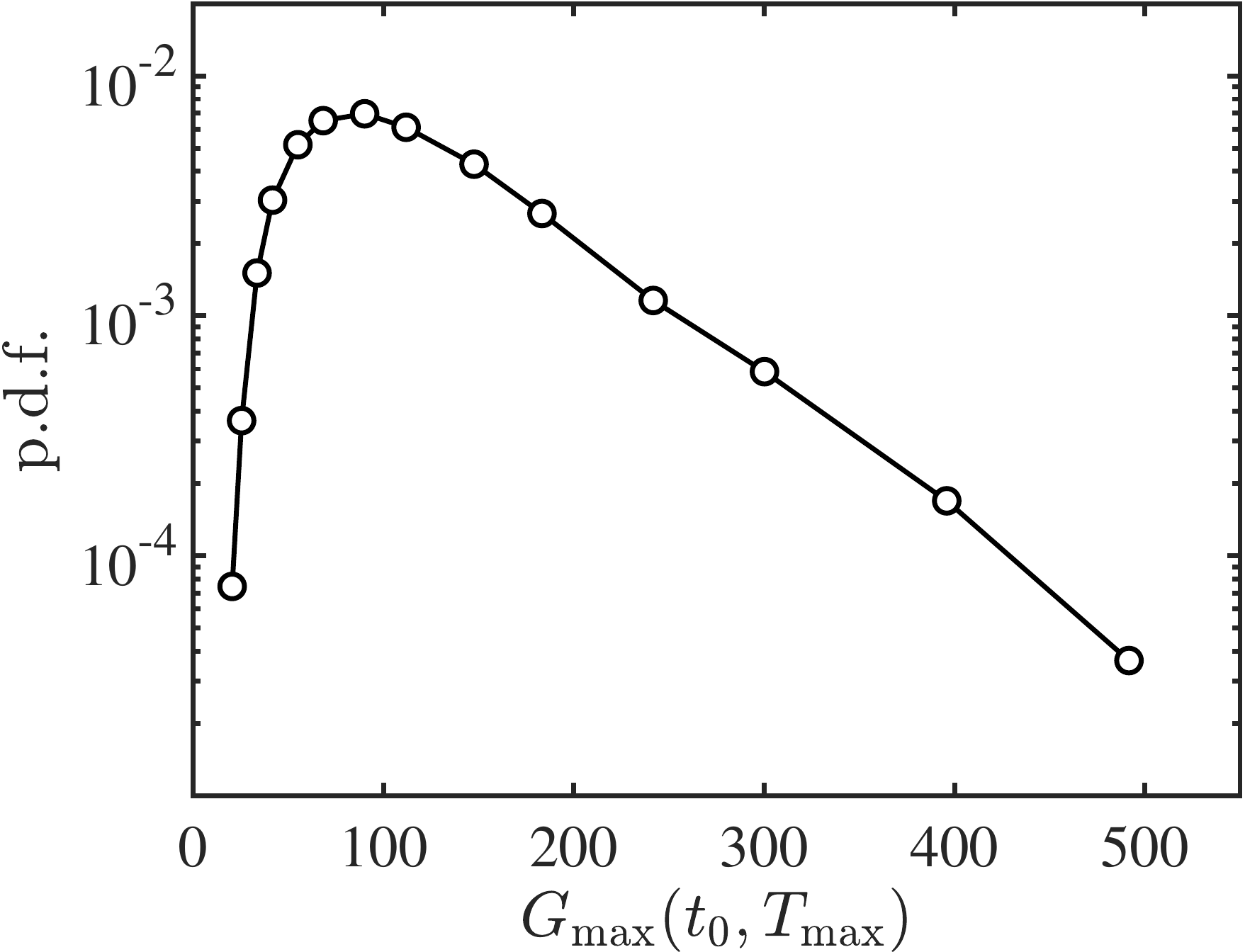}} 
  \end{center}
  \caption{ Energy transfer via transient growth with frozen-in-time
    base flow. (a) The ensemble average of the maximum energy gain
    $G_\mathrm{max}(t_0,T)$ (\solid, see Eq.~\eqref{eq:P_time}) over
    different initial instances $t_0$, as a function of the time
    horizon $T$. Shaded regions denote $\pm$ half standard deviation
    of $G_\mathrm{max}(t_0, T)$ for a given $T$. The vertical dashed
    line denotes $T_{\mathrm{max}} = 0.35 h/ u_\tau$.  The blue dotted
    line is the auto-correlation of $U$, $C_{UU}$ and its values
    appear on the right vertical axis. (b) Probability density
    function of gains $G_\mathrm{max}(t_0,T_{\mathrm{max}})$. Results
    for regular channel R180. \label{fig:gains_regular}}
\end{figure}

The maximum gain $G_\mathrm{max}$ for R180 as a function of the
optimisation time $T$ is shown in
figure~\ref{fig:gains_regular}(a). The values of $G_\mathrm{max}$ also
depend on $t_0$; figure~\ref{fig:gains_regular}(a) features the mean
and the standard deviation of $G_\mathrm{max}$ for more than 1000
uncorrelated instants $t_0$. Figure~\ref{fig:gains_regular}(a) reveals
that non-normality alone is potentially able to produce fluctuation
energy growth of the order of $G_\mathrm{max} \approx 100$. On
average, the time-horizon for maximum gain is attained at
$T_\mathrm{max} \approx 0.35 h/u_\tau$.  Thus, the maximum non-normal
energy gain is obtained at a similar time-scale as the bursting time,
$T_p$ (see \S~\ref{sec:regular}).  For an elapsed time of
$T_\mathrm{max}$, the auto-correlation of the base flow,
\beq\label{eq:corr}
C_{UU} \defn \Big \langle
\left[U(y,z,t)-\langle U \rangle_{t}\right]\left[U(y,z,t+T)-\langle U \rangle_{t}\right] \Big \rangle_{yzt},
\eeq
has a value of 0.7, as shown in figure~\ref{fig:gains_regular}(a),
which is reasonably high to justify the `frozen-base-flow' assumption
underlying the calculation of $G$. The p.d.f.~of $G_\mathrm{max}$ at
$T_\mathrm{max}$ (figure~\ref{fig:gains_regular}b) shows that
$U(y,z,t_0)$ at certain times can support gains as high as 300.

The results here support the hypothesis of transient growth of the
``frozen'' mean streamwise flow $U(y,z,t_0)$ as a tenable candidate to
sustain wall turbulence. It is worth noting that the maximum gain
$G_\mathrm{max}$ obtained with a streaky base flow $U(y,z,t_0)$ is
considerably larger than the limited gains of around 10 reported in
previous studies focused in the buffer layer \citep{Delalamo2006a,
  Pujals2009, Cossu2009}. In these works, the base flow selected was
$\langle u \rangle_{xzt}$, which lacks any spanwise $z$-structure and,
hence supports much lower gains compared to $U(y,z,t_0)$.
%

\begin{figure}
  \begin{center}
  \vspace{0.1cm}
  \includegraphics[width=\textwidth]{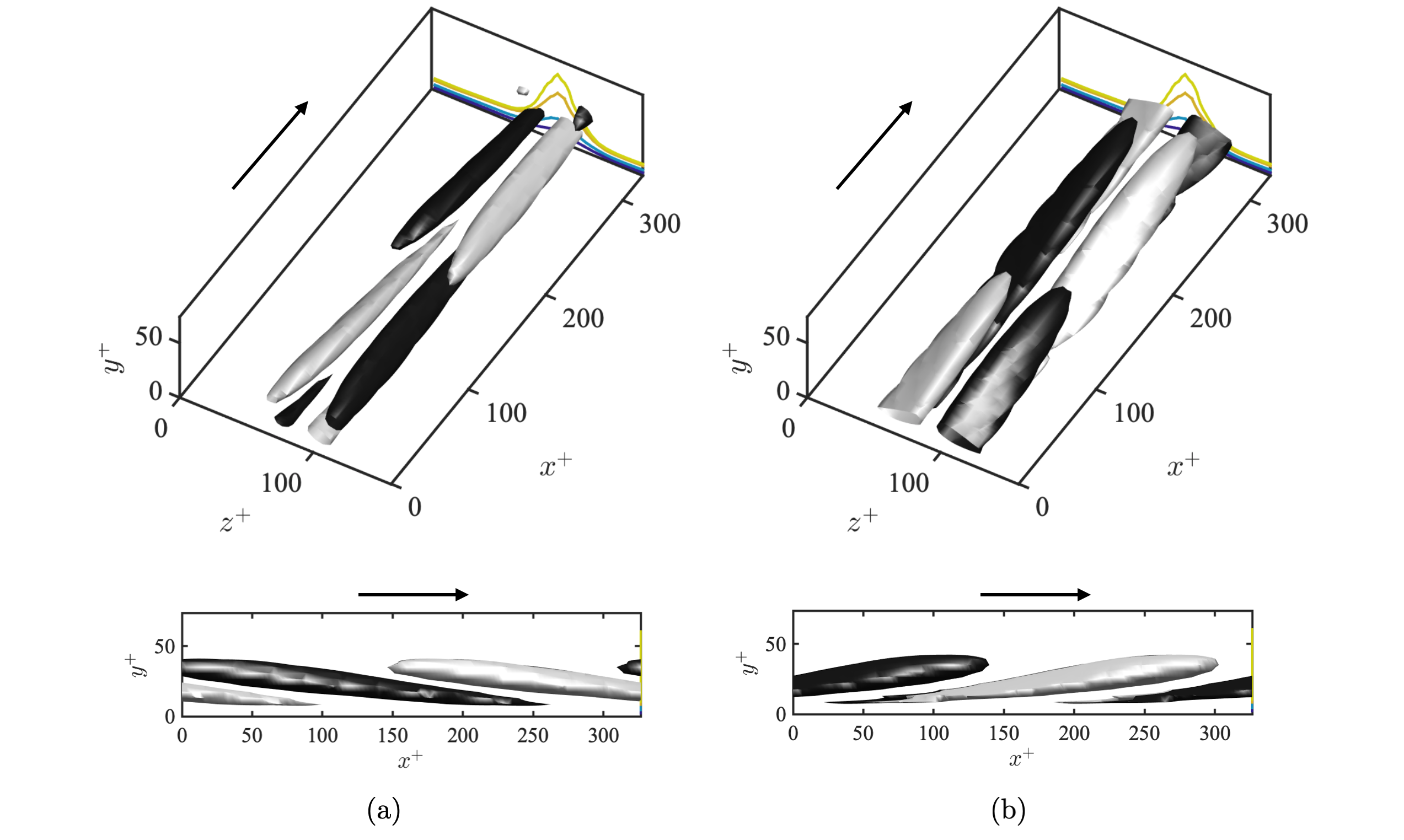} 
  \end{center}
 \caption{Representative sinuous input and output modes associated
   with transient growth of the streak. Isosurfaces of (a) the input
   and (b) the output wall-normal velocity mode associated with the
   largest singular value of $\tilde{\Phi}_{t_0 \rightarrow t_0+T}$
   from (\ref{eq:G_0}) at $T= 0.35h/u_\tau$. The isosurface are $-0.5$
   (dark) and $0.5$ (light) of the maximum wall-normal velocity. The
   gain is $G_{\mathrm{max}}=136$. The coloured lines at $x=L_x$ are
   0.2, 0.4, 0.6, and 0.7 of the maximum velocity of the base
   flow. The result is for the regular channel
   R180. \label{fig:TG_example} }
\end{figure}

Figure~\ref{fig:TG_example} provides an example of the input and
output modes associated with the maximum optimal gain for one selected
instant $t_0$. The flow displays a sinuous backwards-leaning
perturbation (input mode) that is being tilted forward by the mean
shear over the time $T$ (output mode).  The process is reminiscent of
the linear Orr/lift-up mechanism driven by continuity and wall-normal
transport of momentum characteristic of the bursting process and
streak formation \citep{Orr1907, Ellingsen1975, Kim2000, Jimenez2013,
  Encinar2020}. Unlike the studies that used the base flow $\langle
u\rangle_{xzt}$, our choice of a spanwise-varying base flow
$U(y,z,t)=\langle u \rangle_{x}$ limits both the spanwise extent and
location of the input and output modes, which are controlled by the
spanwise location of the streak.
%

\subsection{Energy transfer via transient growth with time-varying base flow} 
\label{subsec:theories_parametric}

In the previous section we have considered frozen-in-time base
flows. We now relax this assumption such that the (stabilised) linear
operator is time-dependent $\tilde\L(U(y,z,t))$. The propagator
$\tilde{\Phi}^t_{t_0 \rightarrow t_0+T}$ (now with superscript $t$),
is given by the Peano-Baker series \citep{Rugh1996},
\begin{equation}\label{eq:P_time}
  \tilde{\Phi}_{t_0 \rightarrow t_0+T}^t 
  =\mathcal{I}+\int_{t_0}^{t_0+T}{\tilde{\mathcal{L}}}(t_{1})\,\mathrm{d}t_{1}
  +\int_{t_0}^{t_0+T}{\tilde{\mathcal{L}}}(t_{1})\int_{t_0}^{{t_{1}}}{\tilde{\mathcal{L}}}(t_2)\,
  \mathrm{d} t_{2}\,\mathrm{d} t_{1} + ...,
\end{equation}
where $\mathcal{I}$ is the identity matrix and we have simplified the
notation to $\tilde{\mathcal{L}}(t) =
\tilde{\mathcal{L}}(U(y,z,t))$. The propagator $\tilde{\Phi}^t_{t_0
  \rightarrow t_0+T}$ represents the cumulative effect of $U(y,z,t)$
from $t_0$ to $t_0+T$ accounting for time-variations in the base
flow. The energy gain of (\ref{eq:P_time}) is
\begin{equation}\label{eq:G_1}
  G^t(t_0,T, \bu'_0) 
  = \frac{\Big ( \bu'_0 \, , \,
    (\tilde{\Phi}_{t_0 \rightarrow t_0+T}^t)^\dagger\, (\tilde{\Phi}_{t_0 \rightarrow t_0+T}^t)\,  \bu'_0 \Big )}
  {\big ( \bu'_0 \, , \, \bu'_0 \big )}.
\end{equation}
In contrast with the frozen-base-flow propagator $\tilde{\Phi}_{t_0
  \rightarrow t_0+T}$ in~\eqref{eq:G_0}, the time variations of the
operator $\tilde{\mathcal{L}}(U)$ can either weaken or enhance the
energy transfer from $\bU$ to $\bu'$. Another difference is that the
$G^t$ now admits a finite value at $T\rightarrow\infty$, despite that
$\tilde{\mathcal{L}}(U)$ is modally stable at all instances.  One
potential route to enhance the gain for short times and/or achieve
finite gains for long times is the parametric instability of the
streak discussed in the introduction \citep{Farrell2012}. However, it
is shown below that none of these effects seem to be at play.
%
\begin{figure}
  \begin{center}
  \hspace{0.1cm}
  \subfloat[]{\includegraphics[width=0.475\textwidth]{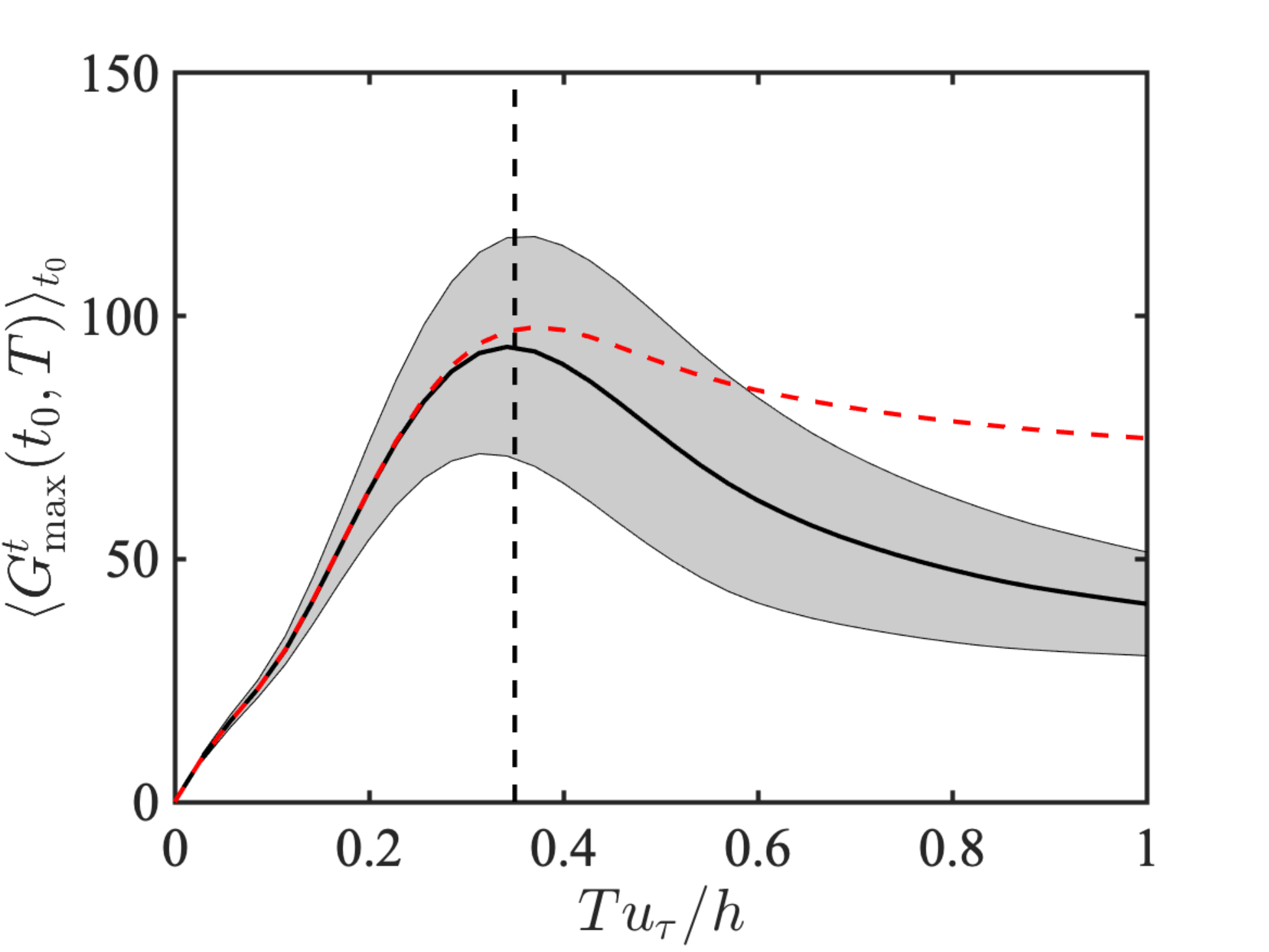}} 
  \hspace{0.1cm}
  \subfloat[]{\includegraphics[width=0.435\textwidth]{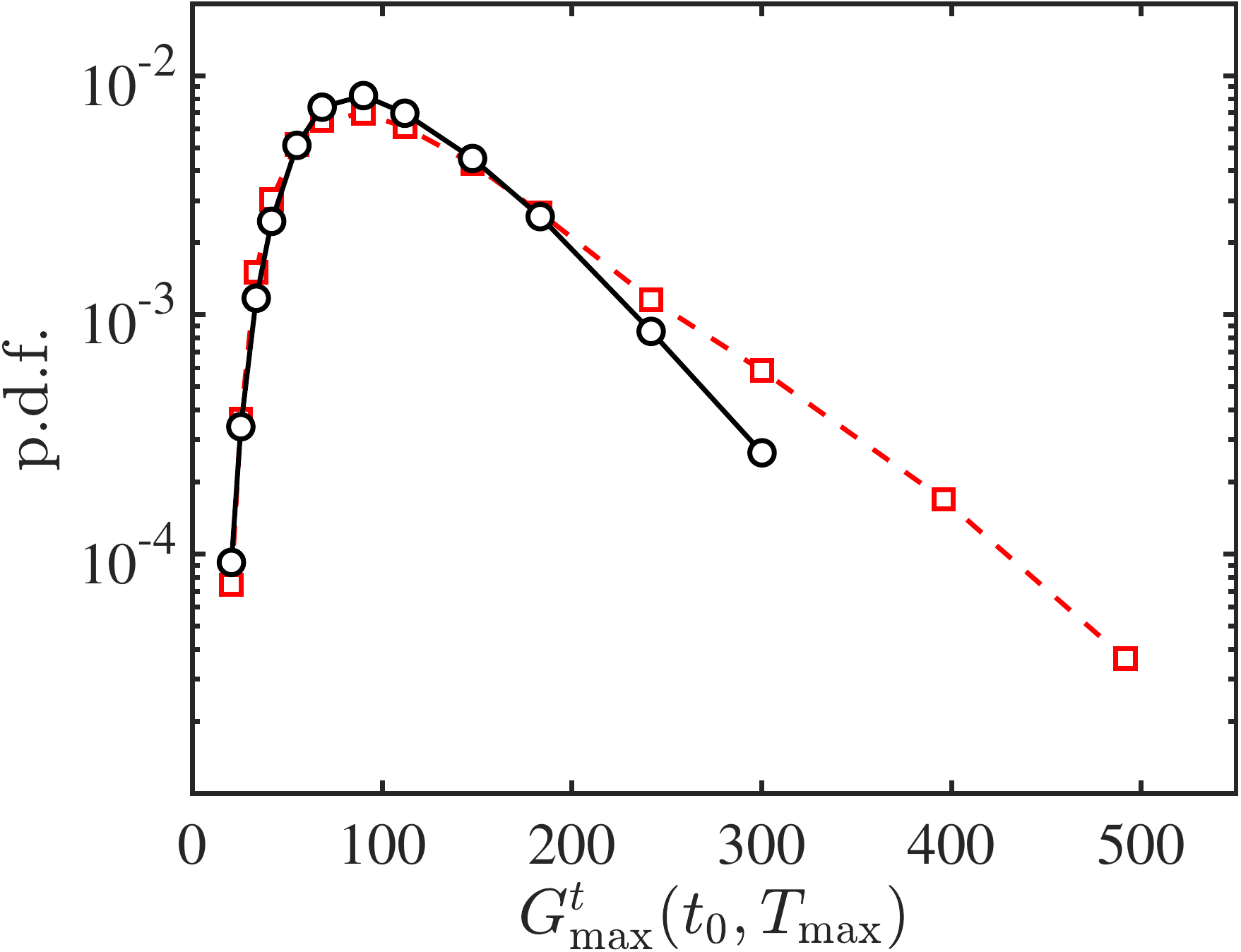}} 
  \end{center}
 \caption{ Energy transfer via transient growth with time-varying base
   flow. (a) The ensemble average of the maximum energy gain
   $G^t_\mathrm{max}(t_0,T)$ (\solid, see Eq.~\eqref{eq:P_time}) and
   $G_\mathrm{max}(t_0,T)$ (\textcolor{red}{\dashed}, see
   Eq.~\eqref{eq:tildePhi0}) over different initial instances $t_0$,
   as a function of the time horizon $T$. Shaded regions denote $\pm$
   half standard deviation of $G^t_\mathrm{max}(t_0, T)$ for a given
   $T$. The vertical dashed line denotes $T_{\mathrm{max}} = 0.35 h/
   u_\tau$.  (b) Probability density function of gains for
   $G^t_\mathrm{max}(t_0,T_{\mathrm{max}})$ (\solid) and
   $G_\mathrm{max}(t_0,T_{\mathrm{max}})$
   (\textcolor{red}{\dashed}). Results for regular channel
   R180. \label{fig:gains_regular_t}}
\end{figure}

To evaluate the transient growth with time-varying base flows, we
reconstruct the propagator without exponential instabilities
$\tilde{\Phi}^t_{t_0 \rightarrow t_0+T}$ for case R180. In virtue of
the property $\tilde{\Phi}_{t_0 \rightarrow t_0+T}^t =
\tilde{\Phi}_{t_0 \rightarrow t_1}^t \tilde{\Phi}_{t_1 \rightarrow
  t_0+T}^t$ for $t_0\le t_1 \le t_0+T$, the propagator is numerically
computed by the ordered product of exponentials under the assumption of
small $\Delta t$ as
\begin{equation}\label{eq:P_time_discrete}
\tilde{\Phi}_{t_0 \rightarrow t_0+T}^t \approx \exp \left[
  \tilde{\mathcal{L}}(t_0+(n-1)\Delta t)\Delta t \right]\dotsb  \exp\left[
  \tilde{\mathcal{L}}(t_0+\Delta t) \Delta t \right] \exp\left[
  \tilde{\mathcal{L}}(t_0)          \Delta t \right],
\end{equation}
where $T = n\Delta t$, with $n$ a positive integer.  We saved the
time-history of $U(y,z,t)$ from R180 at all time steps and used it to
compute $\tilde{\Phi}_{t_0 \rightarrow t_0+T}^t$
via~\eqref{eq:P_time_discrete}.  We take $\Delta t^+ \approx 0.05$,
which is the time step used to integrate the equations of motion. The
maximum gain supported by $\tilde{\Phi}_{t_0 \rightarrow t_0+T}^t$ is
compared with its frozen-base-flow counterpart $\tilde{\Phi}_{t_0
  \rightarrow t_0+T}$ in figure~\ref{fig:gains_regular_t}. The results
reveal that energy growth with time-varying base flows is almost
identical to the energy growth under the frozen-base-flow assumption
up to $T \approx T_{\mathrm{max}}=0.35 h/u_\tau$, which also
corresponds to the time for maximum gain for $\tilde{\Phi}^t_{t_0
  \rightarrow t_0+T}$. For longer times $T > T_{\mathrm{max}}$, the
gain with time-varying base flows is depleted with respect to that of
the frozen-base-flow, and tends to zero for $T \rightarrow \infty$
(not shown). The results show that accounting for time-variations of
the base flow has a negligible effect on energy transfers for short
times, but gains for frozen base-flows are over estimated for long
times otherwise.

The propagator $\tilde{\Phi}^t_{t_0 \rightarrow t_0+T}$ can also be
analysed in terms of input and output modes.  The input and output
modes for the time-varying base flow are again a backwards-leaning
perturbation (input mode) that is being tilted forward by the mean
shear (output mode), very similar to the example shown in
figure~\ref{fig:TG_example}, but not shown here for brevity.

\section{Cause-and-effect discovery with interventions}
\label{sec:interventions}

The analysis above was performed \emph{a priori} by interrogating
the data from R180 in a non-intrusive manner. This provides a valuable
insight about the energy injection into the fluctuations but hinders
our ability to faithfully assess cause-and-effect links between linear
mechanisms and their actual impact on the fully nonlinear
system. 

The most intuitive definition of causality relies on
\emph{interventions}: manipulation of the causing variable leads to
changes in the effect~\citep{Eichler1997, Pearl2009}.  More precisely,
to describe the causal effect that a process $A$ (e.g, exponential
growth of instability) exerts on another process $B$ (e.g., growth of
turbulent kinetic energy), we consider the intervention in the
governing equations of the system that sets $A$ to a modified value
$A_i$ and observe the post-intervention consequences.  How to identify
the intervention $A_i$ that best unveils the causality from $A$ to $B$
is not trivial and relies on our knowledge of the system and
shrewdness to modify it~\citep{Eberhardt2007, Hyttinen2013}. When we
do not have any prior knowledge of how $A$ might affect $B$, we need
to resort to randomised interventions for discovering causal
relationships~\citep{Fisher1936}.  In following sections, the reader
will notice that many of the conclusions drawn on ($A$ causes $B$) are
often framed as a result of a negation, which is justified by the
duality: ($A$ causes to $B$) $\equiv$ (no $B$ implies no $A$). Thus,
we can assess the causality from $A$ to $B$ using either of the two
hypothesis.

As turbulence is a high-dimensional chaotic system, we are concerned
with the statistical alterations in the system after the intervention
rather than changes in individual events. The probability distribution
of the process $B$ for the non-intervened system is measured by
$\mathbb{P}(B)$.  The causal effect of $A$ on $B$ can be quantified by
any functional of the post-intervention distribution $\mathbb{P}( B |
A = A_i)$, where $\mathbb{P}( \ \cdot \ | A = A_i)$ is the probability
of the intervened system. The most commonly used measure of the
statistical effect of $A$ on $B$ is the mean causal effect defined as
the average increase or decrease in value caused by the
intervention.

In the next section, we follow this approach to assess the relevance
of different linear mechanisms on the energy transfer from the base
flow $\bU$ to the fluctuations $\bu'$. The starting point is the R180
system~\eqref{eq:R180}, which is sensibly modified to suppress a
targeted linear mechanism.

\section{Linear theories of self-sustaining wall turbulence: cause-and-effect analysis}
\label{sec:constrain}

\subsection[Wall turbulence without explicit feedback
  from u' to U]{Wall turbulence without explicit feedback from $\bu'$
  to $\bU$}
\label{sec:nofeeback}

In previous sections, we have acted as if 
\begin{gather} \label{eq:NS_again}
\frac{\partial\bu'}{\partial t} =  
\mathcal{L}(U)\bu'+ \bN(\bu'),
\end{gather}
is linear in the term $\mathcal{L}(U)\bu'$.  This is obviously not
true because $U(y,z,t)$ depends on $\bu'$ via the nonlinear feedback
term $-\langle \bu' \cdot \bnabla \bu' \rangle_x$ (see the base-flow
evolution equation~\eqref{eq:R180_2}).

Prior to investigating the cause-and-effect links of linear mechanisms
in $\mathcal{L}(U)$, we derive a surrogate system in which the energy
injection is strictly linear by preventing the explicit feedback from
$\bu'$ to $\bU$. To achieve this, we proceed as follows:
\begin{enumerate}
  \item We perform a simulation of R180 for $600 h/u_\tau$ units of
    time (after transients) with a constant time step.
  \item We store the base flow at all time steps.  We denote the
    time-series of this base flow as $U_0 = U(y,z,t)$ from case R180.
  \item We time-integrate the system
   \begin{gather} \label{eq:NF180}
   \frac{\partial\bu'}{\partial t} = \mathcal{L}(U_0)\bu'+ \bN(\bu'), \\
   U_0 = U(y,z,t) \ \mathrm{from \ case \ R180}. 
   \end{gather}
\end{enumerate}
Equation~\eqref{eq:NF180} is initialised from a random, incompressible
velocity field and it is integrated for $600 h/u_\tau$ units of time
using the same time step as in R180. Equation~\eqref{eq:NF180} is akin
to the Navier--Stokes equations, in which the equation of motion of
$\bU$ is replaced by $\bU = (U_0, 0, 0)$. We refer to this case as
``channel flow with no-feedback'' or NF180 for short. Note that the
base flow $U_0$ has no \emph{explicit} feedback from $\bu'$
in~\eqref{eq:NF180}, although it has been \emph{implicitly} `shaped'
by the velocity fluctuations of R180 and, as such, it contains dynamic
information of actual wall turbulence. The key difference
in~\eqref{eq:NF180} is that the term $\mathcal{L}(U_0)\bu'$ is now
strictly linear while preserving the modal and non-modal properties of
$\mathcal{L}(U)$ in R180.

\begin{figure}
 \begin{center}
   \subfloat[]{\includegraphics[width=0.46\textwidth]{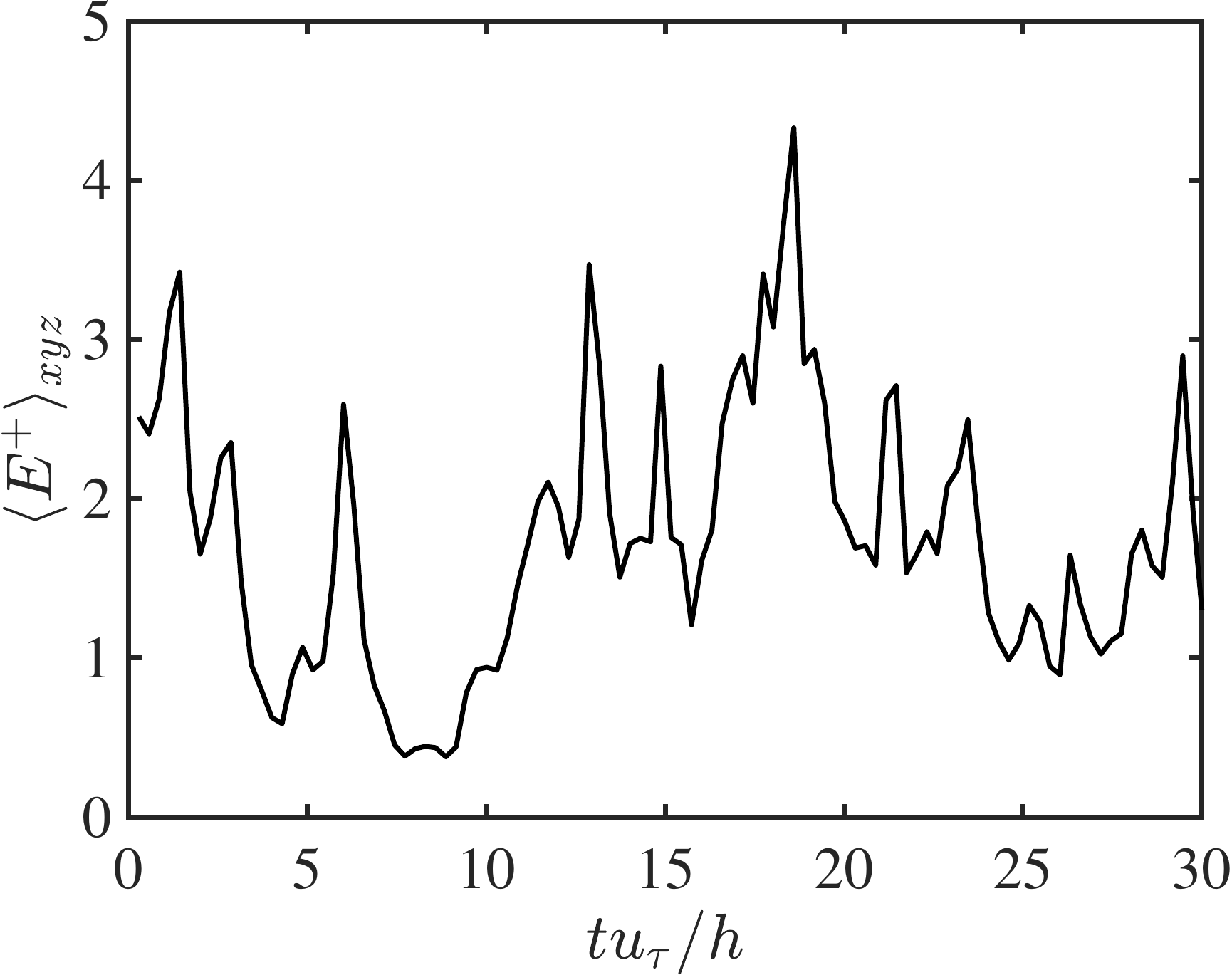}} 
   \hspace{0.1cm}
   \subfloat[]{\includegraphics[width=0.50\textwidth]{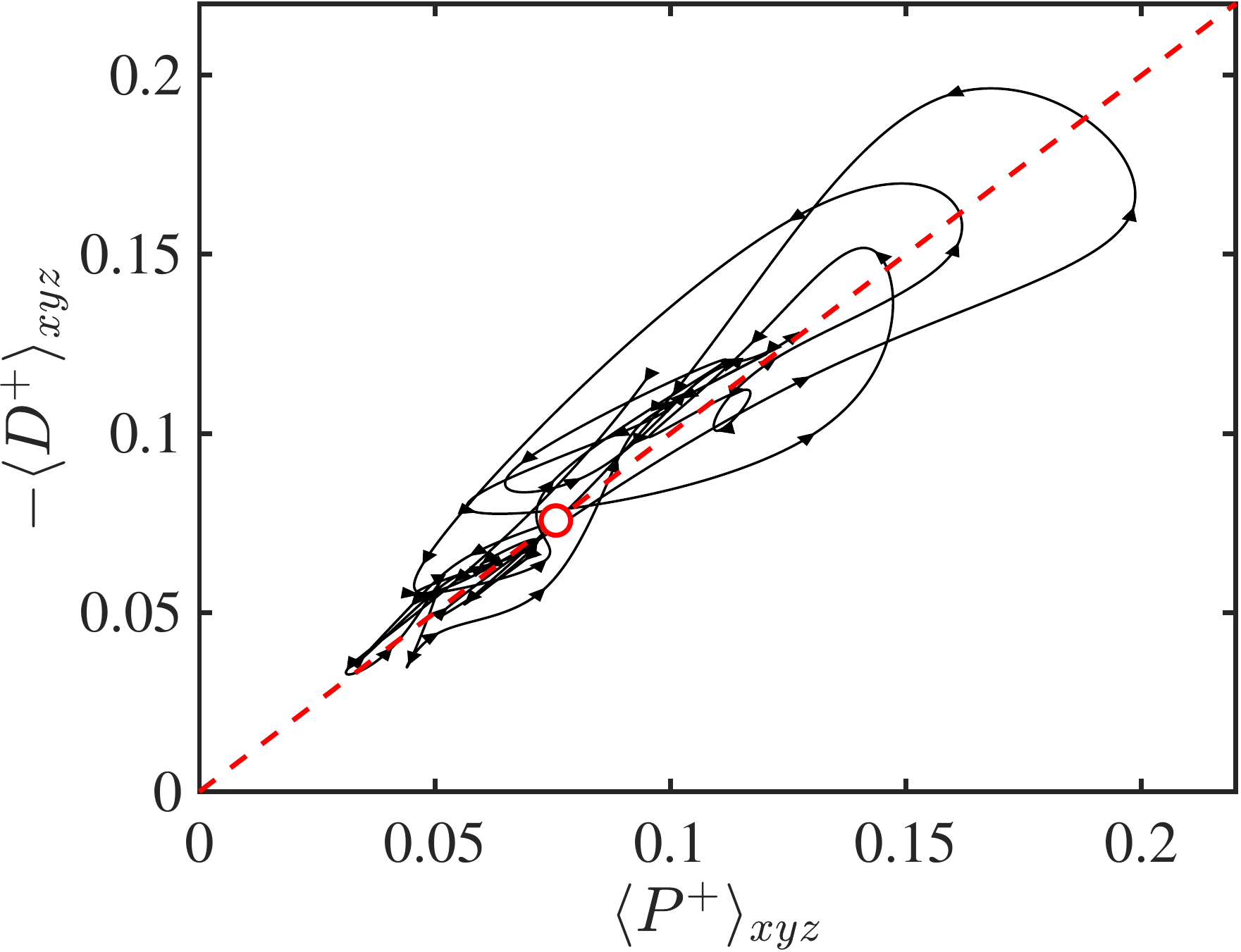}} 
 \end{center}
\caption{(a)~The history of the domain-averaged turbulent kinetic
  energy of the fluctuations $\langle E \rangle_{xyz}$. Note that only
  $30 h/u_\tau$ units of time are shown in the panel but the
  simulation was carried out for $600 h/u_\tau$. (b)~Projection of the
  flow trajectory onto the average production rate $\langle P
  \rangle_{xyz}$ and dissipation rate $\langle D \rangle_{xyz}$
  plane. The arrows indicate the time direction of the trajectory,
  which on average rotates counter-clockwise. The red dashed line is
  $\langle P \rangle_{xyz} = -\langle D \rangle_{xyz}$ and the red
  circle $\langle P \rangle_{xyzt} = -\langle D \rangle_{xyzt}$. The
  trajectory projected covers $15 h/u_\tau$ units of time. The results
  are for NF180.
 \label{fig:TKE_Upred}}
\end{figure}

The flow sustained in NF180 is turbulent as revealed by the history of
the turbulent kinetic energy in
figure~\ref{fig:TKE_Upred}(a). Moreover, the footprint of the flow
trajectory projected onto the $\langle P\rangle_{xyz}$--$\langle
D\rangle_{xyz}$ plane in figure~\ref{fig:TKE_Upred}(b) also exhibits a
similar behaviour to R180: the flow is organised around $\langle P
\rangle_{xyz} = -\langle D \rangle_{xyz}$ with excursions into the
high/low dissipation and production regions with predominantly
counter-clockwise motions. This assessment is merely qualitative and
some differences are expected between the $\langle
P\rangle_{xyz}$--$\langle D\rangle_{xyz}$ trajectories in R180 and
NF180.
%

\begin{figure}
  \begin{center}
  \hspace{0.05cm}
  \subfloat[]{\includegraphics[width=0.32\textwidth]{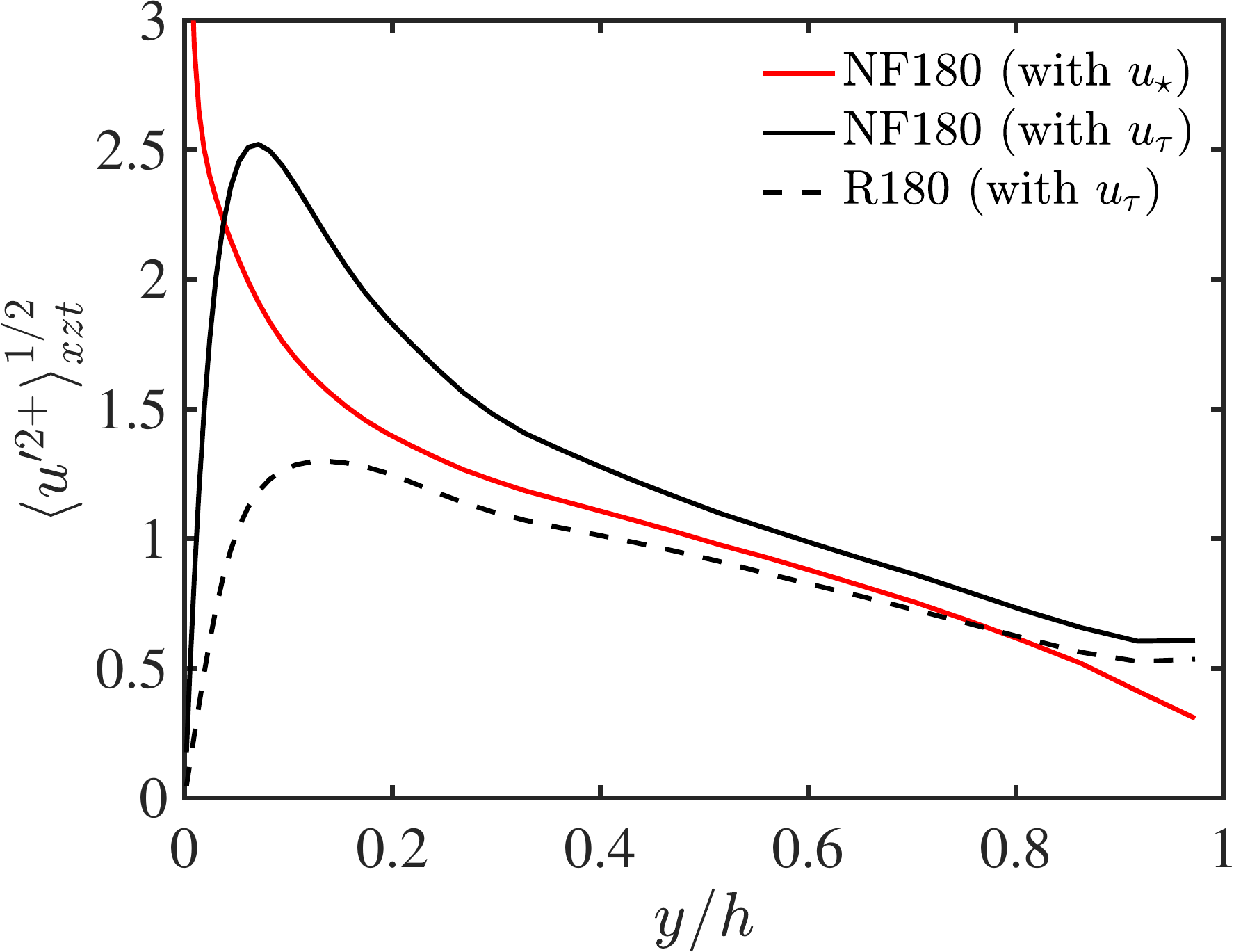}}  
  \hspace{0.05cm}
  \subfloat[]{\includegraphics[width=0.32\textwidth]{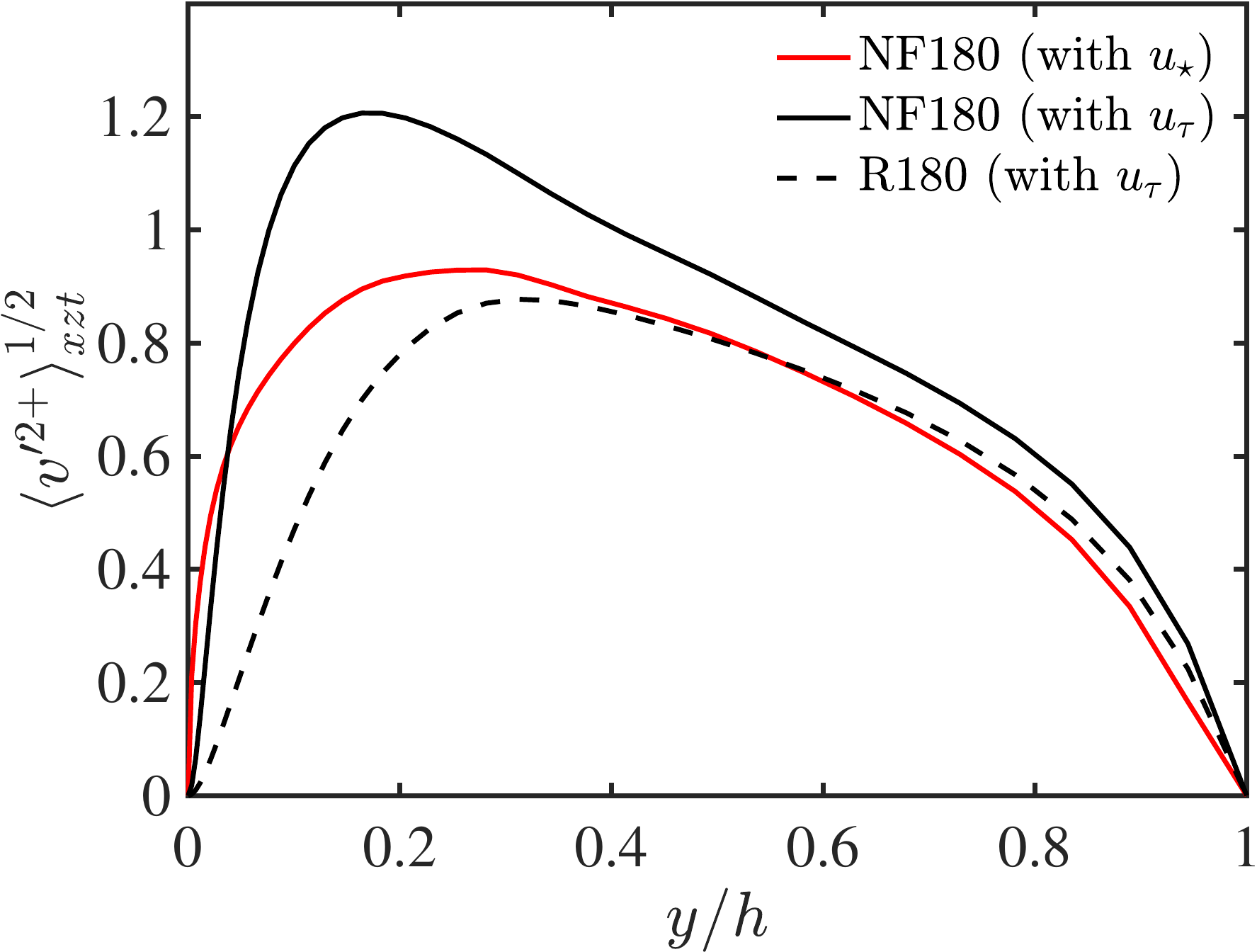}}  
  \hspace{0.05cm}
  \subfloat[]{\includegraphics[width=0.32\textwidth]{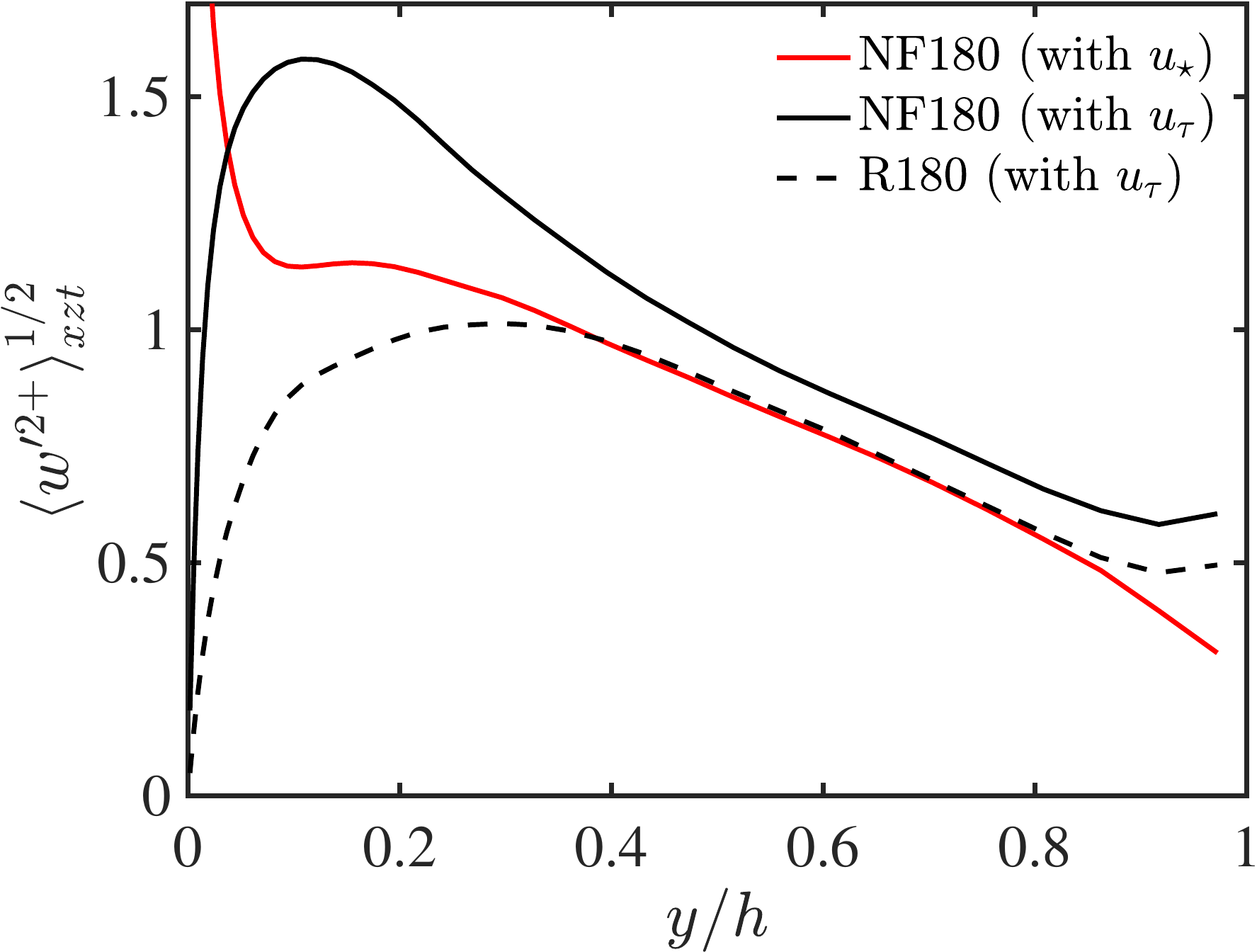}}  
  \end{center}
\caption{ (a) Streamwise, (b) wall-normal, and (c) spanwise mean
  root-mean-squared fluctuating velocities as a function of the
  wall-normal distance for case R180 normalised by $u_\tau$ (\dashed),
  case NF180 normalised by $u_\tau$ (\solid), and NF180 normalised by
  $u_\star$ (\textcolor{red}{\solid}).
\label{fig:stats_Upred}}
\end{figure}

The mean turbulence intensities for NF180 are shown in
figure~\ref{fig:stats_Upred}.  Statistics are collected once the
system reaches the statistically steady state. The mean velocity
profile is omitted as it is identical to that of R180 in
figure~\ref{fig:PD_stats_regular}(b). For comparison,
figure~\ref{fig:stats_Upred} includes one-point statistics for R180
(previously shown in figure~\ref{fig:PD_stats_regular}(c,d,e)).  The
main consequence of precluding the non-linear feedback from $\bu'$ to
$\bU$ is an increase of the level of the turbulence intensities, i.e.,
the feedback mechanism counteracts the growth of fluctuating
velocities in R180.  Despite these differences, we can still argue
that the turbulence intensities in NF180 are alike those in R180 by
noting that the friction velocity $u_\tau$ is no longer the
appropriate scaling velocity for NF180. The traditional argument for
$u_\tau$ as the relevant velocity-scale for the energy-containing
eddies is that the turbulence intensities equilibrate to comply with
the mean integrated momentum balance,
\begin{equation}\label{eq:uv_balance}
  -\langle u v \rangle \approx u_\tau^2 (1- y/h),
\end{equation}
after viscous effects are neglected \citep{Townsend1976, Tuerke2013}.
As a result, $u_\tau \approx \sqrt{-\langle u v \rangle/(1-y/h)}$
stands as the characteristic velocity for all wall-normal distances.
However, we have altered the momentum equation for case NF180, which
renders the balance in (\ref{eq:uv_balance}) invalid.  A more general
argument was made by \cite{Lozano2019} by which the characteristic
velocity of the energy-containing eddies, $u_\star$, is controlled by
the characteristic production rate of turbulent kinetic energy,
$P_\star \sim u_\star^2/t_\star$, where $t_\star$ is the time-scale to
extract energy from the mean shear
\begin{equation}
t_\star \sim \frac{1}{\sqrt{ (\partial U/\partial y)^2+ (\partial U/\partial
  z)^2}}.
\end{equation}
Taking as characteristic production rate
\begin{equation}
P_\star \sim \sqrt{ \left(u'v'\frac{\p U}{\p y}\right)^2 + \left(u'w'\frac{\p U}{\p z}\right)^2 },
\end{equation}
a characteristic velocity-scale is constructed as
\begin{equation}
  u_\star(y) \defn 
     \sqrt{ \frac{\left\langle P_\star t_\star \right\rangle_{xzt}}{1-y/h}},
\end{equation}
which generalises the concept of friction velocity. The factor
$1/\sqrt{1-y/h}$ is introduced for convenience in analogy with
$u_\tau$ in~\eqref{eq:uv_balance} so that $u_\star$ reduces to
$u_\tau$ for the regular wall turbulence.
%

Figure~\ref{fig:stats_Upred} shows that the turbulence intensities,
when scaled with $u_\star$, resemble those of R180, at least for
$y>0.1h$ where viscous effects are negligible. This suggests that the
underlying flow dynamics of NF180 is of similar nature as the regular
channel case R180 under the proper scaling.  Thus, hereafter we
utilise NF180 as reference case for comparisons as we have shown that
it exhibits similar dynamics to regular wall turbulence while being
truly linear in $\tilde{\mathcal{L}}(U)\bu'$. Occasionally, we allow
back the feedback $\bu'\rightarrow \bU$.

\subsection{Wall turbulence without exponential instability of the streaks}
\label{subsec:nonmodal_Upred}

We modify the operator $\mathcal{L}(U_0)$ so that all the unstable
eigenmodes are rendered stable at all times. We refer to this case as
the ``non-feedback channel with suppressed exponential instabilities''
(NF-SEI180) and we inquire whether turbulence is sustained under those
conditions. The approach is implemented by replacing
$\mathcal{L}(U_0)$ at each time-instance by its exponentially-stable
counterpart $\tilde{\mathcal{L}}(U_0)$, introduced
in~\eqref{eq:Atilde}. The governing equations for the channel with
suppressed exponential instabilities are
\begin{gather} \label{eq:NF-SEI180} 
\frac{\partial\bu'}{\partial t} =  \tilde{\mathcal{L}}(U_0)\bu'+ \bN(\bu'),\\
 U_0 = U(y,z,t) \ \mathrm{from \ case \ R180}. 
\end{gather}

The stable counterpart of $\mathcal{L}(U_0)$ given by
$\tilde{\mathcal{L}}(U_0)$ guarantees an exponentially stable wall
turbulence with respect to the base flow at all times, while leaving
other linear mechanisms almost intact. Note that the analysis \S
\ref{subsec:theories_modal} was performed \emph{a priori} using
data from R180, while in the present case the nonlinear dynamical
system~\eqref{eq:NF-SEI180} is actually integrated in time.  The
simulation was initialised using a flow field from R180, from which
the unstable and neutral modes are projected out, and integrated in
time for $300h/u_\tau$ after transients. It was assessed that
initialising the equation with a random velocity field yields exactly
the same conclusions.

It is useful to note that the stabilisation of $\L(U_0)$
in~\eqref{eq:NF-SEI180} can be interpreted as the addition of a
forcing term to the right-hand side of~\eqref{eq:NF-SEI180} by
considering the approximation to $\tilde{\mathcal{L}}(U_0)$
\begin{equation}\label{eq:Lhat}
 \hat{\mathcal{L}}(U_0) =
\mathcal{L}(U_0) - \sum_{j=1}^n 2 \lambda_j \boldsymbol{\mathcal{U}}_j
\boldsymbol{\mathcal{U}}^\dagger_j  \approx \tilde{\mathcal{L}}(U_0),
\end{equation}
where $\boldsymbol{\mathcal{U}}_j$ is the eigenmode of
$\mathcal{L}(U_0)$ associated with eigenvalue $\lambda_j>0$, and $n$
is the total number of unstable eigenvalues. The factor 2 on the
right-hand size of \eqref{eq:Lhat} transforms $\lambda_j>0$ for
$\mathcal{L}(U_0)$ into $-\lambda_j$ for $\hat{\mathcal{L}}(U_0)$ in
analogy with the stabilised operator $\tilde{\mathcal{L}}$;
see~\eqref{eq:Atilde}. Equation~\eqref{eq:Lhat} is approximate, as
$\mathcal{L}(U_0)$ is highly non-normal. However, we confirmed that
the largest eigenvalues and eigenmodes of $\hat{\mathcal{L}}(U_0)$ and
$\tilde{\mathcal{L}}(U_0)$ are almost identical most of the time (see
discussion in Appendix~\ref{sec:appendix_forcing}).  In virtue of
\eqref{eq:Lhat}, the modification of $\mathcal{L}(U_0)$
in~\eqref{eq:Lhat} is easily interpretable: stabilising
$\mathcal{L}(U_0)$ is equivalent to introducing a linear drag term,
$-\mathcal{F}\bu'$, in which the drag coefficient depends on the base
flow $U(y,z,t)$, i.e.,
\begin{equation}\label{eq:Fdrag}
\mathcal{F}(U) =\sum_{j=1}^{n} 2 \lambda_j
\boldsymbol{\mathcal{U}}_j \boldsymbol{\mathcal{U}}^\dagger_j,
\end{equation}
that counteracts the growth of the unstable modes at a rate
proportional to the growth rate of the mode itself.

\begin{figure}
  \begin{center}
   \subfloat[]{\includegraphics[width=0.425\textwidth]{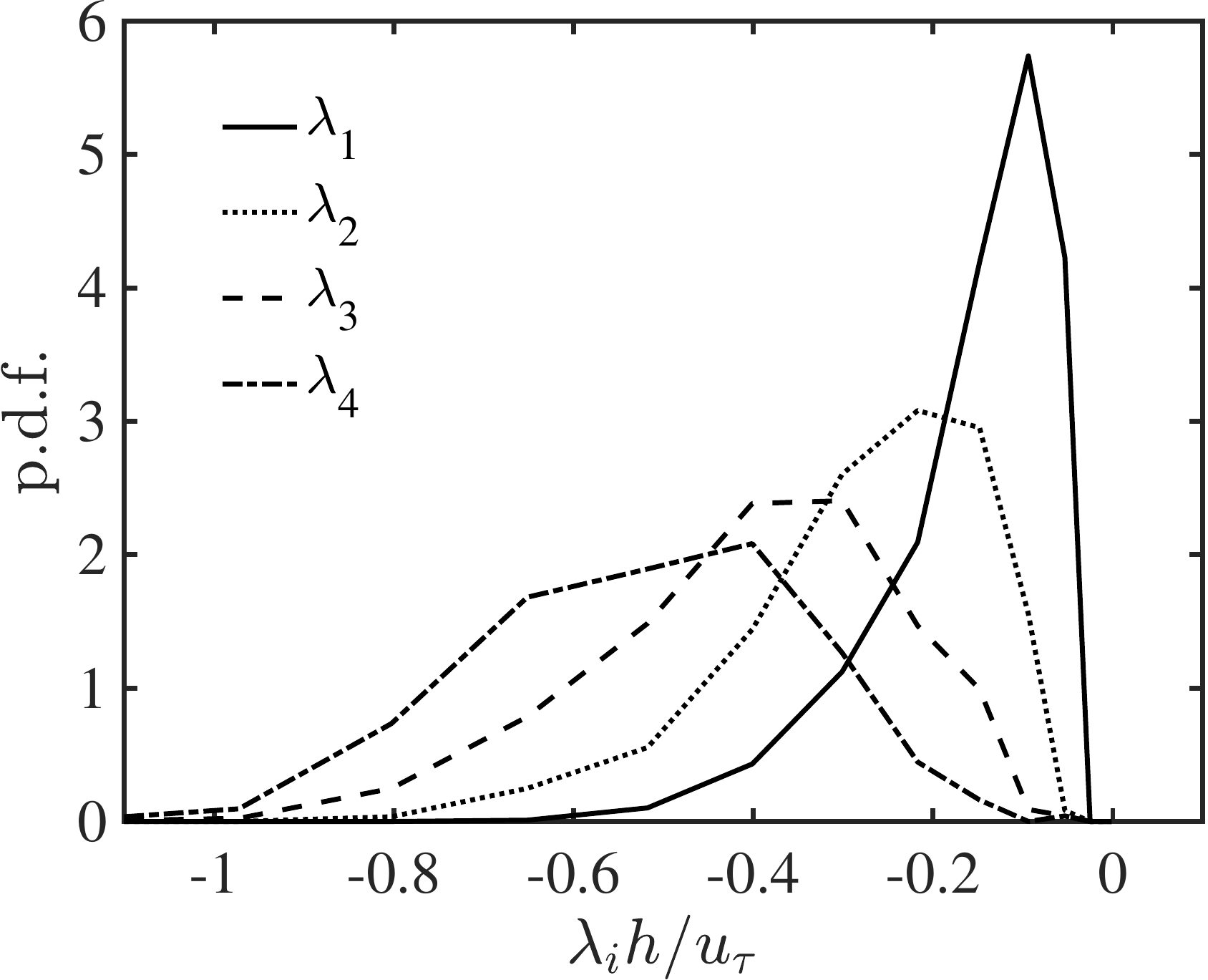}} 
   \hspace{0.2cm}
   \subfloat[]{\includegraphics[width=0.45\textwidth]{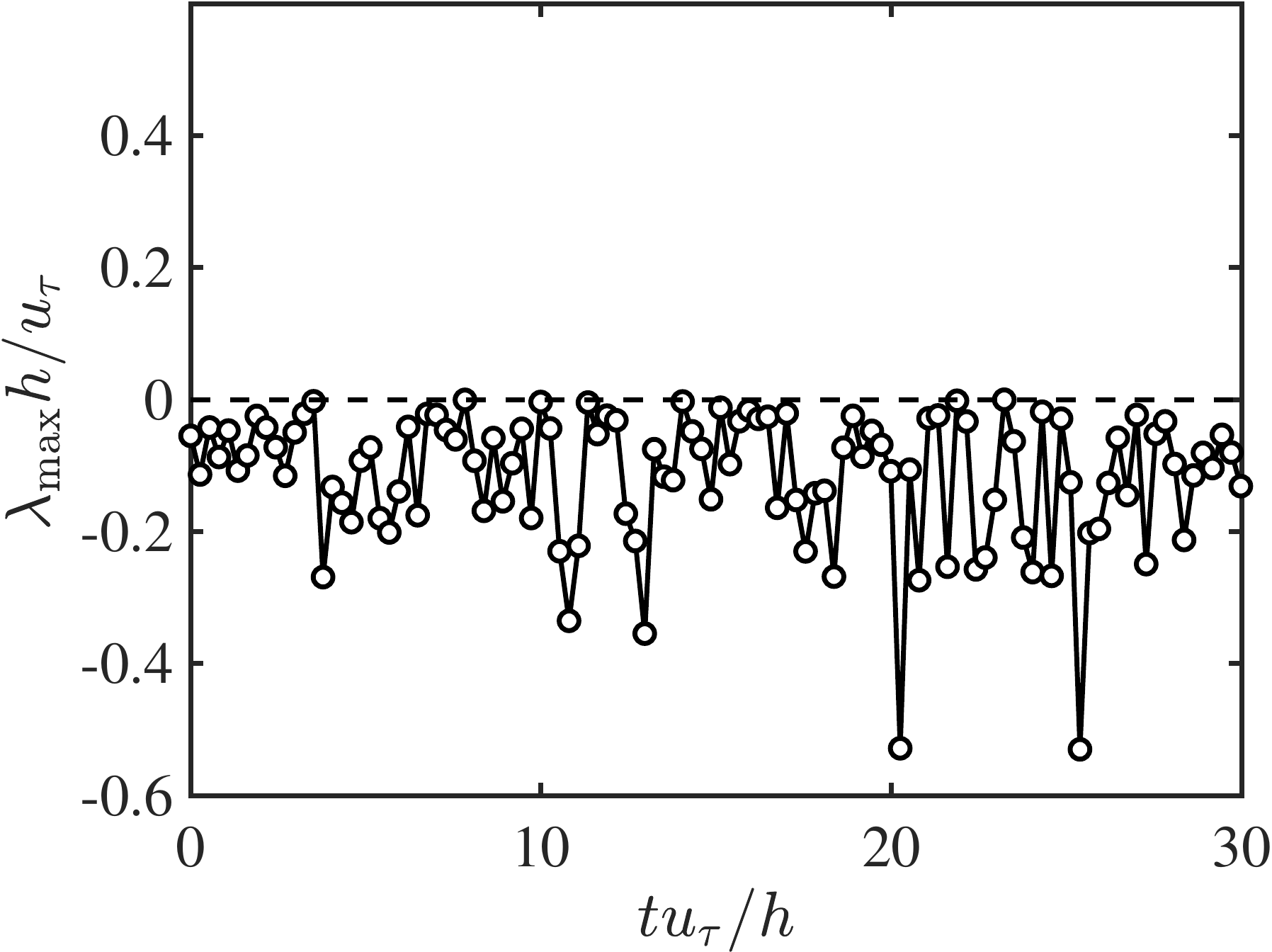}} 
 \end{center}
  \caption{ (a) Probability density functions of the growth rate of
    the four least stable eigenvalues of $\tilde{\mathcal{L}}(U_0)$,
    $\lambda_1>\lambda_2>\lambda_3>\lambda_4$. (b) The history of the
    most unstable eigenvalue~$\lambda_{\rm max}$ of
    $\tilde{\mathcal{L}}(U_0)$. Results are for the channel with
    suppressed modal instabilities
    NF-SEI180. \label{fig:P_eig_nonmodal_Upred}}
\end{figure}

\begin{figure}
 \begin{center}
   \subfloat[]{\includegraphics[width=0.45\textwidth]{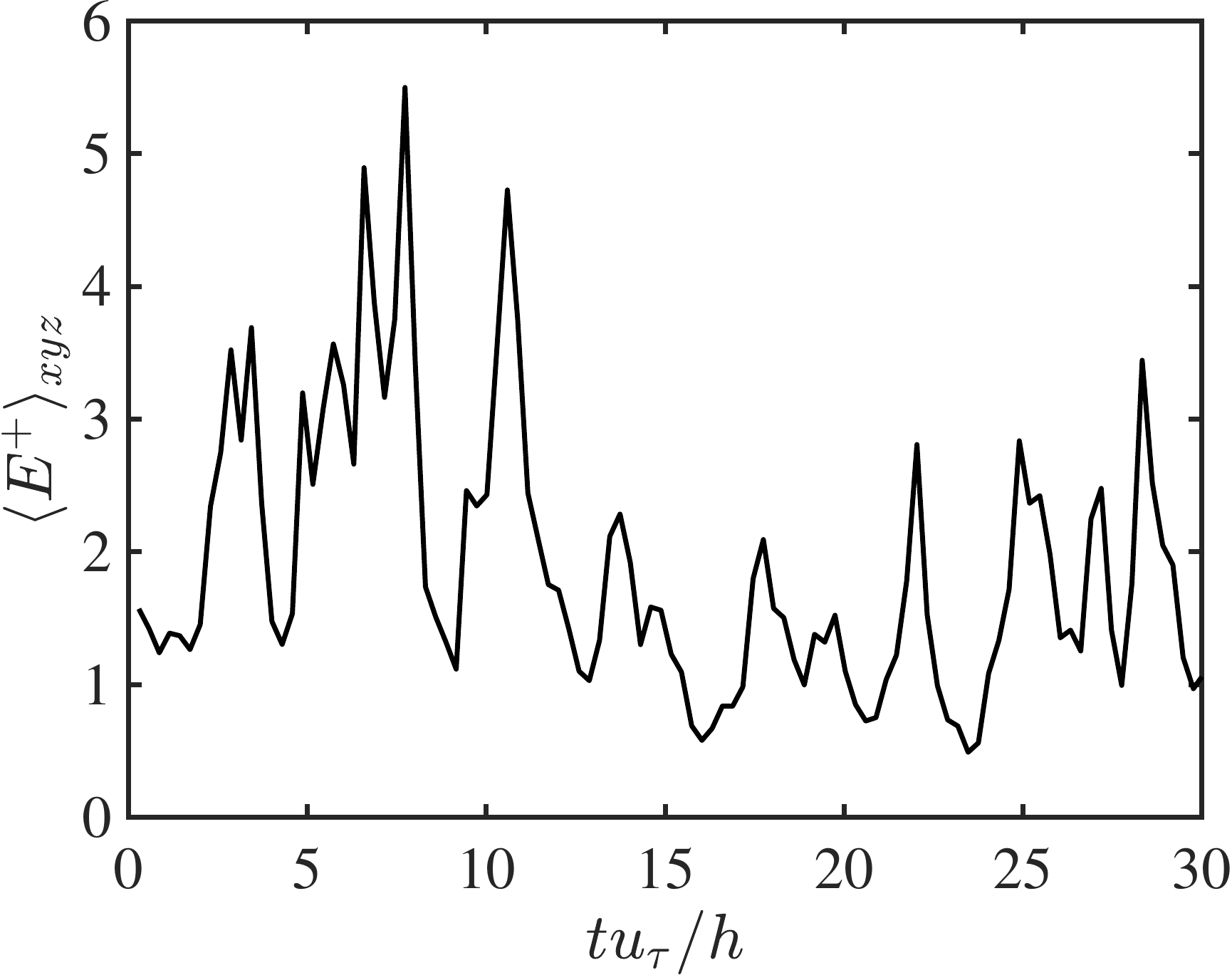}} 
   \hspace{0.1cm}
   \subfloat[]{\includegraphics[width=0.45\textwidth]{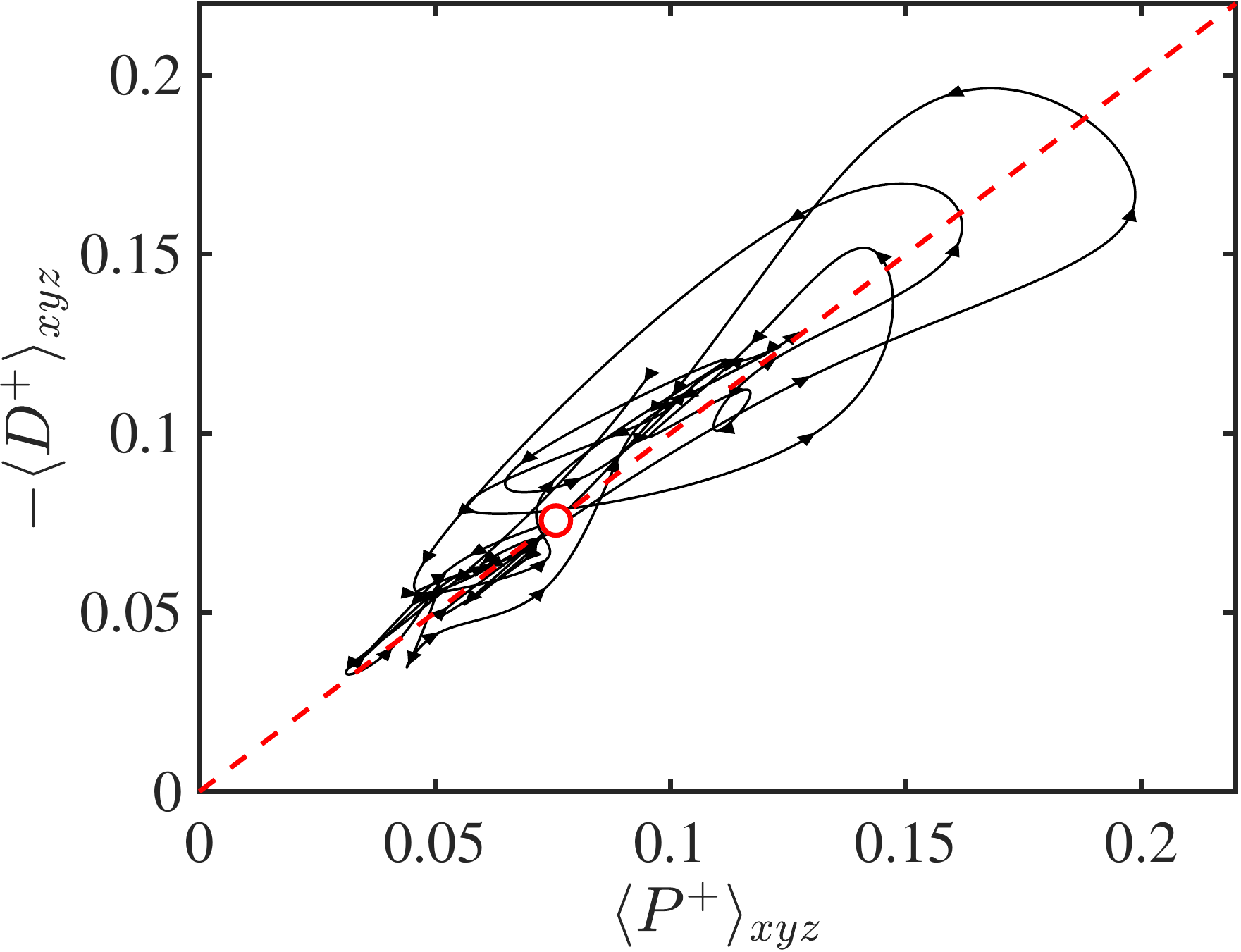}}  
 \end{center}
 \caption{ (a)~The history of the domain-averaged turbulent kinetic
   energy of the fluctuations $\langle E \rangle_{xyz}$. Note that only
   $30 h/u_\tau$ units of time are shown in the panel but the
   simulation was carried out for more than $300
   h/u_\tau$. (b)~Projection of the flow trajectory onto the average
   production rate $\langle P \rangle_{xyz}$ and dissipation rate
   $\langle D \rangle_{xyz}$ plane. The arrows indicate the time
   direction of the trajectory, which on average rotates
   counter-clockwise. The red dashed line is $\langle P \rangle_{xyz}
   = -\langle D \rangle_{xyz}$ and the red circle $\langle P
   \rangle_{xyzt} = -\langle D \rangle_{xyzt}$. The trajectory
   projected covers $15 h/u_\tau$ units of time. Results are for the
   channel with suppressed modal instabilities NF-SEI180.
 \label{fig:TKE_nonmodal_Upred}}
\end{figure}

The results of integrating (\ref{eq:NF-SEI180}) are presented in
figures~\ref{fig:P_eig_nonmodal_Upred} and
\ref{fig:TKE_nonmodal_Upred}.  The p.d.f.s~of $\lambda_j$ and a
segment of the time-series of the maximum modal growth rate of
$\tilde{\mathcal{L}}(U_0)$ are shown in
figure~\ref{fig:P_eig_nonmodal_Upred}, which confirms that the system
is successfully stabilised.
%

Figure~\ref{fig:TKE_nonmodal_Upred}(a) shows the history of the
turbulent kinetic energy for NF-SEI180 after initial transients. The
result verifies that turbulence persists when $\mathcal{L}(U_0)$ is
replaced by $\tilde{\mathcal{L}}(U_0)$. The patterns of the flow
trajectories projected onto the production--dissipation plane
(figure~\ref{fig:TKE_nonmodal_Upred}b)  exhibits features similar
to those discussed above for R180 and NF180.
%

\begin{figure}
  \begin{center}
  \subfloat[]{\includegraphics[width=0.32\textwidth]{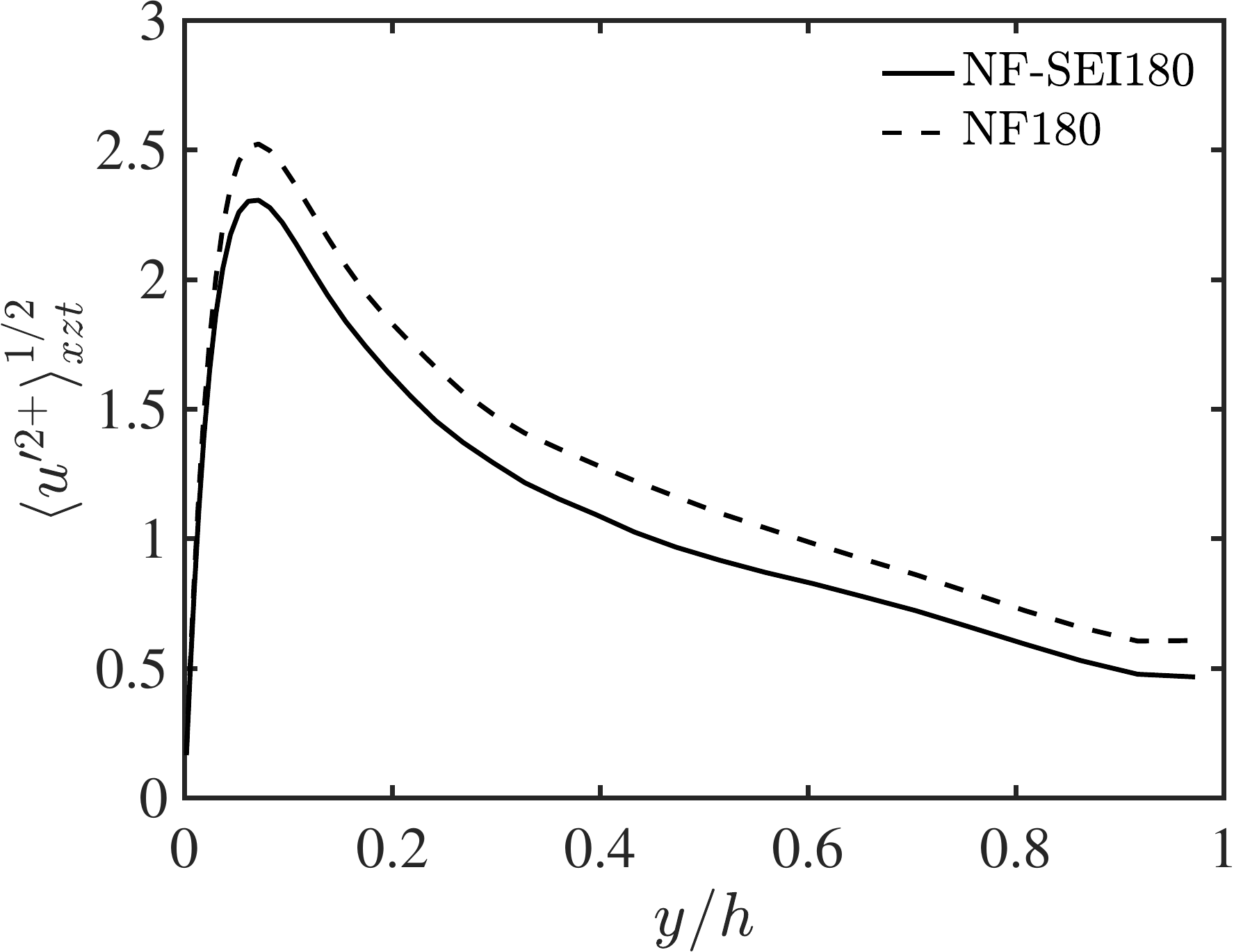}} 
  \hspace{0.05cm}
  \subfloat[]{\includegraphics[width=0.32\textwidth]{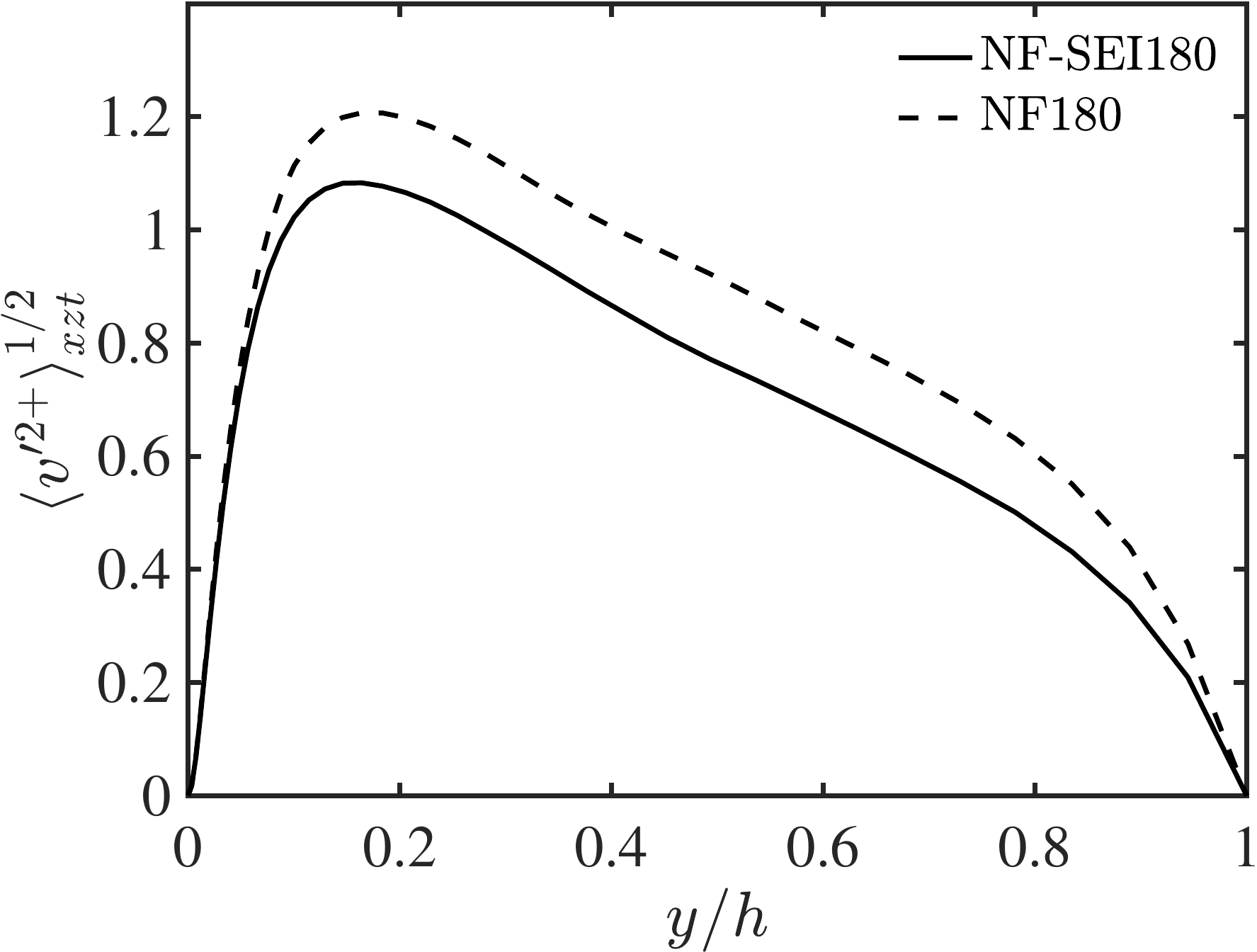}} 
  \hspace{0.05cm}
  \subfloat[]{\includegraphics[width=0.32\textwidth]{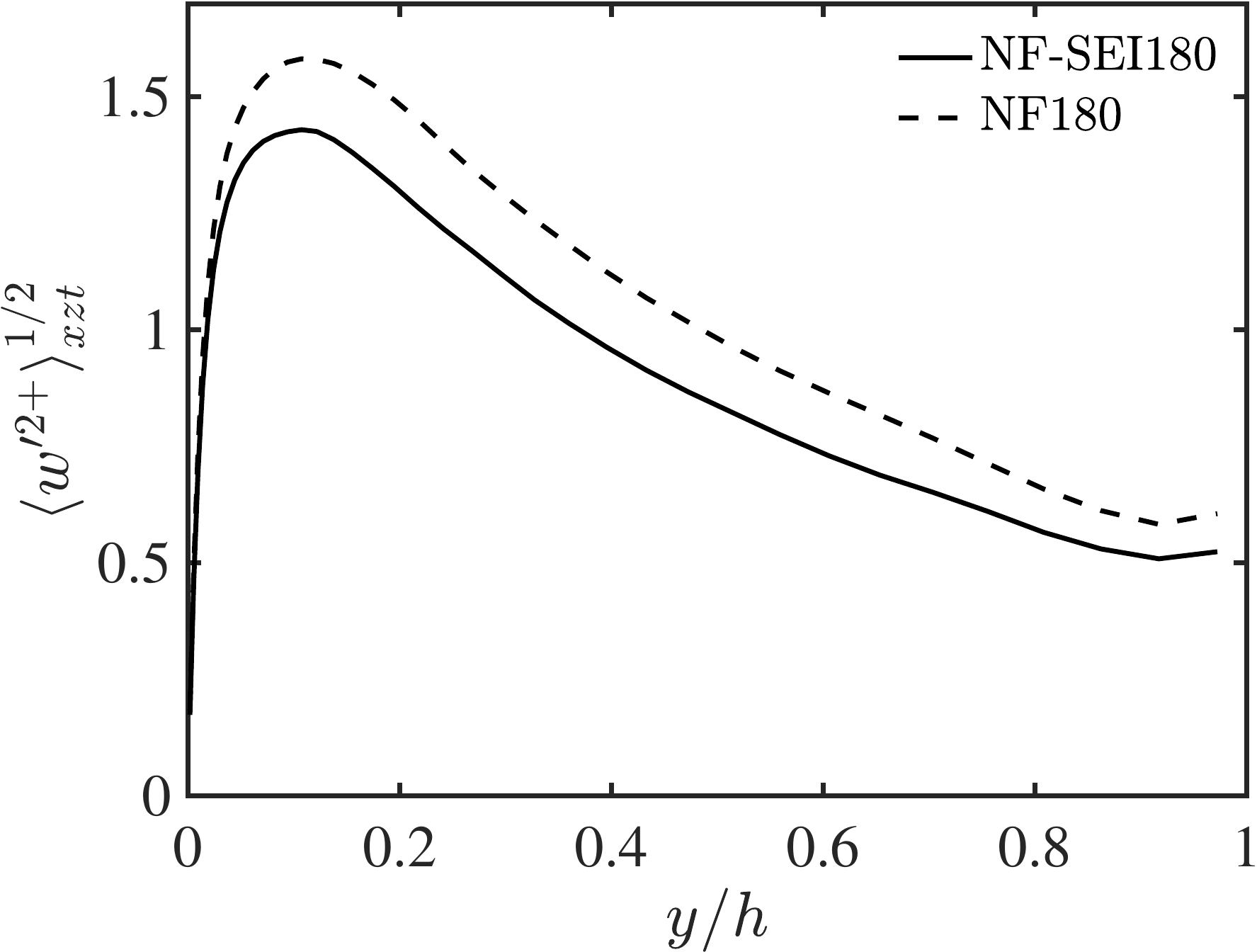}}  
  \end{center}
\caption{ (a) Streamwise, (b) wall-normal, and (c) spanwise
  root-mean-squared fluctuating velocities as a function of the
  wall-normal distance for the non-feedback channel NF180 (\dashed)
  and the non-feedback channel with suppressed exponential
  instabilities NF-SEI180 (\solid).
\label{fig:stats_nonmodal_Upred}}
\end{figure}

The turbulence intensities for NF-SEI180 are presented in
figure~\ref{fig:stats_nonmodal_Upred} and compared with those for
NF180.  The mean profile is the same as R180 (not shown).  Notably,
the channel flow without exponential instabilities is capable of
sustaining turbulence.  The new flow equilibrates at a state with
fluctuations depleted by roughly 10\%--20\%.  The outcome demonstrates
that, even if the linear instabilities of the streak manifest in the
flow, they are not required for maintaining wall turbulence.
%

It is worth emphasising that, according to the post-processing
analysis in \S \ref{sec:theories}, modal instabilities stand as a
viable mechanism to explain the energy transfer form $\bU$ to
$\bu'$. Yet, we have demonstrated here that turbulence remains almost
unchanged in their absence. This illustrates very vividly the risks of
evaluating linear (and other) theories \emph{a priori} without
accounting for the cause-and-effect relations in the actual flow.

\subsubsection{Case with explicit feedback
  from $\bu'$ to $\bU$ allowed}
\label{subsec:regular_nonmodal}

It was shown above that turbulence is sustained despite the absence of
exponential instabilities. This was demonstrated for NF-SEI180, in
which the nonlinear feedback from $\bu'$ to $\bU$ was excluded. We
have seen in \S \ref{sec:nofeeback} that inhibiting the feedback from
$\bu'$ to $\bU$ actually enhances the turbulence intensities with
respect to $u_\tau$. This may cast doubts on whether the `weaker'
fluctuations from R180 can be sustained when modal instabilities are
also cancelled out. To clarify this point, we resolve a channel with
suppressed exponential instabilities in which the feedback from $\bu'$
to $\bU$ is allowed. The equations of motion in this case are:
\begin{subequations} \label{eq:R-SEI180}
\begin{gather} \label{eq:R-SEI180_1}
\frac{\partial\bu'}{\partial t} = 
\tilde{\mathcal{L}}(U)\bu'+ \bN(\bu'),\\
  \frac{\partial \bU}{\partial t} =
 -\bU \cdot
 \bnabla \bU
 -\mathcal{D}\langle \bu' \cdot
  \bnabla \bu' \rangle_x
 - \frac{\mathcal{D}}{\rho}\bnabla \langle p \rangle_x + \nu \nabla^2
 \bU + \boldsymbol{f}, \quad 
 \bnabla \cdot \bU = 0. \label{eq:R-SEI180_2}
\end{gather}
\end{subequations}
We refer to this case as ``regular channel with suppressed exponential
instabilities'' or R-SEI180. Note that the only difference
of~\eqref{eq:R-SEI180} from the original Navier--Stokes
equations~\eqref{eq:NS_original} is the modally stable $\tilde\L(U)$
instead of $\L(U)$. The base flow $\bU$ is now dynamically coupled to
$\bu'$ via the nonlinear term $-\mathcal{D}\langle \bu' \cdot \bnabla
\bu' \rangle_x$ in~\eqref{eq:R-SEI180_2}.  A similar experiment was
done by \cite{Farrell2012} for Couette flow at low Reynolds
numbers. We initialise simulations of~\eqref{eq:R-SEI180} from a flow
field of R180 after projecting out the unstable and neutral modes from
this initial condition. It was checked that using random velocities as
initial condition yields the same results.

\begin{figure}
  \begin{center}
   \subfloat[]{\includegraphics[width=0.32\textwidth]{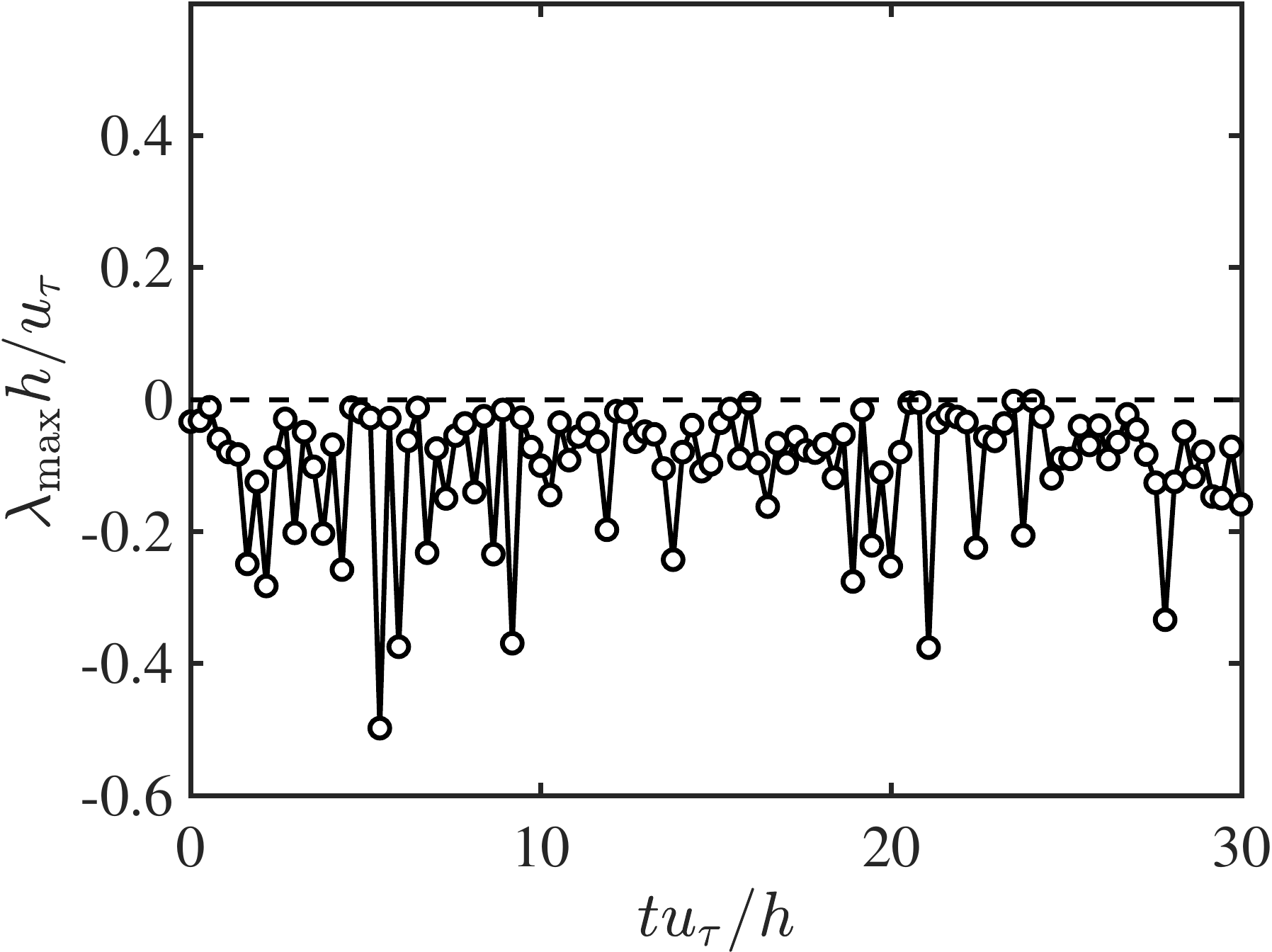}} 
   \hspace{0.02cm}
   \subfloat[]{\includegraphics[width=0.32\textwidth]{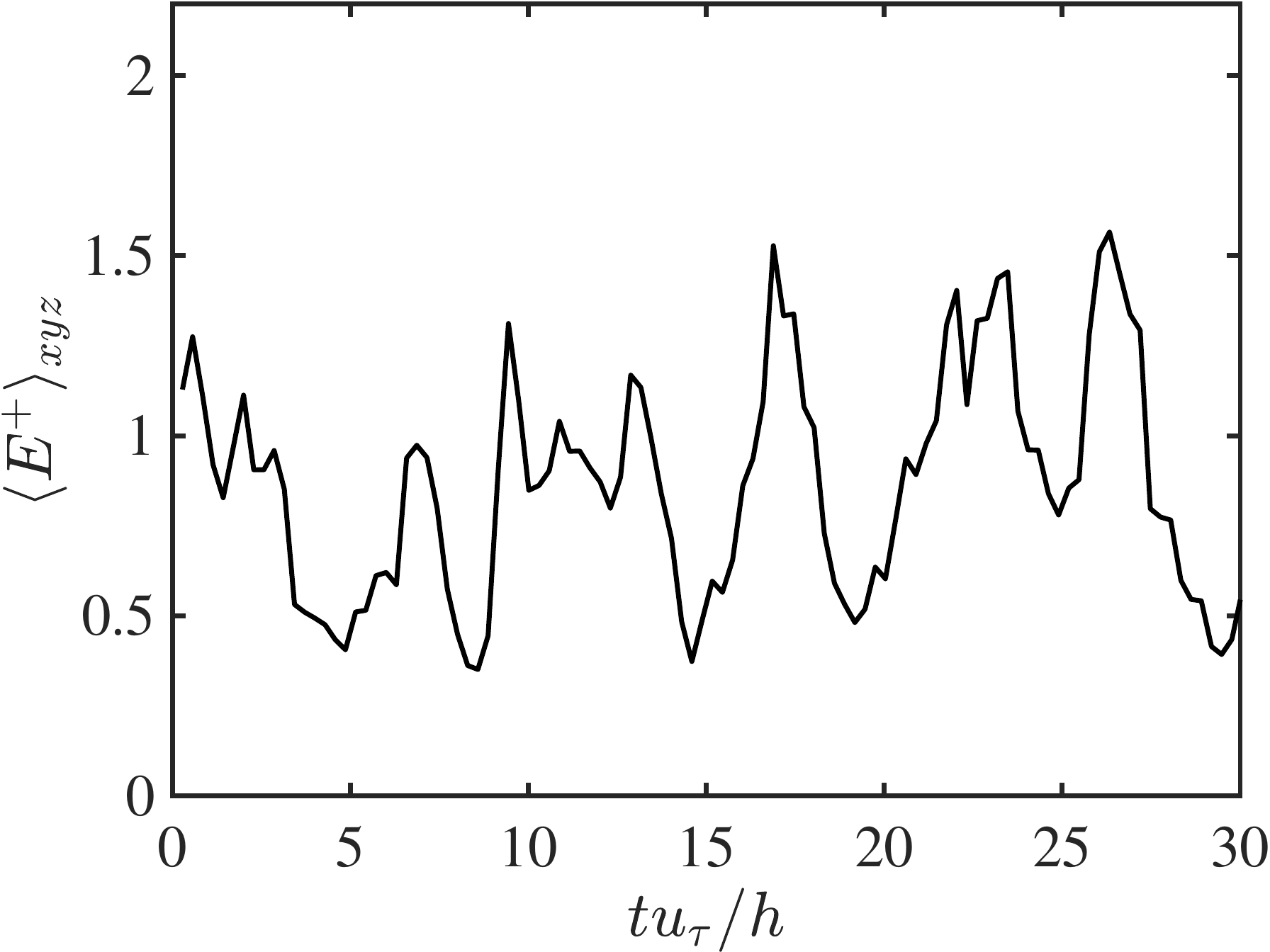}} 
   \hspace{0.02cm}
   \subfloat[]{\includegraphics[width=0.315\textwidth]{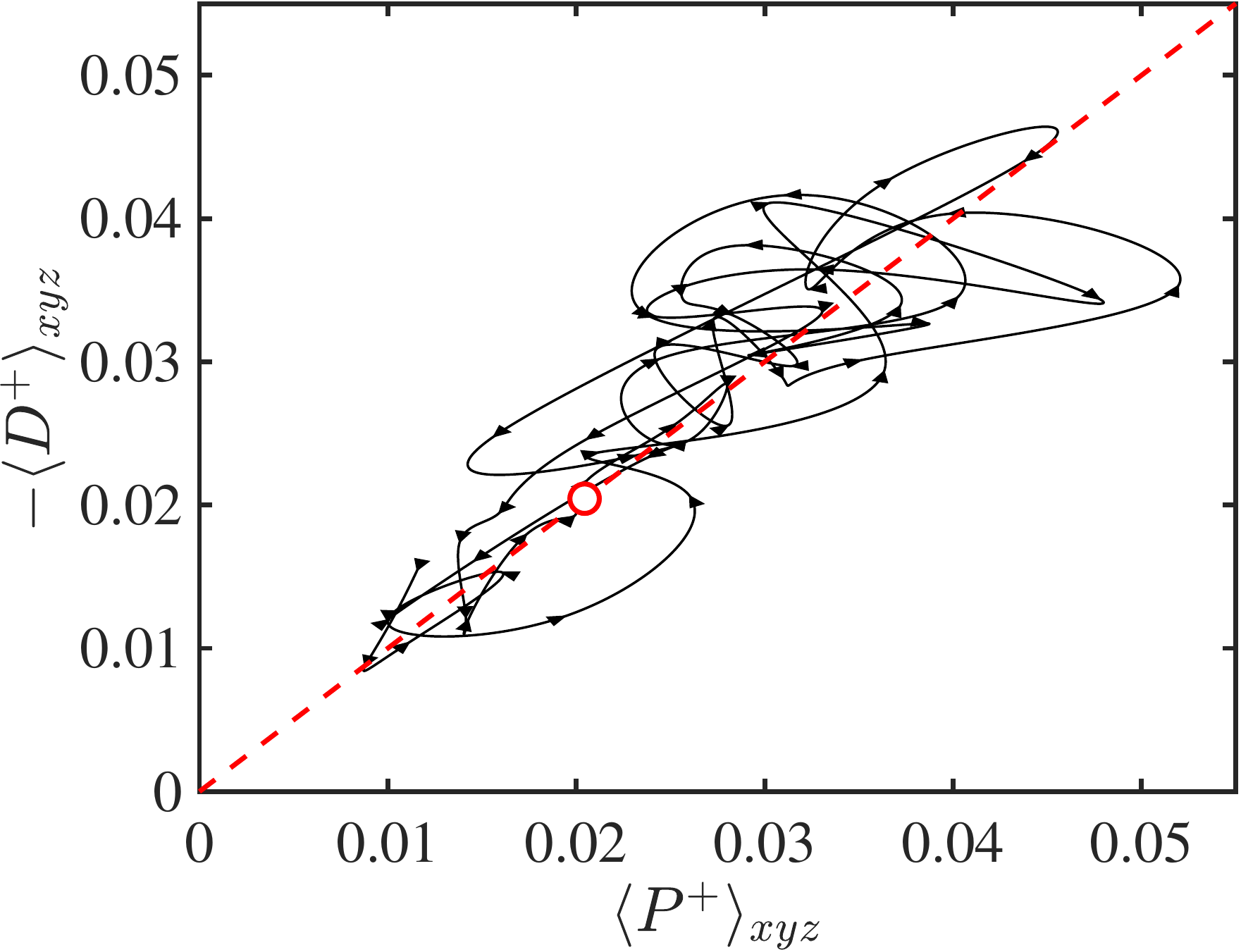}} 
 \end{center}
  \caption{ (a) The history of the most unstable
    eigenvalue~$\lambda_{\rm max}$ of
    $\tilde{\mathcal{L}}(U)$. Results for channel with suppressed
    modal instabilities and feedback from $\bu'$ to $\bU$ allowed
    (R-SEI180).  (b)~The history of the turbulent kinetic energy of
    the fluctuation energy~$E=\tfrac1{2} \bu' \cdot \bu'$ averaged
    over the channel domain.  Note that  only $30 h/u_\tau$ units
    of time are shown in the panels~(a)~and~(b) but the simulation was
    carried out for more than $300 h/u_\tau$. (c)~Projection of the
    flow trajectory onto the average production rate $\langle P
    \rangle_{xyz}$ and dissipation rate $\langle D \rangle_{xyz}$
    plane. The arrows indicate the time direction of the trajectory,
    which on average rotates counter-clockwise. The red dashed line is
    $\langle P \rangle_{xyz} = -\langle D \rangle_{xyz}$ and the red
    circle $\langle P \rangle_{xyzt} = -\langle D \rangle_{xyzt}$. The
    trajectory projected covers $15 h/u_\tau$ units of time. Results
    for channel with suppressed modal instabilities but with feedback
    from $\bu'$ to $\bU$ allowed (R-SEI180).
    \label{fig:nonmodal_regular}}
\end{figure}

The history of $\lambda_\mathrm{max}$ for $\tilde{\mathcal{L}}(U)$,
shown in figure~\ref{fig:nonmodal_regular}(a), confirms that modal
instabilities are successfully
removed. Figure~\ref{fig:nonmodal_regular}(b) contains the evolution
of the turbulent kinetic energy and shows that turbulence persists
under the stabilised linear dynamics of~\eqref{eq:R-SEI180_1}.  The
flow trajectories projected onto the production--dissipation plane
(figure~\ref{fig:nonmodal_regular}c) also exhibit similar features to
those discussed above for R180 and NF-SEI180.
%

\begin{figure}
  \begin{center}
  \hspace{0.1cm}
  \subfloat[]{\includegraphics[width=0.45\textwidth]{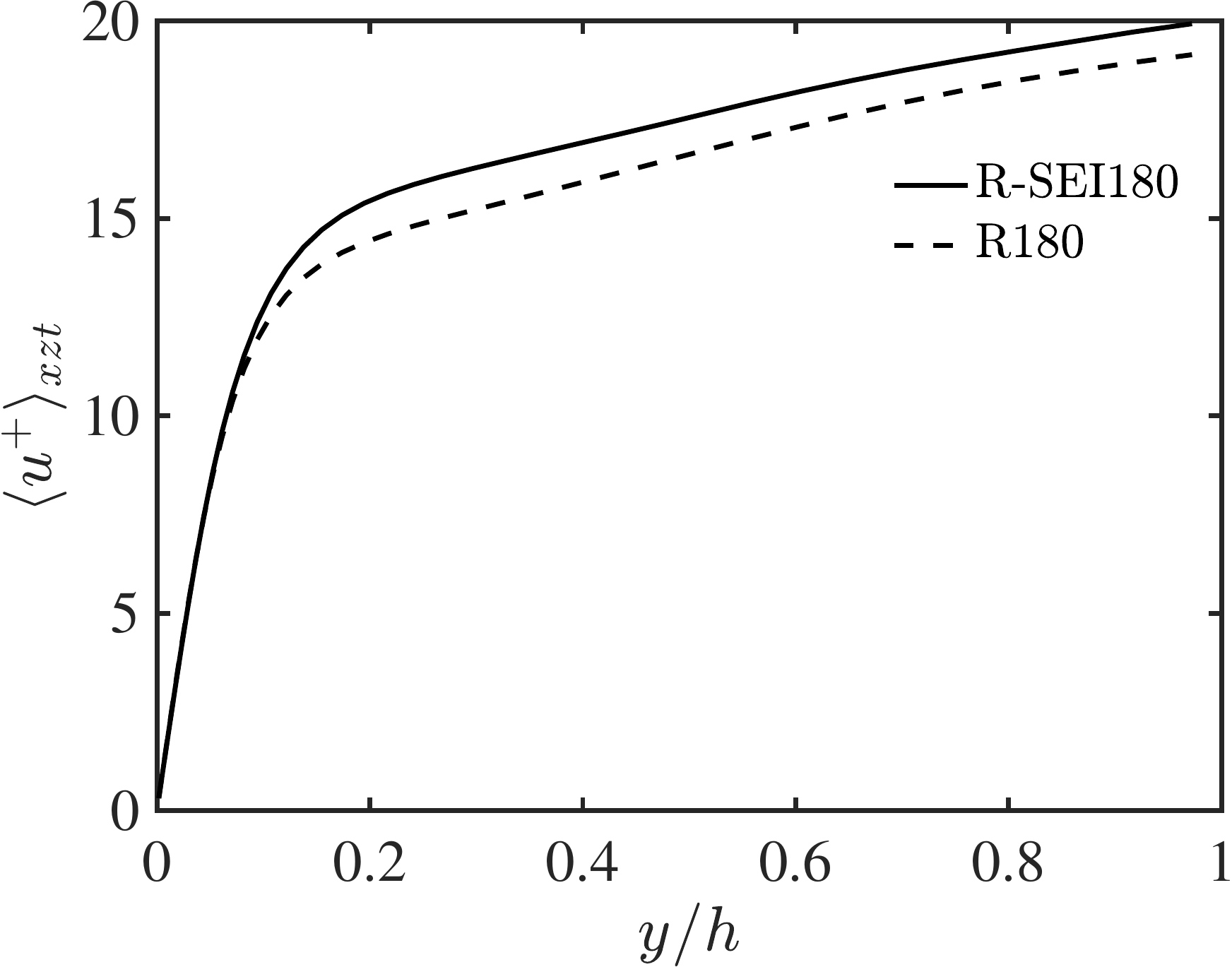}} 
  \hspace{0.1cm}
  \subfloat[]{\includegraphics[width=0.45\textwidth]{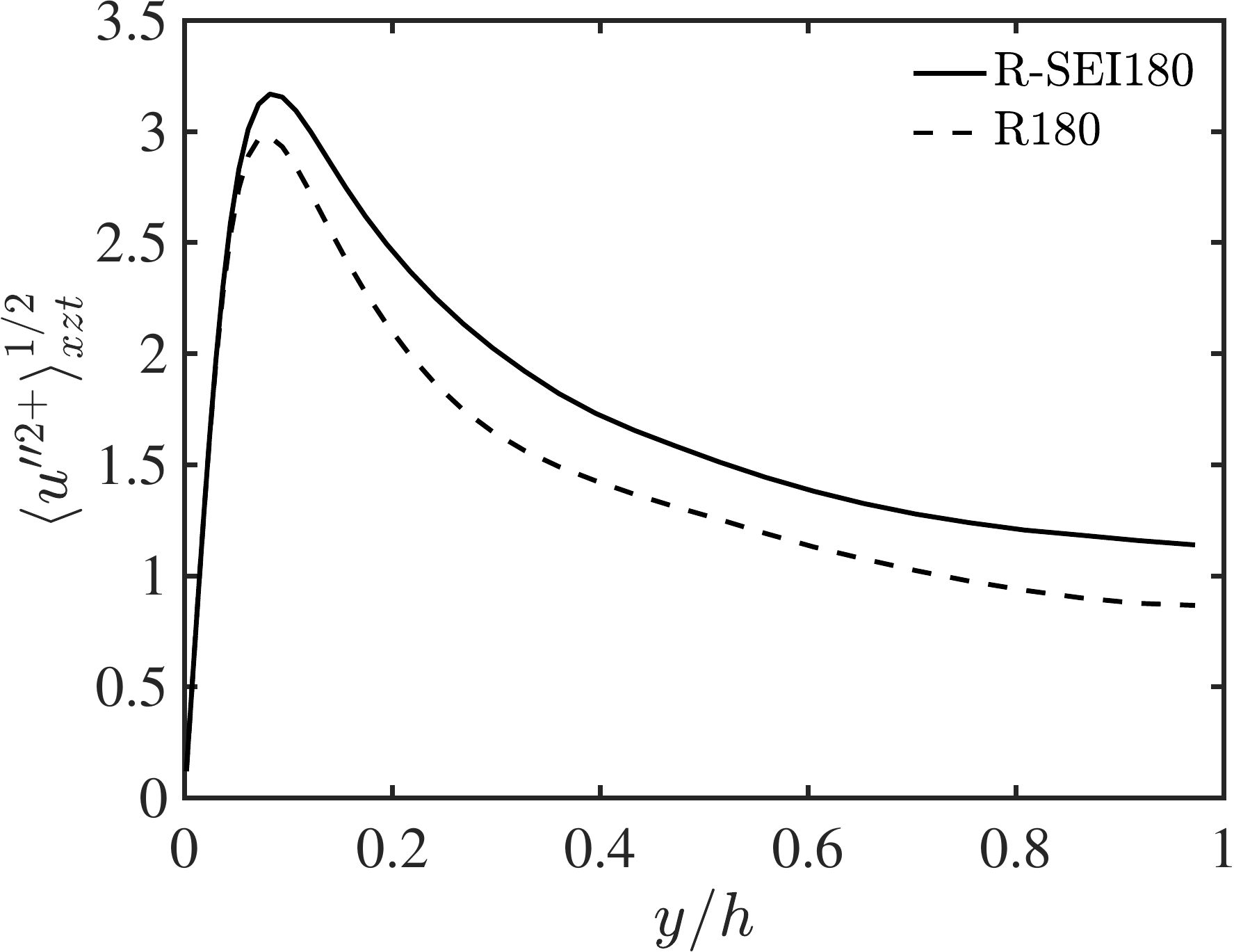}} 
  \end{center}
  \begin{center}
    \subfloat[]{\includegraphics[width=0.32\textwidth]{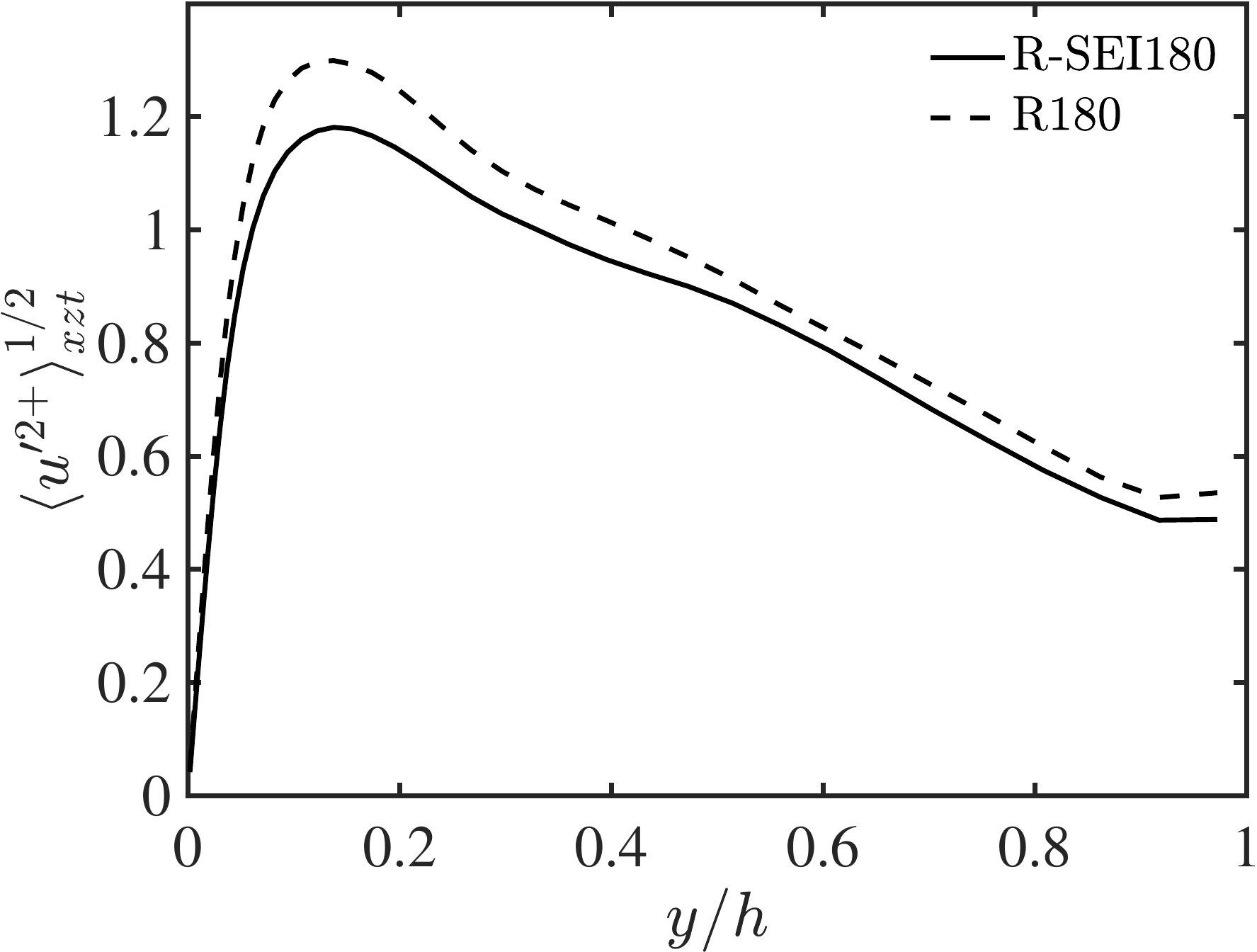}} 
    \hspace{0.05cm}
    \subfloat[]{\includegraphics[width=0.32\textwidth]{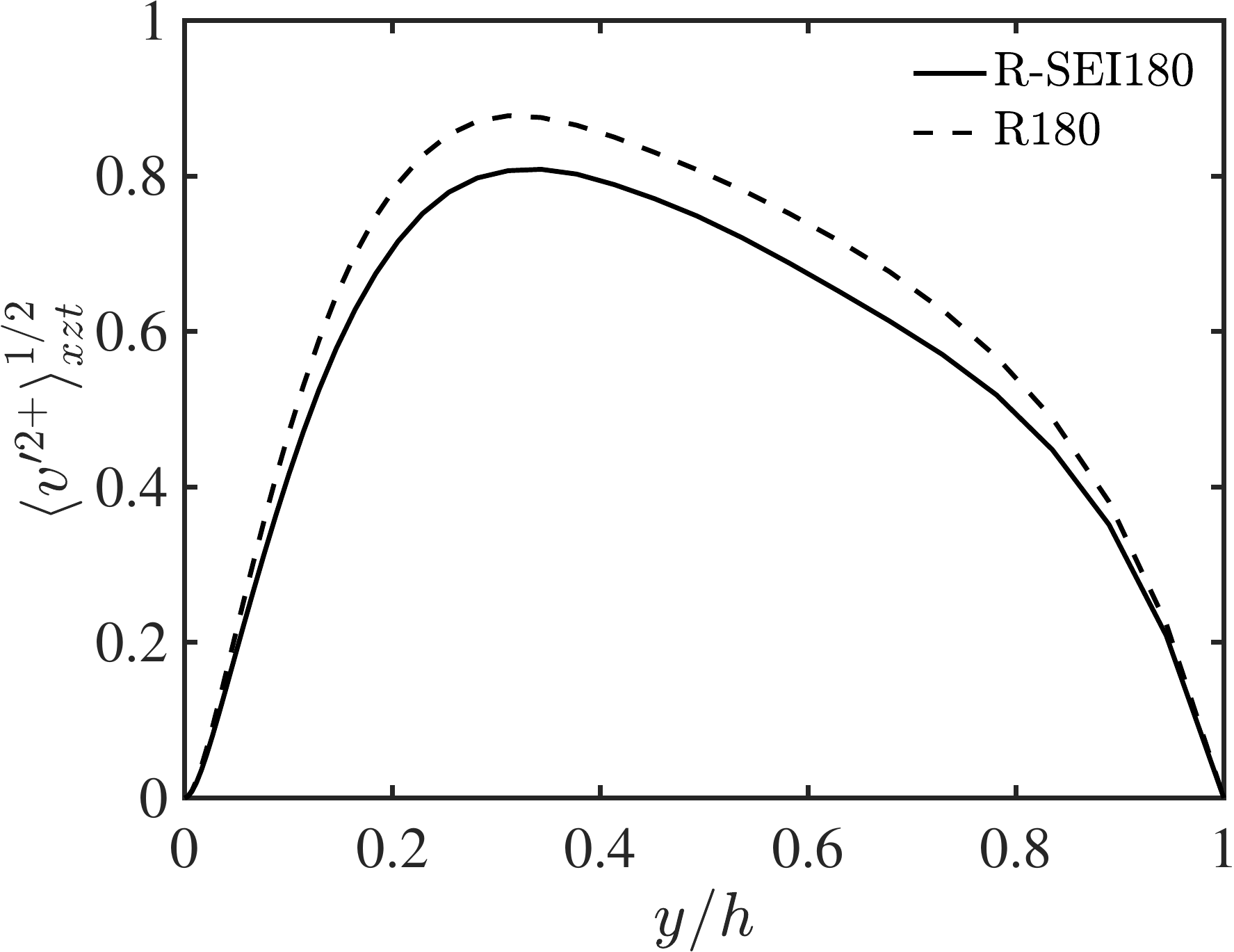}} 
    \hspace{0.05cm}
    \subfloat[]{\includegraphics[width=0.32\textwidth]{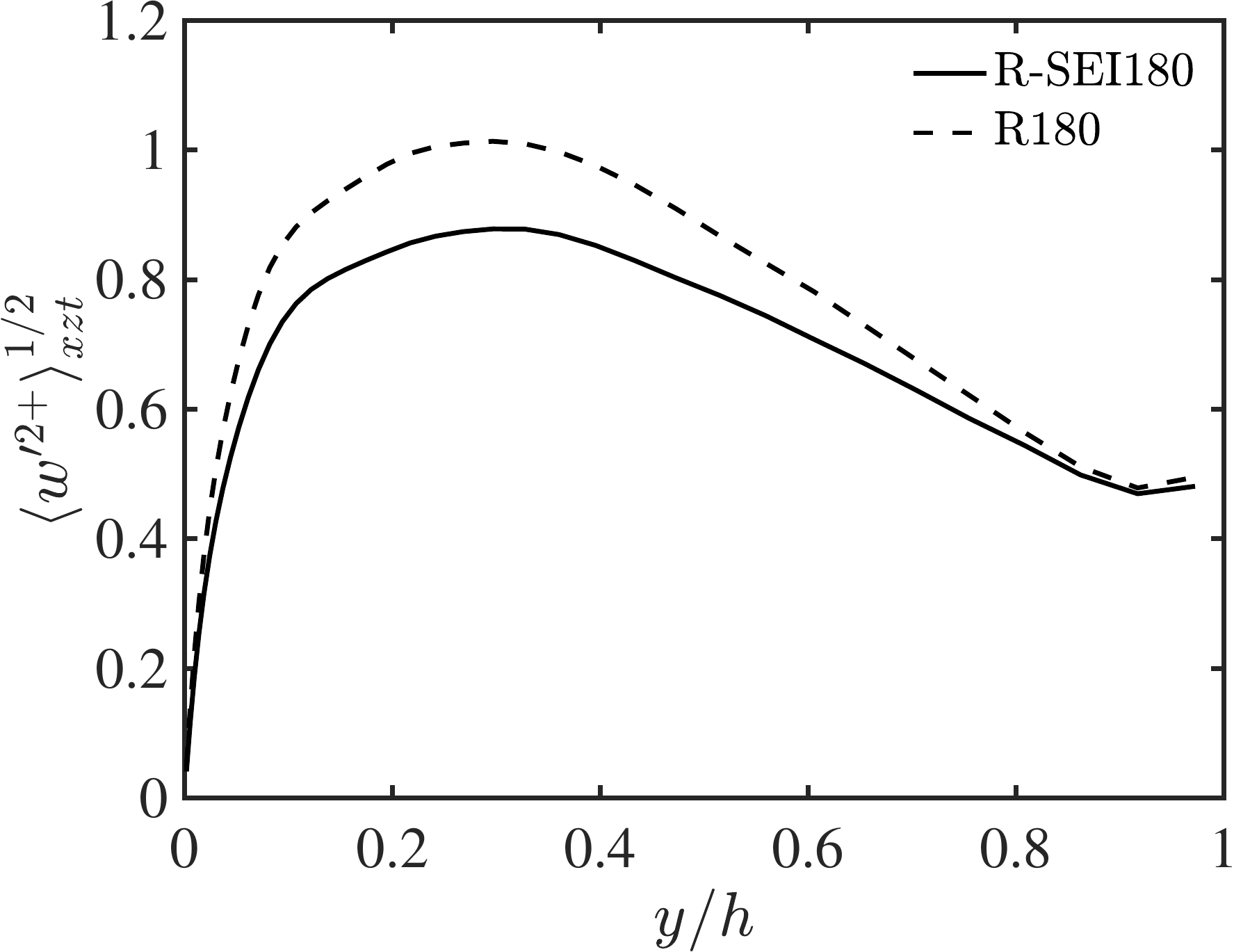}} 
  \end{center}
\caption{ (a) Streamwise mean velocity profile, and (b, c) streamwise,
  (d) wall-normal, and (e) spanwise mean root-mean-squared fluctuating
  velocities as a function of the wall-normal distance for the regular
  channel (R180) (\dashed) and the channel with suppressed exponential
  instabilities but with the feedback from $\bu'$ to $\bU$ allowed
  (R-SEI180) (\solid).  Note that the streamwise fluctuating velocity
  in panel (b) is defined as $u'' = u - \langle u \rangle_{xzt}$,
  while in panel (c) is defined as $u'=u-U$.
\label{fig:stats_nonmodal_regular}}
\end{figure}

\begin{figure}
 \begin{center}
  \includegraphics[width=1\textwidth]{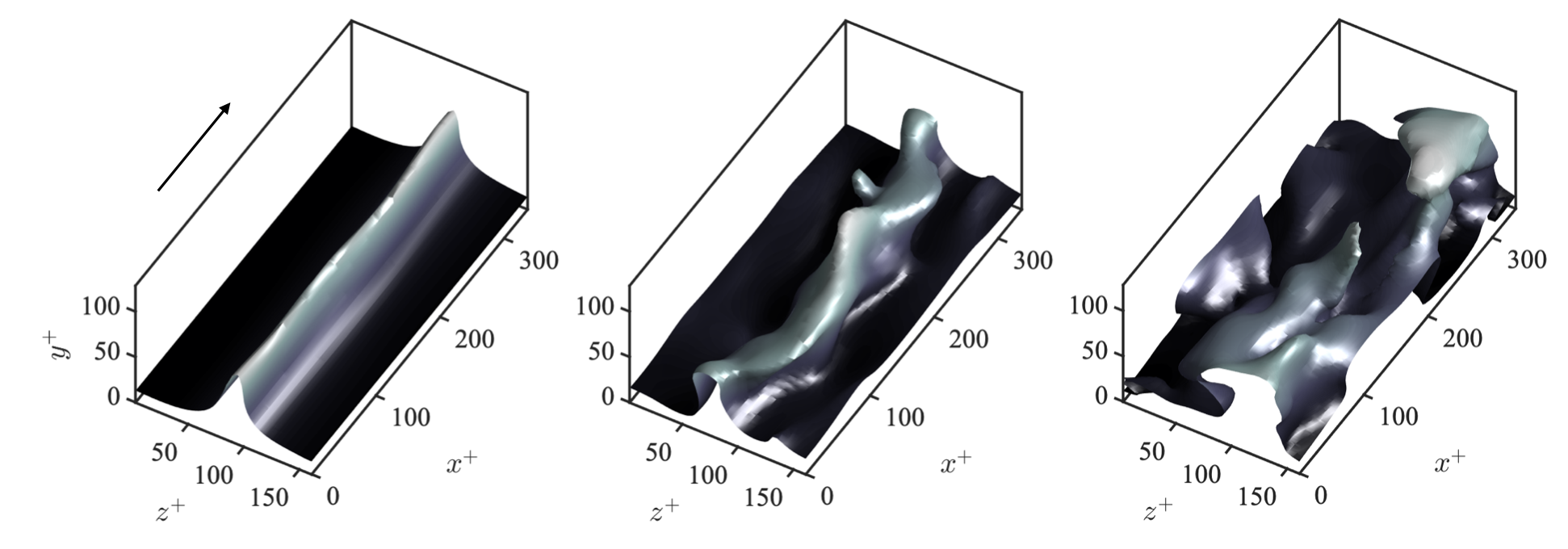} 
 \end{center}
\caption{ Instantaneous isosurface of the streamwise velocity at
  different times for R-SEI180. The value of the isosurface is 0.65 of
  the maximum streamwise velocity.  Shading represents the distance to the
  wall from dark ($y=0$) to light ($y=h$). The arrow indicates the mean flow
  direction.
 \label{fig:snapshots_R-SEI180}}
\end{figure}

The mean velocity profiles and turbulence intensities for R-SEI180 and
R180 are shown in figure~\ref{fig:stats_nonmodal_regular}.  The
results are consistent with the trends reported in
figure~\ref{fig:stats_nonmodal_Upred} for NF-SEI180 and NF180
simulations: turbulence without modal instabilities is sustained
despite allowing the feedback from $\bu'$ to $\bU$. As in NF-SEI180,
the resulting velocity fluctuations are diminished by roughly 10\%.
%
%
Figure~\ref{fig:snapshots_R-SEI180} portrays snapshots of the
streamwise velocity at three different instants for the R-SEI180
simulation.  As in R180 (cf.~figure~\ref{fig:snapshots_regular}), the
spatial organisation of the streak cycles through different stages of
elongated straight motion, meandering and breakdown, although the
first two states (panels~(a)~and~(b)) occur more frequently than in
R180. Indeed, if we consider the common definition for the streamwise
velocity fluctuation $u''=u - \langle u \rangle_{xzt}$, which contains
part of the streaky flow, the new flow in R-SEI180 attains an augmented
streak intensity as clearly depicted in
figure~\ref{fig:stats_nonmodal_regular}(b). The outcome is consistent
with the occasional inhibition of the streak meandering or breakdown
via exponential instability, which enhances $u''$, whereas wall-normal
($v''=v'$) and spanwise ($w''=v'$) turbulence intensities are
diminished due to a lack of vortices succeeding the collapse of the
streak (namely, mechanism (i) discussed in the introduction). This
behaviour was also observed in many drag reduction investigations
\citep{Jung1992, Laadhari1994, Choi2001, Ricco2008}.
%

As a final comment, \cite{Lozano_brief_2018b} showed in a preliminary
work that turbulence was not sustained when $\mathcal{L}(U)$ was
stabilised by $\mathcal{L}(U)-\mu\mathcal{I}$, where $\mu>0$ is a
damping parameter and $\mathcal{I}$ is the identity operator. However,
it can be shown that introducing the linear drag $-\mu\bu'$ reduces
the gains supported by $\mathcal{L}(U)$ by a factor of $\exp(- 2 \mu
T)$, with $T$ the optimisation time. Hence, stabilising
$\mathcal{L}(U)$ via a linear drag term $-\mu\bu'$ also disrupts the
transient growth mechanism severely and this was the cause for the
lack of sustained turbulence in \cite{Lozano_brief_2018b}. Conversely,
we have shown that $\tilde{\mathcal{L}}(U)$ is physically
interpretable as the stabilisation of $\mathcal{L}(U)$ via a linear
forcing directed toward modal instabilities
($\tilde{\mathcal{L}}(U)\bu' \approx
\mathcal{L}(U)\bu'-\mathcal{F}\bu'$, see (\ref{eq:Fdrag})). This
entails a much gentler modification which leaves almost intact the
transient growth mechanisms of $\mathcal{L}(U)$, as opposed to
$\mathcal{L}(U)-\mu \mathcal{I}$.

\subsection{Wall turbulence exclusively supported by transient growth}
\label{subsec:TG}

The effect of non-modal transient growth as the main source for energy
injection from~$\bU$ into~$\bu'$ is now assessed by ``freezing'' the
base flow $U(y,z,t_i)$ at the instant~$t_i$. In order to steer clear
of the potential effect of exponential instabilities, the numerical
experiment here is performed using the stabilised linear operator
$\tilde{\mathcal{L}}(U(y,z,t_i))$. For a given $U(y,z,t_i)$, we refer
to this case as ``channel flow with modally-stable, frozen-in-time
base-flow'', or NF-TG180$_i$, with $i$ an index indicating the
case number.  Let us denote the flow for NF-TG180$_i$ as $\bu_{\{i\}}$
(and similarly for other flow quantities). The governing equations for
NF-TG180$_i$ are
\begin{subequations}
 \label{eq:NF-TG180}
\begin{gather}
\label{eq:NF-TG180_1}
\frac{\partial\bu'_{\{i\}}}{\partial t} = \tilde{\mathcal{L}}_{\{i\}}\,\bu'_{\{i\}}+ \bN(\bu'_{\{i\}}),\\
U_{\{i\}} = U(y,z,t_i) \ \mathrm{from \ case \ R180}, \label{eq:NF-TG180_2}
\end{gather}
\end{subequations}
where $\tilde{\mathcal{L}}_{\{i\}} = \tilde{\mathcal{L}}(U(y,z,t_i))$.
The set-up in~\eqref{eq:NF-TG180} disposes of energy transfers that
are due to both modal and parametric instabilities, while allowing
the transient growth of fluctuations. For a given
  $t_i$, the simulation is initialised from NF-SEI180 at $t=t_i$
  (projecting out neutral and unstable modes), and continued for
  $t>t_i$. We performed more than 500 simulations using different
  frozen base flows $U(y,z,t_i)$ extracted from R180.

\begin{figure}
  \begin{center}
  \subfloat[]{\includegraphics[width=0.32\textwidth]{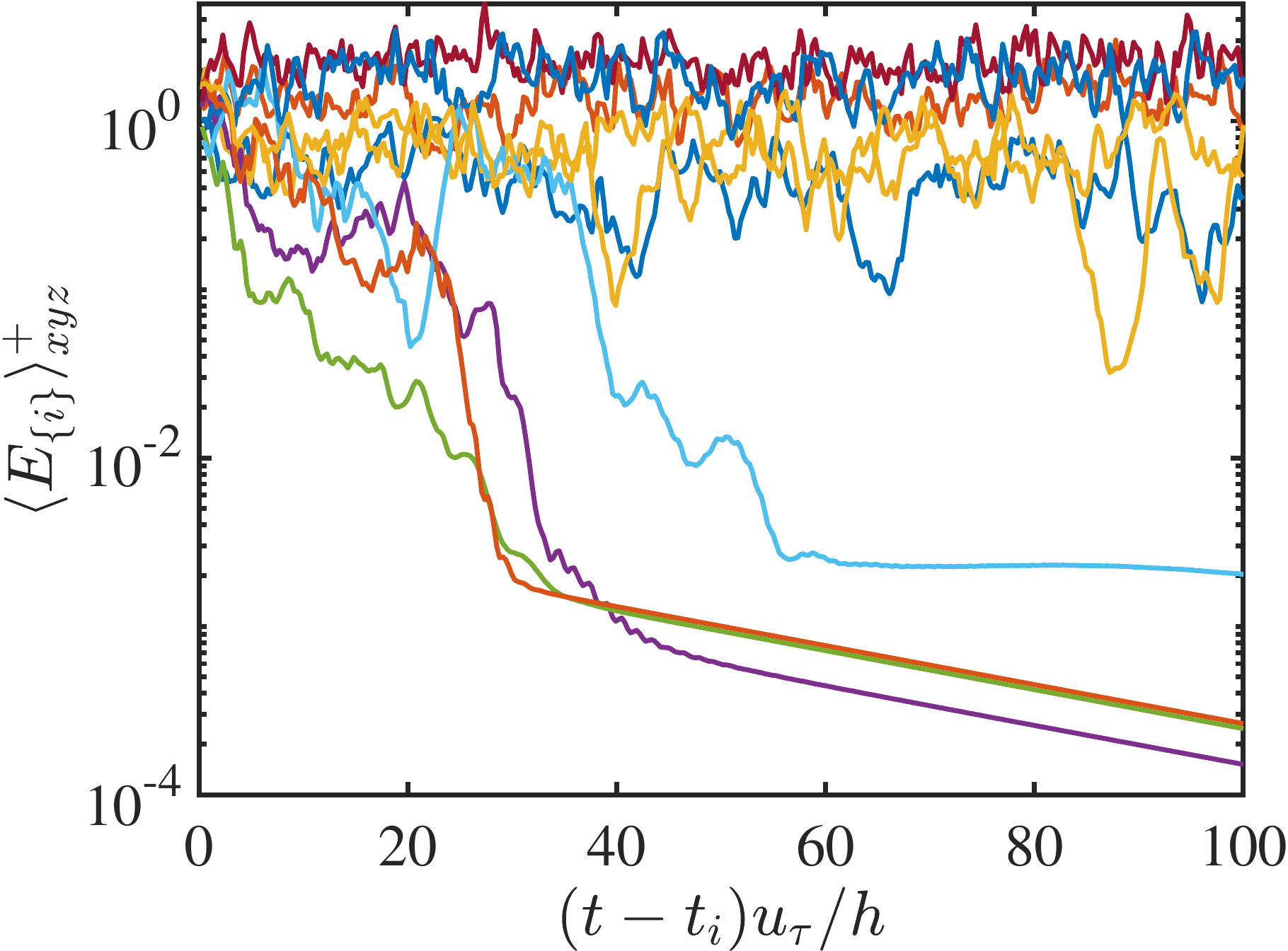}}   
  \subfloat[]{\includegraphics[width=0.32\textwidth]{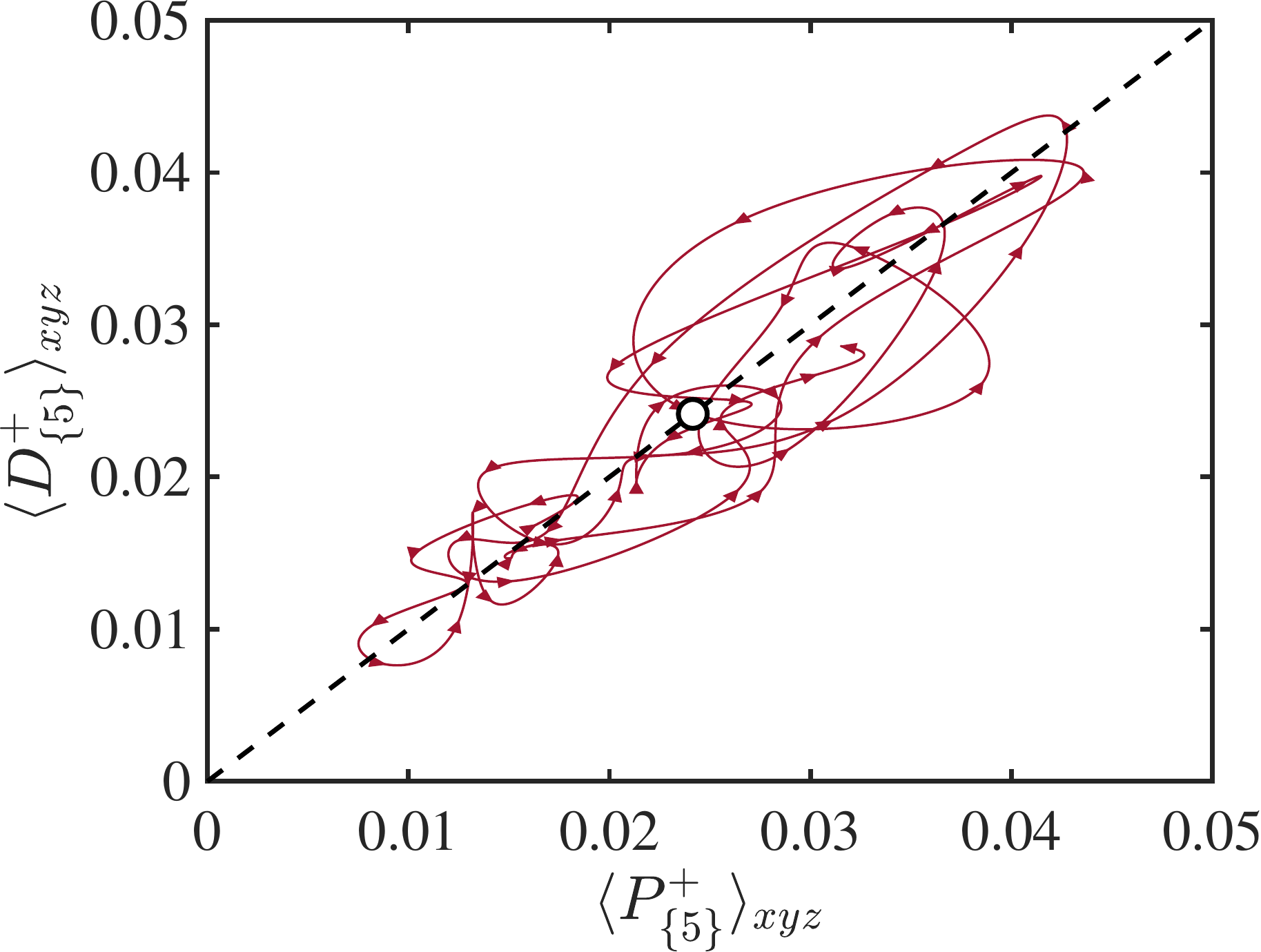}}   
  \subfloat[]{\includegraphics[width=0.31\textwidth]{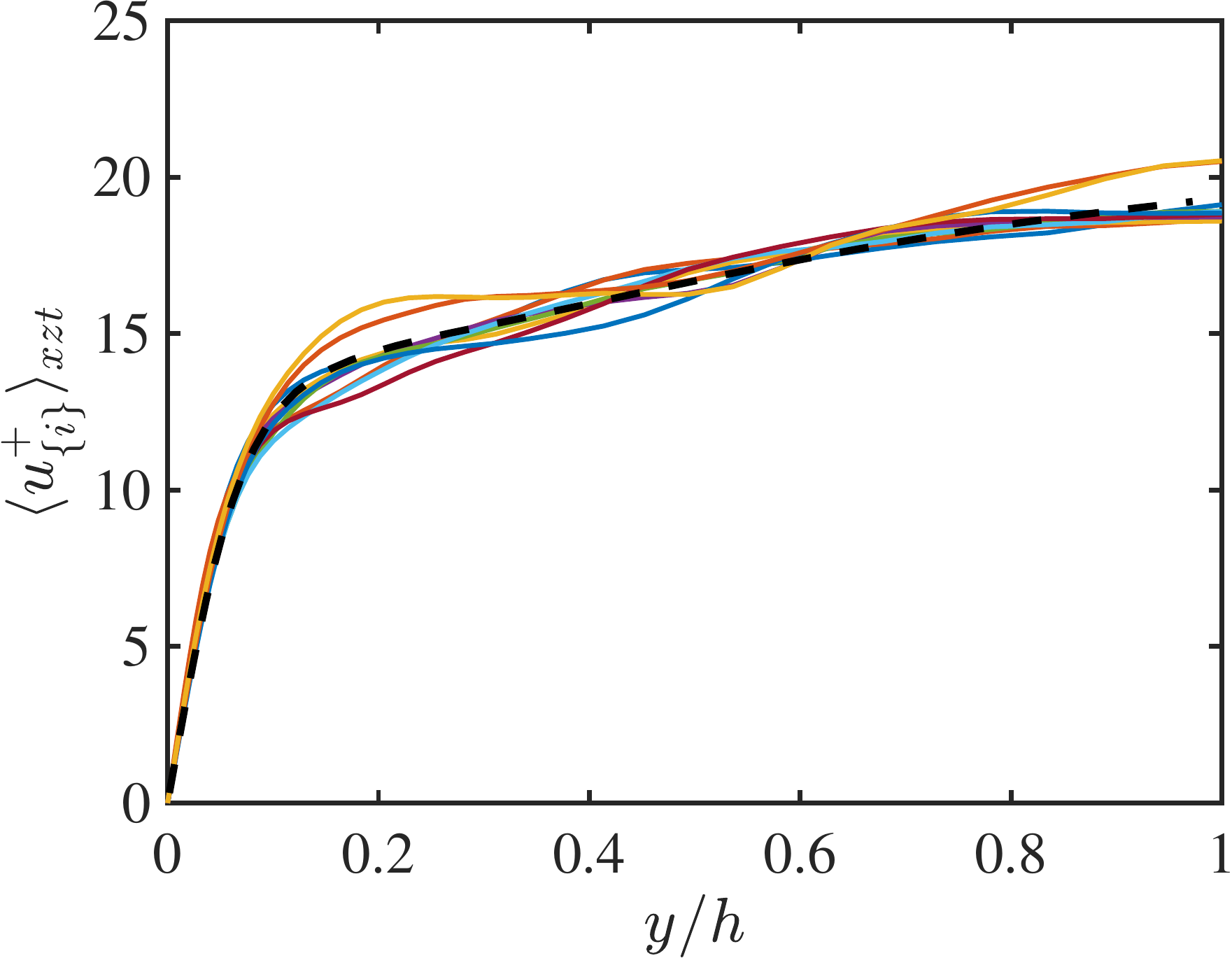}} 
  \end{center}
  \begin{center}
  \subfloat[]{\includegraphics[width=0.32\textwidth]{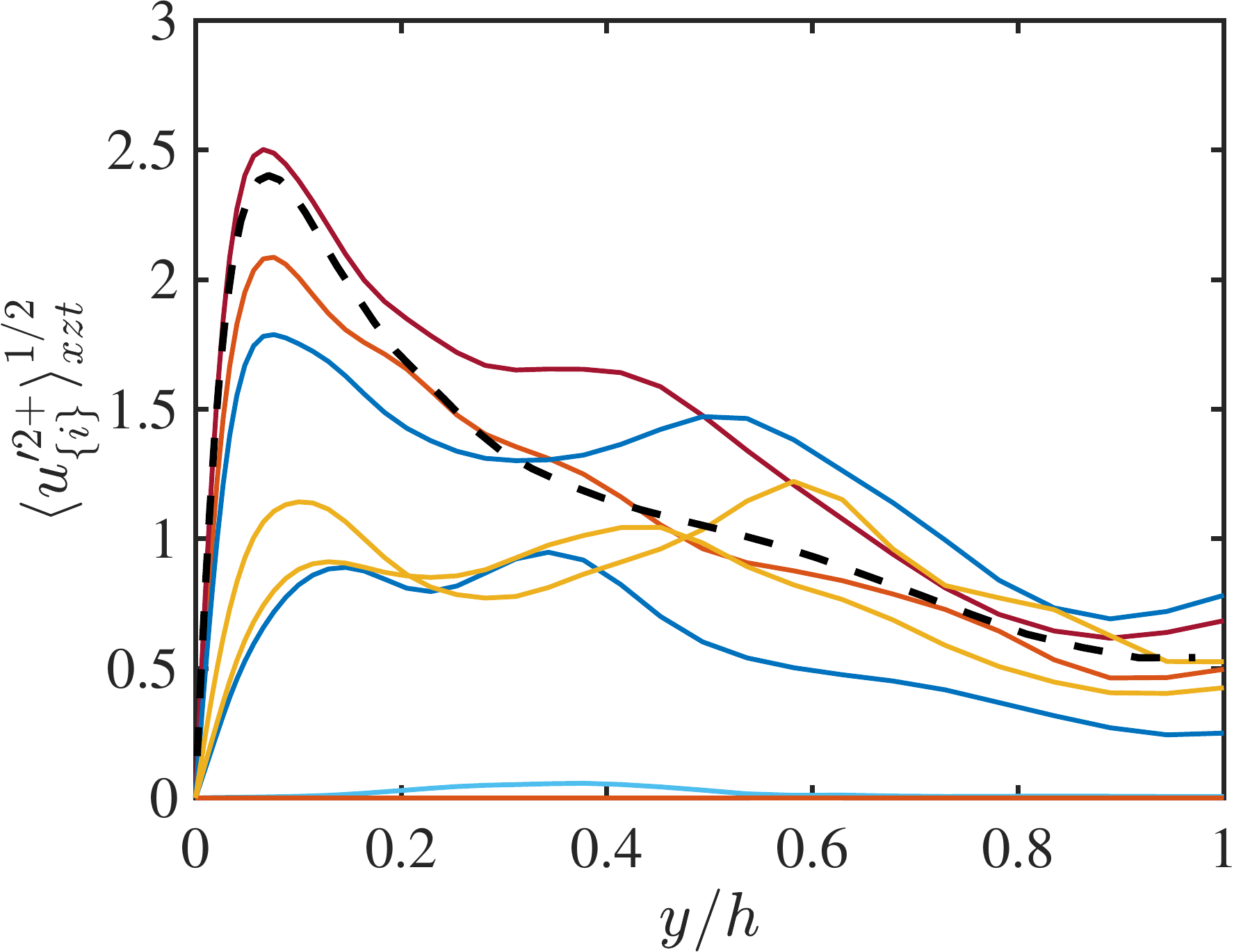}}  
  \subfloat[]{\includegraphics[width=0.32\textwidth]{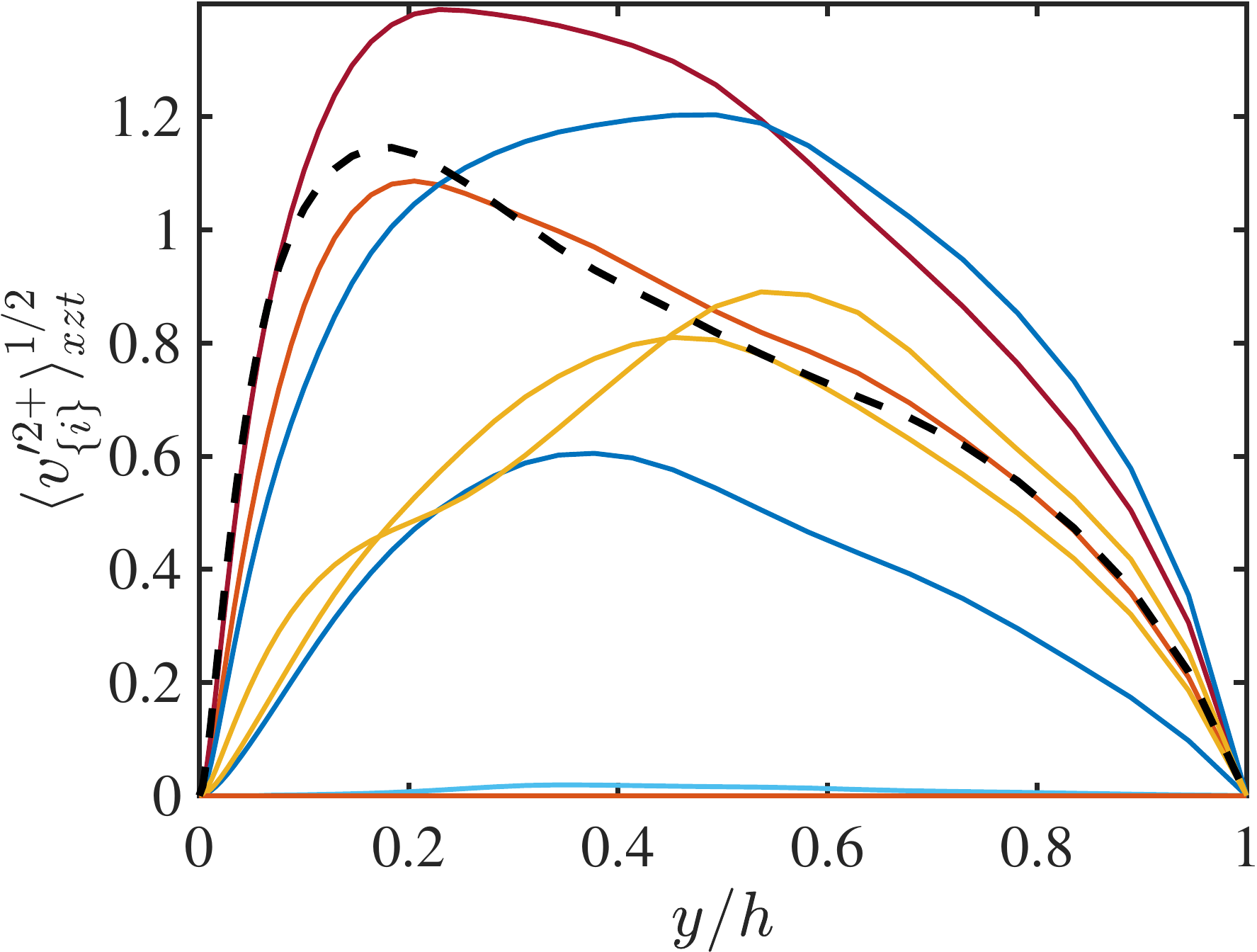}}  
  \subfloat[]{\includegraphics[width=0.32\textwidth]{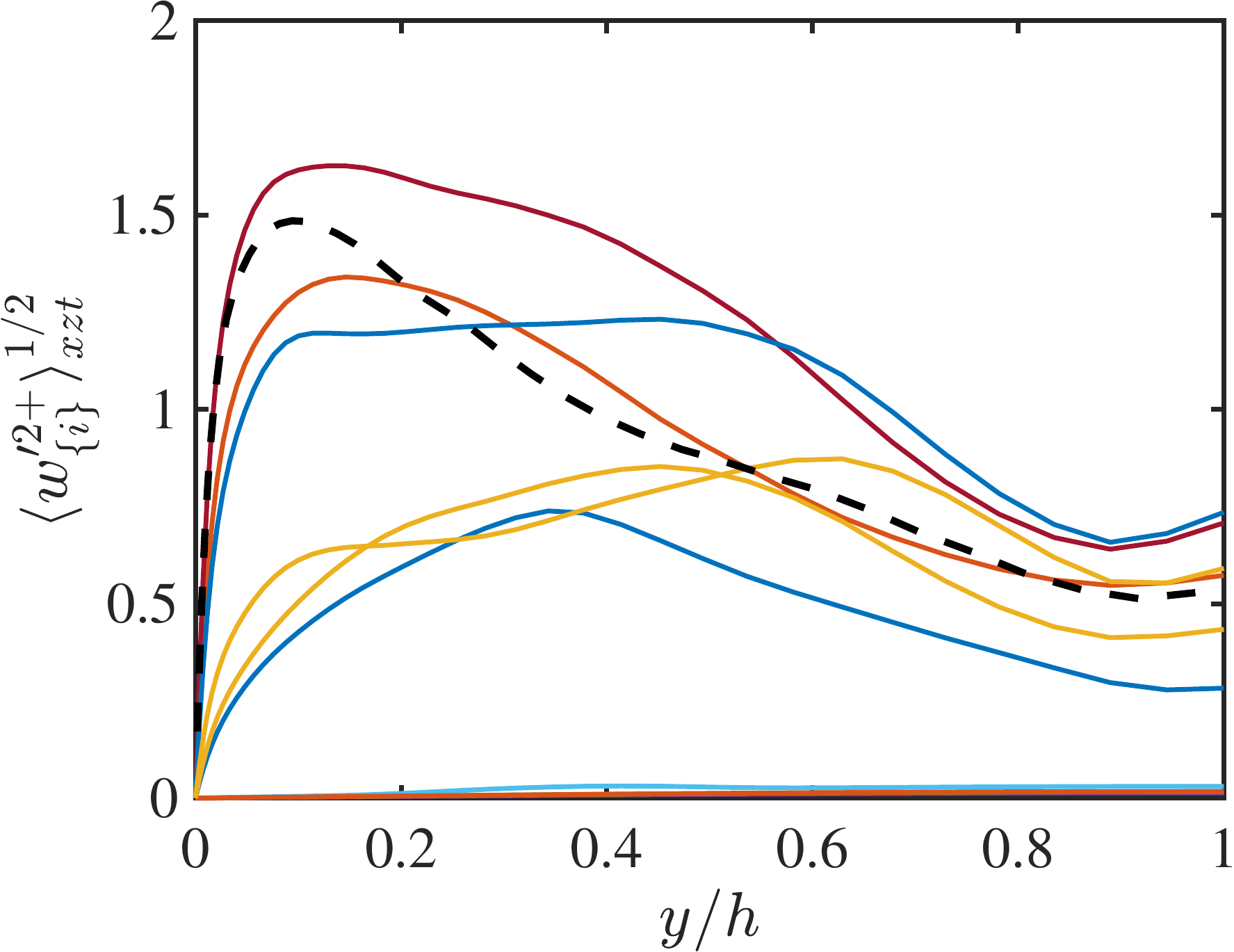}}  
  \end{center}
  \caption{ (a)~The history of the domain-averaged turbulent kinetic
    energy of the fluctuations $\langle E \rangle_{xyz}$.  Different
    colours denote various case of NF-TG180$_i$ for $i=1,...,10$. The
    time $t_i$ is the instant at which the mean flow is frozen in
    time. (b)~Projection of the flow trajectory onto the average
    production rate $\langle P_{\{5\}} \rangle_{xyz}$ and dissipation
    rate $\langle D_{\{5\}} \rangle_{xyz}$ plane for NF-TG180$_5$. The
    arrows indicate the time direction of the trajectory, which on
    average rotates counter-clockwise. The red dashed line is $\langle
    P_{\{5\}} \rangle_{xyz} = -\langle D_{\{5\}} \rangle_{xyz}$ and
    the red circle $\langle P_{\{5\}} \rangle_{xyzt} = -\langle
    D_{\{5\}} \rangle_{xyzt}$. The trajectory projected covers $15
    h/u_\tau$ units of time. (c) Mean velocity profile, and (d)
    root-mean-squared streamwise, (e) wall-normal, and (f) spanwise
    fluctuating velocities for ten cases NF-TG180$_i$,
    $i=1,...,10$. The dashed line is for
    NF-SEI180.  \label{fig:stats_frozen_example} }
\end{figure}

The evolution of the turbulent kinetic energy is shown in
figure~\ref{fig:stats_frozen_example}(a) for ten cases NF-TG180$_i$,
$i=1,...,10$. After freezing the base flow at $t_i$, most of the cases
remain turbulent, while some others decay before $40 h/u_\tau$.
Turbulence was sustained in 80\% of the NF-TG180$_i$ simulations. In
the cases for which turbulence persists, the projection of the flow
trajectory onto the $\langle P \rangle_{xyz}$--$\langle D
\rangle_{xyz}$ plane is reminiscent of the self-sustaining cycle for
R180; an example is shown in figure~\ref{fig:stats_frozen_example}(b).
Since $\tilde{\mathcal{L}}_{\{i\}}$ is modally stable, a key
ingredient to sustain turbulence in NF-TG180$_i$ is the scattering and
generation of new disturbances by $\bN(\bu'_{\{i\}})$. Indeed, we
verify in Appendix~\ref{sec:appendix_linear} that the
system~\eqref{eq:NF-TG180} decays when the nonlinear term
$\bN(\bu'_{\{i\}})$ is discarded.
%

\begin{figure}
 \begin{center}
  \includegraphics[width=0.95\textwidth]{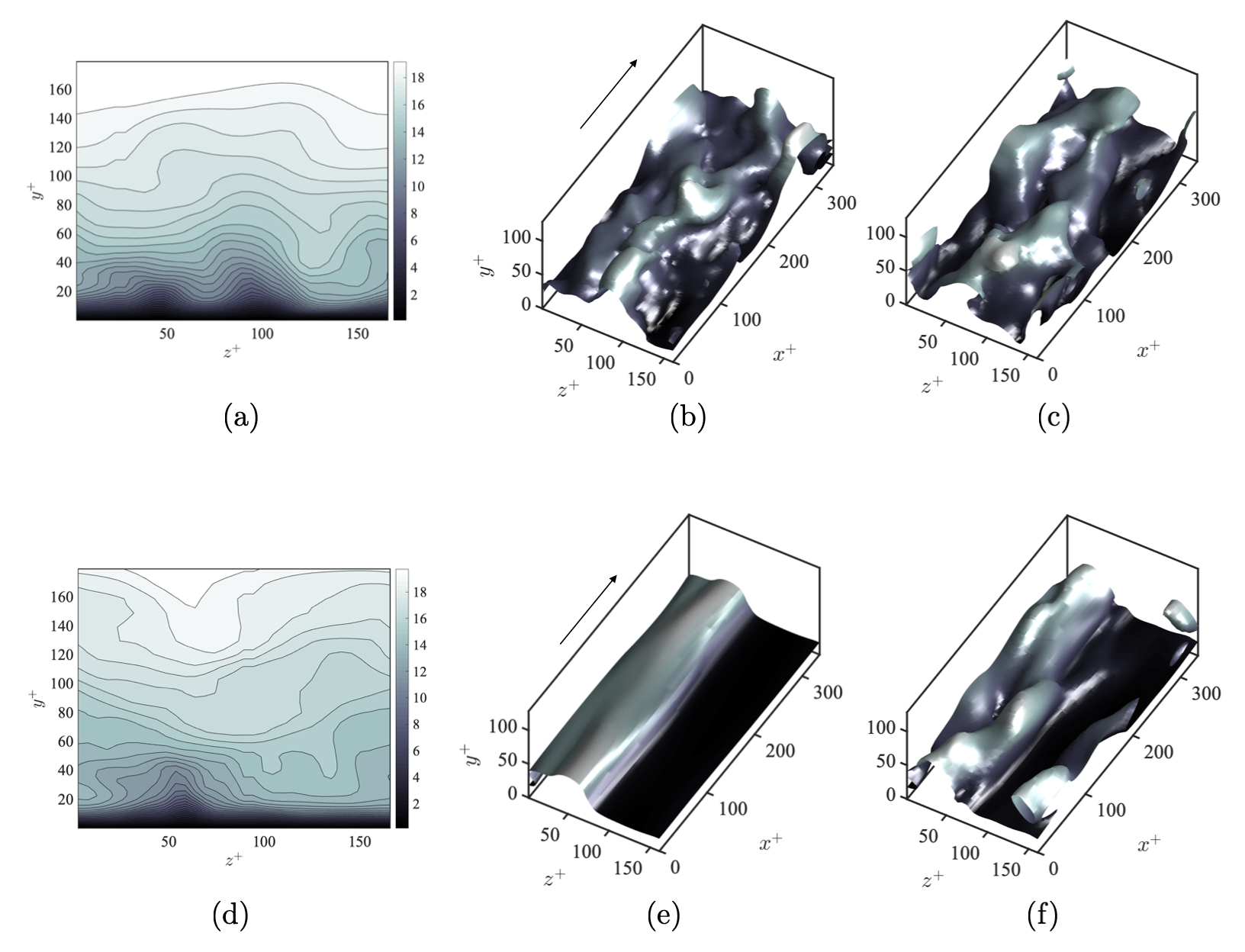} 
 \end{center}
\caption{ Examples of base flows (a,d) and instantaneous isosurfaces
  of the streamwise velocity at different times (b,c,e,f). Top panels
  are for NF-TG180$_5$, which is representative of a state with
  intensified turbulence intensities. Bottom panels are for
  NF-TG180$_{10}$, which is representative of a state with weakened
  turbulence. In panels (b,c,e,f), the value of the isosurfaces is
  0.65 of the maximum streamwise velocity and colours represent the
  distance to the wall located at $y=0$. The arrow indicates the mean
  flow direction.
 \label{fig:snapshots_TG180}}
\end{figure}

The one-point statistics for each NF-TG180$_i$ vary for different
$U(y,z,t_i)$.  To illustrate the differences among cases,
figures~\ref{fig:stats_frozen_example}(c,d,e,f) contain the mean
velocity profiles and fluctuating velocities for NF-TG180$_i$,
$i=1,...,10$. Note that this is only a small sample; the total number
of cases simulated is above 500. In some occasions, $U(y,z,t_i)$ is
such that the system equilibrates in a state of intensified turbulence
with respect to NF-SEI180 (i.e., $\langle E \rangle_{xyzt}$ for
NF-TG180$_i$ larger than for NF-SEI180), while other base flows result
in weakened turbulence. Figure~\ref{fig:snapshots_TG180} shows
instances of the streamwise velocity for representative cases with
intensified (top panels) and weakened (bottom panels) turbulent
states.  The intensified turbulence features a highly disorganised
state akin to a broken streak, whereas the weakened turbulence
resembles the quiescent stages of wall turbulence with a well-formed
persistent streak.
%
%
%

\begin{figure}
  \begin{center}
  \includegraphics[width=0.45\textwidth]{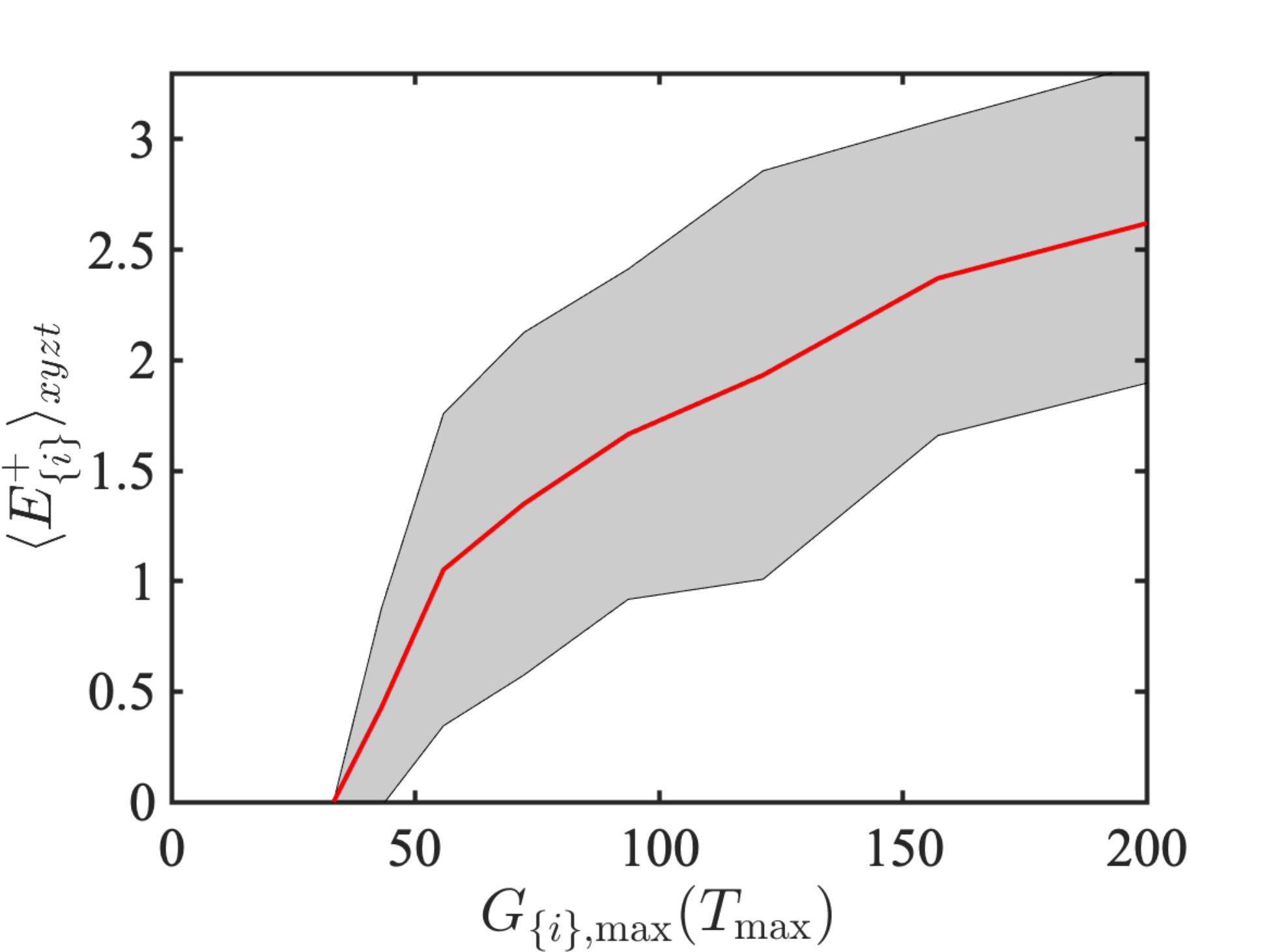} 
  \end{center}
 \caption{ Mean turbulent kinetic energy conditioned to the
   maximum gain $G_{{\{i\}},\mathrm{max}}$ at $T=T_{\mathrm{max}}$
   compiled over NF-TG180$_{i}$; (\textcolor{red}{\solid}) represents
   the mean value; the shaded area denotes $\pm$ one standard deviation.
 \label{fig:gain_frozen}}
\end{figure}

Figure~\ref{fig:gain_frozen} shows the average turbulent kinetic
energy of a given case NF-TG180$_i$ as a function of the maximum gain
$G_{\{i\},\mathrm{max}}$ at $T=T_\mathrm{max}$. The results reveal
that turbulence is not maintained when $G_{\{i\},\mathrm{max}}
\lessapprox 50$, although this critical gain might be Reynolds number
dependent. The trend also suggests that the level of
  the turbulence intensities for NF-TG180$_i$ increases with the
  amount of transient growth supported by $U(y,z,t_i)$ and scales
  approximately as,
\begin{equation}\label{eq:E_observation}
  \langle E_{\{i\}} \rangle_{xyzt} \sim G_{\{i\},\mathrm{max}}.
\end{equation}
We attempt to explain this observation by invoking the severe
assumption that $\bN(\bu'_{\{i\}})$ acts as a time-varying forcing
whose net effect is independent of $\bu'_{\{i\}}$,
i.e. $\bN(\bu'_{\{i\}}) \approx \bcalN_{\{i\}}(t)$ \citep[see, for
  instance][]{Farrell1993b, Zare2020, Jovanovic2020}. Under those
conditions, the solution to~\eqref{eq:NF-TG180_1} is obtained via the
Green's function as:
\begin{equation}\label{eq:approx_sol}
  \bu'_{\{i\}}(t) \approx
 \tilde\Phi_{\{i\}}(t-t_i)\bu'_{\{i\}}(t_i)
  + \int_{t_i}^t  \tilde\Phi_{\{i\}}(t-\tau)  \bcalN_{\{i\}}(\tau) \mathrm{d}\tau,
\end{equation}  
with $\tilde\Phi_{\{i\}}(t) = \exp \left[ \tilde{
    \mathcal{L}}_{\{i\}}(t)\right]$. The turbulent kinetic energy
of~\eqref{eq:approx_sol}, after transients, is
\begin{align}\label{eq:E_model_pre}
  \langle E_{\{i\}} \rangle_{xyzt} & = \frac1{2} \left\langle \Big ( \int_{t_i}^t
  \tilde\Phi_{\{i\}}(t-\tau) \bcalN_{\{i\}}(\tau) 
  \mathrm{d}\tau \, , \, \int_{t_i}^t \tilde\Phi_{\{i\}}(t-\tau) 
  \bcalN_{\{i\}}(\tau) \mathrm{d}\tau \Big ) \right\rangle_t.
\end{align}
After considering the singular-value decomposition on
$\tilde\Phi_{\{i\}}= M_{\{i\}} \Sigma_{\{i\}} N_{\{i\}}^\dagger$ then,
\begin{equation}\label{eq:E_model}
  \langle E_{\{i\}} \rangle_{xyzt} \sim  \sigma_{\{i\},\mathrm{max}}^2 \sim G_{\{i\},\mathrm{max}},
\end{equation}
which establishes a link between the level of turbulent kinetic energy
and the non-normal energy gain provided by the linear dynamics as
anticipated by figure~\ref{fig:gain_frozen}. Nonetheless, the scatter
of the data in figure~\ref{fig:gain_frozen} is still large and the
relation between $\langle E_{\{i\}} \rangle_{xyzt}$ and
$G_{\{i\},\mathrm{max}}$ is not perfectly linear. This is not
surprising as the actual mechanism regulating the intensity of
turbulence does not depend exclusively on $G_{\{i\},\mathrm{max}}$ but
also on the replenishment of fluctuations given by the projection of
$\bN(\bu'_{\{i\}})$ onto $\tilde\Phi_{\{i\}}$.
%

\begin{figure}
  \begin{center}
  \subfloat[]{\includegraphics[width=0.45\textwidth]{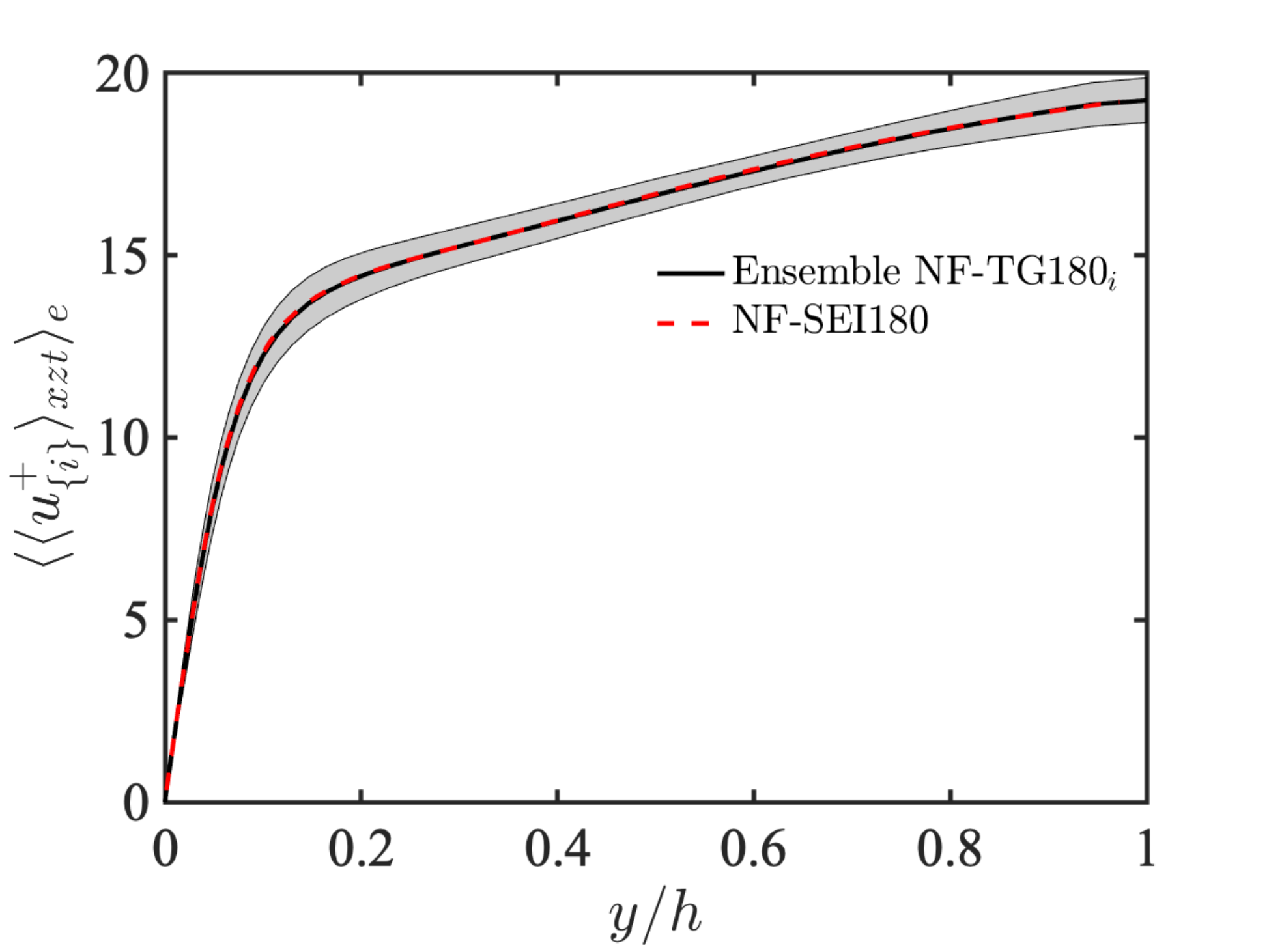}} 
  \hspace{0.05cm}
  \subfloat[]{\includegraphics[width=0.45\textwidth]{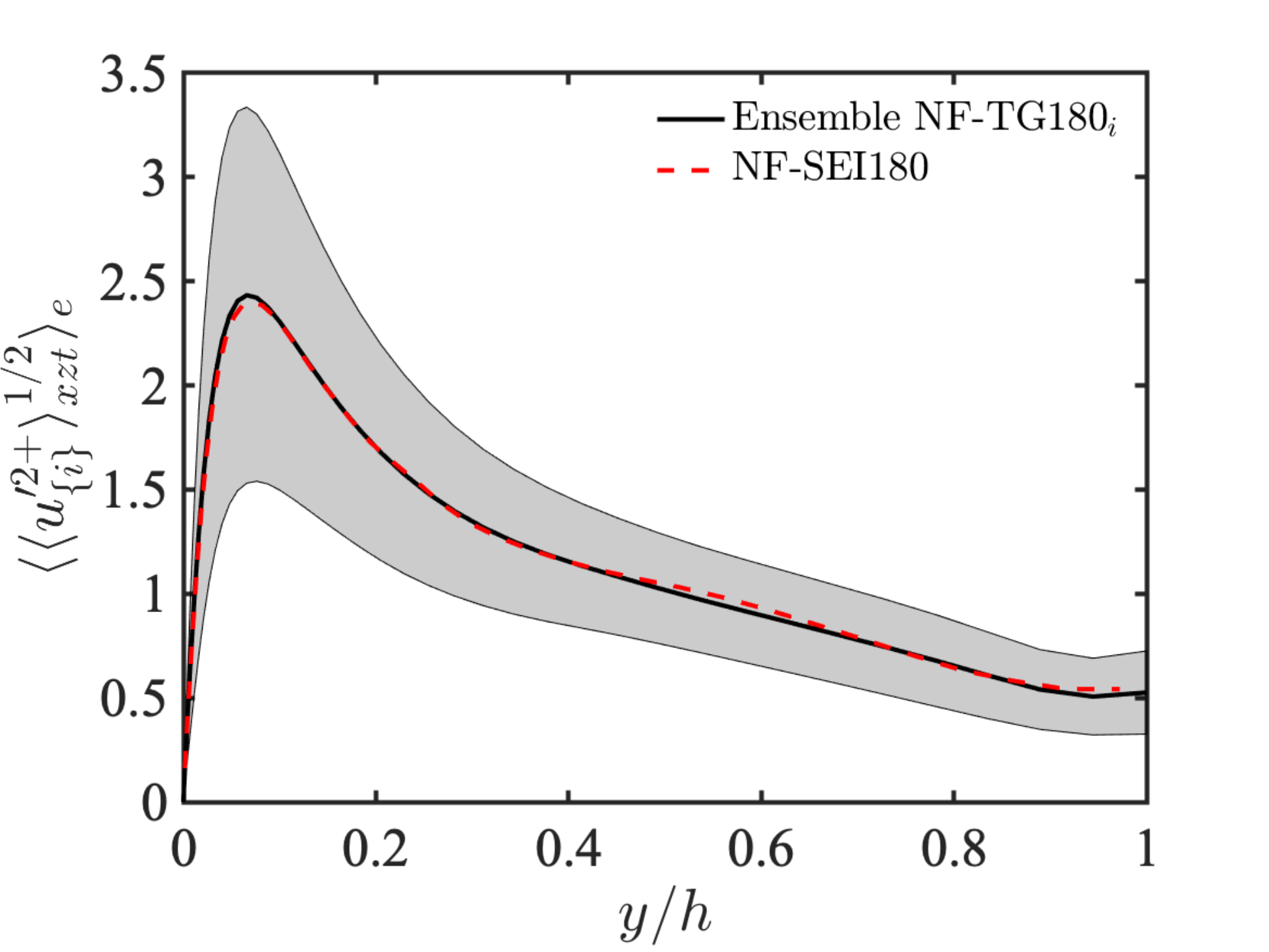}} 
  \end{center}
  \begin{center}
  \subfloat[]{\includegraphics[width=0.45\textwidth]{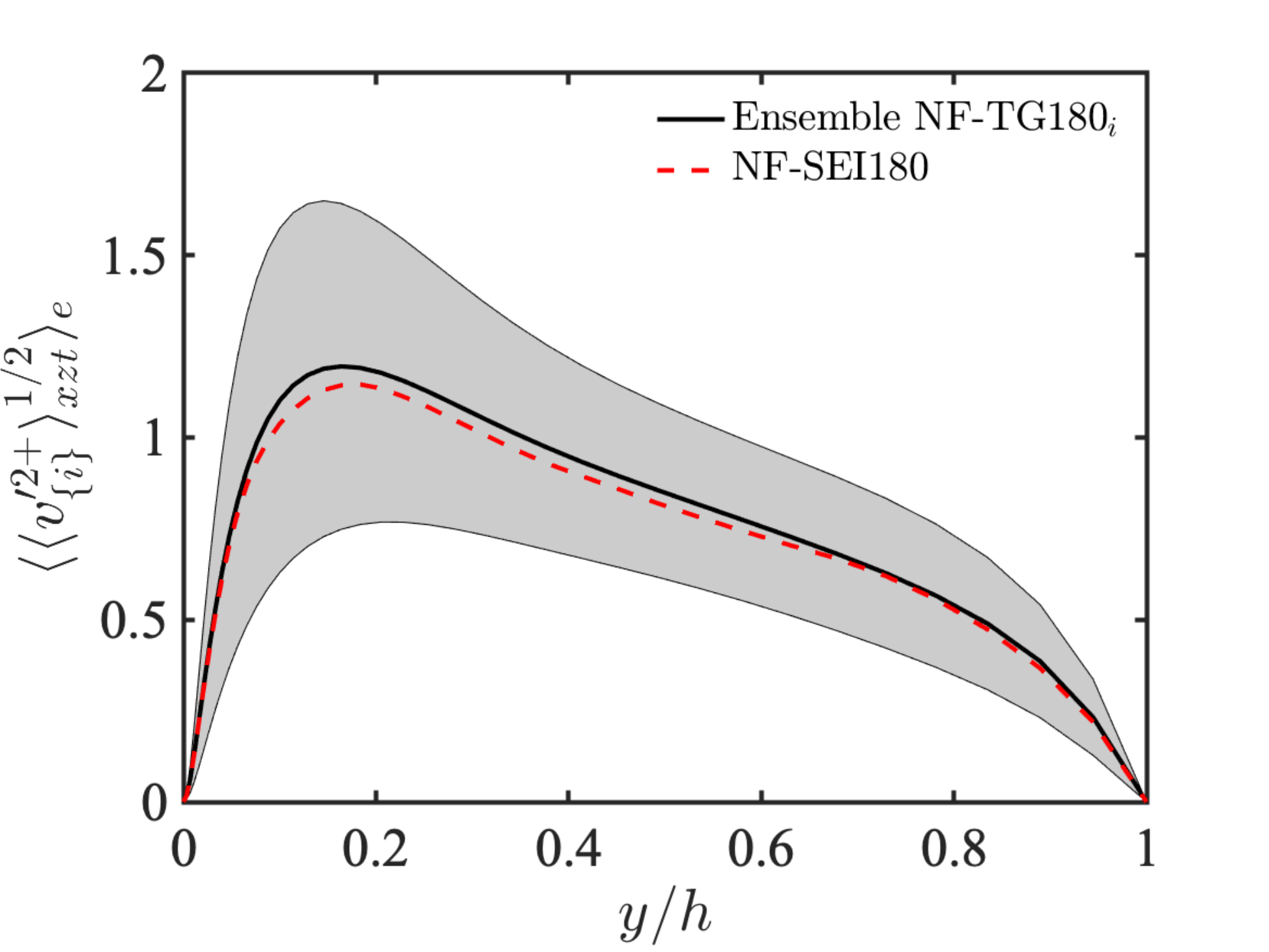}} 
  \hspace{0.05cm}
  \subfloat[]{\includegraphics[width=0.45\textwidth]{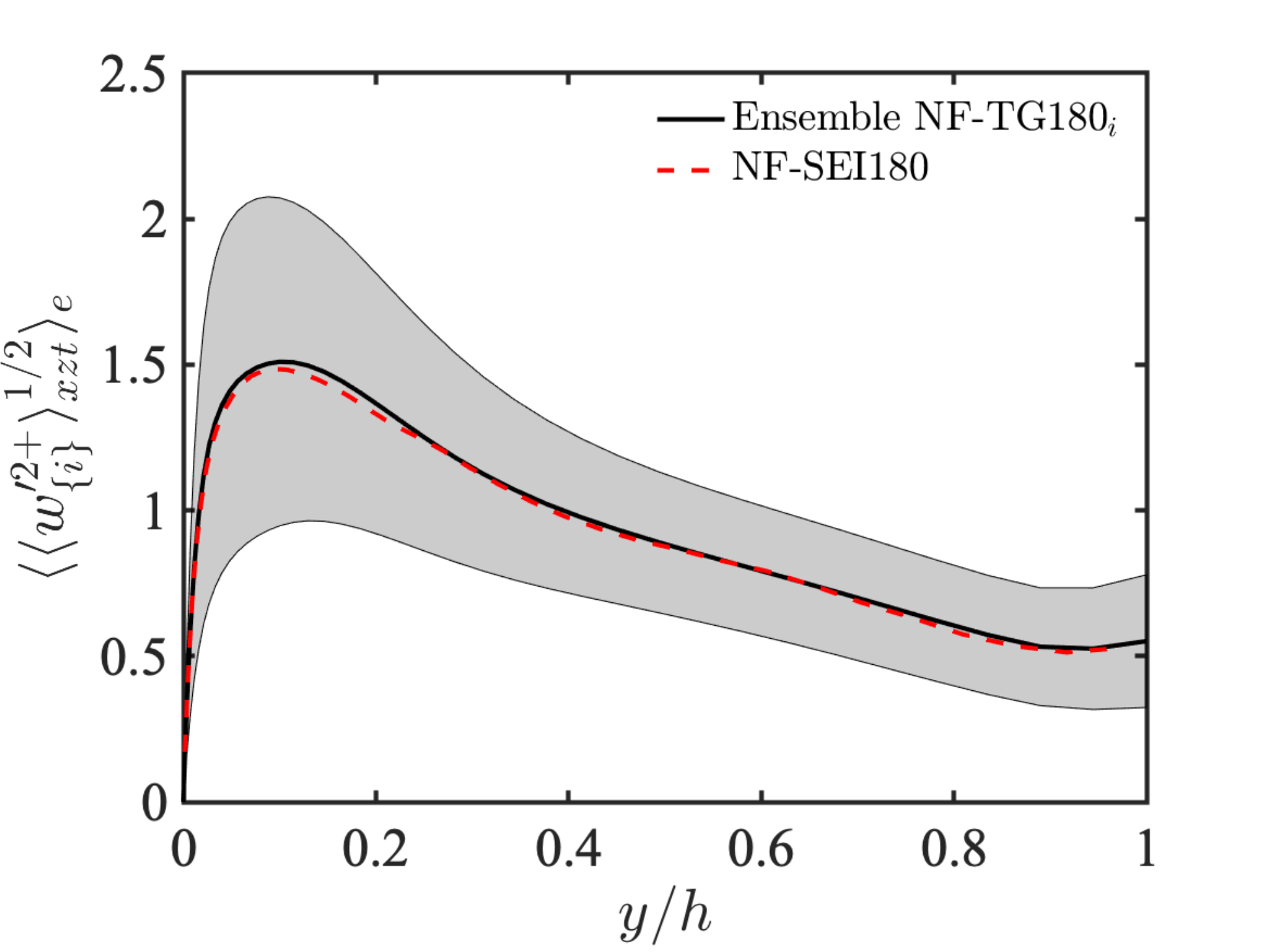}} 
  \end{center}
 \caption{ (a) Mean velocity profile, (b) root-mean-squared
   streamwise, (c) wall-normal, and (d) spanwise fluctuating
   velocities: (\solid), the ensemble average of turbulent cases
   NF-TG180$_{i}$, namely, $\langle \langle u_{\{i\}} \rangle_{xzt}
   \rangle_e$, $\langle \langle u'^2_{\{i\}} \rangle_{xzt}^{1/2}
   \rangle_e$, $\langle \langle v'^2_{\{i\}} \rangle_{xzt}^{1/2}
   \rangle_e$, and $\langle \langle w'^2_{\{i\}} \rangle_{xzt}^{1/2}
   \rangle_e$; the shaded region denotes $\pm$ one standard deviation with
   respect to the ensemble average operator $\langle \ \cdot
   \ \rangle_e$; (\textcolor{red}{\dashed}) is $\left \langle u'^2
   \right\rangle_{xzt}^{1/2}$, $\left \langle v'^2
   \right\rangle_{xzt}^{1/2}$, and $\left \langle w'^2
   \right\rangle_{xzt}^{1/2}$ for NF-SEI180.  \label{fig:stats_frozen}
 }
\end{figure}

To evaluate the compound result of NF-TG180$_i$, we define the
ensemble average of a quantity $\phi_{\{i\}}$ over cases NF-TG180$_i$
as
\begin{equation}
  \langle \phi_{\{i\}} \rangle_e = \sum_{i=1}^N \frac{\phi_{\{i\}}}{N},
\end{equation}
where ${1,...,N}$ is the collection of cases NF-TG180$_i$ which remain
turbulent.  The ensemble average of the mean and rms fluctuating
velocities are presented in figure~\ref{fig:stats_frozen}. The results
are compared with those from NF-SEI180, which is similar to
NF-TG180$_i$ but with time-varying $U$. The outcome is striking: the
ensemble averages over NF-TG180$_i$ cases (black solid lines) coincide
almost perfectly with the one-point statistics for NF-SEI180 (dashed
red lines). Given that the current setup is composed of `static' base
flows, $\partial U/\partial t$ ($=0$) does not play any role in the
flow dynamics of NF-TG180$_i$. Thus, we conclude that energy transfer
via parametric instabilities (intimately related to $\partial
U/\partial t$) is not required to sustain the flow.  Time-variations
of $U$ are only necessary to sample the phase space of `regular'
turbulence with different non-normal gains so that the ensemble of
NF-TG180$_i$ results in nominal wall turbulence statistics.
%

The wall-normal behaviour of the turbulence intensities for
NF-TG180$_i$ are determined by the fluctuation energy balance
\begin{equation}\label{eq:balance_i}
  \left\langle \bu'_{\{i\}} \cdot \tilde{\mathcal{L}}_{\{i\}}\bu'_{\{i\}} + \bu'_{\{i\}} \cdot  \bN(\bu'_{\{i\}}) \right\rangle_{xzt} = 0.
\end{equation}
Similarly, the average turbulent kinetic energy for NF-SEI180 is
dictated by the balance
\begin{equation}\label{eq:balance_NF-SEI180}
\left\langle \bu' \cdot \tilde{\mathcal{L}}(U) \bu' + \bu' \cdot \bN(\bu') \right\rangle_{xzt}= 0,
\end{equation}
where $\tilde{\mathcal{L}}(U)$ and $\bu'$ are now the linear operator
and velocity vector, respectively, for case NF-SEI180. The excellent
agreement between NF-SEI180 and the ensemble average over NF-TG180$_{i}$
suggests that
\begin{eqnarray}\label{eq:equivalence}
 \left\langle  \bu'_{\{i\}} \cdot \tilde{\mathcal{L}}_{\{i\}}
  \bu'_{\{i\}} \right\rangle_{xzte} &\approx &
  \left\langle \bu' \cdot \tilde{\mathcal{L}} \bu' \right\rangle_{xzt}.
\end{eqnarray}
An interpretation of~\eqref{eq:equivalence} (and of
figure~\ref{fig:stats_frozen}) is that the collection of linear
transient-growth events due to frozen $U(y,z,t_i)$ at different
instances $t_i$ provides an accurate representation of the actual
time-varying energy transfer from $\bU$ to $\bu'$ in NF-SEI180.  From
a dynamical-systems viewpoint: the sampling of the phase-space under
the time-varying $U$ is statistically equivalent to an ensemble
average of solutions in equilibrium with frozen instances of $U$. This
is an indication that the nonlinear dynamics supported by $\bN$ are in
quasi-equilibrium with $\mathcal{L}(U)$, i.e., the way the energy is
input into the system changes slowly in time. The latter argument can
be posed in terms of the time scales of the base flow $T_U$ and
turbulent kinetic energy $T_E$. Defining $T_U$ as the time at which
the auto-correlation of $U$ (see (\ref{eq:corr})) decays to 0.5
(similarly for $T_E$ from the auto-correlation of $E$), the ratio
$T_U/T_E$ is found to be roughly 10. Therefore, changes in the base
flow are ten times slower than changes in the turbulent kinetic
energy, which is consistent with the discussion above.

As a final note, in a preliminary work \cite{Lozano_madrid_2020}
noticed that turbulent channel flows decayed when freezing the base
flow, which may initially seem inconsistent with the present results.
However, a main difference is that in the present work we are imposing
the base flow from actual wall turbulence (R180), while
\cite{Lozano_madrid_2020} imposed a base flow from modified
turbulence. The statistical sample used in the present work is also
far larger than that used by \cite{Lozano_madrid_2020}.

\subsection{Distilling the transient growth mechanisms}
\label{sec:TG_more}

We have shown above that transient growth is the simplest linear model
to explain self-sustaining turbulence.  In this section we further
dissect the relevance of different transient growth mechanisms and the
implications for the streaky structure of the base flow. We turn out
attention back to (\ref{eq:NF180}), which can be written as
\begin{subequations}
  \label{eq:linear_details}
  \begin{gather}
    \frac{\p u'}{\p t} = -U_0\frac{\p u'}{\p x} -v'\frac{\p U_0}{\p y} - w'\frac{\p U_0}{\p z} - \frac{1}{\rho}\frac{\p p'}{\p x} + \nu\nabla^2 u' + N_{u'}, \label{eq:linear_details_1}\\
\frac{\p v'}{\p t} = -U_0\frac{\p v'}{\p x} -\frac{1}{\rho}\frac{\p p'}{\p y} + \nu\nabla^2 v'  + N_{v'}, \label{eq:linear_details_2}\\
\frac{\p w'}{\p t} = -U_0\frac{\p w'}{\p x} -\frac{1}{\rho}\frac{\p p'}{\p z} + \nu\nabla^2 w'  + N_{w'}, \label{eq:linear_details_3}\\
    0 = \frac{\p u'}{\p x} + \frac{\p v'}{\p y}+\frac{\p w'}{\p z},\label{eq:linear_details_4}
  \end{gather}
\end{subequations}
where we have explicitly expanded the linear components, and $N_{u'},
N_{v'}$ and $N_{w'}$ stand for the remaining nonlinear terms. The
baseflow is $U_0 = U(y,z,t)$ from case R180 and no feedback from
$\bu'$ to $U$ is allowed.  Equations (\ref{eq:linear_details}) allow
for exponential and parametric instabilities, but we have shown that
these are inconsequential for sustaining the flow. Hence, we admit the
possibility of both instabilities for the sake of reducing the
computational cost of solving (\ref{eq:linear_details}).

We define the streak flow by $U_{\mathrm{streak}} = U_0(y,z,t) - \langle
u \rangle_{xz}$, and proceed to examine three transient growth mechanisms:
\begin{itemize}
\item[i)] Linear lift-up of the streak by $-v' \p U_{\mathrm{streak}}/\p y$ in (\ref{eq:linear_details_1}).
\item[ii)] Linear push-over of the streak by $-w'\p U_{\mathrm{streak}}/\p z$ in (\ref{eq:linear_details_1}).
\item[iii)] Linear Orr of the streak by $-\p p'/\p y$ in (\ref{eq:linear_details_2})
  and continuity in (\ref{eq:linear_details_4}).
\end{itemize}
In mechanisms i) and ii), velocities perpendicular to the base-shear
$(0, \p U_0/\p y, \p U_0/\p z)$ extract energy from the latter to
energise the streamwise perturbations, which persist after transients
\citep{Ellingsen1975}. For perturbations in the form of $v'$, the
active linear term for energy transfer is $-v' \p
U_{\mathrm{streak}}/\p y$, which is referred to as lift-up effect.
For perturbations in the form of $w'$, the energy is transferred
through $-w'\p U_{\mathrm{streak}}/\p z$. We label the latter as
`push-over effect' to make a clear distinction from the lift-up
mechanism, as one relies on spanwise shear of the base flow and the
other on the wall-normal shear. The terminology varicose and sinuous
is commonly used to refer to the perturbations from mechanism i) and
mechanism ii), respectively.  In mechanism iii), velocities
perpendicular to the base-shear are amplified when backwards-leaning
perturbations are tilted forward until they are roughly normal to the
base-shear, and are damped as they continue to be tilted past that
point. Mechanisms i) and ii) occur concurrently with mechanisms iii),
but the amplification in the latter is guided by continuity. In
mechanisms iii), the pressure inhibits the cross-shear velocities when
the structures are strongly tilted, and releases the inhibition when
they are closer to vertical~\citep{Orr1907, Jimenez2013,
  Chagelishvili2016}.

We perform three additional experiments each aiming to suppress one of
the linear mechanisms discussed above. In the first experiment, we
modify (\ref{eq:linear_details_1}) to suppress the linear lift-up of
the streak,
\begin{equation}
  \frac{\p u'}{\p t} = -U_0\frac{\p u'}{\p x} - \cancel{v'\frac{\p U_{\mathrm{streak}}}{\p y}} - v'\frac{\p \langle u \rangle_{xz} }{\p y}
  - w'\frac{\p U_{\mathrm{streak}}}{\p z} -\frac{1}{\rho}\frac{\p p'}{\p x} + \nu\nabla^2 u' + N_{u'}, \label{eq:linear_details_1_mod1}
\end{equation} 
while the equations for $v'$, $w'$ and continuity remain intact. We
labelled this case as NF-NFU180 (no-feedback \& no-lift-up). The second
experiment consists of a channel without linear push-over mechanism of
the streak,
\begin{equation}
  \frac{\p u'}{\p t} = -U_0\frac{\p u'}{\p x} - v'\frac{\p U_{\mathrm{streak}}}{\p y} - v'\frac{\p \langle u \rangle_{xz} }{\p y}
  - \cancel{w'\frac{\p U_{\mathrm{streak}}}{\p z}} -\frac{1}{\rho}\frac{\p p'}{\p x} + \nu\nabla^2 u' + N_{u'}, \label{eq:linear_details_1_mod2}
\end{equation} 
which is referred to as NF-NPO180 (no-feedback \& no-push-over). Again
the equations for $v'$, $w'$ and continuity remain the same. We might
note here that both modal and non-modal growth share the physical
source of the fluctuation amplification: $w \partial U/\partial z$ for
sinuous motions and $v\partial U/\partial y$ for varicose motions.
Hence, by modifying the latter terms we are also interfering with the
modal growth of the system, although as argued above we do not worry
about this in this section.

In the third experiment, we modify the linear dynamics of $v'$ to
constrain mechanism iii). The equation dictating the linear Orr is
given by
\begin{equation}
  \frac{\partial v'_{\mathrm{linear}}}{\partial t} + U_0 \frac{\partial
    v'_{\mathrm{linear}}}{\partial x} = -\frac{1}{\rho}\frac{\partial
    p'_{\mathrm{linear}}}{\partial y}, \label{eq:Orr_lin}
\end{equation} 
where we have neglected the viscous effects.  As seen in
(\ref{eq:Orr_lin}), the Orr amplification of $v'_{\mathrm{linear}}$ is
controlled by $p'_{\mathrm{linear}}$, which is the only source term in
the right-hand side of the equation. The linear pressure can be easily
obtained by solving
\begin{equation}
\frac{1}{\rho}\nabla^2 p'_{\mathrm{linear}} = -2 \frac{\p U_0}{\p y} \frac{\p v'}{\p x} - 2\frac{\p U_0}{\p z} \frac{\p w'}{\p x}.
\end{equation} 
If we focus on the linear pressure induced by the streak, then
\begin{equation}
  \frac{1}{\rho}\nabla^2 p'^{s}_{\mathrm{linear}} = -2 \frac{\p U_\mathrm{streak}}{\p y} \frac{\p v'}{\p x}
  - 2\frac{\p U_\mathrm{streak}}{\p z} \frac{\p w'}{\p x}.
\end{equation} 
We inhibit the linear Orr mechanism of the streak by introducing a
forcing term $f_{\mathrm{Orr}}$ in the right-hand-side of the $v'$
equation to counteract the gradient of the linear pressure
\begin{equation}
  f_{\mathrm{Orr}} = -\frac{1}{\rho}\frac{\p p'^{s}_{\mathrm{linear}} }{\p y}.
\end{equation} 
Then, the system
(\ref{eq:linear_details_1})-(\ref{eq:linear_details_4}) is modified by
replacing (\ref{eq:linear_details_2}) by
\begin{equation}\label{eq:linear_details_2_mod}
\frac{\p v'}{\p t} = -U_0\frac{\p v'}{\p x} -\frac{1}{\rho}\frac{\p p'}{\p y} + \nu\nabla^2 v'  + N_{v'} - f_{\mathrm{Orr}}.
\end{equation} 
We refer to this case as NF-NO180 (no-feedback \& no-Orr). 

\begin{figure}
  \begin{center}
  \subfloat[]{\includegraphics[width=0.45\textwidth]{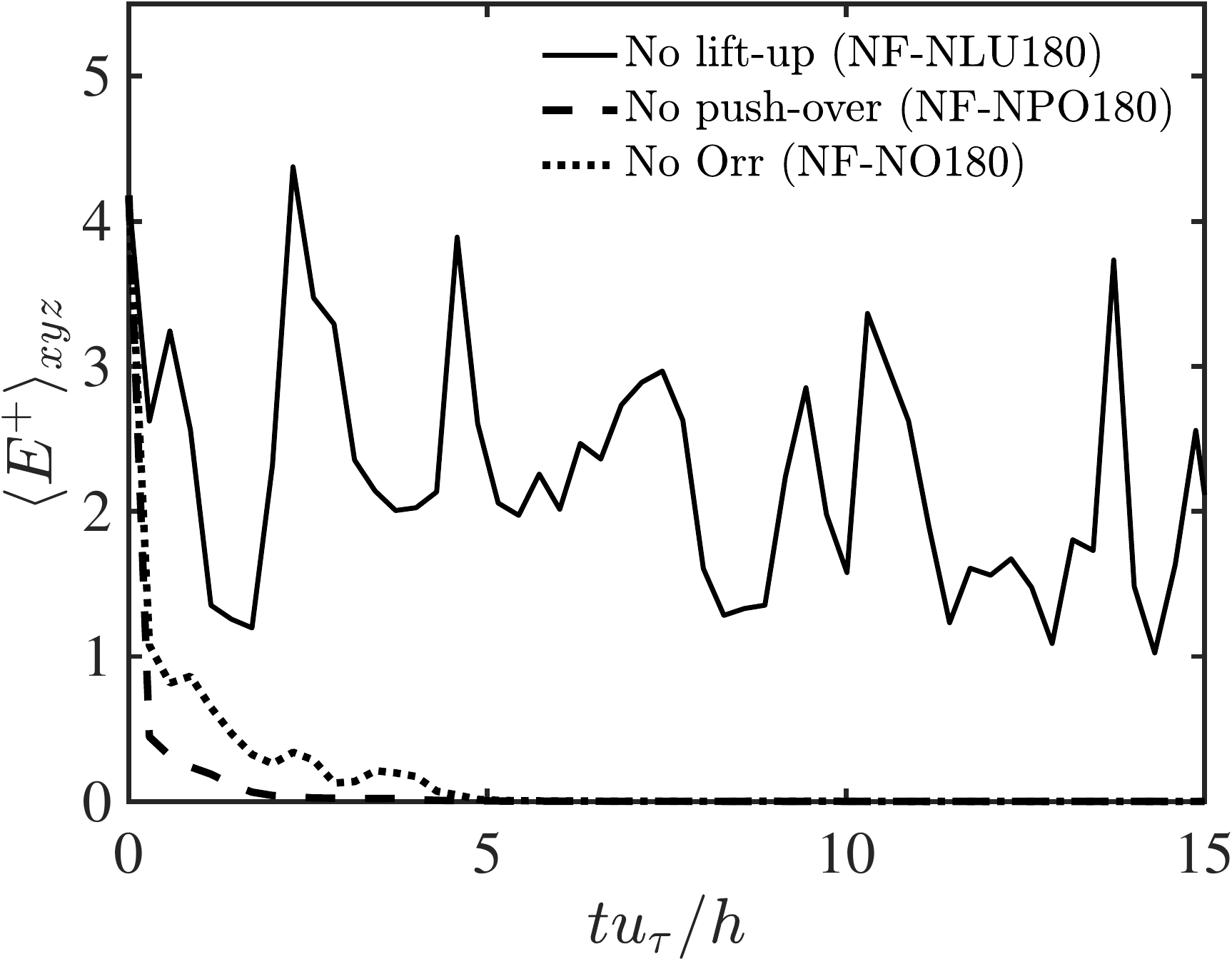}} 
  \hspace{0.5cm}
  \subfloat[]{\includegraphics[width=0.45\textwidth]{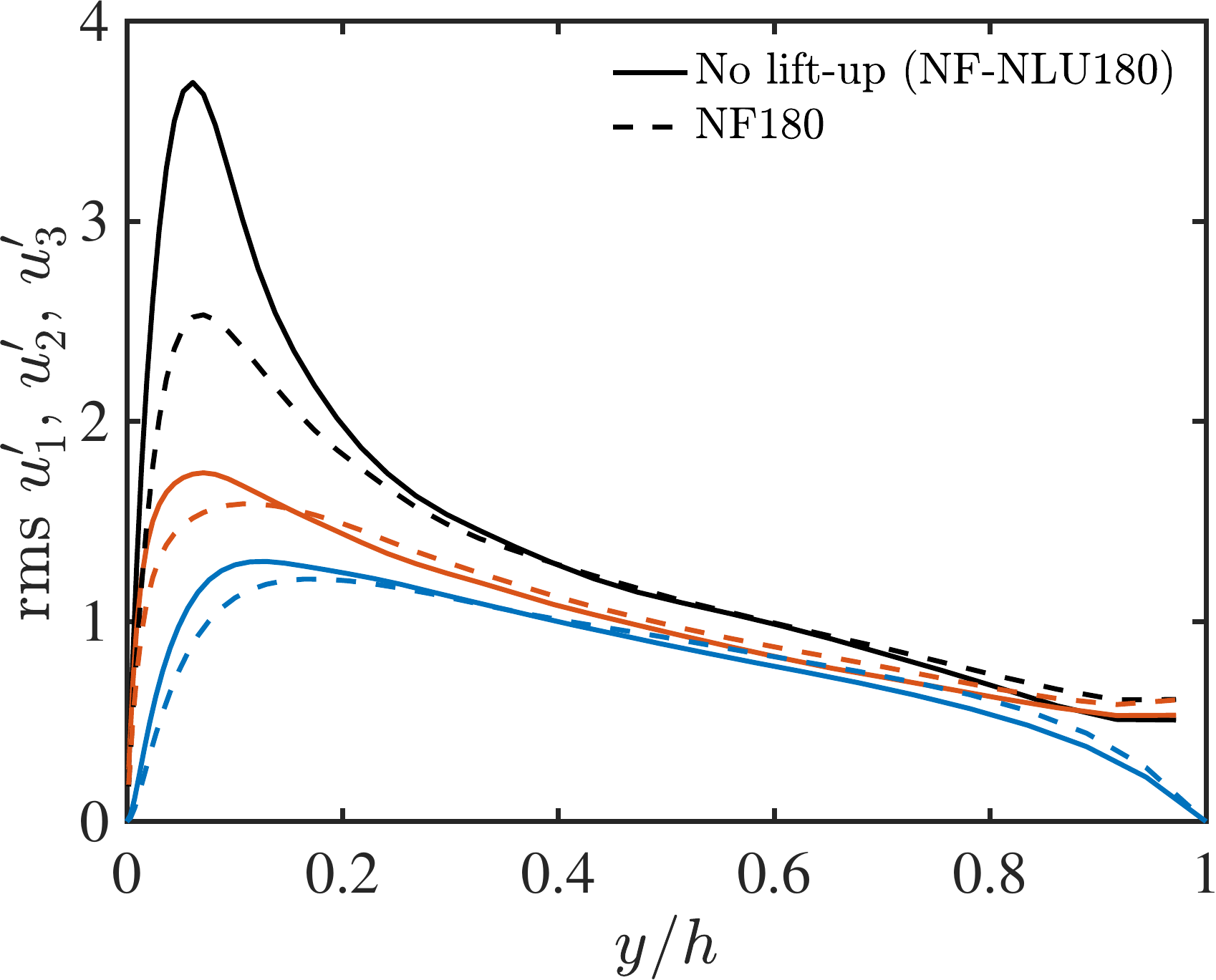}} 
  \end{center}
  \caption{ (a)~The history of the domain-averaged turbulent kinetic
    energy of the fluctuations $\langle E \rangle_{xyz}$ for NF-NLU180
    (\solid); NF-NPO180 (\dashed); and NF-NO180 (\dotted). The time
    $t=0$ is the instant at which the simulation are started.  (a) The
    root-mean-squared streamwise (\solid and \dashed), wall-normal
    (\textcolor{myBlue}{\solid} and \textcolor{myBlue}{\dashed}), and
    spanwise (\textcolor{myRed}{\solid} and
    \textcolor{myRed}{\dashed}) fluctuating velocities for NF-NLU180
    (solid lines) and NF180 (dashed lines).\label{fig:no}}
\end{figure}

The three cases are initialised using flow fields from NF180.
Different initial conditions were tested and the subsequent evolution
of the flow was similar regardless of the details of the initial
velocity.  The evolution of the turbulent kinetic energy for one
initialisation is shown in figure~\ref{fig:no}(a).  The case without
streak lift-up (NF-NLU180) is the only one sustained. The rms velocity
fluctuations after transients for NF-NLU180 are shown in
figure~\ref{fig:no}(b). Interestingly, blocking the streak lift-up
enhances $u'_1$ close to the wall, implying that the wall-normal
variations of the streak provide a stabilising effect.  Conversely,
the cases without streak push-over (NF-NPO180) or Orr mechanisms
(NF-NO180) decay in less than $5 u_\tau/h$. Therefore we conclude that
both streak push-over and Orr amplification are essential transient
growth mechanisms for sustaining the fluctuations. Three more cases
analogous to NF-NLU180, NF-NPO180, and NF-NO180 were conducted by
allowing the feedback of the fluctuations back into the base flow.
The conclusions drawn for these cases are similar to the ones
presented above, and they are not included here for the sake of
brevity.
%

\begin{figure}
  \begin{center}
  \subfloat[]{\includegraphics[width=0.45\textwidth]{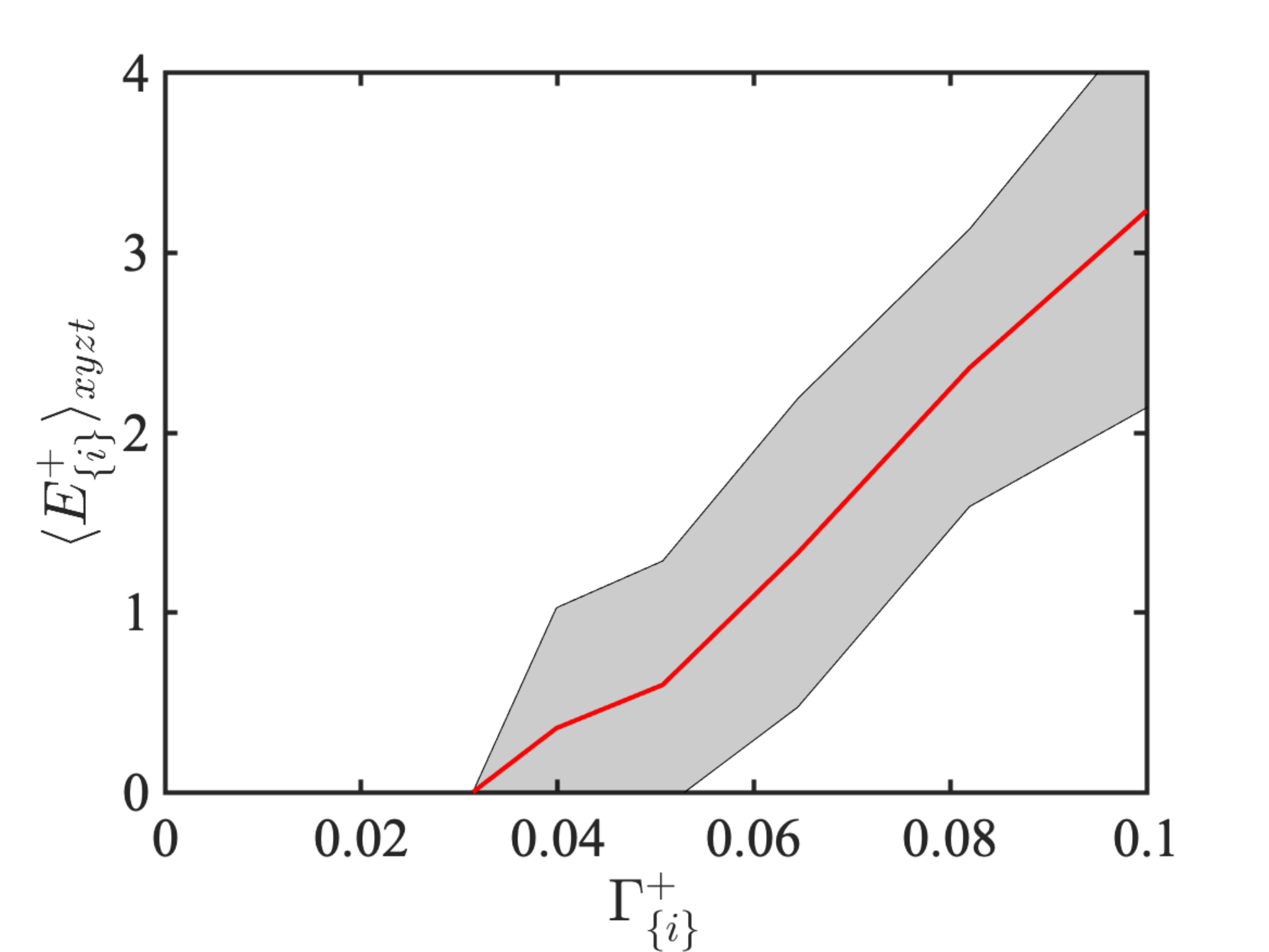}} 
  \hspace{0.5cm}
  \subfloat[]{\includegraphics[width=0.45\textwidth]{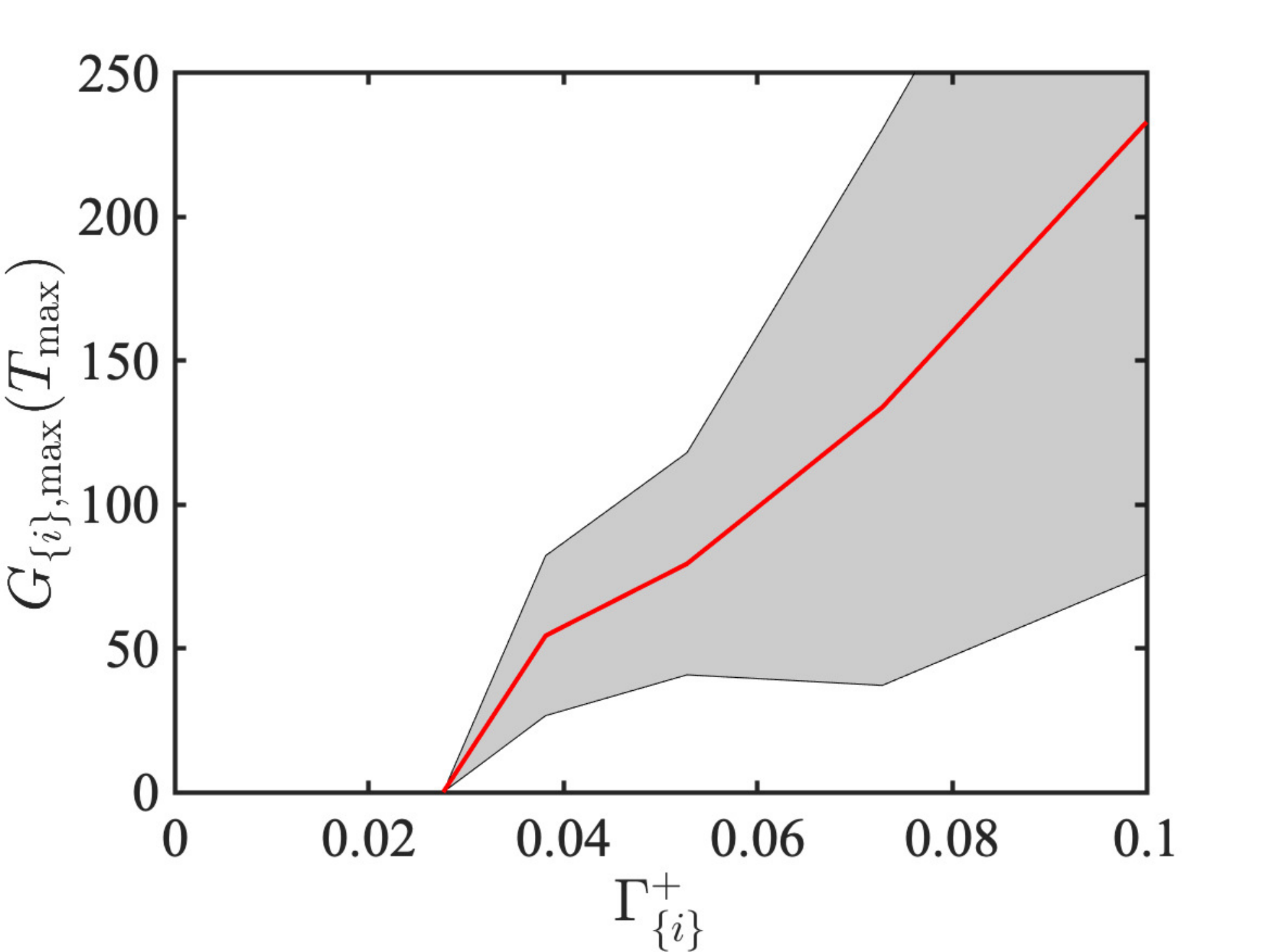}} 
  \end{center}
  \caption{ (a) Mean turbulent kinetic energy and (b) maximum gain
    $G_{{\{i\}},\mathrm{max}}$ at $T=T_{\mathrm{max}}$ conditioned to
    the marker for the spanwise-shear strength $\Gamma_{\{i\}} =
    \langle \left( \p U_0(y,z,t_i)/\p z \right)^2 \rangle_{yzt}^{1/2}$
    compiled over NF-TG180$_{i}$; (\textcolor{red}{\solid}) represents
    the mean value; the shaded area denotes $\pm$ one standard
    deviation.  \label{fig:dUdz}}
\end{figure}

\begin{figure}
  \begin{center}
  \includegraphics[width=1\textwidth]{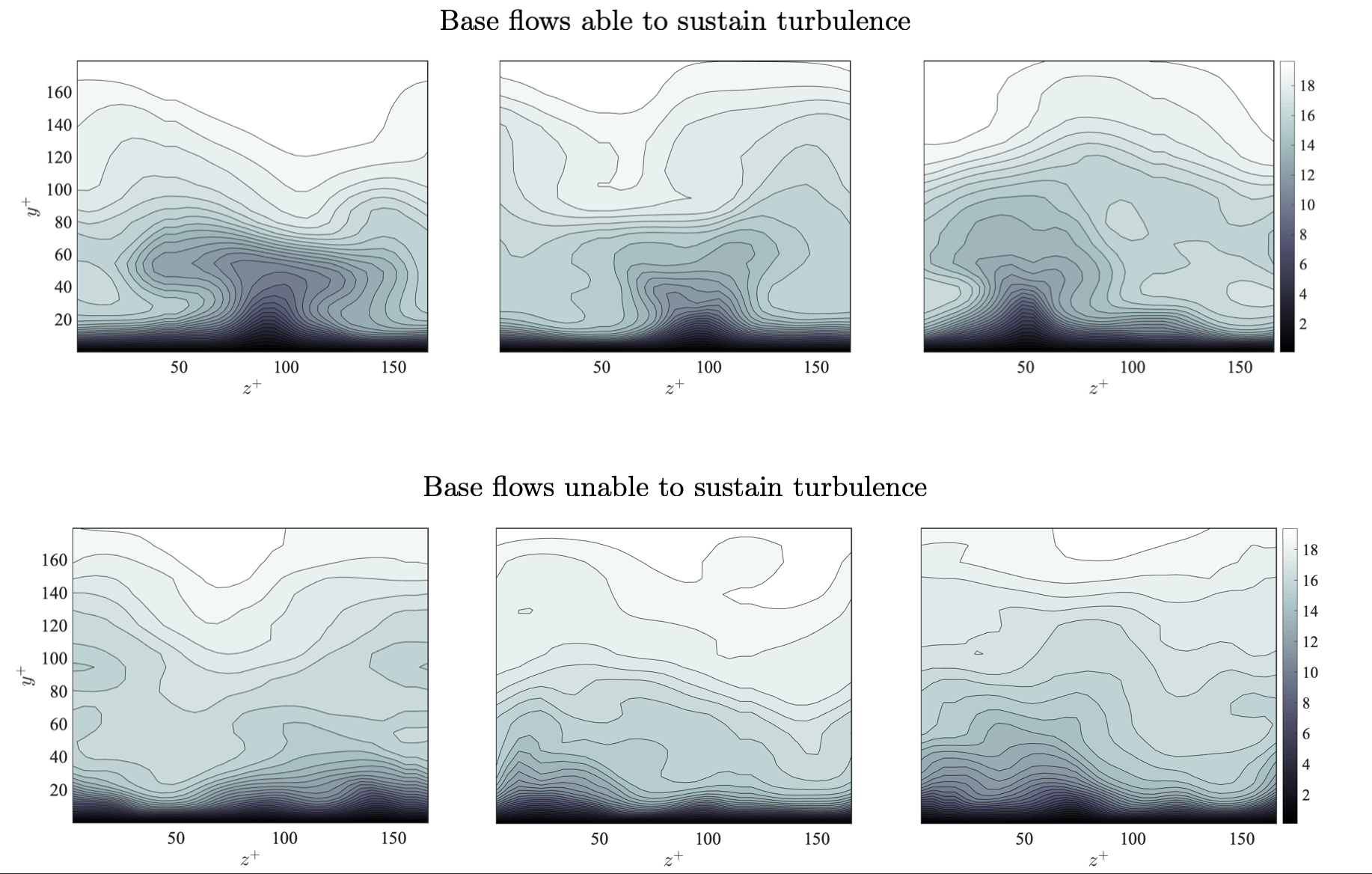} 
  \end{center}
  \caption{ Examples of base flows able to support turbulence (three
    top panels) and unable to support turbulence (three bottom panels)
    from cases NF-TG180$_{i}$ discussed in \S \ref{subsec:TG}. Visual
    inspection of the base flows suggests that spanwise variations of
    $U$ are critical to sustain turbulence. \label{fig:dUdz_examples}}
\end{figure}

The necessity of push-over ($-w'\p U_\mathrm{streak}/\p z$) points at
the spanwise variations of $U_\mathrm{streak}$ (equivalent to spanwise
variations of $U_0$) as a key structural feature to sustain
turbulence. To test this claim, we resort to the cases NF-TG180$_{i}$
presented in \S \ref{subsec:TG} and inspect their mean turbulent
kinetic energy conditioned to the marker for the spanwise-shear
strength
\begin{equation}
\Gamma_{\{i\}} = \left\langle \left( \frac{\p
U_\mathrm{streak}(y,z,t_i)}{\p z} \right)^2 \right\rangle_{yzt}^{1/2}.
\end{equation}
The results, shown in figure~\ref{fig:dUdz}(a), corroborate the
hypothesis that the average kinetic energy of the flow depends on the
strength of $\p U_\mathrm{streak}(y,z,t_i)/\p z$. Additionally, frozen
base flows with $\Gamma_{\{i\}}$ below the critical value of
$\Gamma_{\{i\},c}^+ \approx 0.03$ are too feeble to maintain
turbulence.  This is further supported by figure~\ref{fig:dUdz}(b),
which shows that the maximum gain of the base flow also increases with
$\Gamma_{\{i\}}$. A visual impression of base flows that are either
able or unable to sustain turbulence can be gained from the examples
shown in figure~\ref{fig:dUdz_examples}. The message conveyed by
figure~\ref{fig:dUdz_examples} agrees with the discussion above: base
flows capable of sustaining turbulence are accompanied by strong
spanwise variations, $\p U_\mathrm{streak}/\p z$, while base flows
unable to maintain turbulence have milder $\p U_\mathrm{streak}/\p z$.
%
%

\subsection{Relation with previous studies on secondary linear process}
\label{sec:literature}

Among the studies in table \ref{table:literature} regarding secondary
linear process, those labelled as TG, TG \& EXP, and TG PARA advocate
for the algebraic transient growth of the perturbations as a central
linear mechanism to energise the velocity fluctuations. The work by
Farrell, Ioannou, and co-workers complemented the transient growth
picture with parametric growth of the perturbations resulting from the
time-variability of the base flow. In their view, the growth of
fluctuations is a concatenation of transient growth events that occur
as the base flow varies. We have shown here that time-changes in the
base flow are needed to recover regular turbulence statistics;
however, these time-changes do not enhance the energy transfer from
the base flow to the fluctuations and, thus, they are not strictly
necessary to sustain the flow. The works labelled as TG \& EXP deemed
both transient growth and modal growth relevant for sustaining wall
turbulence, as they are intimately entangled in the equations of
motion. On the contrary, \cite{Schoppa2002} argued that modal
instability was irrelevant to streak breakdown, and that transient
growth driven by the streak profile was the dominant process. The
mechanism in \cite{Schoppa2002} was referred to as STG (Streak
Transient Growth, i.e., secondary linear process) to make a clear
distinction with the more traditional transient growth supported by
$\langle u \rangle_{xzt}$ (i.e., primary linear process).  In this
sense, our conclusions are more aligned with \cite{Schoppa2002}, and
transient growth is identified here to overcome other linear
mechanisms. \citet{Schoppa2002} also analysed the physical process at
play during transient growth in terms of vortex dynamics. They
formulated the problem in the streak-vortex-line coordinate system,
that procures a clear interpretation of the perturbation vorticity
generation. They found that $w'$ perturbations of moderately low
amplitude lead to generation of new vortices and sustained near-wall
turbulence via the `shearing' mechanism. The latter results are also
consistent with our analysis in \S~\ref{sec:TG_more}, where we showed
that the push-over mechanism (represented by $-w' \partial
U_{\mathrm{streak}} /\partial z$) drives the generation of
perturbations. \cite{Schoppa2002} traced back the source of
fluctuations to $\partial \langle u \rangle_{xzt} /\partial n$, where
$n$ is the direction normal to the base-flow vortex lines and includes
contributions of both $\partial \langle u \rangle_{x} /\partial y$ and
$\partial \langle u \rangle_{x} /\partial z$. Here, we have further
demonstrated that the spanwise shear $\partial \langle u \rangle_{x}
/\partial z \equiv \partial U_{\mathrm{streak}} /\partial z$ dominates
over $\partial U_{\mathrm{streak}} /\partial y$.

Despite our main conclusions being consistent with \cite{Schoppa2002},
the causal analysis followed in the present study is fundamentally
different from that adopted in previous works. This is an important
point, as our approach allows to tackle the criticism raised by the
community on transient growth as the prevailing mechanism for the
secondary linear process. In the remainder of this section, we survey
previous criticisms and discuss how our contributions overcome
existing deficiencies. First, it is pertinent to clarify that
transient growth due to high non-normality of the linearised
Navier--Stokes operator is an intrinsic feature of the equations of
motion. Transient growth represents the transport of momentum normal
to the base shear, which is a necessary requirement for mixing of the
flow and hence for turbulence. Therefore, the debate in the community
is not about the ubiquity of transient growth in turbulence, but about
the necessity of additional sorts of instabilities to sustain the
fluctuations (as shown in table \ref{table:literature}).

The scepticism on the ability of transient growth alone to fuel the
generation of $\bu'$ has materialised in different forms:
\begin{enumerate}
\item \cite{Jimenez2018} pointed out that the absence of unstable
  streaks reported by \cite{Schoppa2002} can be interpreted as an
  indication that the instability is important. Quoting
  \cite{Jimenez2018}: ``One may think of the low probability of finding
  upright pencils on a shaking table. Unstable flow patterns would not
  be found precisely because instability destroys them.''
\item \cite{Jimenez2018} also argued that transient growth as
  envisioned by \cite{Schoppa2002} is a property of the streak itself,
  implying that the energy of $\bu'$ is drawn from the energy of
  $U_{\mathrm{streak}}$. The premise is reasonable close to the wall,
  where $1/2\bu'\cdot \bu'$ is a small fraction of streak energy
  $1/2U_{\mathrm{streak}}^2$, but it becomes problematic farther from
  the wall, where both energies are comparable. If the transverse
  velocities had to obtain their energy from the streak, one would
  expect a negative correlation between the two energies, but the
  opposite seems to be true.  Instead, \cite{Jimenez2018} suggested
  that the actual source of energy for $\bu'$ would come from $\langle
  u \rangle_{xzt}$, which is also part of the base flow in
  \citet{Schoppa2002}.
\item Other authors have reasoned, as mentioned above, that it is
  essentially impossible to distinguish between streak transient
  growth and streak modal instability, as both processes can be traced
  back to the same source term in the linearised Navier--Stokes
  equations, namely, $-w' \partial \langle u \rangle_{x} /\partial z -
  v' \partial \langle u \rangle_{x} /\partial y$ \citep{Hopffner2005,
    Giovanetti2017, Cassinelli2017}. Consequently, both transient
  growth and modal instability occur concurrently during the streak
  breakdown and both should be considered responsible for replenishing
  $\bu'$.
\item Another criticism comes from the effect of time-varying base
  flows.  \cite{Schoppa2002} investigated the effect of unfrozen
  streaks on modal instabilities and transient growth. They showed
  that unfrozen (freely diffusing) streaks are still able to support
  transient growth with amplifications of the order of 10, whereas the
  initially unstable base-flow provided only a factor of 2.  However,
  the analysis was performed on freely decaying streaks, while streaks
  in actual wall turbulence are subjected to periods of both growth
  and decay.  Precluding the growth phase of the streaks might have
  important consequences for the growth of perturbations.  For
  example, \cite{Farrell2012} have shown that a potential route to
  enhance the gain for short times and/or achieve finite gains for
  long times is the parametric instability of the streak discussed in
  the introduction~\citep{Farrell1999, Farrell2012, Farrell2016}. In
  contrast with the freely diffusing base flow, alternating periods of
  growth and decay in the base flow can enhance the energy transfer
  from $\bU$ to $\bu'$ and should be taken into consideration.
\item Finally, some authors have criticised or at least found
  questionable the use of linear stability theory to analyse
  time-varying base-flows and base flows defined by an average (for
  example $\langle u \rangle_x$) rather than by a solution of the
  Navier--Stokes equations. Most of the successful and
  well-established results from linear stability theory have been
  derived in the case of laminar-to-turbulence transition, where the
  underlying assumptions are rigorously satisfied. This is obviously
  not the case for turbulent flows~\citep[see discussion
    in][]{Hussain1983, Hussain1986}. One conceptual objection is the
  fact that stability analysis of `frozen' turbulent profiles would be
  appropriate only if the time scales of the actual time-varying
  mean-flow were much smaller than those of the instabilities.
  However, some authors have pointed out that time-changes in the
  turbulent mean-flow could be of the same order as those of the
  instability wave. Thus, the instabilities do not `see' this mean
  flow and their evolution departs noticeably from that predicted from
  linear stability theory. Another recurrent objection is granting the
  status of `perturbations' (assumed to be small, e.g. $<0.1\%$), to
  the turbulent fluctuations, $\bu'$ (which might reach values above
  $10\%$ of the mean flow, especially close to the wall). Hence, the
  evolution of these (not-so-small) $\bu'$ `perturbations' is
  subjected to nonnegligible contributions from nonlinear
  interactions, which might invalidate the predictions from linear
  stability theory.
\end{enumerate}

We have addressed the criticism discussed above by formulating the
problem within a cause-and-effect framework. Whilst there is no such a
thing as a perfect methodology, we have argued that cause-and-effect
analysis entails a substantial leap in the study of turbulence
compared to non-causal analysis by post-processing data. Following the
same order that was used to introduce the criticism above:
\begin{enumerate}
\item We have contributed to settle the debate regarding the role
  played by modal instabilities by showing that these are not
  necessary to sustain wall turbulence. This was achieved by
  completely precluding the possibility of exponential growth from the
  linear Navier--Stokes operator at all times (see
  \S~\ref{subsec:nonmodal_Upred}).
\item We have shown that inhibiting the effect of $\partial
  U_{\mathrm{streak}}/ \partial z$ interrupts the self-sustaining
  cycle. Hence, the extraction of energy from the streak $\partial
  U_{\mathrm{streak}}/ \partial z$ is a necessary condition to
  maintain $\bu'$. Note that we do not imply that $\partial \langle u
  \rangle_{xzt}/ \partial y$ is inconsequential to sustaining
  turbulence, but that the spanwise variations of the streak are also
  an active participant in the self-sustaining cycle of turbulence
  (see \S~\ref{sec:TG_more}).
\item Despite the fact that modal and non-modal growth of the
  fluctuations originate from the same physical term in the
  Navier--Stokes equations, we have established a clear distinction
  between both mechanisms by manipulating the linear Navier--Stokes
  operator. We have shown that it is possible to block modal
  instabilities, while maintaining the transient growth mechanism
  almost intact (see \S~\ref{subsec:TG}).
\item Both post-processing analysis \S~\ref{sec:theories} and
  cause-and-effect analysis in \S~\ref{sec:constrain} comprise
  time-varying base flows which evolve in a realistic manner, as they
  are extracted from actual DNS data. As such, the evolution of the
  base flow experiences periods of both growth and decay consistent
  with actual wall turbulence, which allows for an accurate estimation
  of linear mechanism assisting the growth of $\bu'$.
\item The validity of linear theories for fully-developed turbulence
  is more subtle, and we have commented on this topic in
  \S~\ref{sec:introduction}. Paraphrasing the argument given in the
  introduction, writing the fluctuation equation in the form of
  (\ref{eq:dummy}) does not require invoking linearisation about $\bU$
  nor assuming that $\bu'$ is small.  If the volume integral of
  $\bu'\cdot \bN(\bu')$ is zero, then the only way of sustaining
  $\bu'$ is through the energy injection from $\bu' \cdot
  \mathcal{L}(\bU)\bu'$. Thus, for a partition of the flow $\bU +
  \bu'$, we can always refer to the linear mechanisms supported by
  $\mathcal{L}(\bU)$ and assess their relevance in sustaining
  turbulence regardless of how `good' is the linearisation. This is
  because our cause-and-effect analysis is conducted for the fully
  non-linear equations, instead of only for the linear
  component. Thus, when we inquire about the validity of a particular
  linearisation we are indeed asking about the usefulness of the
  partition $\bU + \bu'$ in explaining the dynamics of $\bu'$ via the
  linear mechanisms supported by $\bU$, which circumvents the problem
  of linearisation.
\end{enumerate} 

We close this section by discussing some discrepancies with
\cite{Schoppa2002}. The stability analysis conducted in
\S~\ref{subsec:theories_modal} reveals that our base flow $U(y,z)$ is
modally unstable 90\% of the time.  On the contrary,
\cite{Schoppa2002} found that most streaks have intensities that are
too low to be modally unstable.  It is unclear what is the root of
such a difference --it might be related to the synthetic base flow
used in \cite{Schoppa2002}, while here we used instantaneous base flow
from DNS, which are more corrugated and prone to modal
instabilities. Other possible explanations are the criteria used by
\cite{Schoppa2002} to quantify unstable streaks via a vorticity-based
inclination angle, or the use of minimal turbulent channel in the
present study. A second difference of our work with \cite{Schoppa2002}
comes from the value of the gains provided by transient growth. In
\S~\ref{subsec:theories_nonmodal}, we have shown that our gains are of
the order of 100, while the gains reported in \cite{Schoppa2002} are
of the order of 10. The cause for this discrepancy is related to the
perturbation chosen by \cite{Schoppa2002}, which differs substantially
from ours. \cite{Schoppa2002} used a physics-motivated perturbation in
$w'$. In our case, we are considering optimal perturbations, which
lead to much higher amplifications.

\section{Conclusions}\label{sec:conclusions}

We have investigated the processes responsible for the energy transfer
in wall turbulence from the streamwise-averaged mean-flow $U(y,z,t)$
to the fluctuating flow $\bu'(x,y,z,t)$. This energy transfer is the
backbone of self-sustaining wall turbulence and a subject of heated
debates.  It has long been hypothesised that the mechanism by which
the energy is transferred from $U$ to $\bu'$ can be captured by the
linearised Navier--Stokes equations and various linear theories stand
as tenable candidates to rationalise this process. The most prominent
theories are exponential instabilities of the base flow, nonlinear
interactions facilitated via neutral modes, non-modal transient
growth, and non-modal transient growth supported by parametric
instability, among others (see table \ref{table:literature}). To date,
a conclusive study regarding the role played by each linear mechanism
has been elusive due to the lack of methodologies designed to unveil
causal inference in the flow.

In the present work, we have used cause-and-effect analysis based on
interventions to assess the role played by different linear mechanisms
in sustaining turbulence. The approach is rooted on the concept that
manipulation of the causing variable leads to changes in the
effect. To that end, we sensibly modified the Navier--Stokes equations
of a turbulent channel flow to preclude one or various linear
mechanisms participating in the energy transfer from $U$ to $\bu'$.
We devised a set of numerical experiments tailored for minimal
turbulent channel units at $Re_\tau\approx 180$ in which the feedback
from $\bu' \rightarrow U$ is blocked to isolate the energy transfer
from $U \rightarrow \bu'$.  The active linear mechanisms for each
numerical experiment and its consequences are summarised in table
\ref{table}. In the first set of the experiments, the linear
Navier--Stokes operator is modified to render any exponential
instabilities of the streaks stable, thus precluding the energy
transfer from $U \rightarrow \bu'$ via exponential growth or
interaction with neutral modes. In the second set of experiments, we
simulated turbulent channel flows with prescribed, frozen-in-time,
exponentially stable base flows, such that both parametric
instabilities as well as exponential instabilities are suppressed.
The last set of experiments is devoted to further pinpoint the process
for energy transfer via transient growth by constraining the linear
Orr, lift-up, or push-over mechanisms, the latter being analogous to
the lift-up effect but in the spanwise direction.

The main contribution of this work is to establish that transient
growth alone is capable of sustaining wall turbulence with realistic
mean velocity and turbulence intensities in the absence of exponential
instabilities, neutral modes, and parametric instabilities. We have
further shown that transient growth originates mostly from the
Orr/push-over mechanisms due to spanwise variations of $U$.  Our
results are obtained for the fully nonlinear Navier--Stokes equations
in which the scattering of fluctuations by the nonlinear term is
required in combination with transient growth.  Exponential
instabilities also manifest in the flow, but they are only responsible
for about 10\% of the turbulent fluctuating velocities, and more
importantly, turbulence persists when they are inhibited. We have also
shown that turbulence persists when disposing of parametric
instabilities by using exponentially-stable frozen-in-time base
flows. In these cases, the statistics of the resulting turbulence
depend on the particular frozen base-flow selected.  However, the
ensemble average of cases with different frozen-in-time base-flows
reproduces the statistics of actual (time-varying-$U$) turbulence with
striking accuracy.  This was justified by showing that the way the
energy is input from $U$ into the system changes slowly compared to
the nonlinear dynamics of $\bu'$. In summary, turbulence statistics
are essentially explained by a collection of linear transient growth
processes in conjunction with nonlinear scattering. The evidence that
the ensemble average over multiple solutions (\S \ref{subsec:TG})
offers a simplified but complete representation of the system also
resembles the dynamical-system viewpoint that a large-enough set of
(invariant) solutions and their manifolds constitute the skeleton of
flow trajectories in turbulence~\citep{Auerbach1987, Cvitanovic1991}.

The outcome of this study is consistent with
\citet{Schoppa2002}. However, as inferred from the literature review
in table \ref{table:literature}, the transient-growth scenario is far
from being widely accepted.  The possibility of turbulence exclusively
supported by transient growth has been long hypothesised
\citep{Trefethen1993}, but its relevance has never been persuasively
shown in the full Navier--Stokes equations using cause-and-effect
analysis. To our best knowledge, our results are the most conclusive
demonstration of transient growth (via Orr/push-over) as a key driving
mechanism of self-sustaining turbulence.  It is important to
emphasise that our conclusions do not imply that other mechanisms are
not active in wall turbulence. Indeed, we have shown that modal
instabilities do manifest in the flow, and to some extent this and
other mechanisms have been observed by previous investigators. We have
also shown that time-variations of $U$ are necessary to sample the
perturbation phase-space and recover the nominal turbulence
statistics.  The picture promoted here is that the linear energy
transfer via transient growth overwhelms other competing mechanisms
and, as such, is able to explain most of the flow statistics.  This
simplifies the conceptual model of wall turbulence and unravels the
linear processes that should be targeted in turbulence modelling and
control.

Our conclusions regarding the dynamics of wall turbulence were drawn
using direct numerical simulations of the Navier--Stokes equations at
low Reynolds numbers representative of the buffer layer. It remains to
establish whether similar conclusions apply to higher $Re_\tau$. The
analysis was also performed in channels computed using minimal flow
units, chosen as simplified representations of naturally occurring
wall turbulence. Yet, we have shown that our results are qualitatively
similar when the domain size is doubled.  We expect that the approach
presented here paves the way for future investigations at
high-Reynolds-numbers turbulence obtained in larger unconstrained
domains, in addition to extensions to different flow configurations in
which the role of linear mechanisms remains unsettled.


\vspace{1em}

This work was also supported by the Coturb project of the European
Research Council (ERC-2014.AdG-669505) during the 2019 Coturb
Turbulence Summer Workshop at the Universidad Polit\'ecnica de Madrid.
M.-A.N. acknowledges the support of the Hellenic Foundation for
Research and Innovation, and the General Secretariat for Research and
Technology (Grant No. 1718/14518).  A.L.-D. acknowledges the support
of the NASA Transformative Aeronautics Concepts Program
(Grant~No.~NNX15AU93A) and the Office of Naval Research
(Grant~No.~N000141712310). N.C.C.~was supported by the Australian
Research Council (Grant~No.~CE170100023). M.K. was supported by the
Air Force Office of Scientific Research under grant FA9550-16-1-0319
and the Office of Naval Research under grant N00014-17-1-2341.  We
thank Jane Bae, Gregory Chini, Brian Farrell, Yongyun Hwang, Philip
Hall, Fazle Hussain, Petros Ioannou, and Javier Jim\'enez for
insightful discussions and comments.

 \section*{Declaration of Interests}
The authors report no conflict of interest.

\appendix

\section{Sensitivity to the size of the computational domain}
\label{sec:appendix_2Lx}

The minimum size of the computational domain required to sustain
turbulence in channels at low Reynolds number was extensively studied
by \citet{Jimenez1991}. The streamwise and spanwise lengths of our
simulations were selected to comply with these minimum requirements.
We also verified that decreasing the streamwise ($L_x$) and spanwise
($L_z$) extends of our domain any further in the current setup leads
to laminarisation of the flow.  In this appendix, we focus on the
sensitivity of our results to increasing $L_x$ and $L_z$. The former
is potentially the most critical length as the base flow is defined by
taking the streamwise average of $u$. We have also seen that the most
dangerous instabilities occur for the wavenumber $k_x=2\pi/L_x$, which
is the harmonic excitation and, hence, might be susceptible to changes
in $L_x$.  Therefore, we centre our attention on $L_x$ and the role of
subharmonic instabilities. Some comments are made at the end of the
appendix on the sensitivity to $L_z$.

Prior to conducting the sensitivity analysis for $L_x$, we can
anticipate that the subharmonic instabilities associated with
$k_x=2\pi/(n L_x)$ with $n>1$ are probably of little relevance for
sustaining the flow. The most obvious reason is that wavenumbers equal
or smaller than $k_x=\pi/L_x$ are not accommodated in the domain (they
simply do not fit in $x$) and subharmonic instabilities cannot
manifest in the flow. Given that our simulations show that turbulence
is sustained with realistic mean profile and fluctuating velocities
for the chosen $L_x$, the importance of the (nonexistent) subharmonic
instabilities must be minor. To ascertain that this is the case, we i)
perform an \emph{a priori} analysis of the stability of $U(y,z,t)$
assuming that instabilities from $k_x=\pi/L_x$ are realisable, and ii)
conduct additional simulations by doubling the streamwise domain.

\begin{figure}
 \begin{center}
   \subfloat[]{\includegraphics[width=0.45\textwidth]{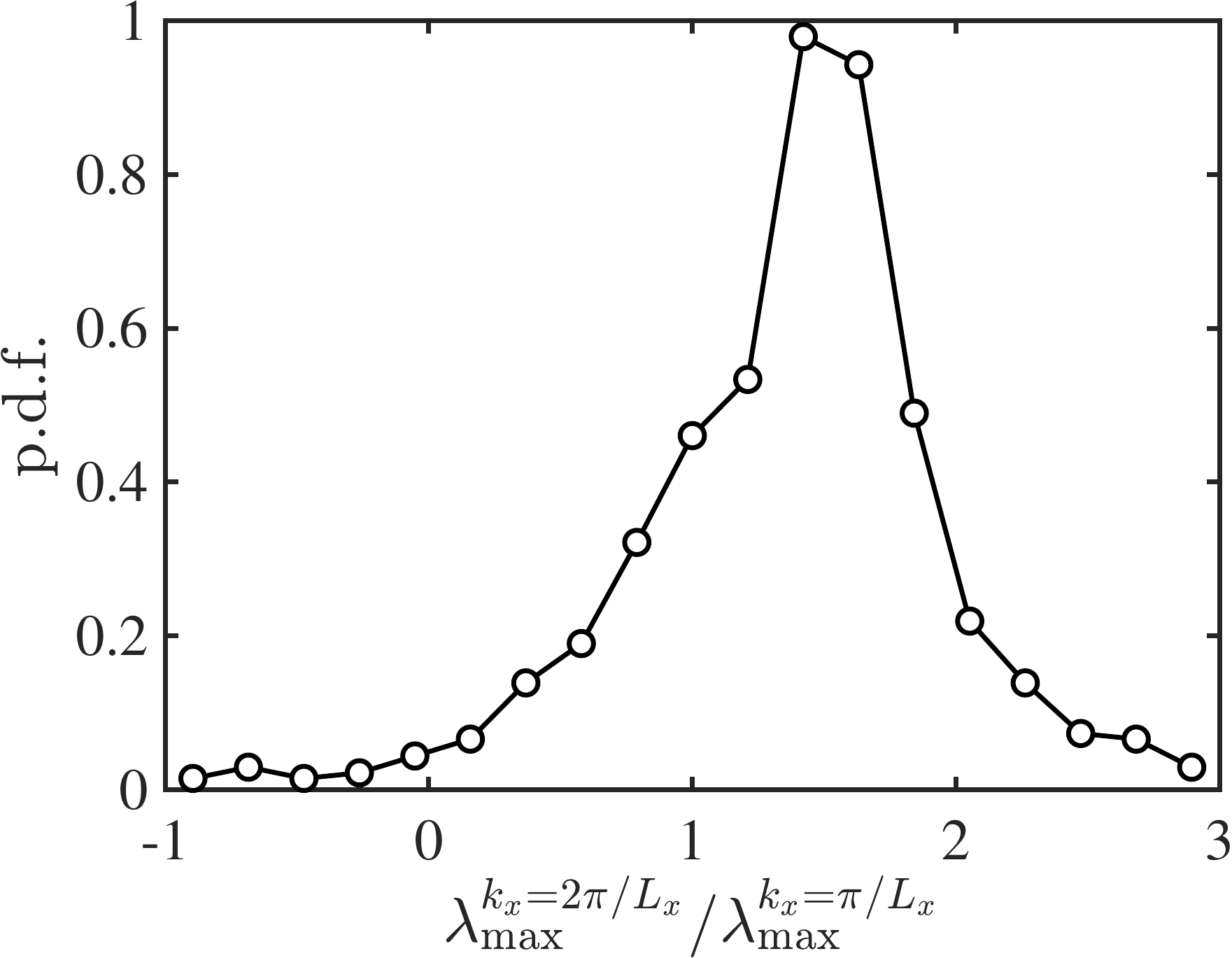}} 
   \subfloat[]{\includegraphics[width=0.45\textwidth]{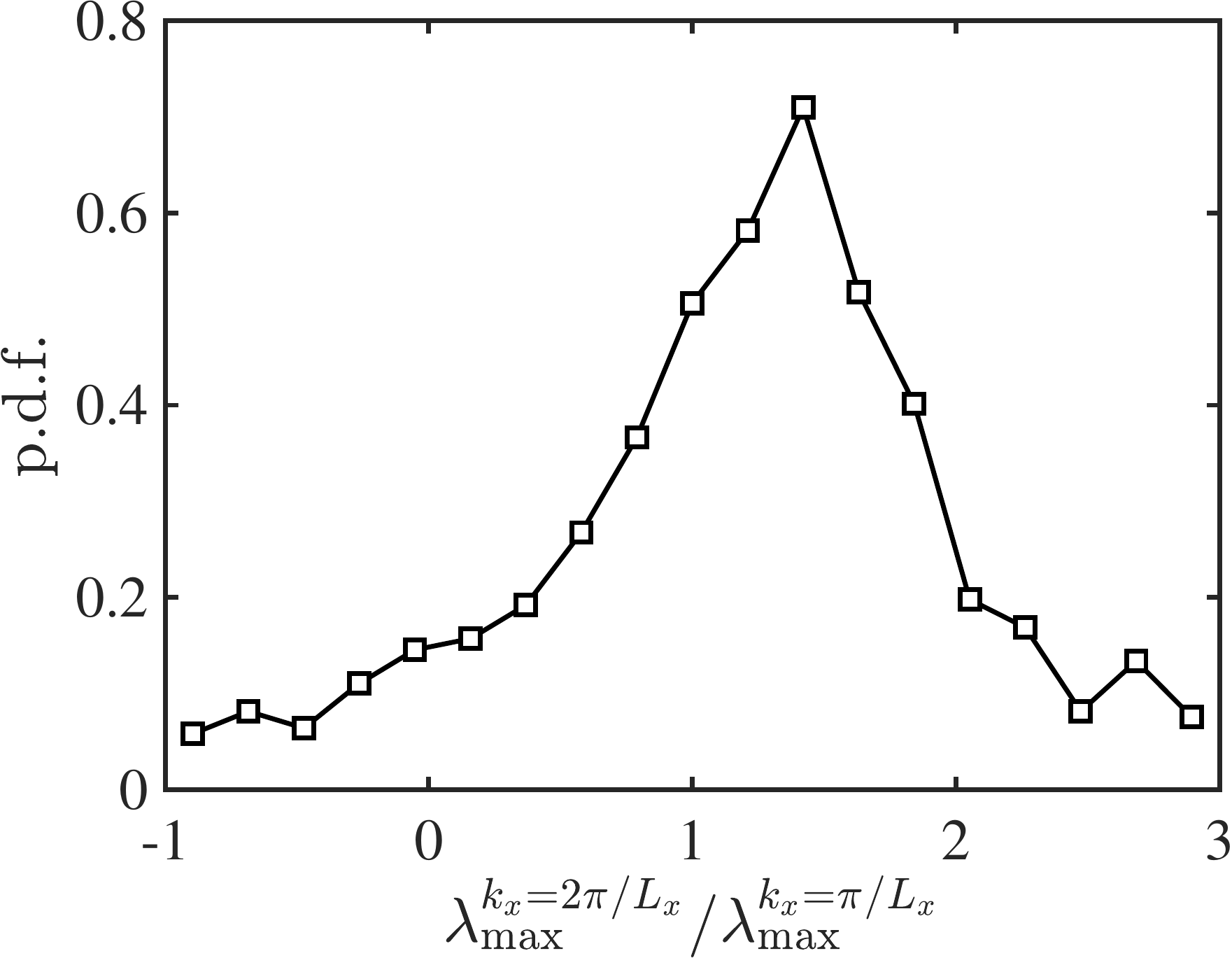}} 
 \end{center}
\caption{ The probability density function of the ratio of the largest
  growth rates $\lambda_{\max}^{k_x=2\pi/L_x} /
  \lambda_{\max}^{k_x=\pi/L_x}$ of a given base flow at time $t_i$,
  $U(y,z,t_i)$ for (a) case R180 and (b) case R-2Lx-180.
 \label{fig:ratio_lambdas_2Lx}}
\end{figure}

We study the stability of $U(y,z,t)$ from case R180, similar to the
analysis in \S \ref{subsec:theories_modal}.  In this occasion, we
focus on the growth rates for perturbations with streamwise
wavenumbers $k_x=\pi/L_x$ and $k_x=2\pi/L_x$ denoted by
$\lambda_{\max}^{k_x=\pi/L_x}$ and $\lambda_{\max}^{k_x=2\pi/L_x}$,
respectively.  It is important to remark that
$\lambda_{\max}^{k_x=\pi/L_x}$ is the hypothetical growth rate of
exponential instabilities with wavenumber $k_x=\pi/L_x$ if they were
allowed to manifest in the flow (which they are not). The p.d.f.  of
the ratio $\lambda_{\max}^{k_x=2\pi/L_x} /
\lambda_{\max}^{k_x=\pi/L_x}$ for a given base flow at time $t_i$,
$U(y,z,t_i)$, is plotted in figure~\ref{fig:ratio_lambdas_2Lx}(a),
and shows that the harmonic instability (which are realisable in the
simulation) prevails over the (hypothetical) subharmonic
instability. The data also reveal that $\lambda_{\max}^{k_x=2\pi/L_x}
>\lambda_{\max}^{k_x=\pi/L_x}$ about 80\% of the time. The result
proves that the base flows for case R180 are more receptive to harmonic
instabilities than they are to subharmonic instabilities.
%

\begin{figure}
 \vspace{1cm}
 \begin{center}
   \includegraphics[width=0.97\textwidth]{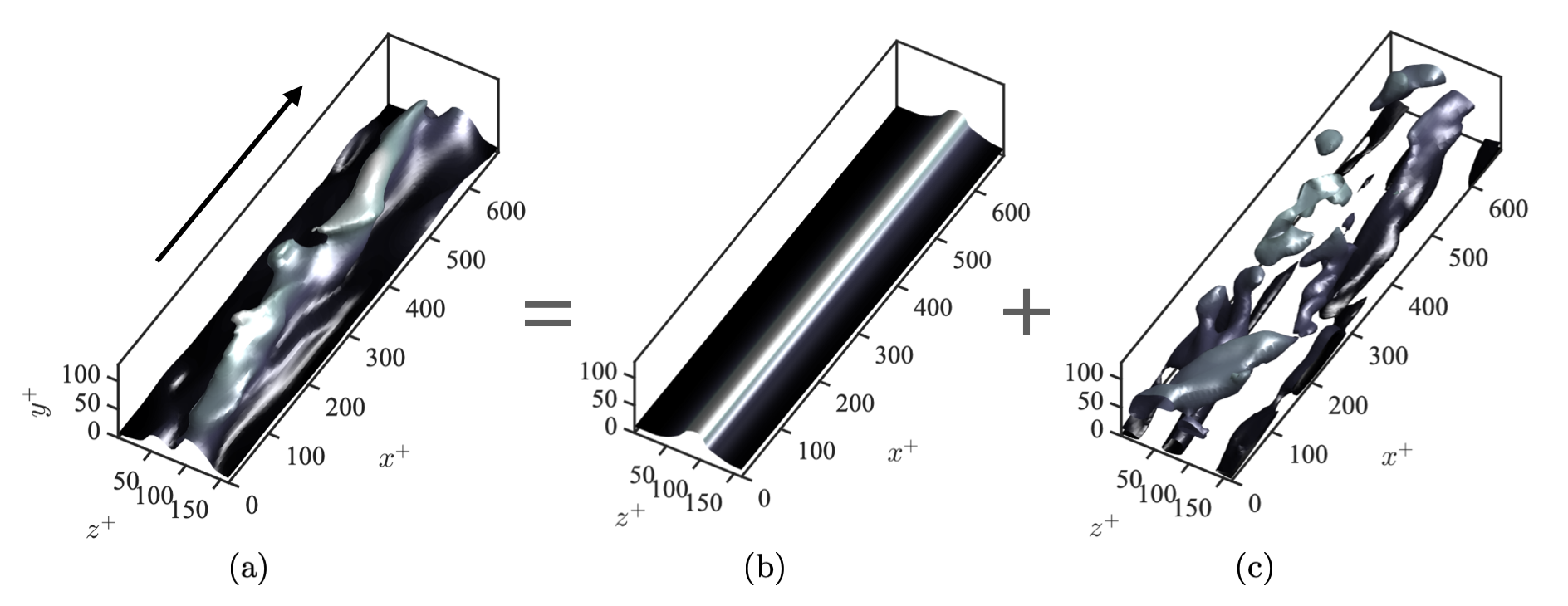} 
 \end{center}
\caption{ Decomposition of the instantaneous flow into a streamwise
  mean base flow and fluctuations for case R-2Lx-180.  Instantaneous
  isosurface of streamwise velocity for (a)~the total flow $u$,
  (b)~the streak base flow $U$, and (c)~the absolute value of the
  fluctuations $|u'|$.  The values of the isosurfaces are 0.6 (a and
  b) and 0.1 (c) of the maximum streamwise velocity.  Shading
  represents the distance to the wall from dark ($y=0$) to light
  ($y=h$). The arrow in panel~(a) indicates the mean flow direction.
  \label{fig:snaphots_2Lx}}
\end{figure}

\begin{figure}
 \vspace{1cm}
  \begin{center}
    \subfloat[]{\includegraphics[width=0.34\textwidth]{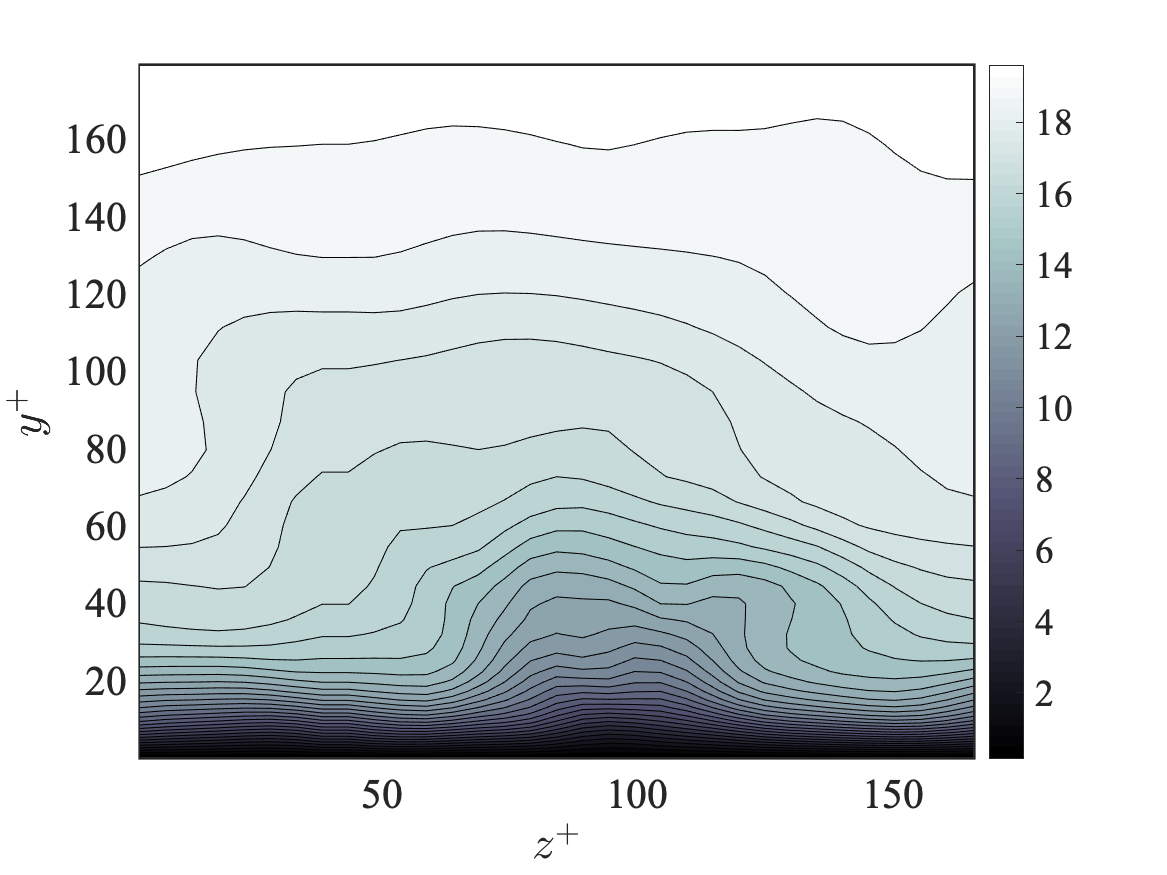}}
    \subfloat[]{\includegraphics[width=0.34\textwidth]{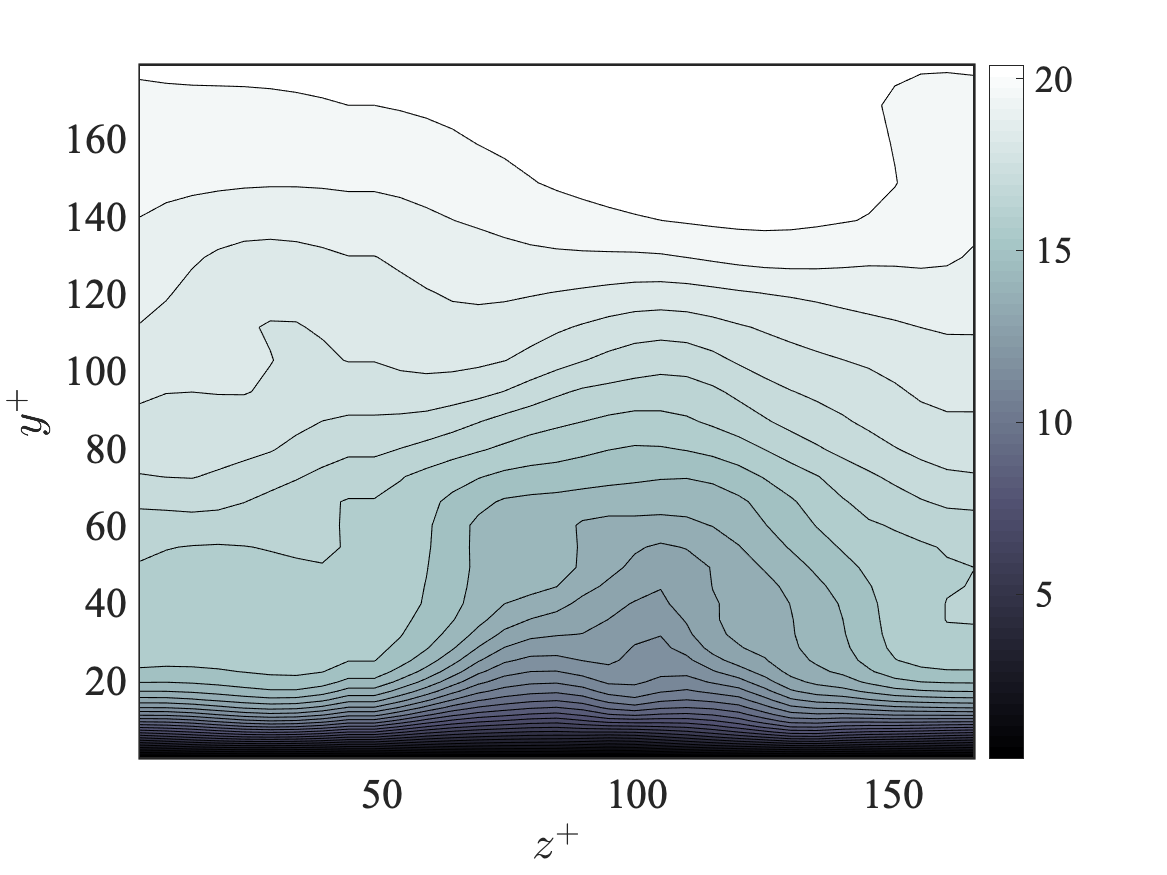}}
    \subfloat[]{\includegraphics[width=0.34\textwidth]{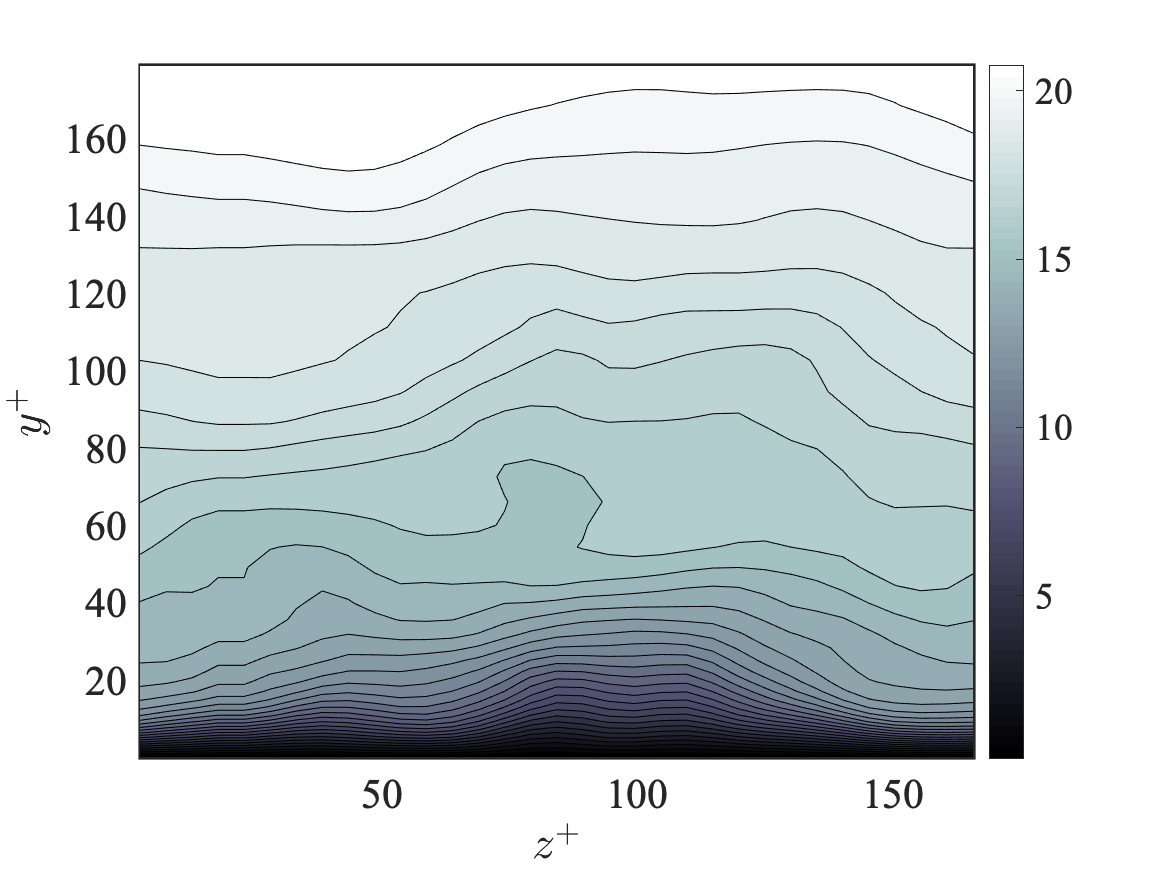}}
 \end{center}
\caption{ Examples of base flow $U(y,z,t)$ for case
  R-2Lx-180.  \label{fig:baseflow_2Lx}}
\end{figure}

The second analysis consists of an actual simulation with streamwise
domain length equal to $2L_x^+ \approx 673$ ($2L_x \approx 3.66h$),
such that the instabilities associated with $k_x=\pi/L_x$ are now
allowed in the flow ($L_x$ and $L_z$ signify the domain size of
R180). We label this case as \mbox{R-2Lx-180}, which is analogous to
R180 but with doubled streamwise domain length.
Figure~\ref{fig:snaphots_2Lx} illustrates the flow decomposition into
base flow and fluctuations for R-2Lx-180 and figure~\ref{fig:baseflow_2Lx} depicts three examples of base
flows. Consistently, the base flows for R-2Lx-180 are defined as
$U(y,z,t) = \langle u \rangle_x = 1/2L_x \int_{0}^{2L_x} u
\mathrm{d}x$.
%
%

\begin{figure}
 \begin{center}
   \subfloat[]{\includegraphics[width=0.33\textwidth]{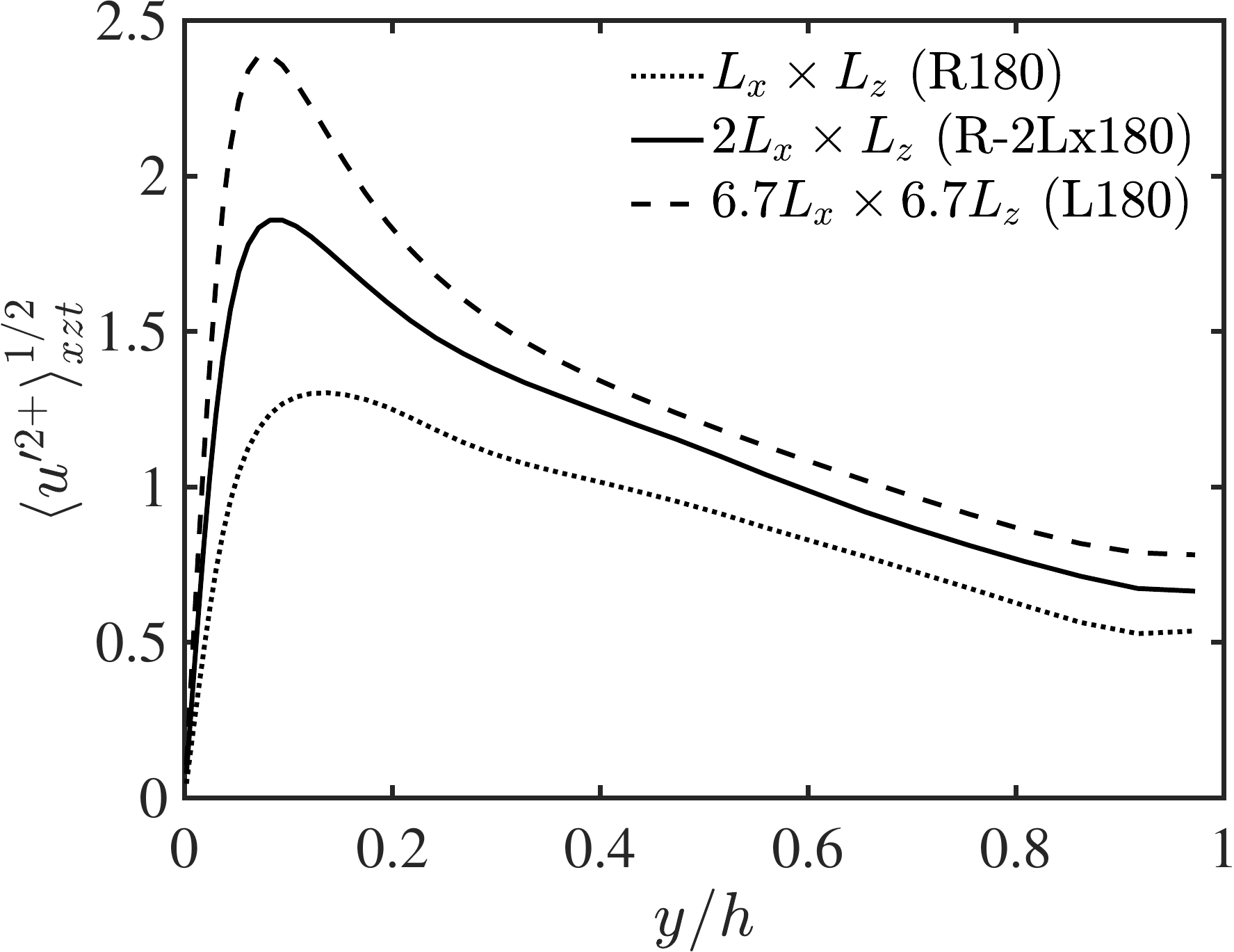}} 
   \subfloat[]{\includegraphics[width=0.33\textwidth]{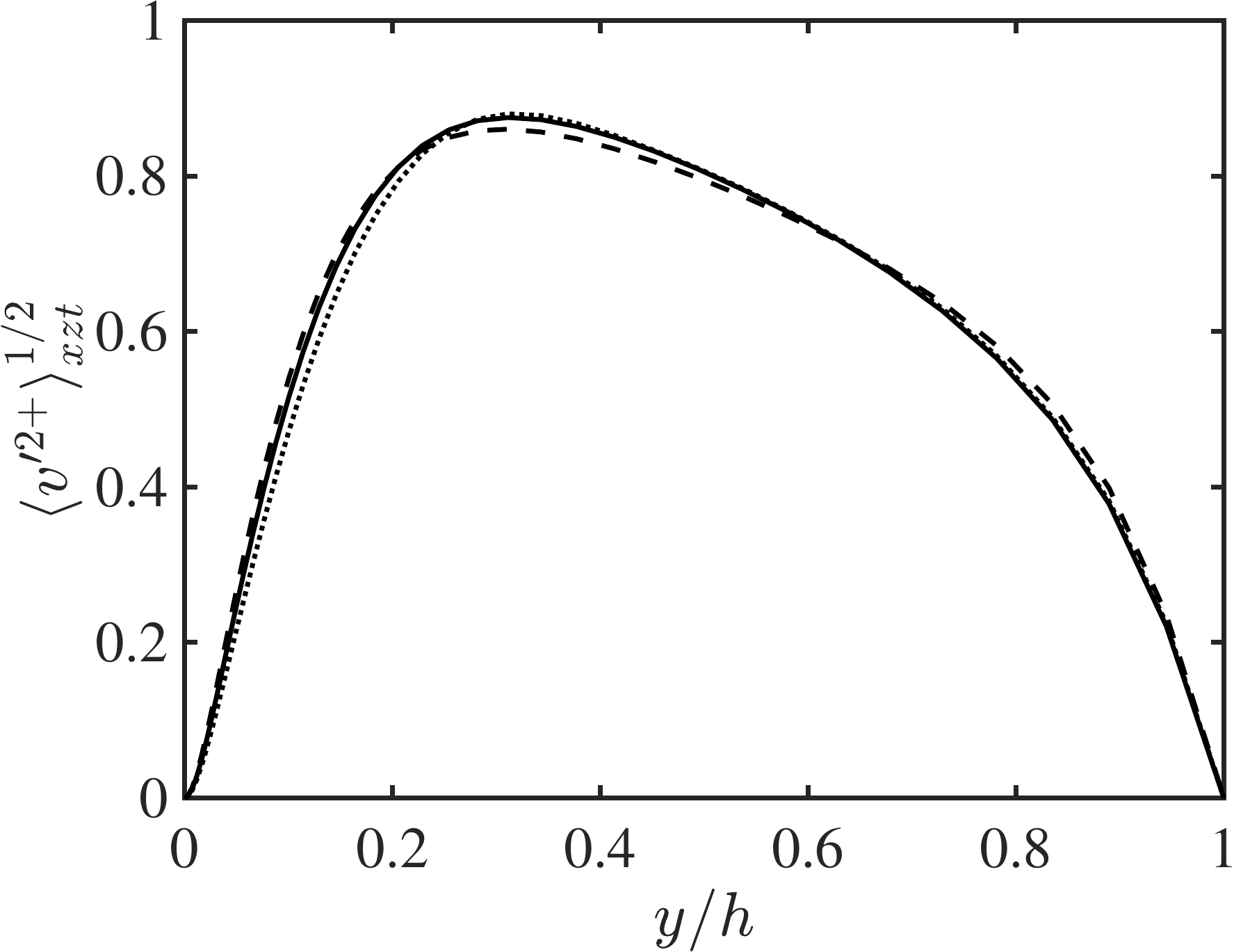}} 
   \subfloat[]{\includegraphics[width=0.33\textwidth]{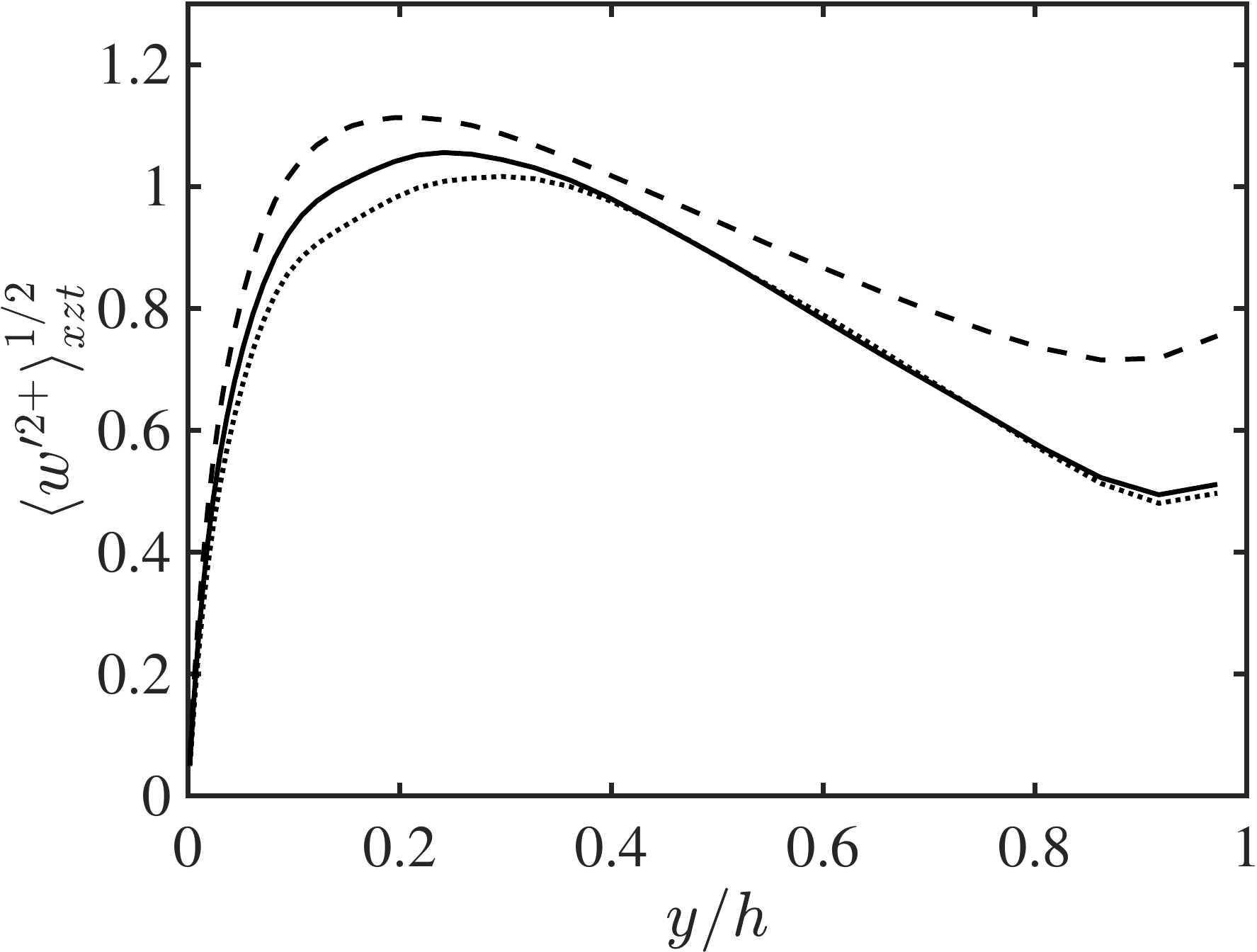}} 
 \end{center}
\caption{ (a) Streamwise (\solid), (c) wall-normal (\dashed), and (c)
  spanwise (\dotted) root-mean-squared fluctuating velocities as a
  function of the wall-normal distance for R-2Lx-180 (\solid) and
  equivalent non-minimal channel (L180) with $6.7L_x^+ \times 6.7L_z^+
  = 2312 \times 1156$ ($\approx 12.5h \times 6.3h$) (\dashed), where
  $L_x$ and $L_z$ signify the channel domain for R180.
 \label{fig:stats_2Lx}}
\end{figure}

Figure~\ref{fig:stats_2Lx} shows the rms fluctuating velocities for
R-2Lx-180 compared with the minimal channel flow R180. The main effect
of enlarging the domain in $x$ is an increase in $u'$, which comes
from the larger scales accommodated by the computational box.  The
energy in $u'$ is now closer to the nominal value in non-minimal
domains, but we have argued in \S \ref{sec:regular} that this
additional energy is not strictly required to sustain turbulence.
%

The stability analysis of $U(y,z,t)$ for R-2Lx-180 and wavenumbers
$k_x = \pi/L_x$ and $k_x = 2\pi/L_x$ is included in figure~\ref{fig:ratio_lambdas_2Lx}(b). The outcome is similar to R180:
$\lambda_{\max}^{k_x=2\pi/L_x}$ prevails over
$\lambda_{\max}^{k_x=\pi/L_x}$ most of the time, and this is true even
if now the streamwise domain is $2L_x$. This suggests that the most
unstable wavelength should be around the length of the minimal channel.

\begin{figure}
 \begin{center}
   \subfloat[]{\includegraphics[width=0.45\textwidth]{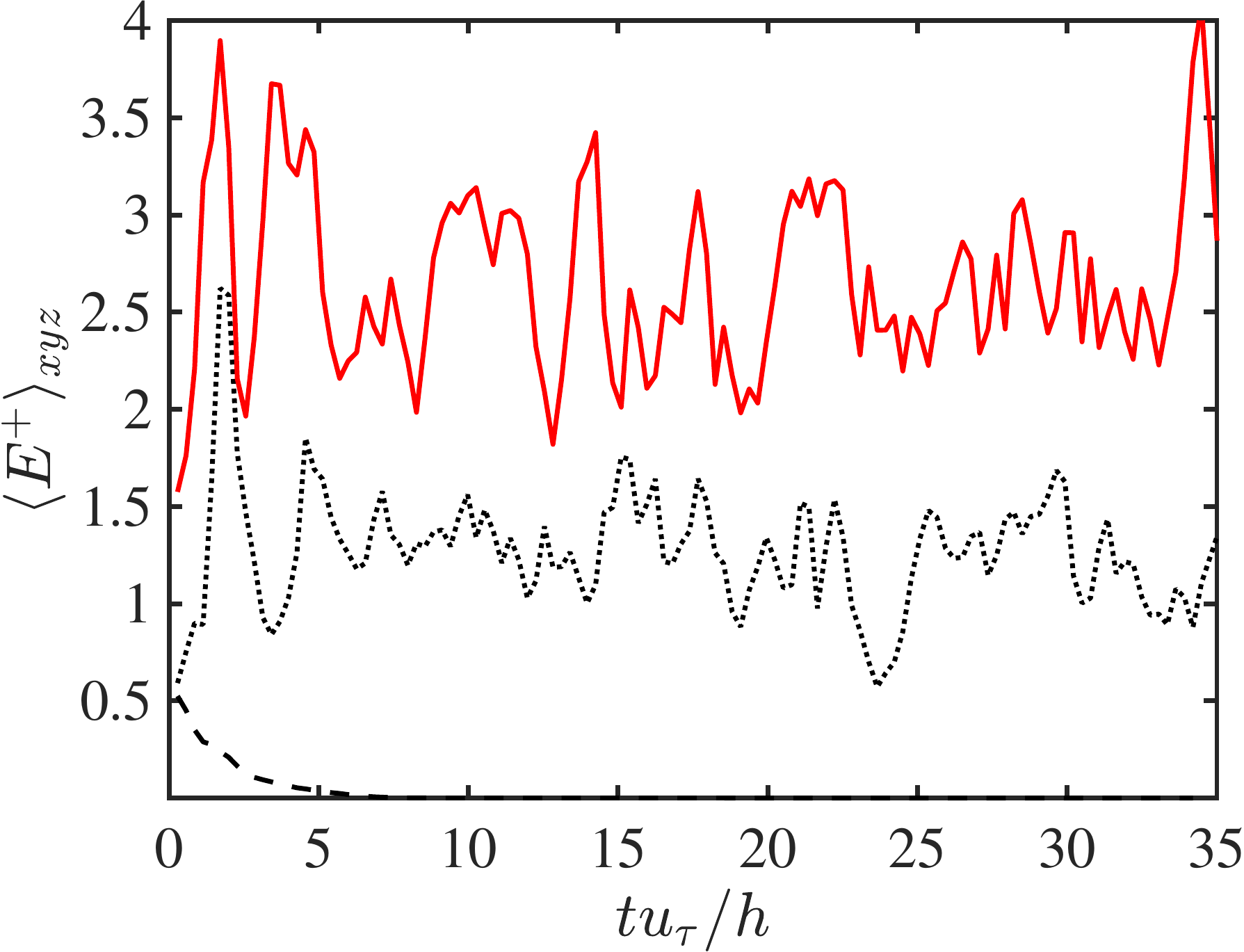}} 
   \subfloat[]{\includegraphics[width=0.45\textwidth]{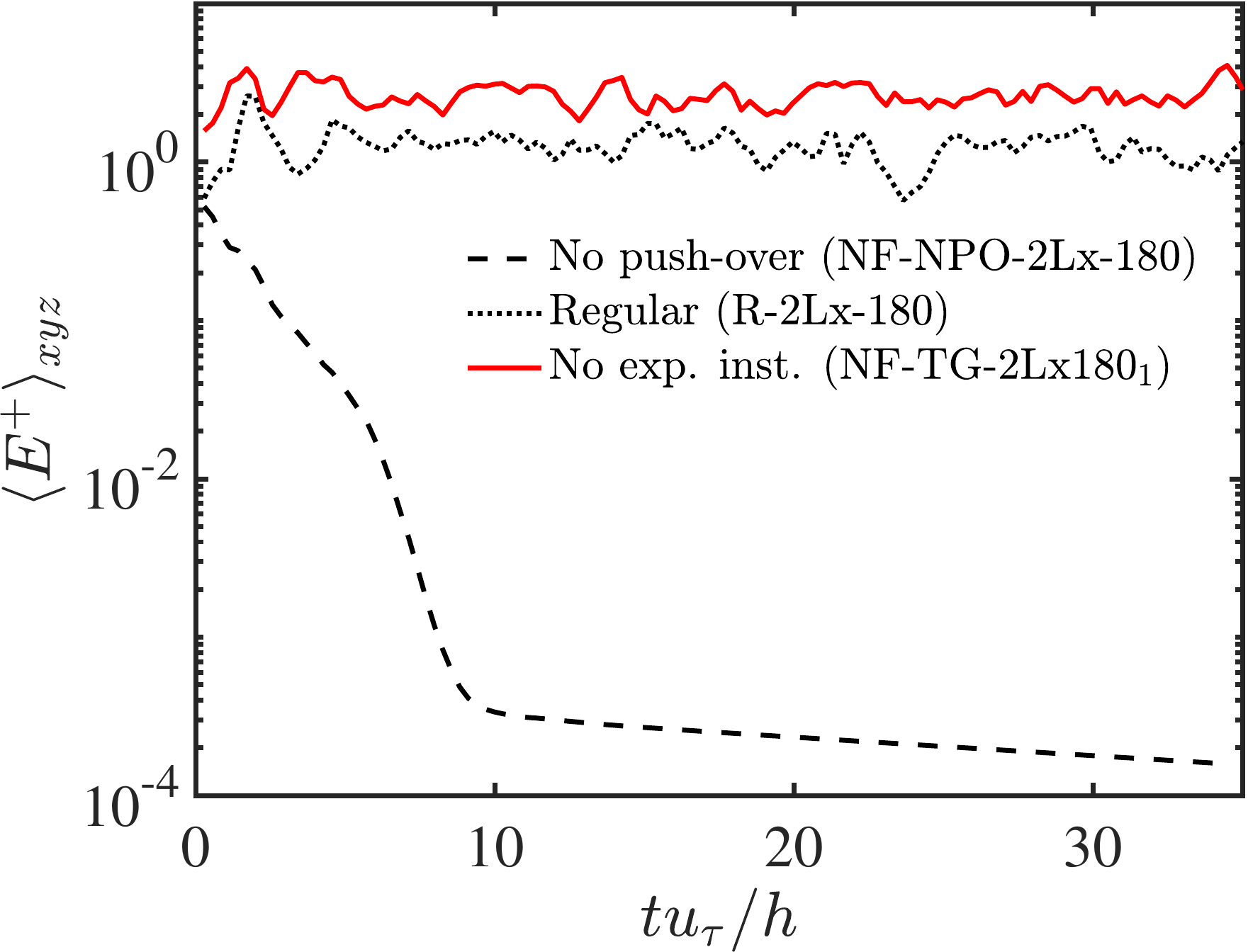}} 
 \end{center}
 \caption{The history of the domain-averaged turbulent kinetic energy
   of the fluctuations $\langle E \rangle_{xyz}$ for the case with
   frozen base flow without exponential instabilities
   (NF-TG-2Lx-180$_{\{1\}}$, \textcolor{red}{\solid}), channel without
   linear push-over (NF-NPO-2Lx-180, \dashed), and regular channel
   flow (R-2Lx-180, \dotted). The vertical axis is linear in panel (a)
   and logarithmic in panel (b).
 \label{fig:TKE_2Lx}}
\end{figure}

To complete the analysis and build confidence in the results presented
in the paper, we perform two simulations with constrained linear
dynamics using R-2Lx-180 as baseline. In the first case, a selected
base flow from R-2Lx-180 is frozen in time and the exponential
instabilities are removed. We denote this case as
NF-TG-2Lx-180$_{\{1\}}$ (similar to cases NF-TG180$_{\{i\}}$ in \S
\ref{subsec:TG}). In the second case, the linear push-over mechanisms
is cancelled out (similar to case NF-NPO180 in \S
\ref{sec:TG_more}). The cases are initialised from R-2Lx-180, although
it was assessed that the conclusions are independent of the initial
condition. The evolution of the turbulence kinetic energy for both
cases is shown in figure~\ref{fig:TKE_2Lx}. For
NF-TG-2Lx-180$_{\{1\}}$, turbulence is maintained despite the lack of
exponential and parametric instabilities. For the particular base flow
chosen in NF-TG-2Lx-180$_{\{1\}}$, the turbulent kinetic energy is on
average larger than that for R-2Lx-180, although other base flows (not
shown) exhibit a lower value.  Conversely, the turbulent kinetic
energy decays for NF-NPO-2Lx-180, where the linear push-over is
suppressed. Therefore, the main conclusions drawn from simulations
with a streamwise domain equal to $2L_x$ are consistent with those
discussed in the paper with streamwise domain $L_x$.
%

Finally, we carried out simulations analogous to those described above
doubling the spanwise length of the domain. Our conclusions remain
unchanged. This is not unexpected as enlarging $L_z$ mostly translates
into an increment on the number of coherent structures contained in
the domain along the $z$ direction. The new channel is not minimal as
it contains more than one elementary flow unit, but the
characteristics of the base flow are barely affected.

\section{Results for base flow $(U,V,W)$}
\label{sec:appendix_UVW}

For completeness, we repeat the analysis in \S \ref{sec:constrain}
using this time the streamwise-averaged $v$ and $w$ as part of our
base flow, i.e.,
\begin{equation}\label{eq:new_base}
  \bU \defn (U,V,W) = ( \langle u \rangle_x, \langle v \rangle_x, \langle w \rangle_x).
\end{equation}
Consequently, the perturbations are now defined as $u' = u - U$, $v' =
v - V$, and $w' = w - W$. We carried out simulations analogous to
those discussed in table \ref{table}. The conclusions drawn using
$(U,V,W)$ as the base flow are similar to those using $(U,0,0)$. Here
we report some of the key results.

The new equation for $\bU$ is obtained by replacing the operator
$\mathcal{D}$ used to set the $y-$ and $z-$components, namely, 
\begin{equation}
\mathcal{D} = \begin{bmatrix}
1 & 0 & 0\\
0 & 0 & 0\\
0 & 0 & 0\\
\end{bmatrix},
\end{equation}
by the identity operator $\mathcal{D} = \mathcal{I}$ such that 
\begin{subequations}
  \begin{gather}
\frac{\partial \bU}{\partial t} + \bU \cdot
 \bnabla \bU =
 -\langle \bu' \cdot
  \bnabla \bu' \rangle_x
 - \frac{1}{\rho}\bnabla \langle p \rangle_x + \nu \nabla^2
 \bU + \boldsymbol{f}, \\
 \bnabla \cdot \bU = 0.
 \end{gather}
\end{subequations}
The equation for the new fluctuating velocity vector is
\begin{subequations}
  \begin{gather}
\frac{\partial\bu'}{\partial t} = \mathcal{L}(\bU)\bu' + \bN(\bu'), \\
  \mathcal{L}(\bU)\bu' =  \mathcal{P}\left[ -\bU\cdot \bnabla \bu' - \bu' \cdot \bnabla \bU
  + \nu \nabla^2 \bu' \right], \\
  \bN(\bu') = \mathcal{P}\left[ -\bu' \cdot \bnabla \bu'
    + \langle \bu' \cdot \bnabla \bu' \rangle_x \right].
  \end{gather}
\end{subequations}
%

\begin{figure}
  \begin{center}
  \hspace{0.05cm}
  \subfloat[]{\includegraphics[width=0.32\textwidth]{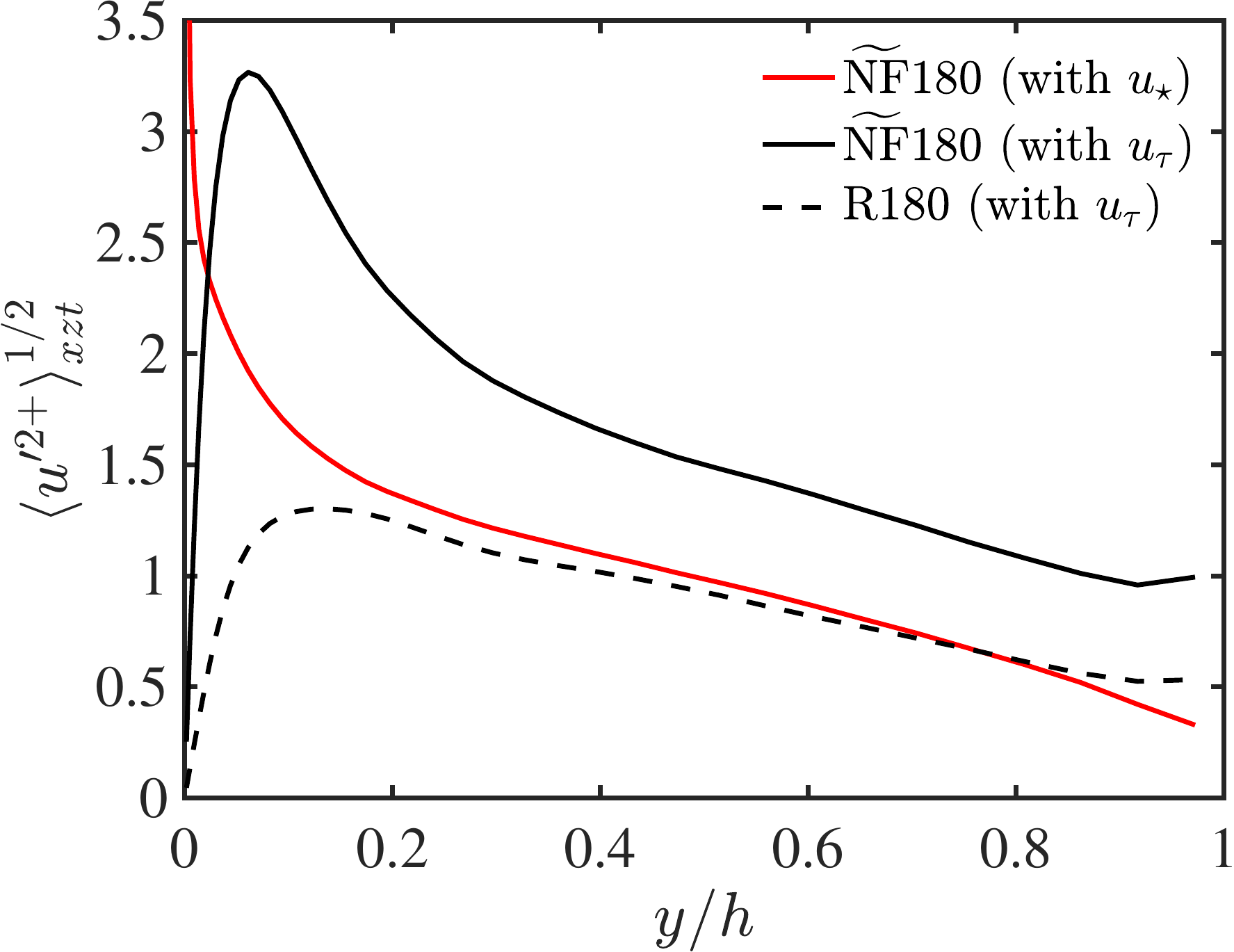}} 
  \hspace{0.05cm} 
  \subfloat[]{\includegraphics[width=0.32\textwidth]{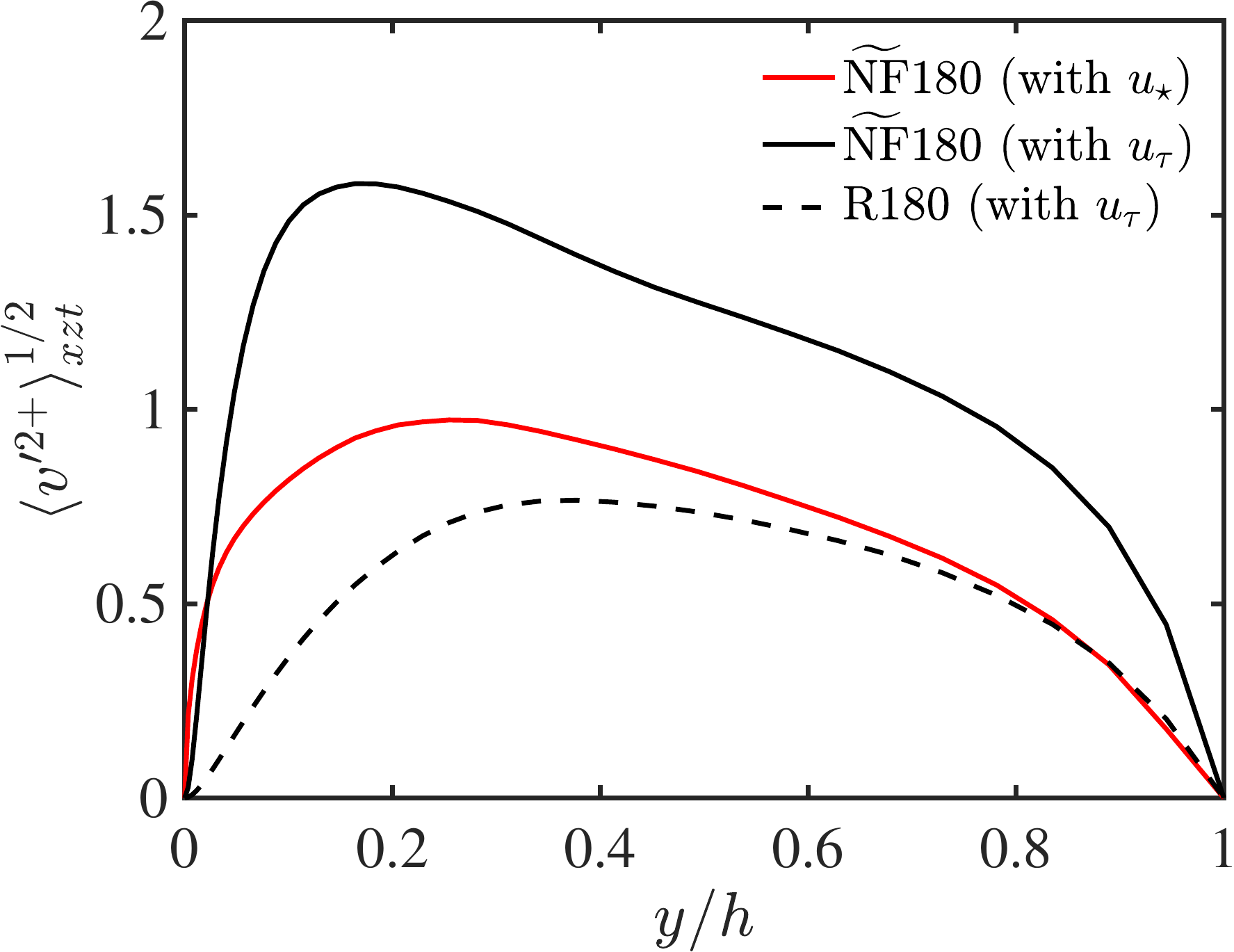}} 
  \hspace{0.05cm}
  \subfloat[]{\includegraphics[width=0.32\textwidth]{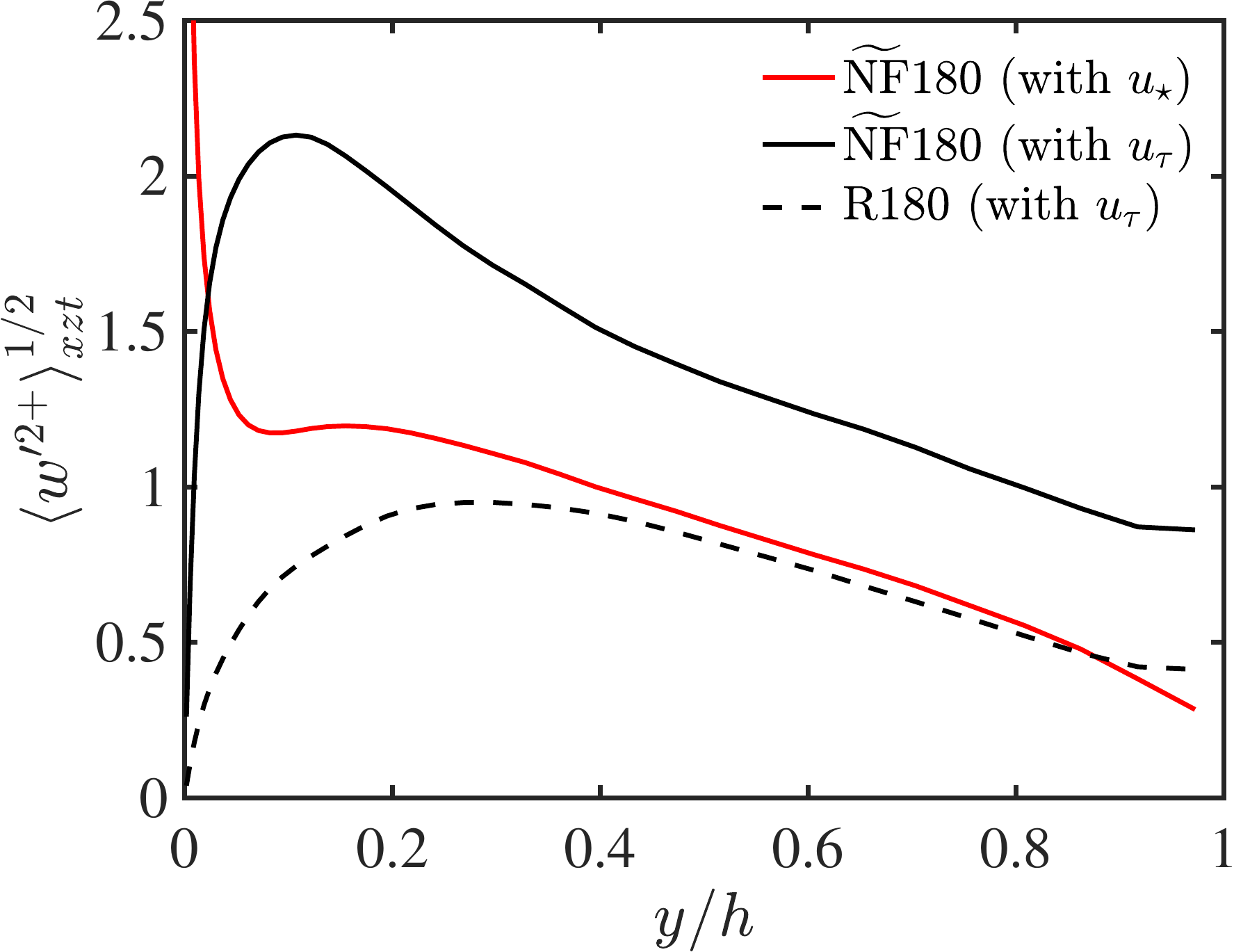}} 
  \end{center}
\caption{ (a) Streamwise, (b) wall-normal, and (c) spanwise mean
  root-mean-squared fluctuating velocities as a function of the
  wall-normal distance for case R180 normalised by $u_\tau$ (\dashed),
  case $\widetilde{\mathrm{NF}}$180 normalised by $u_\tau$ (\solid), and
  $\widetilde{\mathrm{NF}}$180 normalised by $u_\star$
  (\textcolor{red}{\solid}).
\label{fig:stats_UVWpred}}
\end{figure}

First, we remove the explicit feedback from $\bu'$ to $(U,V,W)$ and
refer to this case as $\widetilde{\mathrm{NF}}$180 (analogous to
NF180). Figure~\ref{fig:stats_UVWpred} shows that the effect of
blocking the feedback from $\bu'$ to $(U,V,W)$ is to enhance the
fluctuating velocities, similarly to NF180. The results exhibit an
improved collapse when the velocities are scaled by $u_\star$, which
is more representative of characteristic flow velocity.
%

\begin{figure}
  \begin{center}
  \subfloat[]{\includegraphics[width=0.45\textwidth]{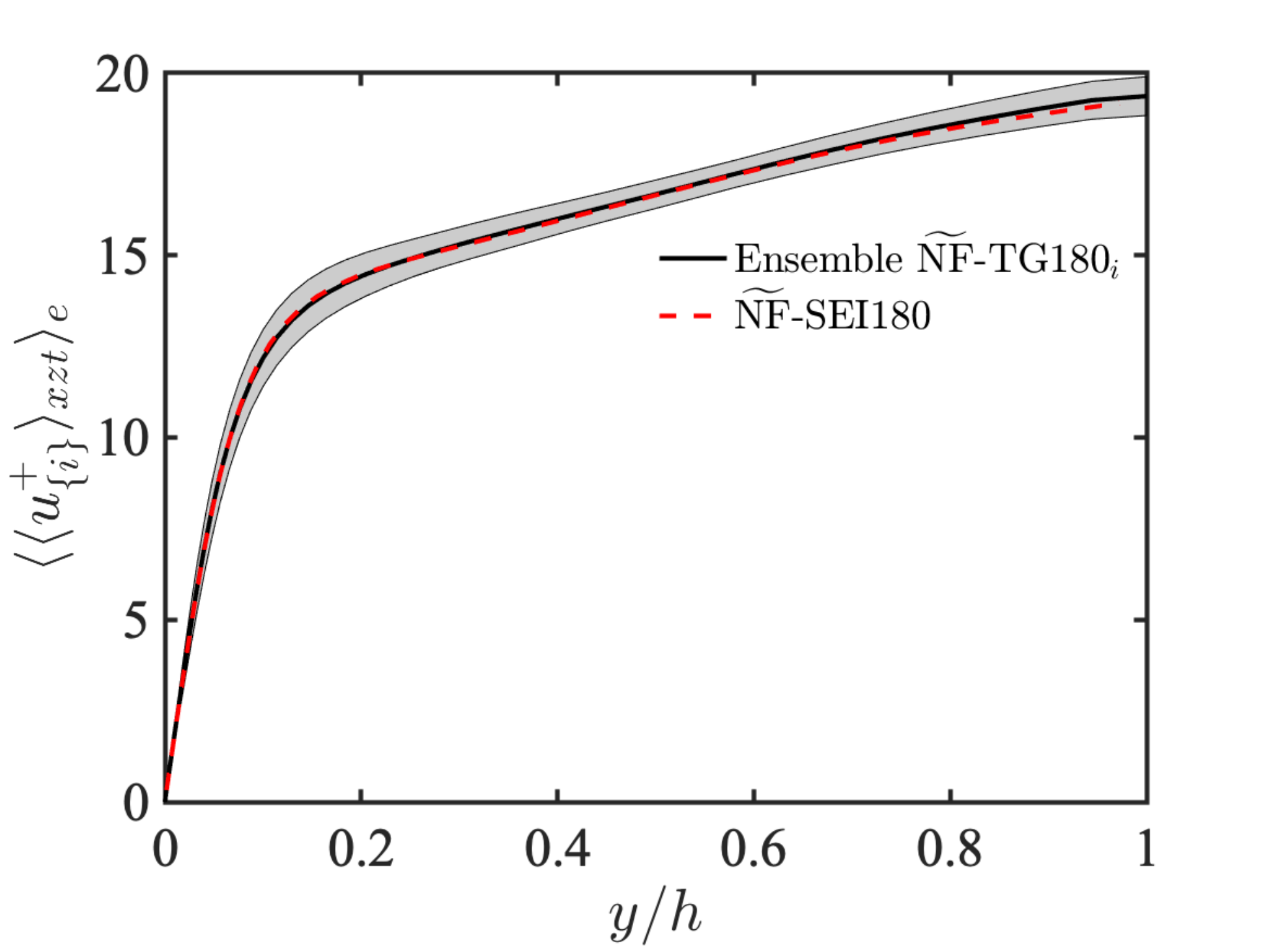}} 
  \hspace{0.05cm}
  \subfloat[]{\includegraphics[width=0.45\textwidth]{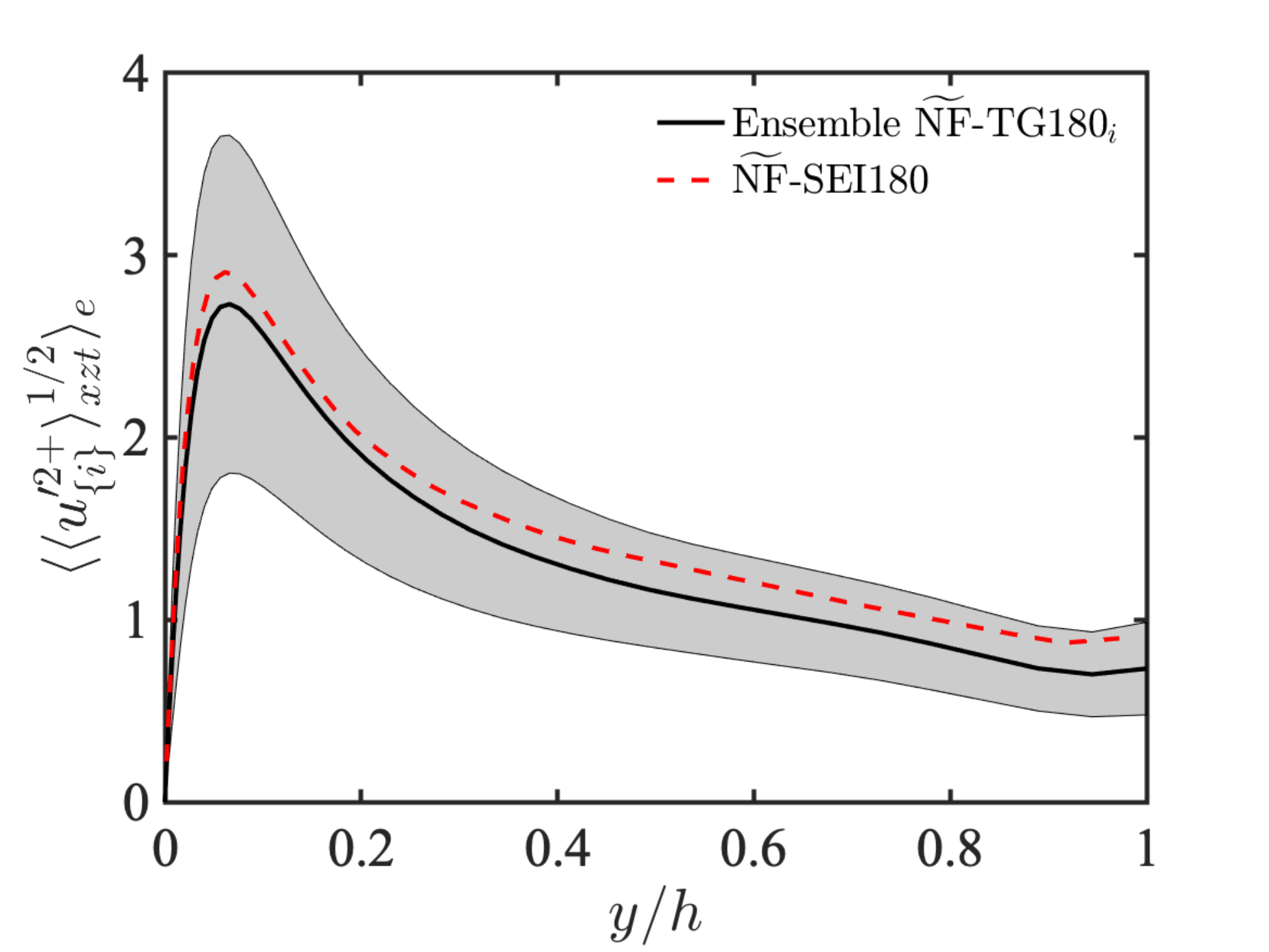}} 
  \end{center}
  \begin{center}
  \subfloat[]{\includegraphics[width=0.45\textwidth]{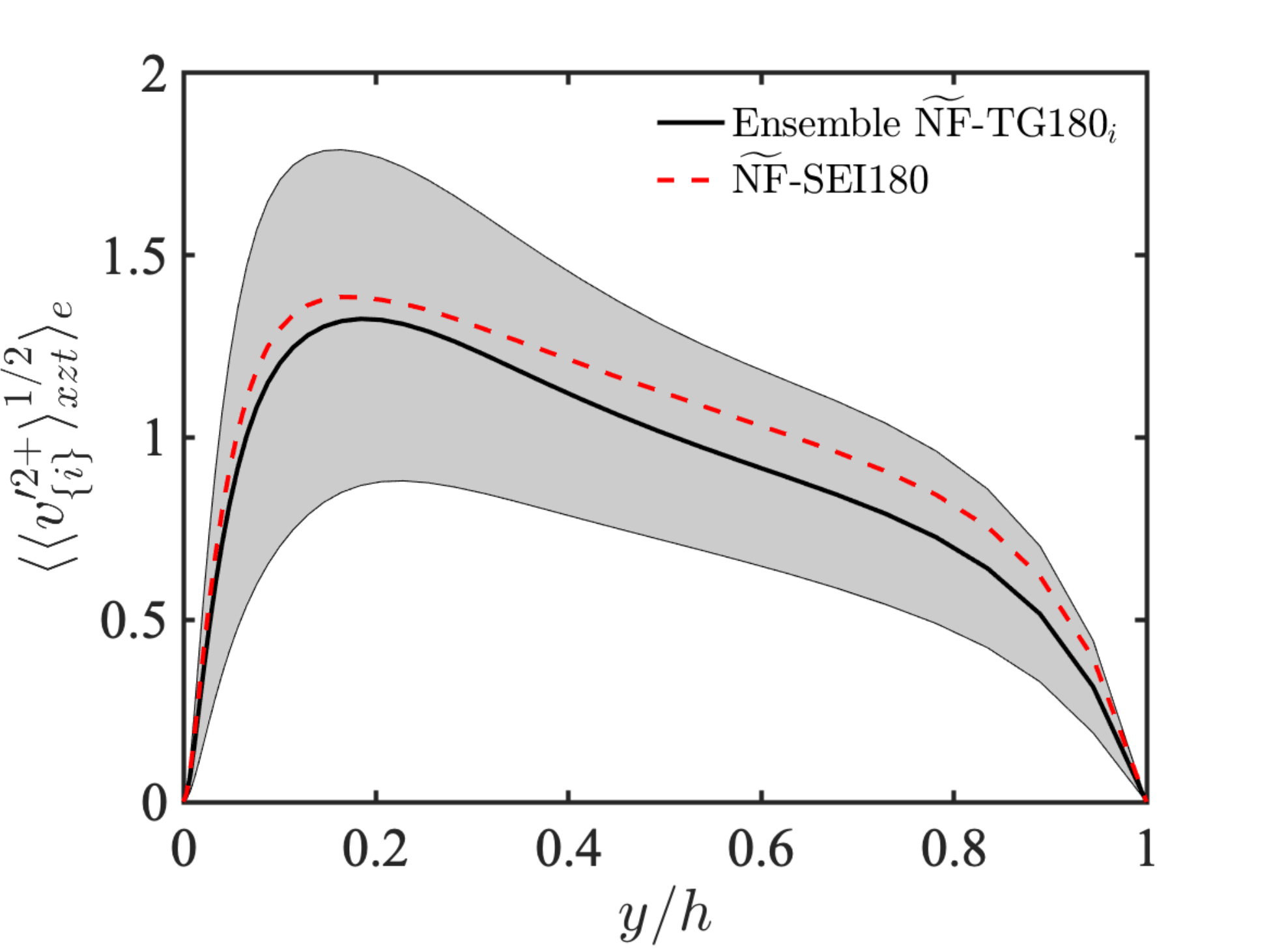}}  
  \hspace{0.05cm}
  \subfloat[]{\includegraphics[width=0.45\textwidth]{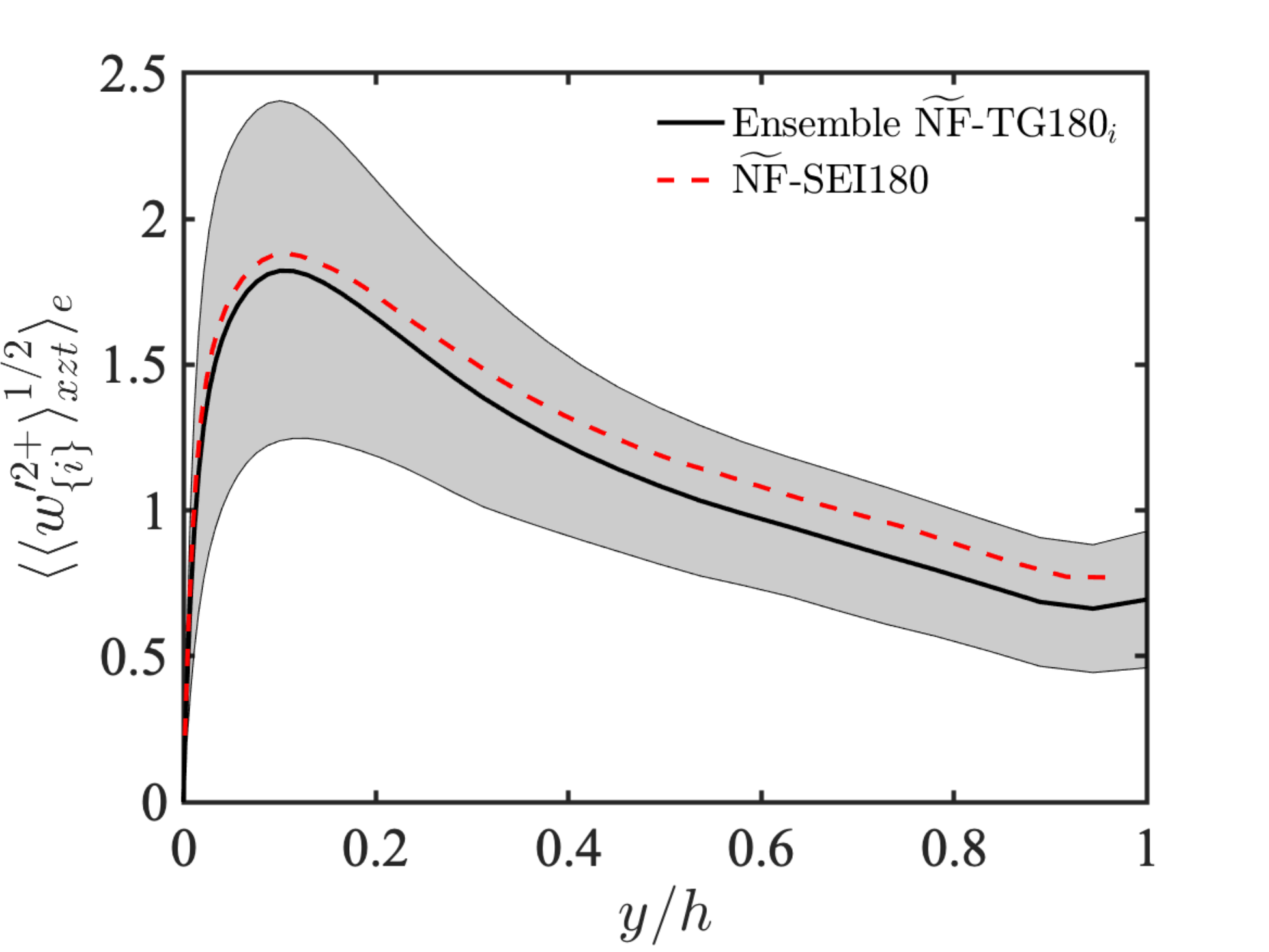}}  
  \end{center}
 \caption{ (a) Mean velocity profile, (b) root-mean-squared
   streamwise, (c) wall-normal, and (d) spanwise fluctuating
   velocities: (\solid), the ensemble average of turbulent cases
   $\widetilde{\mathrm{NF}}$-TG180$_{i}$, namely, $\langle \langle
   u_{\{i\}} \rangle_{xzt} \rangle_e$, $\langle \langle u'^2_{\{i\}}
   \rangle_{xzt}^{1/2} \rangle_e$, $\langle \langle v'^2_{\{i\}}
   \rangle_{xzt}^{1/2} \rangle_e$, and $\langle \langle w'^2_{\{i\}}
   \rangle_{xzt}^{1/2} \rangle_e$; the shaded region denotes $\pm$ one
   standard deviation with respect to the ensemble average operator
   $\langle \ \cdot \ \rangle_e$; (\textcolor{red}{\dashed}) is $\left
   \langle u'^2 \right\rangle_{xzt}^{1/2}$, $\left \langle v'^2
   \right\rangle_{xzt}^{1/2}$, and $\left \langle w'^2
   \right\rangle_{xzt}^{1/2}$ for
   $\widetilde{\mathrm{NF}}$-SEI180.  \label{fig:stats_frozen_UVW}
 }
\end{figure}

In the second experiment, we remove the exponential instability of the
streaks and label the case as $\widetilde{\mathrm{NF}}$-SEI180
(analogous to NF-SE180).  The results, included in
figure~\ref{fig:stats_frozen_UVW} (red dashed lines), show that
turbulence is maintained in the absence of exponential
instabilities. When comparing $\widetilde{\mathrm{NF}}$-SEI180 with
$\widetilde{\mathrm{NF}}$180, the former exhibits a mildly reduced
level of fluctuating velocities (similar to the observation from
NF-SEI180 compared to NF180).

Finally, we perform simulations freezing the base flow in addition to
removing the exponential instabilities as in \S \ref{subsec:TG}. The
cases are denoted as $\widetilde{\mathrm{NF}}$-TG180$_{i}$ (analogous
to NF-TG180$_{i}$). Turbulence is sustained in 90\% of the cases. From
figure~\ref{fig:stats_frozen_UVW}, we conclude that the discussion in
\S \ref{subsec:TG} is broadly applicable to the base flow $(U,V,W)$:
wall turbulence exclusively supported by transient growth is sustained
and able to produce realistic flow statistics.
%

\section{Details of the stability analysis}\label{sec:appendix_details}

In this appendix we describe the linear stability analysis of a base
flow, $U(y, z, t)$, which is inhomogeneous in two spatial directions
\citep[e.g.,][]{Karp2014}. At given time $t=t_0$, we assume the following
velocity field
\beq \label{eq:ansantz}
\bu=\left(U(y,z,t_0),0,0\right)+\e\bu',\quad 0<\e\ll 1,
\eeq
where the base flow $U$ is assumed parallel, steady, and streamwise
independent, and $\bu'$ is the disturbance. Substituting~\eqref{eq:ansantz} into 
the incompressible Navier--Stokes equations~\eqref{eq:NS_original}, neglecting
terms of order $\e^2$ and higher, and gathering the terms at order~$\e$,
we obtain the linearised equations for the disturbances:
\bse\label{ns}
\begin{align}
\frac{\p u'}{\p x}+\frac{\p v'}{\p y}+\frac{\p w'}{\p
z} & =0, \\
\frac{\p u'}{\p t} + U\frac{\p u'}{\p x}+v'\frac{\p U}{\p
y} + w'\frac{\p U}{\p z} &= -\frac{1}{\rho}\frac{\p p'}{\p
x} + \nu\nabla^2 u', \\
\frac{\p v'}{\p t} + U\frac{\p v'}{\p x} &= -\frac{1}{\rho}\frac{\p p'}{\p
y} + \nu\nabla^2 v', \\
\frac{\p w'}{\p t} + U\frac{\p w'}{\p x} &= -\frac{1}{\rho}\frac{\p p'}{\p
z} + \nu\nabla^2 w'. \end{align}
\ese
The boundary conditions are no slip at the wall and free slip for $u'$
and $w'$ and impermeability for $v'$ at the top.  Homogeneity in $x$
and $t$ allows us to assume that all flow fields for the disturbances
take the form, e.g.,
\beq \label{eq:eigenanstantz}
	u'=\hat{u}'(y, z)\,e^{(\lambda+i\omega)t + i k_x x},
\eeq
where $k_x$ is the streamwise wavenumber, and $\lambda+i\omega$ is the temporal 
complex eigenvalue. (Similarly for $v'$, $w'$, and $p'$.)
 
Substituting~\eqref{eq:eigenanstantz} into the linearised equations~\eqref{ns}, they can
be rearranged as a generalised eigenvalue problem,
\beq \begin{pmatrix}
\mathsfbi{D_x}~&\mathsfbi{D_y}~&\mathsfbi{D_z}~&\mathsfbi{O}\\
\mathsfbi{C}~&\mathsfbi{U_y}~&\mathsfbi{U_z}~&\mathsfbi{D_x}\\
\mathsfbi{O}~&\mathsfbi{C}~&\mathsfbi{O}~&\mathsfbi{D_y}\\
\mathsfbi{O}~&\mathsfbi{O}~&\mathsfbi{C}~&\mathsfbi{D_z}
\end{pmatrix} \begin{pmatrix}
\ba{c}\tilde{u}'\\\tilde{v}'\\\tilde{w}'\\\tilde{p}'
\ea \end{pmatrix} = (\lambda+i\omega) \begin{pmatrix}\mathsfbi{O}~&\mathsfbi{O}~&\mathsfbi{O}~&\mathsfbi{O}\\
-\mathsfbi{I}~&\mathsfbi{O}~&\mathsfbi{O}~&\mathsfbi{O}\\
\mathsfbi{O}~&-\mathsfbi{I}~&\mathsfbi{O}~&\mathsfbi{O}\\
\mathsfbi{O}~&\mathsfbi{O}~&-\mathsfbi{I}~&\mathsfbi{O} \end{pmatrix} 
\begin{pmatrix}\tilde{u}'\\\tilde{v}'\\\tilde{w}'\\\tilde{p}'
\end{pmatrix}. \eeq
Here, $\mathsfbi{I}$ is the identity matrix, $\mathsfbi{O}$ is a zero
matrix, $\tilde{u}'$ is a one-dimensional
representation of a two-dimensional vector
\beq
\tilde{u}' \defn \big(\hat{u}'(y_1,z_1), \dots, \hat{u}'(y_1,z_{N_z}), \dotsb,
\hat{u}'(y_{N_y},z_1), \dots, \hat{u}'(y_{N_y},z_{N_z})\big)^\mathsf{T}, 
\eeq
and similarly for $\tilde{v}'$, $\tilde{w}'$, and
$\tilde{p}'$. Furthermore, the matrices $\mathsfbi{C}$,
$\mathsfbi{U_y}$, $\mathsfbi{U_z}$, $\mathsfbi{D_x}$,
$\mathsfbi{D_y}$, and $\mathsfbi{D_z}$ are given by
\bse\begin{align}
\mathsfbi{C} &= i k_x\;\textnormal{diag}\left(\mathsfbi{U}\right) -
\nu\left(\mathsfbi{\bar{I}_z}\otimes\mathsfbi{\bar{D}^2_y}
+ \mathsfbi{\bar{D}^2_z}\otimes\mathsfbi{\bar{I}_y} -
k_x^2\mathsfbi{\bar{I}_z}\otimes\mathsfbi{\bar{I}_y}\right),\\
\mathsfbi{U_y} &=\textnormal{diag}\left\{\left(
\mathsfbi{\bar{I}_z}\otimes\mathsfbi{\bar{D}_y}\right)\mathsfbi{U}\right\},\\
\mathsfbi{U_z} &=\textnormal{diag}\left\{\left(
\mathsfbi{\bar{D}_z}\otimes\mathsfbi{\bar{I}_y}\right)\mathsfbi{U}\right\},\\
\mathsfbi{D_x} &=i k_x \;\mathsfbi{\bar{I}_z}\otimes\mathsfbi{\bar{I}_y},\\
\mathsfbi{D_y} &=\mathsfbi{\bar{I}_z}\otimes\mathsfbi{\bar{D}_y},\\
\mathsfbi{D_z} &=\mathsfbi{\bar{D}_z}\otimes\mathsfbi{\bar{I}_y},
\end{align} \ese
where $\otimes$ is the Kronecker product and $\mathsfbi{U}$ is the
one-dimensional representation of $U$ (similarly to $\tilde{u}'$). The
matrices $\mathsfbi{\bar{I}_y}$ and $\mathsfbi{\bar{I}_z}$ are the
identity matrices of dimensions $N_y\times N_y$ and $N_z\times N_z$,
respectively, and $\mathsfbi{\bar{D}_y}$ and $\mathsfbi{\bar{D}_z}$
are the matrices that represent differentiation in $y$ and $z$
directions, respectively.  The eigenvalue problem is solved numerically for 
all streamwise wavenumbers~$k_x$ on-the-fly during the simulations.

\section{Validation of eigenvalue calculation}\label{sec:appendix_validation}

\begin{figure}
  \begin{center}
  \subfloat[]{\includegraphics[width=0.45\textwidth]{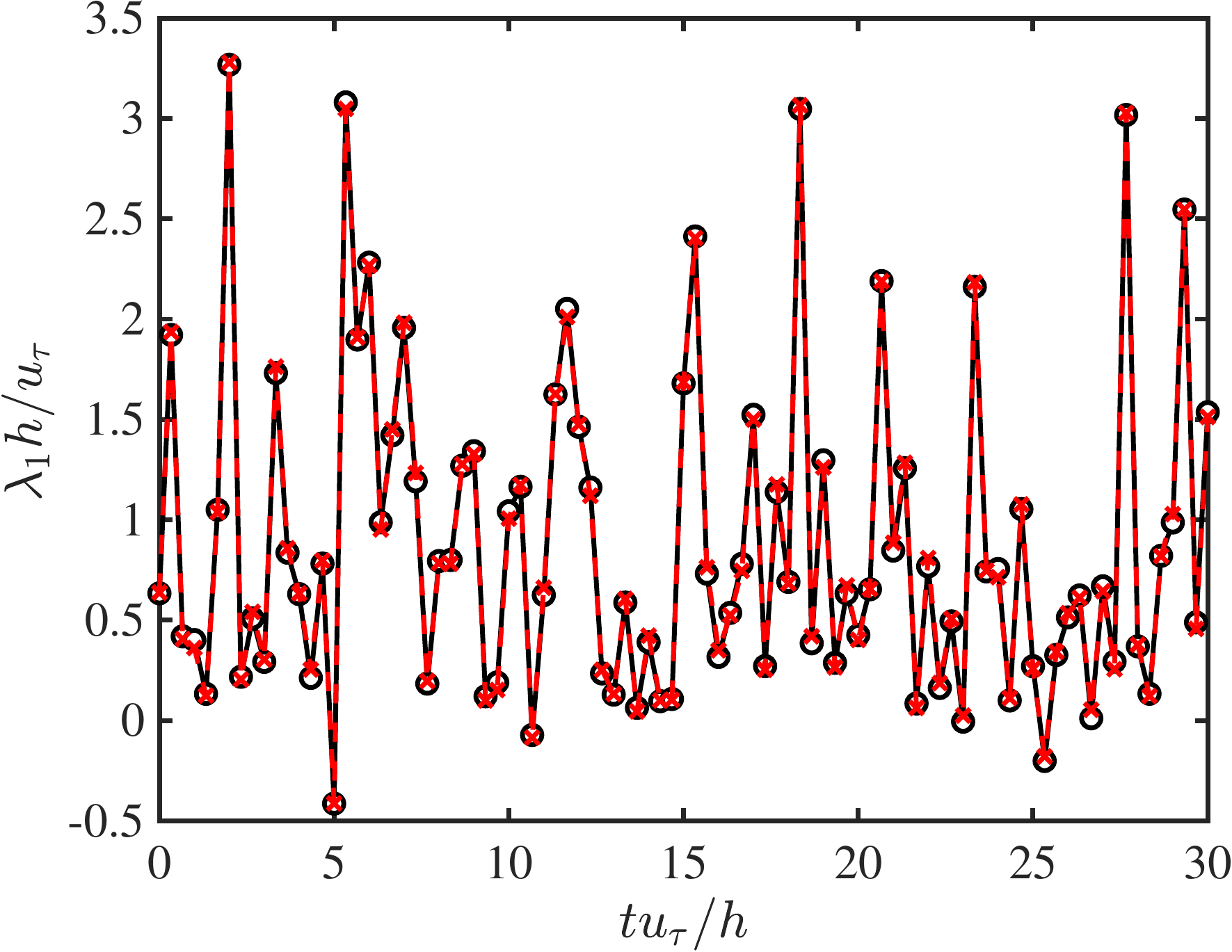}} 
  \hspace{0.1cm}
  \subfloat[]{\includegraphics[width=0.45\textwidth]{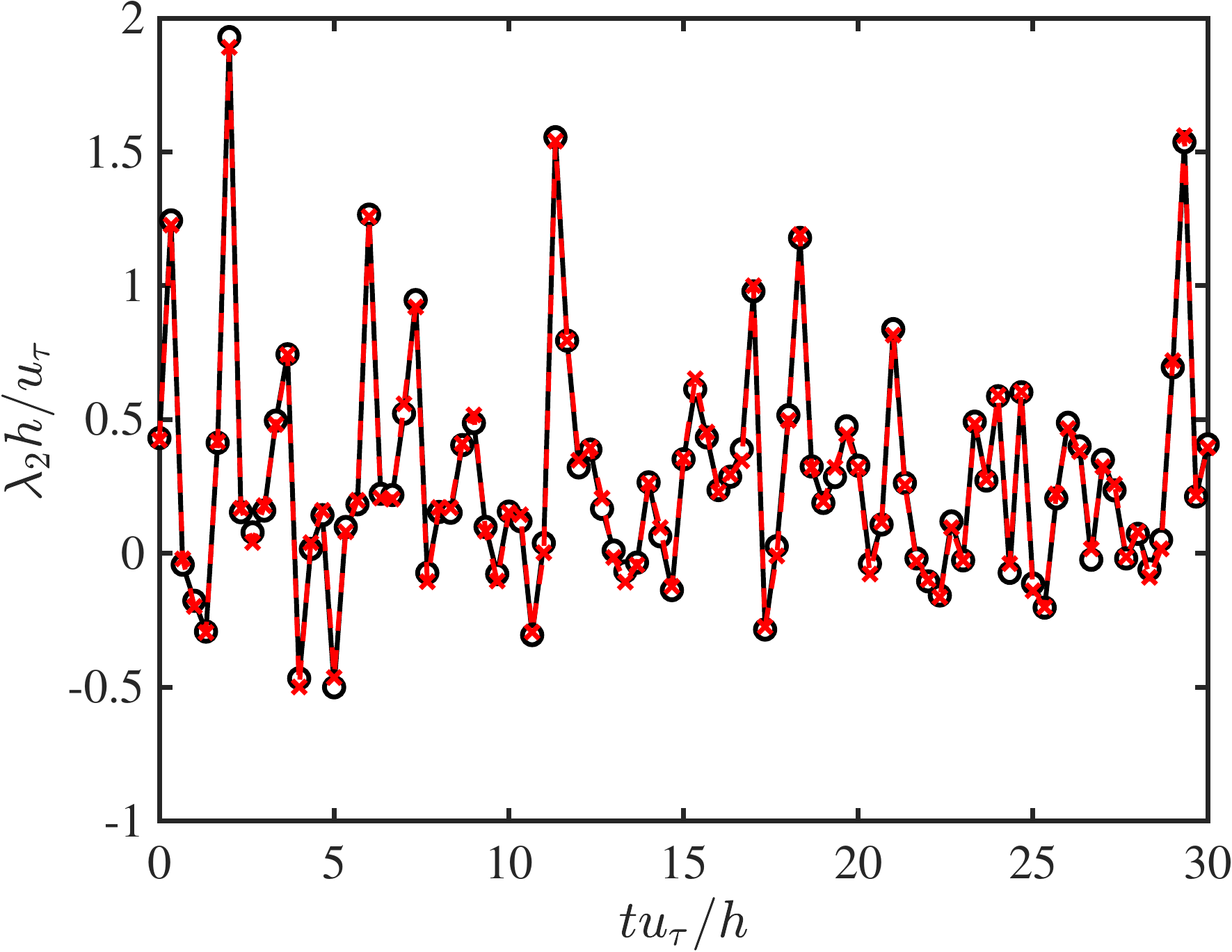}} 
  \end{center}
 \caption{The history of the real part of the two most unstable
   eigenvalues (a) $\lambda_1$, $\breve{\lambda}_1$ and (b) $\lambda_2$,
   $\breve{\lambda}_2$ of $\mathcal{L}(U)$ computed on-the-fly by the solver
   which integrates the equations of motion for the fluctuating
   velocities ($-\circ-$) and computed \emph{a posteriori} by a second
   independent solver
   (\textcolor{red}{$-\times-$}). \label{fig:appendixB} }
\end{figure}

The eigenvalue calculation described in Appendix
\ref{sec:appendix_details} was numerically implemented in the code
which solves the equations of motion such that, at a given time $t$,
the eigenvalues of $\mathcal{L}(U(y,z,t))$ are computed on-the-fly. To
verify the implementation, a second independent solver was used, which
takes as input the base flows $U(y,z,t)$ stored from the
simulation. The second algorithm solves the eigenvalues problem in the
$y$-vorticity--Laplacian of $v$ formulation discretised with
first-order finite differences in a collocated grid. We have referred
to the real part of the eigenvalues computed by the first solver as
$\lambda_j$.  Let us denote the eigenvalues computed by the second
solver as $\breve{\lambda}_j$.  Figure~\ref{fig:appendixB} shows the
history of the real part of the two most unstable eigenvalues
$\lambda_1$ and $\lambda_2$, and $\breve{\lambda}_1$ and
$\breve{\lambda}_2$.  On average, the error $|\lambda_j -
\breve{\lambda}_j|/|\lambda_j|$ for all unstable eigenvalues is of the
order of 0.1\%.  These small differences are  expected, as the
numerical details of the two solvers differ. Yet, the errors are small
enough to provide confidence in the calculation of the modal
instabilities.  An additional validation is presented in
Appendix~\ref{sec:appendix_linear}.
%

\section{Approximate calculation of $\tilde{\mathcal{L}}(U)$ using linear forcing}
\label{sec:appendix_forcing}

The exponential instabilities in $\mathcal{L}(U)$ were rigorously
removed to obtained $\tilde{\mathcal{L}}(U)$ via eigendecomposition
(see \ref{eq:eigen_LU}). This approach might obscure the
interpretation of $\tilde{\mathcal{L}}(U)$ and, at the same time, it
entails a rather costly procedure.  In this appendix, we present an
alternative approach to suppress exponential instabilities, which aids
the interpretation of the stabilisation of $\mathcal{L}(U)$ and is
computationally more affordable.

In general, the operator $\mathcal{L}$ is stabilised by subtracting
the eigenspaces that correspond to eigenvalues with positive real part
$\lambda_j>0$, $j=1,...,N$:
\begin{equation} \label{eq:Ahat_bi}
\tilde{\mathcal{L}} = \mathcal{L} - \sum_{j=1}^N a
\lambda_j\, \boldsymbol{\mathcal{U}}_j
\boldsymbol{\mathcal{V}}^\dagger_j,
\end{equation}
where $a$ is a real coefficient $a>1$, $\boldsymbol{\mathcal{U}}_j$ is
the $j$-th eigenmode of $\mathcal{L}$, and
$\boldsymbol{\mathcal{V}}_j$ is the $j$-th eigenmode of the adjoint
operator $\mathcal{L}^\dagger$, appropriately normalised so that
$\boldsymbol{\mathcal{V}}_i^\dagger \boldsymbol{\mathcal{U}}_j =
\delta_{ij}$. Hence, the approach to project out the manifolds
associated with particular eigenvector from an operator whose
eigenbasis is not orthogonal involves the biorthogonal eigenbasis of
its adjoint operator.

\begin{figure}
  \begin{center}
  \subfloat[]{\includegraphics[width=0.32\textwidth]{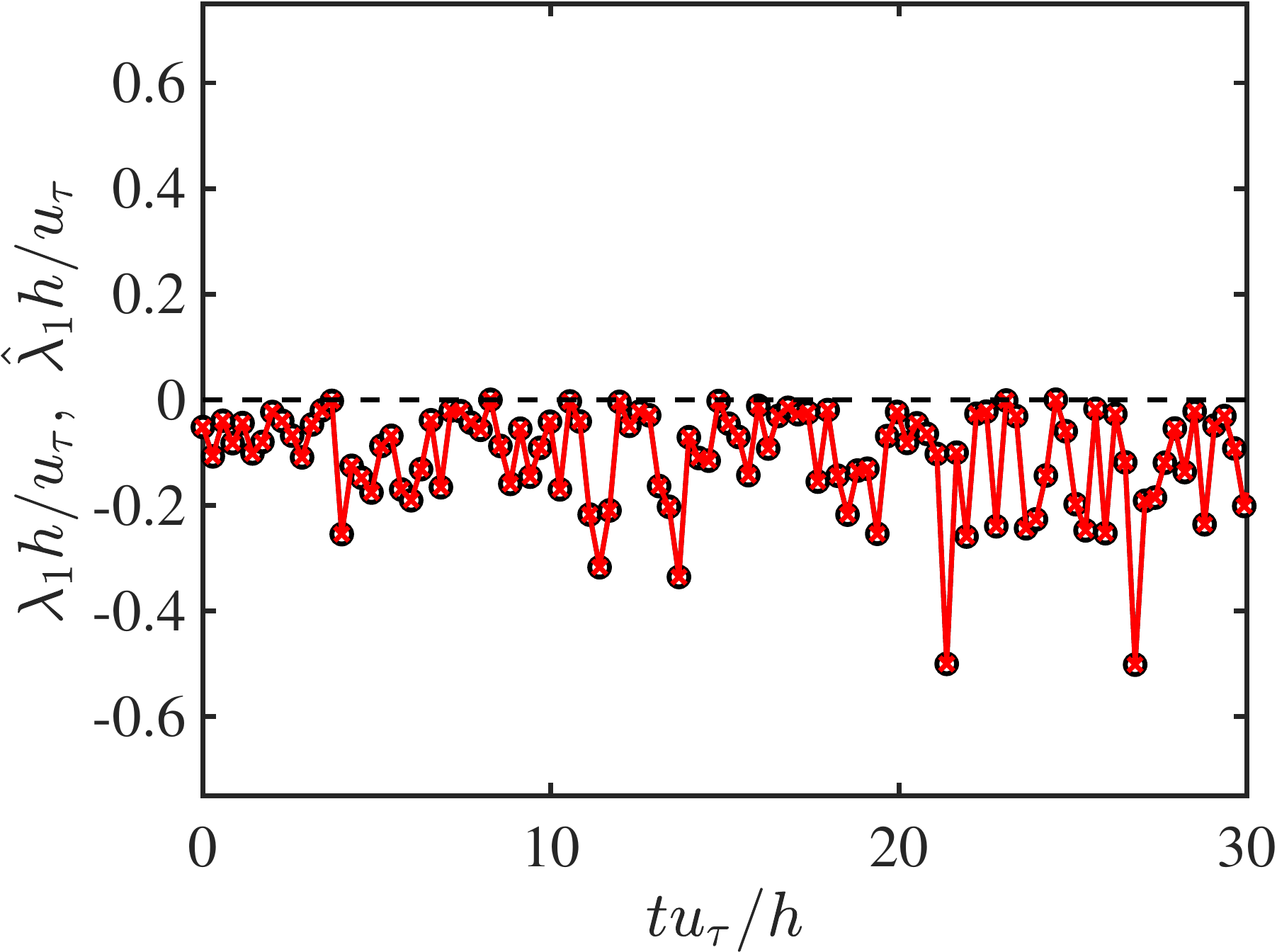}} 
  \hspace{0.1cm} 
  \subfloat[]{\includegraphics[width=0.32\textwidth]{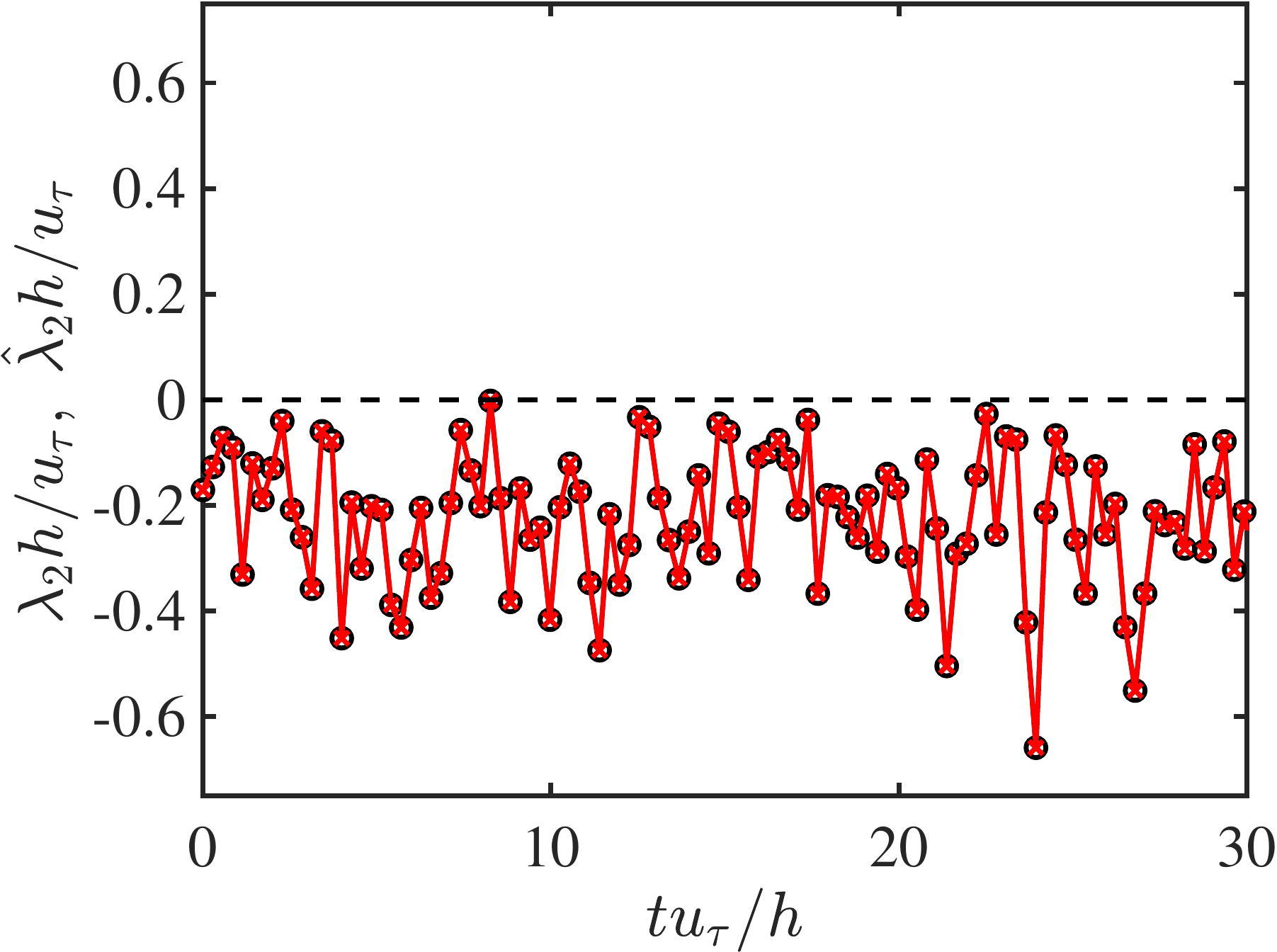}} 
    \hspace{0.1cm}
  \subfloat[]{\includegraphics[width=0.32\textwidth]{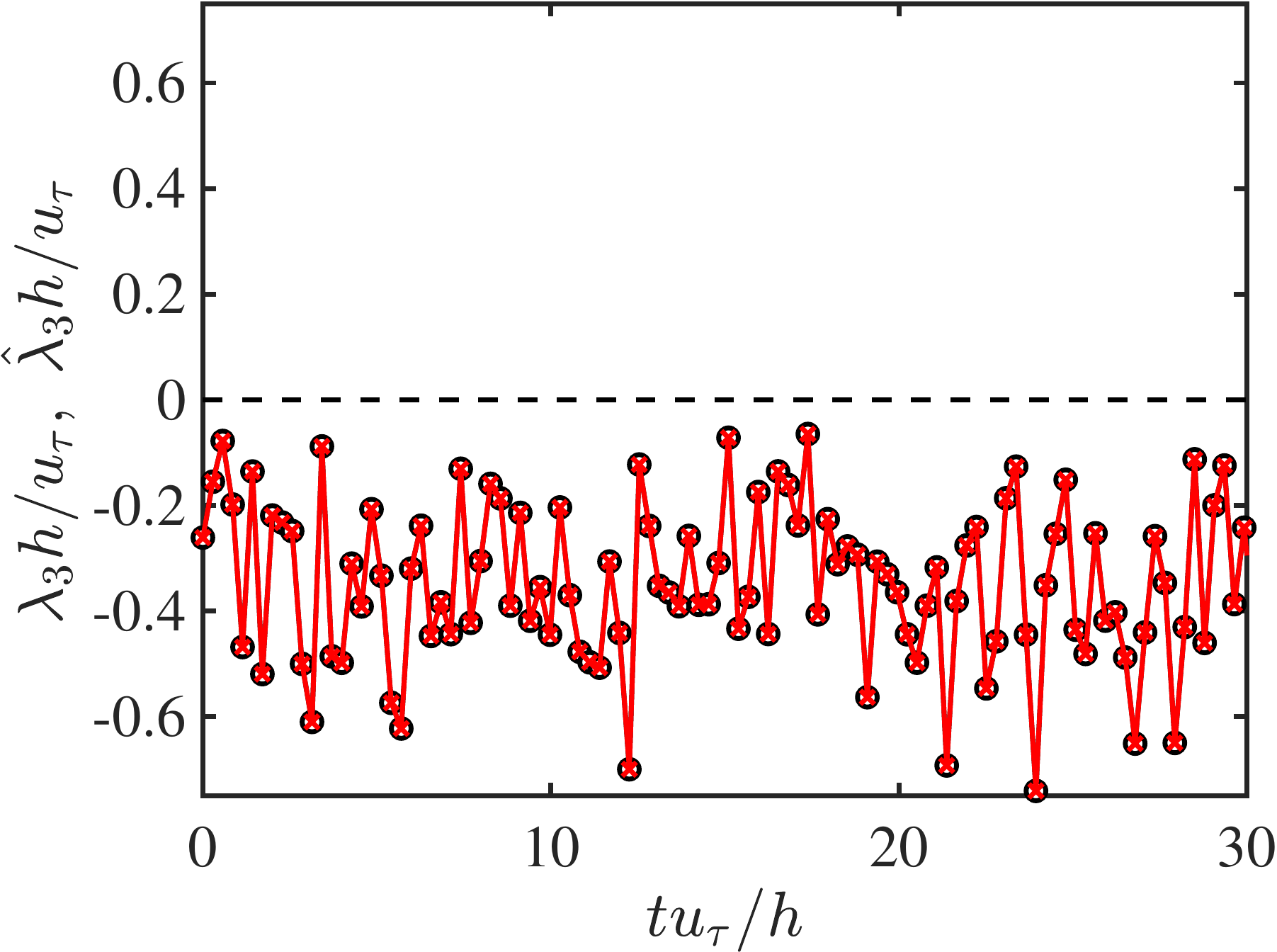}} 
  \end{center}
 \caption{The history of the real part of the three most unstable
   eigenvalues (a) $\lambda_1$, $\hat\lambda_1$, (b) $\lambda_2$,
   $\hat\lambda_2$, (c) $\lambda_3$, $\hat\lambda_3$ of
   $\tilde{\mathcal{L}}(U)$ ($-\circ-$) and $\hat{\mathcal{L}}(U)$
   (\textcolor{red}{$-\times-$}). \label{fig:appendix_forcing} }
\end{figure}

If $\L$ was normal, its eigenvectors and those of its adjoing would
coincide. Therefore, stabilisation would be simplified as
\begin{equation} \label{eq:Ahat_uni}
\hat{\mathcal{L}} = \mathcal{L} - \sum_{j=1}^N a
\lambda_j\, \boldsymbol{\mathcal{U}}_j
\boldsymbol{\mathcal{U}}^\dagger_j,
\end{equation}
for $\lambda_j>0$, $j=1,...,N$.  Under the assumption that the most
unstable eigenspace of $\L$ can be suppressed considering $\L$ as
normal, then we can use the approximation
\begin{equation}\label{eq:hatL_appex}
  \tilde{\L}(U) \approx \hat{\L}(U) = \L(U) - \sum_{j=1}^N 2 \lambda_j
  \boldsymbol{\mathcal{U}}_j \boldsymbol{\mathcal{U}}^\dagger_j,
\end{equation}
where we have chosen $a=2$. It is well-known that the operator $\L$
from the Navier--Stokes equations is highly nonnormal and an
approximation like~\eqref{eq:hatL_appex} is not guaranteed to
stabilise $\L$.  Nonetheless, we show here that it works reasonably
well. Figure~\ref{fig:appendix_forcing} compares the real part of the
three most unstable eigenvalues of the properly stabilised
$\tilde{\L}(U)$ (denoted by $\lambda_i$) and those of $\hat{\L}(U)$
(denoted by $\hat\lambda_i$). The approximate method $\hat{\L}(U)$
succeeds in stabilising $\L(U)$ with the largest eigenvalues (now
stable) obtained within to more than 0.1\% accuracy when compared to
$\tilde{\mathcal{L}}(U)$.
%

In addition to providing an intuitive interpretation of the
stabilisation, the approximate approach is also included here given
its easier implementation using the power method.  The power method
solves the linearised Navier--Stokes equations rescaling the velocity
field amplitude at each time step to track the most unstable mode. The
process can be repeated iteratively to obtain approximations of the
first, second, third,... most unstable modes and eigenvalues.  The
advantage of the power method is that it does not require constructing
$\mathcal{L}$ explicitly nor performing the eigendecomposition of the
operator, which might be beneficial in those cases where computing
$\mathcal{L}(U)$ is numerically impractical. We repeated cases
NF-SEI180 and R-SEI180 using $\hat{\mathcal{L}}(U)$ and tested that
our conclusions remains the same.

\section{Linear analysis of channel flow with modally-stable base-flow}
\label{sec:appendix_linear}

\begin{figure}
  \begin{center}
  \subfloat[]{\includegraphics[width=0.45\textwidth]{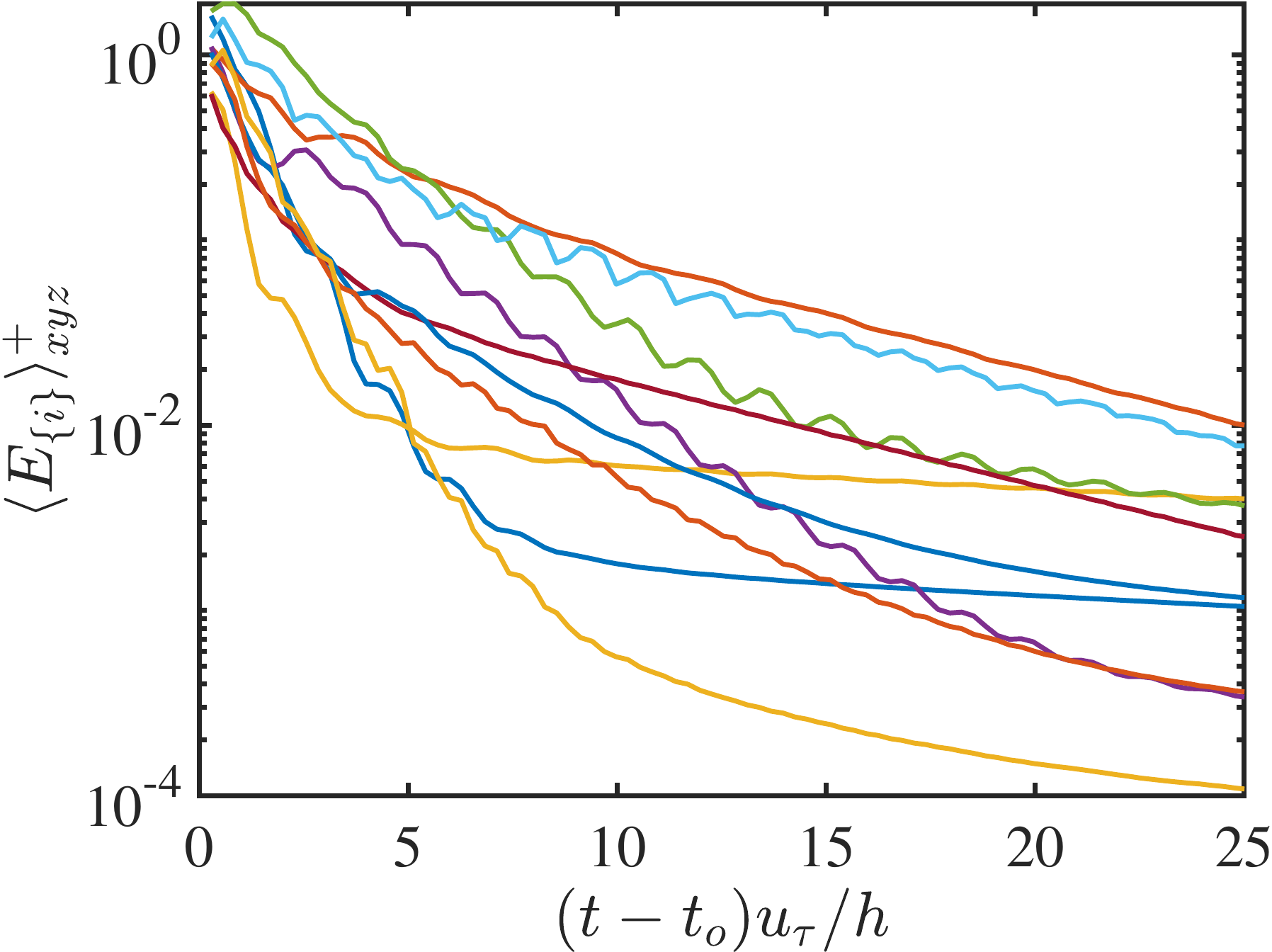}} 
  \hspace{0.2cm}
  \subfloat[]{\includegraphics[width=0.45\textwidth]{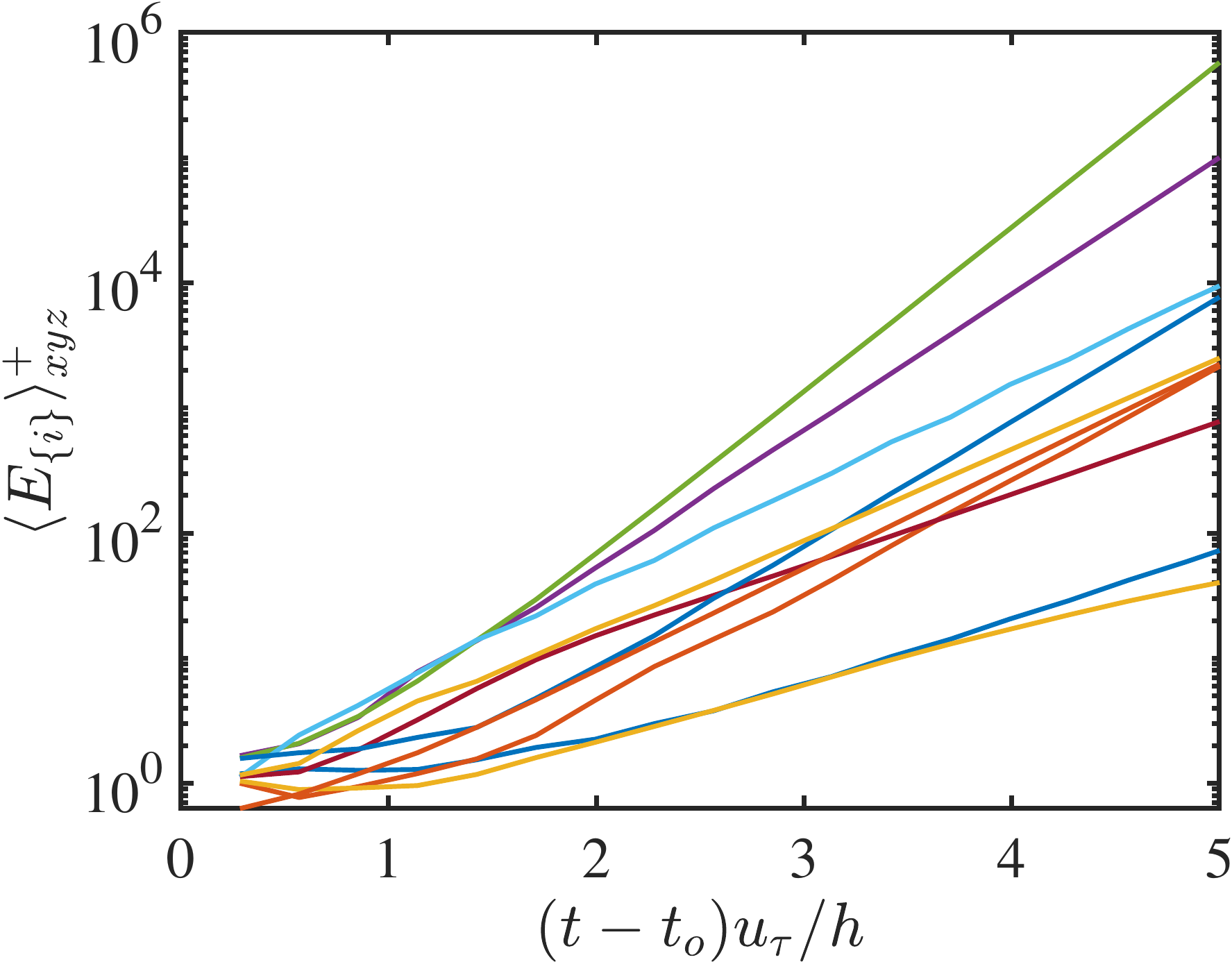}} 
  \end{center}
  \caption{The history of the domain-averaged turbulent kinetic energy
    of the fluctuations $\langle E \rangle_{xyz}$.  Different colours
    are for cases for (a)~modally-stable system
    \eqref{eq:frozen_linear} and (b)~modally-unstable system
    \eqref{eq:frozen_linear_exp}. $t_0$ is initial time to integrate
    the system. \label{fig:stats_frozen_appendix} }
\end{figure}

We consider the governing equations for the linear
channel flow with modally-stable frozen base-flow
\begin{subequations} \label{eq:frozen_linear}
\begin{gather}
\frac{\partial\bu'}{\partial t} = \tilde{\mathcal{L}}(U)\bu',\\
\bU = (U(y,z,t_0),0,0) \ \mathrm{from \ case \ R180},
\end{gather}
\end{subequations}
where we have disposed of the nonlinear term $\bN(\bu')$. We repeat
the simulations in \S \ref{subsec:TG} using the same set-up. As an
example, the evolution of the turbulent kinetic energy for ten cases
is shown in figure~\ref{fig:stats_frozen_appendix}(a). Given that
$\tilde{\mathcal{L}}(U)$ is modally stable, the turbulent kinetic
energy decays without exception. We verified that this was the case
for all the simulations considered in \S~\ref{subsec:TG} once
$\bN(\bu')$ is set to zero. Conversely, if we consider the system
\begin{subequations}\label{eq:frozen_linear_exp}
\begin{gather} 
\frac{\partial\bu'}{\partial t} = \mathcal{L}(U)\,\bu',\\
\bU = (U(y,z,t_0),0,0) \ \mathrm{from \ case \ R180},
\end{gather}
\end{subequations}
in which modal instabilities are allowed, the turbulent kinetic energy
grows exponentially as seen in
figure~\ref{fig:stats_frozen_appendix}(b) given that the ten cases
considered are all such that $\mathcal{L}(U)$ is modally unstable. It
was also verified that the growth rate obtained by
integrating~\eqref{eq:frozen_linear_exp} coincides with the growth
rate $\lambda_1$ of the most unstable mode as predicted by the
eigenvalue analysis of $\mathcal{L}(U)$. The present appendix serves
as validation of the successful suppression of modal instabilities in
$\mathcal{L}(U)$, and complements the results in
figure~\ref{fig:P_eig_nonmodal_Upred} and the analysis in Appendix
\ref{sec:appendix_validation}.
%

Finally, we consider the governing equations for the linear channel
flow with modally-stable time-varying base-flow
\begin{subequations} \label{eq:linear_timevaryingU}
\begin{gather}
\frac{\partial\bu'}{\partial t} = \tilde{\mathcal{L}}(U)\bu',\\
\bU = (U(y,z,t),0,0) \ \mathrm{from \ case \ R180},
\end{gather}
\end{subequations}
where base flow is now allowed to change in time.  The system
(\ref{eq:linear_timevaryingU}) is supported by transient growth
potentially assisted by parametric instabilities. Thus, turbulence
could survive even if $\tilde{\mathcal{L}}(U)$ is modally stable at
all instances. However, we found that this is not the case and
(\ref{eq:linear_timevaryingU}) is unable to sustain turbulence: $\bu'$
decays after a few eddy turnover times.


\end{document}